\documentclass[12pt, a4 paper, twoside, oneside]{book}
\usepackage[latin1]{inputenc}
\usepackage{simpler-wick}
\usepackage[italian, english]{babel}
\usepackage{cite}
\usepackage{mathtools, stackrel}
\usepackage{amsmath, amsthm, bbm, slashed}
\usepackage{bbm}

\usepackage{array}
\usepackage{longtable}
\usepackage{booktabs} 
\usepackage{multirow}
\usepackage{dsfont}
\newcolumntype{C}[1]{>{\centering\let\newline\\ \arraybackslash\hspace{0pt}}m{#1}}

\usepackage{amssymb, mathabx}

\usepackage{graphicx}

\usepackage[pdftex,plainpages=false,colorlinks,hyperindex,bookmarksopen,linkcolor=black,citecolor=blue,urlcolor=blue]{hyperref}

\usepackage{epstopdf}

\usepackage{braket}

\usepackage{appendix}

{\cleardoublepage \fncyblank \null \vfill \begin{center}%
\bfseries \abstractname \end{center}}%
{\vfill \null}

\usepackage{setspace}
\usepackage[top=30mm,bottom=30mm,right=30mm,left=30mm]{geometry}
\usepackage{titlesec}
\usepackage{fancyhdr}
\usepackage{titling}

\usepackage{titlesec}
\titleformat{\chapter}[display]
  {\normalfont\sffamily\huge\mdseries\rmfamily}
  {\chaptertitlename\ \thechapter}{30pt}{\Huge}
\titleformat{\section}
  {\normalfont\sffamily\Large\mdseries\rmfamily}
  {\thesection}{1em}{}
 \titleformat{\subsection}
 {\normalfont\sffamily\Large\mdseries\rmfamily}
  {\thesubsection}{1em}{}
\titleformat{\title}[display]
  {\normalfont\sffamily\huge\mdseries\rmfamily}
  {\chaptertitlename\ \thechapter}{30pt}{\Huge}

\title{Higher Order Corrections To The Lifetime Of Heavy Hadrons}
\author{
Maria Laura Piscopo
}
\date{}

\begin{document}

\pagestyle{empty}

\begin{center}

{\bf \huge Higher Order Corrections To The \\[4mm] Lifetime Of Heavy Hadrons}

\normalsize

\vspace{70pt}

{ \textsc{DISSERTATION}} \\[1mm]
{\large zur Erlangung des akademischen Grades eines Doktors\\
der Naturwissenschaften }

\vspace{90pt}

{\large vorgelegt von} \\[2mm]
\textbf{\Large M.Sc. Maria Laura Piscopo}

\vspace{185pt}
{\large eingereicht bei der Naturwissenschaftlich-Technischen~Fakult\"at \\
der Universit\"at Siegen \\}
\vspace{60pt}
{\large Siegen \\[1mm]
September 2021}

\normalsize
\end{center}

\newpage
\thispagestyle{empty}
\vspace*{4cm}
\begin{center}
Betreuer und erster Gutachter\\[2mm]
Prof. Dr. Alexander Lenz \\[2mm]
Universit\"at Siegen\\[1cm]
Zweiter Gutachter\\[2mm]
Prof. Dr. Thomas Mannel\\[2mm]
Universit\"at Siegen\\[1cm]
Weitere Mitglieder der Promotionskommission\\[5mm]
Prof. Dr. Markus Cristinziani\\[2mm]
Universit\"at Siegen\\[3mm]
und\\[3mm]
Prof. Dr. Alexey Petrov\\[2mm]
Wayne State University\\[4cm]
Tag der m\"undlichen Pr\"ufung\\[2mm]
27. Oktober 2021
\end{center}
%


\pagestyle{plain}

\chapter*{Abstract}

In this work we discuss the theoretical status for the study of the lifetime of heavy hadrons. 
After presenting some introductory topics like the effective weak Hamiltonian and the heavy quark effective theory (HQET), we describe the construction of the heavy quark expansion (HQE), which constitutes the theoretical framework to systematically compute the total decay width of heavy hadrons, in terms of an expansion in inverse powers of the heavy quark mass. The structure of the HQE is discussed in detail, and the computation of the lowest dimensional contributions, explicitly outlined. Particular emphasis is put in describing the expansion of the quark propagator in the external gluon field using the Fock-Schwinger (FS) gauge, which represents a fundamental ingredient of the calculation. Moreover, the main result is the computation of the dimension-six contribution due to the Darwin operator, only recently determined and found to have a sizeable effect.  Finally, we consider two phenomenological applications of the HQE in the charm sector, namely the study of the lifetime of charmed mesons and the analysis of the Glashow-Iliopoulos-Maiani (GIM) cancellations in neutral $D$-meson mixing. By comparing our results with recent measurements performed by the LHCb, Belle-II and BESIII collaborations, we conclude that the HQE is able to reproduce, within large theoretical uncertainties, the experimental pattern for the lifetimes of charmed mesons and we discuss a potential solution for the discrepancy of previous theoretical determinations of $D$-mixing with data.

\chapter*{Zusammenfassung}
 
In dieser Arbeit er\"{o}rtern wir den theoretischen Status der Untersuchung von Lebensdauern schwerer Hadronen.
Nach einigen einf\"{u}hrenden Themen wie dem schwachen effektiven  Hamiltonian und der effektiven Theorie
f\"{u}r schwere Quarks (HQET) beschreiben wir die Konstruktion der Heavy Quark Expansion (HQE), die den theoretischen
Rahmen f\"{u}r die systematische Berechnung der totalen Zerfallsbreite schwerer Hadronen in Form einer Entwicklung in
inversen Potenzen der schweren Quarkmasse bildet. Die Struktur der HQE wird im Detail diskutiert und die Berechnung
der niedrigstdimensionalen Beitr\"{a}ge wird explizit dargestellt. Ein besonderes Augenmerk wird auf die Beschreibung der
Entwicklung des Quark-Propagators in einem externen Gluon-Feld unter Verwendung der Fock-Schwinger-Eichung (FS) gelegt,
die einen grundlegenden Bestandteil unserer Rechnungen darstellt. Dar\"{u}ber hinaus ist das Hauptergebnis die Berechnung
des Beitrags des Darwin Terms mit der Massendimension sechs, der erst vor kurzem von uns erstmals bestimmt wurde und
numerisch bedeutend ist.
Schlie{\ss}lich stellen wir zwei ph\"{a}nomenologische Anwendungen der HQE im Charm-Sektor vor, n\"{a}mlich die Untersuchung
der Lebensdauern von Charm Mesonen und  die Analyse der Glashow-Iliopoulos-Maiani (GIM)-Kanzellierungen in der
Mischung neutraler $D$ Mesonen.
Durch den Vergleich unserer Ergebnisse mit den j\"{u}ngsten Messungen der Kollaborationen LHCb, BelleII und BesIII kommen wir zu dem Schluss,
dass die HQE in der Lage ist, innerhalb gro{\ss}er theoretischer Unsicherheiten, die experimentellen Resultate f\"ur die Lebensdauern
von Charm Mesonen zu reproduzieren, und wir diskutieren eine m\"{o}gliche L\"{o}sung f\"{u}r die Diskrepanz zwischen fr\"{u}heren theoretischen
Bestimmungen von $D$-Mischung und den Daten.
 
\newpage
 
\vspace*{6cm}
\hspace*{10cm}{Alla mia famiglia: }\\
\hspace*{9.7cm}{mamma, pap\`a e Beniamino}
\thispagestyle{empty}

 \newpage
 
\vspace*{6cm}
{\centering
``If the doors of perceptions were cleansed, \\
everything would appear to man as it is, Infinite." \\[4mm]
\hspace*{4cm} William Blake, The Marriage of Heaven and Hell. 
}
\thispagestyle{empty}

\newpage
\tableofcontents

\chapter*{Introduction}
\addcontentsline{toc}{chapter}{Introduction}

The standard model of particle physics (SM) describes our knowledge 
about the fundamental constituents of nature, quarks and leptons, and the
interactions among them \cite{Weinberg:1967tq,Glashow:1961tr,Salam:1968rm}
and it is confirmed by numerous measurements to an astonishing
precision, see e.g.\
textbooks like \cite{Langacker:2010zza}. With the discovery of the Higgs
boson \cite{Higgs:1964pj,Englert:1964et,Guralnik:1964eu} by the experimental collaborations 
ATLAS \cite{ATLAS:2012yve} and CMS \cite{CMS:2012qbp} at the Large Hadron
Collider (LHC) at CERN in 2012, the spectrum of the SM particles is complete. 
\\
Despite the enormous success, the SM leaves many important questions
open, in fact, e.g.\ it is not able to explain the existence of ordinary
matter in the Universe or that of dark matter. According to
the Sakharov criteria  \cite{Sakharov:1967dj}, the fundamental theory of
nature must incorporate C and CP violation, baryon number violating
processes and a strong first order phase transition in the early Universe, to potentially explain the existence of ordinary matter. C violation is implemented by construction in the SM and CP
violation is present in the Cabibbo-Kobayashi-Maskawa matrix 
\cite{Cabibbo:1963yz,Kobayashi:1973fv}, although 
typically, the amount of CP violation contained in the CKM matrix is considered to
be too small to explain the matter-antimatter asymmetry
\cite{Gavela:1994dt}, see, however Ref.~\cite{Alonso-Alvarez:2021qfd} for a
counter example. Baryon number is violated in the SM via sphalerons
\cite{Klinkhamer:1984di}, but a strong first order phase transition could only
occur for Higgs masses below 70 GeV \cite{Kajantie:1996mn}, which 
is not realised in nature \cite{ATLAS:2012yve,CMS:2012qbp} \footnote{In Ref.~\cite{Piscopo:2019txs}, we have developed a method to solve differential equations using neural networks, applied then to the study of cosmological phase transitions in the early Universe.}.
\\
Because of this, the SM is typically considered to be
an effective theory, see e.g.\ the textbook \cite{Petrov:2016azi}, extended at higher energies
with contributions that might explain some of
the open questions. Numerous possible extensions of the SM have been studied in the literature, one of the simplest predicts the existence of a second Higgs doublet, see e.g.\ the review
\cite{Branco:2011iw}, which could provide the missing amount
of CP violation and also a strong first order phase transition, see e.g.\ Ref.~\cite{Atkinson:2021eox}. 
Another 
example is the framework presented in Ref.~\cite{Alonso-Alvarez:2021qfd}, in which it is investigated the possibility to explain the existence of
matter and dark matter, through new sources of CP violation in mixing of
neutral $B$ mesons and new couplings of the $B$ mesons with light dark matter
particles.
\\
The search for beyond standard model (BSM) effects in particle physics, can be direct and
indirect. With the former, new heavy resonances
can be produced in particle collisions
 by
increasing the center of mass energy, however, apart from the discovery of the Higgs boson, direct searches have not been successful so far at
the LHC. On the other side, with indirect searches, measurements of observables  with high precision, are compared with the corresponding SM predictions. 
In this case a robust control over the theoretical uncertainties is 
crucial, and the bottleneck is represented by the strong
interaction, which either requires the calculation of higher order perturbative corrections or the use of non perturbative methods.
Since the LHC will continue
running for several years with increased luminosity and in the upcoming future there will not be a new particle accelerator with higher center of mass, 
in recent years there has been a progressive shift from direct to indirect searches, see
e.g.\ Ref.~\cite{Dainese:2019rgk}.
\\
Quark flavour physics is particularly well suited for indirect searches of
BSM effects due to several reasons.
First, many experiments are  providing precise flavour
data, e.g.\ LHCb, Belle II, BESIII, ATLAS, CMS and formerly  BaBar, Belle
and many more, see e.g.\ the extensive HFLAV report for a list of the
numerous measurements \cite{HFLAV:2019otj}. 
Second, the theoretical description of quark flavour observables is
theoretically very advanced and enables a control of the hadronic effects, 
see e.g.\ the textbooks \cite{bookAK, Buras:2020xsm,Manohar:2000dt}.
The computation of higher order perturbative corrections can be systematically improved, see the 
recent N$^3$LO-QCD calculation for the semileptonic
$b \to c \ell \nu_\ell $ decay \cite{Fael:2020tow}.
Moreover, many heavy flavour observables can be 
expressed in terms of a series in inverse powers of the heavy quark mass,
see e.g.\ the review \cite{Lenz:2014jha}, and again higher order power corrections can be systematically determined, see e.g.\ the computation of the contributions up to order $1/m_b^5$ for semileptonic $b$-decays \cite{Mannel:2010wj}.
Non perturbative effects can be determined with theoretical tools like 
light-cone sum rules (LCSR) \cite{Shifman:1978by, Shifman:1978bx, Balitsky:1989ry}
or lattice QCD \cite{Wilson:1974sk}, which can also be systematically improved, in order to match the increasing experimental precision.
Third, CP violating effects are large in the $B_d$-system and they are well
studied, see e.g.\ Ref.~\cite{Bigi:2000yz}. 
Conversely, they are expected to be very small in the charm sector, see e.g.\ the review \cite{Lenz:2020awd}, and in the $B_s$ system, see e.g.\ the review \cite{Artuso:2015swg}, and can then provide a useful null-hypothesis test of the SM, since any
measurement of a sizeable amount of CP violation could be a clear signal
for BSM effects. 
Finally, we currently witness a significant number of deviations between
experiments and SM predictions for quark flavour observables.
The most famous are the so-called ``flavour anomalies" \cite{Albrecht:2021tul},
observed in semileptonic loop-level decays, induced by the $b \to s \ell \ell $, with $\ell = \mu, e$, transitions and semileptonic tree-level decays, induced 
by the  $b \to c \ell \nu$, with $\ell = \mu, \tau$, transitions. A combined
statistical analysis of these anomalies points at deviations of the order of  
six to seven standard deviations, see e.g.\ Ref.~\cite{Alguero:2021anc}.
\\
The above arguments show that indirect BSM searches with quark flavour 
observables represent a very interesting and promising field for future
investigations in elementary particle physics. However, in order to be able to unequivocally identify the signals of BSM effects, it is of primary importance to further improve the control over the theoretical predictions. In this respect, the work here presented   constitutes a detailed study of the theoretical status for the determination of the lifetime of heavy hadrons, like the $B$- and the $D$-mesons. In particular, we analyse the structure of the heavy quark expansion (HQE), which provides a consistent framework to compute the total decay width of heavy hadrons in terms of a series in inverse powers of the heavy quark mass, and discuss the recent computation of higher power corrections of dimension-six. Specifically, the content presented in this work is divided into four major parts. In Chapter~\ref{ch:theory-bg} we introduce the main theoretical ingredients required for the computation, and in particular describe the construction of the HQE. In Chapter~\ref{ch:HQE-ex} we present the explicit calculation of the lowest dimensional contributions to the HQE of a $B$-meson, namely due to two-quark operators up to order $1/m_b^2$ and to four-quark operators up to order $1/m_b^4$. In Chapter~\ref{ch:Darwin} we outline in detail the computation of the contribution of order $1/m_b^3$ due to the Darwin operator for the case of arbitrary non-leptonic decay modes of the $b$-quark, which has only recently been determined and found to be sizeable. Moreover, particular emphasis is put in describing the mixing between four-quark operators and the Darwin operator at dimension-six, that ensures the cancellation of the infrared divergences, arising from the emission of a soft gluon from a light quark propagator, otherwise present in the coefficients of the Darwin operator. In Chapter~\ref{ch:pheno} we consider two phenomenological applications of the HQE in the charm-sector, specifically, we perform a comprehensive study of the inclusive decay width of charmed mesons and propose a possible solution to explain the large discrepancy between the theoretical determination of mixing of neutral $D$-mesons and the corresponding experimental data. Finally we conclude with a discussion of the results. 


\chapter*{Notations}
\label{ch:Notations}
\addcontentsline{toc}{chapter}{Notations}
Here we list some of the notations adopted throughout this work, mostly following the textbooks \cite{Itzykson:1980rh, ellis_stirling_webber_1996}. We use the natural system of units, i.e.\ $c = \hbar = 1$. Indices representing all four components of a four-vector are always  labelled by Greek letters e.g.\ $\mu = 0, 1,2,3$, while indices corresponding only to the three space components are labelled by Latin letters e.g.\ $k = 1,2,3$. Summation over repeated indices is understood unless otherwise stated. The four-dimensional Minkoswki metric tensor is $g_{\mu \nu} = diag (1, -1, -1, -1)$, so that the invariant product between two four-vectors $x^\mu$ and $y^\mu$ is given by $x \cdot y \equiv g_{\mu \nu} x^\mu y^\nu = x^0 y^0 - \bf x \cdot \bf y $, with three-vectors denoted in bold type. Moreover the differential operator reads  \\ 
\begin{equation}
\partial_\mu = \frac{\partial}{\partial x^\mu} = \left(\partial_t, \bf \nabla \right)\,, \qquad \partial^\mu = \frac{\partial}{\partial x_\mu} = \left(\partial_t, - \bf \nabla \right)\,.
\end{equation}\\
The Pauli matrices are the three hermitean $2 \times 2$ matrices\\
\begin{equation}
\sigma_1 = 
\begin{pmatrix}
0 & 1 \\
1 & 0 
\end{pmatrix}\,,
\quad 
\sigma_2 = 
\begin{pmatrix}
0 & - i \\
i & 0 
\end{pmatrix}\,,
\quad 
\sigma_3 = 
\begin{pmatrix}
1 & 0 \\
0 & -1 
\end{pmatrix}\,,
\end{equation}\\
satisfying $\sigma_j \sigma_k = \delta_{jk} + i \epsilon_{jkl} \sigma_l$\,, with $\epsilon_{123} = 1$. 
The four-dimensional gamma matrices $\gamma^\mu$, in the standard representation, are respectively given by\\
\begin{equation}
\gamma^0 = 
\begin{pmatrix}
\mathds{1}_2 & 0 \\
0 & - \mathds{1}_2 
\end{pmatrix}\,,
\quad 
\gamma^k = 
\begin{pmatrix}
0 & \sigma_k \\
- \sigma_k & 0
\end{pmatrix}\,,
\end{equation}\\
with
\begin{equation}
\{ \gamma^\mu, \gamma^\nu \} = 2 g^{\mu \nu} \mathds{1}_4\,, \qquad \gamma^{\mu \dagger} = \gamma^0 \gamma^\mu \gamma^0\,.
\end{equation} \\
The commutator of two gamma matrices is\\
\begin{equation}
\sigma^{\mu \nu} = \frac{i}{2}\,[\gamma^\mu, \gamma^\nu ]\,,
\end{equation}
while the fifth gamma matrix is defined as 
\begin{equation}
\gamma_5 = \gamma^5 = i \gamma^0 \gamma^1 \gamma^2 \gamma^3\,.
\label{eq:gamma5-def}
\end{equation}
Regarding the convention for the four-dimensional Levi-Civita tensor $\epsilon^{\mu \nu \rho \sigma}$, we use  $\epsilon^{0123} = 1 = - \epsilon_{0123}$. With the above definitions for $\gamma_5$ and $\epsilon^{\mu \nu \rho \sigma}$, it follows that the tensor decomposition of three gamma matrices reads\\
\begin{equation}
\gamma^\mu \gamma^\nu \gamma^\rho = g^{\mu \nu} \gamma^\rho - g^{\mu \rho} \gamma^\nu + g^{\nu \rho} \gamma^\mu + i \epsilon^{\mu \nu \rho \sigma}\gamma_\sigma \gamma_5\,,
\label{eq:tensor-decomposition-gamma-mat}
\end{equation}\\
and that the trace of four gamma matrices and one $\gamma_5$ is\\
\begin{equation}
{\rm Tr} \left[\gamma^\mu \gamma^\nu \gamma^\rho \gamma^\sigma \gamma_5 \right] = - 4 i \epsilon^{\mu \nu \rho \sigma}\,.
\end{equation}\\
Quantum Chromodynamics (QCD) is a non-abelian gauge theory with the symmetry group $SU(N_c)$ and number of colours $N_c = 3$. Colour indices of fields in the adjoint representation are indicated by $a,b,c = 1, \ldots, (N_c^2 -1)$, whereas $i,j,k  = 1,\ldots, N_c$, are used to label fields in the fundamental representation. The generators in the fundamental and in the adjoint representation are respectively denoted by $t^a$ and $T^a$. They satisfy the following commutation relations\\
\begin{equation}
\big[ t^a, t^b \big] = i  f^{abc}\, t^c \,, \qquad   \big[ T^a, T^b \big] = i f^{abc} T^c\,, \qquad  \left( T^a \right)_{bc} = - i f^{abc}\,, 
\label{eq:commut-t}
\end{equation}\\
where $f^{abc}$ are the structure constants of the group. From the normalisation choice\\
\begin{equation}
{\rm Tr} [t^a t^b] = \frac12 \delta^{ab}\,,
\label{eq:norm-t}
\end{equation}\\
it follows that\\
\begin{equation}
\big( t^a \cdot t^a \big)_{ij} = C_F\, \delta_{ij}\,,
\qquad
f^{abc} f^{dbc} = C_A \, \delta^{ad}\,,
\label{eq:ta-properties}
\end{equation}\\
with $C_F = (N_c^2 - 1)/2N_c$, and $C_A = N_c$.
The Feynman rules for a perturbative analysis of QCD are derived from the Lagrangian \footnote{Note that this is an abuse of notation, it actually corresponds to the Lagrangian density.}\\
\begin{equation}
 {\cal L}_{classical} + {\cal L}_{gauge-fixing} + {\cal L}_{ghost}\,,
\end{equation}\\
here, without discussing ${\cal L}_{gauge-fixing}$ and ${\cal L}_{ghost}$, required for the renormalisation of the theory, the classical Lagrangian reads\\
\begin{equation}
{\cal L}_{classical} = - \frac14 G^a_{\mu \nu}(x) G^{a \mu \nu}(x) + \sum_{flavours} \bar q^i(x) \left( i \slashed D - m_q \right)_{ij} q^j(x)\,.
\label{eq:L-classical}
\end{equation}\\
In Eq.~(\ref{eq:L-classical}), the gluon field strength tensor is \\
 \begin{equation}
 G^a_{\mu \nu}(x) = \partial_\mu A^a_\nu(x) - \partial_\nu A^a_\mu(x) + g_s f^{abc} A_\mu^b(x) A_\nu^c(x)\,,
 \end{equation}\\
$A_\mu^a(x)$ denotes the corresponding gauge field and $g_s$ is the strong coupling. Moreover, we use $G_{\mu \nu} = G_{\mu \nu}^a t^a$ and $A_\mu = A_\mu^a t^a$. Acting respectively on fields in the fundamental and adjoint representation, the covariant derivative takes the form\\
\begin{equation}
(D_\mu)_{ij} =  \partial_\mu \delta_{ij}  - i g_s A^a_\mu(x) (t^a)_{ij}\,,
\label{eq:D-fun}
\end{equation}\\
and 
\begin{equation}
(D_\mu)_{ab} =  \partial_\mu  \delta_{ab} - i g_s A_\mu^c(x) (T^c)_{ab}\,.
\label{eq:D-adj}
\end{equation}\\
Finally, the gluon field strength tensor can be expressed in terms of the commutator of two covariant derivatives in the fundamental representation, as \\
 \begin{align}
 G_{\mu \nu} &= G_{\mu \nu}^a t^a= 
 \frac{i}{g_s} \left[D_\mu, D_\nu \right]
 \nonumber \\[3mm]
 & = \frac{i}{g_s} \left[ \partial_\mu  - i g_s A^b_{\mu}(x) t^b, \partial_\nu - i g_s A^c_{\nu}(x) t^c  \right]
 \nonumber \\[3mm]
 & = \left(\partial_\mu A^a_\nu(x) -\partial_\nu A^a_{\mu}(x)\right) t^a - i g_s A^b_{\mu}(x) A^c_{\nu}(x) \left[t^b, t^c\right]
 \nonumber \\[3mm]
 & = \left(\partial_\mu A^a_\nu(x)  -\partial_\nu A^a_{\mu}(x) + g_s f^{abc} A^b_\mu (x) A^c_\nu (x)\right) t^a
 \,,
 \label{eq:G-munu}
 \end{align}\\
while from Eqs.~(\ref{eq:D-adj}), (\ref{eq:D-fun}), and (\ref{eq:commut-t}), we obtain that a covariant derivative acting on the gluon field strength tensor i.e.\ $ D_\rho G_{\mu \nu} = ( D_\rho)_{ab} G^b_{\mu \nu} t^a $, see also e.g.\  Ref.~\cite{Shifman:1980ui}, can be written~as 
 \begin{align}
 D_\rho G_{\mu \nu}  & = \partial_\rho  \delta_{ab}   G_{\mu \nu}^b t^a - g_s   A^c_\rho G^b_{\mu \nu}  f^{cab} t^a
 \nonumber \\[3mm]
 & = \partial_\rho G_{\mu \nu} - i g_s \left[ t^c, t^b \right] A^c_\rho G^b_{\mu \nu} 
 \nonumber \\[3mm]
 & = \partial_\rho G_{\mu \nu} - i g_s \left[ A_\rho ,G_{\mu \nu} \right]  
 \nonumber \\[3mm]
  & =[D_\rho, G_{\mu \nu}]
 =\frac{i}{g_s} \left[ D_\rho, \left[ D_\mu, D_\nu \right] \right] \,,
 \label{eq:DG-munu}
 \end{align}\\
where the covariant derivative on the l.h.s.\ of Eq.~(\ref{eq:DG-munu}) is in the adjoint representation and those in the last line of Eq.~(\ref{eq:DG-munu}) in the fundamental.
However, in the following, the coupling constant will be mostly absorbed in the definition of the gluon field i.e.\ $A_\mu(x) \to 1/ g_s A_\mu (x)$, and $G_{\mu \nu} \to g_s G_{\mu \nu}$, so that in this case \\
\begin{equation}
G_{\mu \nu} = i  \left[D_\mu, D_\nu \right]\,, \qquad D_\rho G_{\mu \nu} = i \left[ D_\rho,  \left[D_\mu, D_\nu \right] \right]\,.
\label{eq:G-munu-DG}
\end{equation}
%
\thispagestyle{plain}


\chapter{The Theoretical Framework}
\pagestyle{fancy}
\label{ch:theory-bg}

In this first chapter we present four of the fundamental theoretical tools necessary for the upcoming discussions and computations. Specifically, in Section~\ref{sec:Heff}, we introduce the effective weak Hamiltonian, which provides the appropriate framework to study processes like $b$- and $c$-quark decays, that happen at energy scales much lower than the $W$-boson mass. We then briefly describe in Section~\ref{sec:HQET}, the construction of the heavy quark effective theory (HQET), which represents an approximation of QCD, valid in the case of heavy quarks $Q$ with mass $m_Q \gg \Lambda_{QCD}$, where $\Lambda_{QCD}$ characterises the onset of the non perturbative regime of the strong coupling $\alpha_s$. Particular emphasis is put in deriving in Section~\ref{sec:FS}, the expansion of the quark-propagator in the external gluon field using the Fock-Schwinger (FS) gauge, a key ingredient for the calculations presented in the subsequent chapters. Finally, in Section~\ref{sec:HQE}, we introduce the general framework in which all of the computations and results obtained in the present work are embedded, namely, the heavy quark expansion (HQE).

\section{The effective weak Hamiltonian}
\label{sec:Heff}

The study of hadronic weak decays defines a typical multi-scale problem in which the mass of the $W$-boson, $m_W$, the one of the decaying constituent quark, $m$, and the hadronic non perturbative scale $\Lambda_{QCD}$, lead to the hierarchy $m_W \gg m \gg \Lambda_{QCD}$. The construction of effective field theories (EFTs) provides a general way to deal with multi-scale problems, as it allows to reduce them to a combination of simpler and single-scale ones. 

In order to derive the effective Hamiltonian needed to describe weak decays of $B$- and $D$-hadrons in the sequent chapters, we consider as a paradigmatic example the $c \to s \bar d u$ decay. We stress that the content of this section closely follows the one of the reviews \cite{Buras:1998raa, Buchalla:1995vs,Buchalla:2002pd,Grozin:2013hra}, to which we refer for a comprehensive introduction to the effective Hamiltonians for weak decays as well as for further references on the topic.

\begin{figure}[t]
\centering
\includegraphics[scale=0.5]{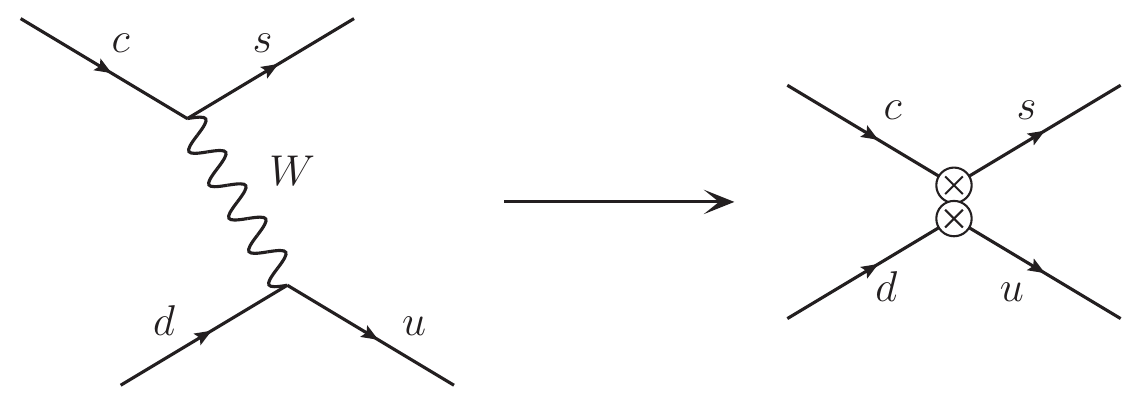}
\caption{By expanding in powers of $1/m_W^2$, the non-local amplitude in the full theory (left), results in a local interaction in the effective theory (right). The crosses in vertices denote the insertion of the effective four-quark operator.}
\label{fig:full-to-eff}
\end{figure}
The tree-level flavour changing transition
$c \to s \bar d u$, proceeds through the exchange of a $W$-boson  
between  the $(c s)$ and $(u d)$ left-handed quark currents, as it is diagrammatically shown in the left diagram of Figure~\ref{fig:full-to-eff}. The amplitude for process is given by \\
\begin{equation}
i T =  \left( i \frac{g}{2 \sqrt 2}\right)^{\! 2} V_{cs}^* V_{ud} \, \langle  \int d^4 x \int d^4 y\, 
\bar s^i(x) \, \Gamma^\mu  c^i(x) \, i D_{\mu \nu}(x,y) \, \bar u^j(y) \,\Gamma^\nu  d^j(y) \rangle \,,
\label{eq:ampl-full}
\end{equation}\\
where $g$ is the coupling corresponding to the $SU(2)_L$ symmetry group, 
$V_{q_1 q_2}$ the elements of the quark-mixing Cabibbo-Kobayashi-Maskawa (CKM) matrix, and we have introduced the short-hand notation for the Lorentz structure
$\Gamma_\mu = \gamma_\mu (1 - \gamma_5)$, and for the matrix element between external quark states $\langle \ldots \rangle$. Moreover, in Eq.~(\ref{eq:ampl-full}) the propagator of the $W$-boson $ D_{\mu \nu}(x,y)$, in the unitary gauge, admits the Fourier representation, see e.g.\ the textbook~\cite{Itzykson:1980rh} \\
\begin{equation}
D_{\mu \nu} (x,y) = \int \frac{d^4 k}{(2 \pi)^4}  \frac{-1}{k^2 - m_W^2} \left(g_{\mu \nu} - \frac{k_\mu k_\nu}{m_W^2} \right) e^{-i k \cdot (x-y)}\,.
\label{eq:W-prop}
\end{equation}\\
Because of momentum conservation, the integral in Eq.~(\ref{eq:W-prop}) is saturated by values of $k$ of the order of the decaying quark mass $m_c$, much smaller than $m_W$. It follows that by expanding in powers of $1/m_W^2$, the expression of the $W$-propagator reduces to \\
\begin{equation}
D_{\mu \nu} (x,y) = \int \frac{d^4 k}{(2 \pi)^4} \left[ \frac{g_{\mu \nu}}{m_W^2} + {\cal O}\left( \frac{k^2}{m_W^4}\right) \right] e^{- i k \cdot (x-y)} \approx \frac{g_{\mu \nu}}{m_W^2} \delta^{(4)}(x-y) \,.
\label{eq:W-prop-exp}
\end{equation}\\
Substituting Eq.~\eqref{eq:W-prop-exp} into Eq.~\eqref{eq:ampl-full}, and performing the integration over the variable $y^\mu$, we obtain that the transition amplitude for the tree-level process, up to corrections suppressed by powers of $ k^2/m_W^2 $ in the $W$-propagator, reads
\begin{equation}
i T =  - i \frac{G_F}{\sqrt 2} V_{cs}^* V_{ud} \, \langle \int d^4 x \, \bar s^i(x) \, \Gamma^\mu  c^i(x) \bar u^j(x) \,\Gamma_\mu  d^j(x) \rangle \,,
\label{eq:ampl-eff}
\end{equation}
where the Fermi constant $G_F$ is defined as\\
\begin{equation}
\frac{G_F}{\sqrt 2} = \frac{g^2}{8 m_W^2}\,.
\end{equation}\\
The amplitude in Eq.~\eqref{eq:ampl-eff}, valid at energy scales much lower than  $m_W$, could have been equivalently derived  
starting from the following effective Hamiltonian \
\begin{equation}
{\cal H}_{eff} (x) = 
\frac{G_F}{\sqrt 2} V_{cs}^* V_{ud} \, \bar s^i(x) \, \Gamma^\mu  c^i(x) \bar u^j(x)\,\Gamma_\mu  d^j(x)\,.
\label{eq:Heff-LO-QCD}
\end{equation} \\
\begin{figure}[t]
\centering 
\includegraphics[scale=0.4]{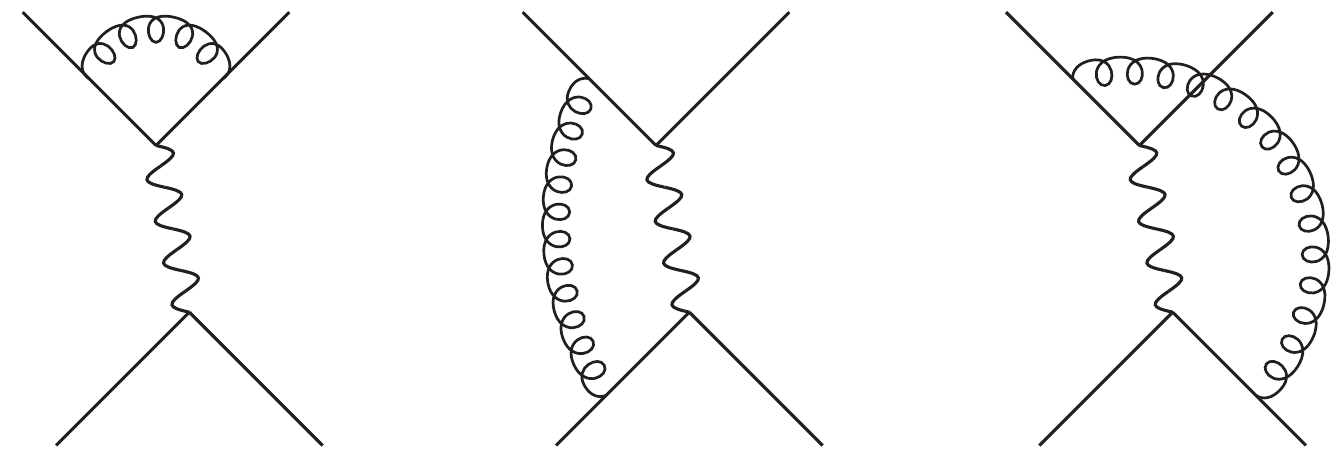}
\caption{Diagrams describing the decay $c \to s \bar d u$, at NLO-QCD in the full theory.
Left-right and up-down reflected diagrams are not shown.}
\label{fig:NLO-full}
\end{figure}
We see that, by exploiting the hierarchy $m_W \gg m_c$, the non-local product of two currents, namely the non-local operator in Eq.~(\ref{eq:ampl-full}), has been expressed in Eq.~(\ref{eq:ampl-eff}) in terms of a local operator weighted by an effective coupling. This is schematically shown in Figure~\ref{fig:full-to-eff} and represents a basic illustration of the Wilson operator product expansion (OPE) \cite{Wilson:1969zs, Wilson:1973jj}.
The next step is to include perturbative QCD corrections to the tree-level 
transition $c \to s \bar d u$, as schematically shown in Figure~\ref{fig:NLO-full}. In this case the effective Hamiltonian must be modified as\\
\begin{equation}
{\cal H}_{eff}(x) = \frac{G_F}{\sqrt 2} V_{cs}^* V_{ud} \, \Big( C_1 \, Q_1(x) + C_2 \, Q_2(x) \Big)\,,
\label{eq:Heff}
\end{equation}\\ 
where the local effective four-quark operators are given by
\begin{align}
Q_1 (x)
& = 
\Big( \bar s^i(x)  \Gamma^\mu c^i (x) \Big) \Big( \bar u^j(x) \Gamma_\mu d^j(x) \Big)\,,
\\[3mm]
Q_2  (x)
& = 
\Big( \bar s^i (x) \Gamma^\mu c^j (x) \Big) \Big( \bar u^j(x) \Gamma_\mu d^i(x) \Big) \,.
\end{align}
We see that in addition to the operator $Q_1(x)$ \footnote{Note that we do not adopt the convention historically used in the literature, see e.g.\ Ref.~\cite{Buchalla:1995vs}, and instead denote by $Q_1$ the colour-singlet operator.}, already obtained in the case of tree-level transition, there is a new operator $Q_2(x)$, with different contractions of the colour indices, which arises due to the fact that the exchange of a gluon leads to two possible colour structures, because of the completeness property of the $SU(3)_c$ generators $t^a$, i.e. \\
\begin{equation}
t^a_{ik} \, t^a_{jl} = \frac12 \left(  \delta_{il} \, \delta_{jk} 
-\frac{1}{N_c} \delta_{ik} \, \delta_{jl} \right)\,.
\end{equation}\\
In Eq.~\eqref{eq:Heff}, $C_1$ and $C_2$ denote the corresponding Wilson coefficients (WCs) of the effective operators $Q_1$ and $Q_2$.
From the result in Eq.~\eqref{eq:Heff-LO-QCD}, it follows that in the absence QCD corrections, it is $C_1 = 1$ and $C_2 = 0$. 

\begin{figure}[t]
\centering 
\includegraphics[scale=0.4]{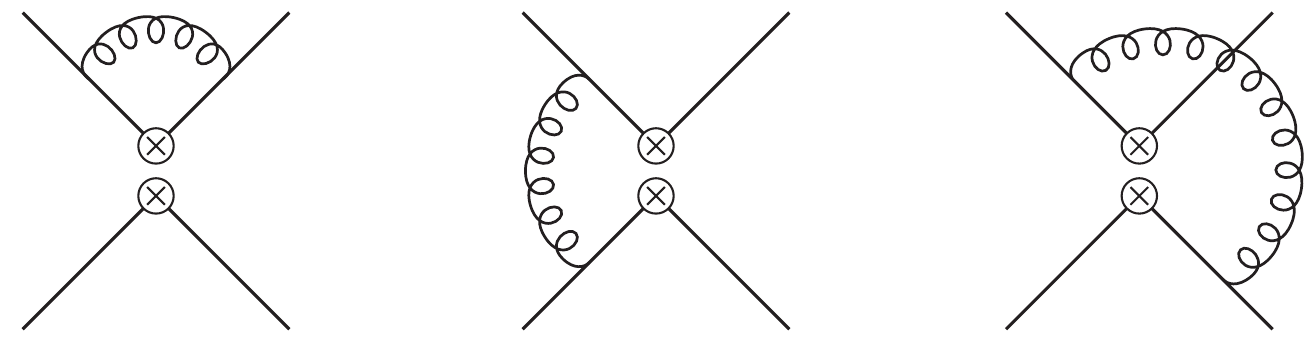}
\caption{Diagrams describing the $c \to s \bar d u$ at NLO-QCD in the effective theory.
Again, left-right and up-down reflected diagrams are not shown.}
\label{fig:NLO-eff}
\end{figure}

The general prescription to determine the expression of the Wilson coefficients is to require that 
the amplitude in the full theory is reproduced by the corresponding one in the effective theory, which reads\\
\begin{equation}
i T = - i \frac{G_F}{\sqrt 2} V_{cs}^* V_{ud} \, \Big( C_1 \, \langle Q_1 \rangle  + C_2 \, \langle Q_2 \rangle \Big)\,.
\label{eq:matching}
\end{equation}\\
By computing, on one side, QCD corrections to the amplitude $i T$ in the full theory, see Figure~\ref{fig:NLO-full}, and on the other side, 
the matrix elements of the effective operators $\langle Q_1 \rangle$ and $\langle Q_2 \rangle$, at the same order in $\alpha_s$, see Figure~\ref{fig:NLO-eff}, we can obtain the corresponding expressions  for the Wilson coefficients by equating the two results and by taking into account Eq.~\eqref{eq:matching}. 
This procedure is called matching of the full theory onto the effective theory.
Omitting the explicit calculation, we only show the final result for the renormalised amplitude in the full theory up to NLO-QCD corrections. This is  \\
\begin{align}
i T
& =  - i \frac{G_F}{\sqrt 2} V_{cs}^* V_{ud} 
\biggl[ 
\left(
1 + 2 \, C_F \frac{\alpha_s}{4 \pi} \log \left(\frac{\mu^2}{-p^2} \right)
\right) \langle Q_1 \rangle_{ tree}
\nonumber \\[3mm]
&
+ \, \frac{3}{N_c} \frac{\alpha_s}{4 \pi} \, \log \left(\frac{m_W^2}{-p^2} \right)
\langle Q_1 \rangle_{ tree}
- 3 \frac{\alpha_s}{4\pi} \log \left(\frac{m_W^2}{-p^2} \right)
\langle Q_2 \rangle_{ tree}
\biggr]\,,
\label{eq:A-full}
\end{align}\\
where $\langle Q_{1,2} \rangle_{tree}$ denote the tree level matrix elements 
of the operators $Q_1$ and $Q_2$. The expression in Eq.~(\ref{eq:A-full}) has been obtained in dimensional regularisation \cite{tHooft:1972tcz, Bollini:1972ui, Cicuta:1972jf, Ashmore:1972uj }, with $D = 4 - 2 \epsilon$ space-time dimensions, using the Feynman gauge for the gluon propagator, massless external quark states and an off-shell momentum $p$, see Ref.~\cite{Buchalla:1995vs}.
Note that Eq.~\eqref{eq:A-full} includes only logarithmic corrections of the type $\alpha_s \cdot \log$ and constant terms of order ${\cal O} (\alpha_s)$ have been neglected, which corresponds to the leading logarithmic approximation.
Moreover, the renormalisation of the quark fields has been already implemented and has resulted in the explicit $\mu$ dependence.
\\[2mm]
Similarly, by computing the diagrams within the effective theory shown in Figure~\ref{fig:NLO-eff}, leads to the following results for the unrenormalised matrix elements of the operators $Q_1$ and $Q_2$, up to NLO-QCD corrections, namely\\
\begin{align}
\langle Q_1 \rangle^{(0)} 
& =  
\left[
1 + 2 \, C_F \frac{\alpha_s}{4 \pi} 
\left(\frac{1}{\epsilon} + \log \left(\frac{\mu^2}{-p^2} \right) \right)
\right] \langle Q_1 \rangle_{ tree}
\nonumber \\[3mm]
& 
\, + \frac{3}{N_c} \frac{\alpha_s}{4 \pi} 
\left(\frac{1}{\epsilon} + \log \left(\frac{\mu^2}{-p^2} \right) \right)
\langle Q_1 \rangle_{ tree}
- 3 \frac{\alpha_s}{4 \pi} 
\left(\frac{1}{\epsilon} + \log \left(\frac{\mu^2}{-p^2} \right) \right)
\langle Q_2 \rangle_{ tree} \,,
\label{eq:ME-Q1-0}
\end{align}\\
and\\
\begin{align}
\langle Q_2 \rangle^{(0)} 
& = 
\left[
1 + 2 \, C_F \frac{\alpha_s}{4 \pi} 
\left(\frac{1}{\epsilon} + \log \left(\frac{\mu^2}{-p^2} \right) \right)
\right] \langle Q_2 \rangle_{ tree}
\nonumber \\[3mm]
&  
\, + \frac{3}{N_c} \frac{\alpha_s}{4 \pi} 
\left(\frac{1}{\epsilon} + \log \left(\frac{\mu^2}{-p^2} \right) \right)
\langle Q_2 \rangle_{ tree}
- 3 \frac{\alpha_s}{4 \pi} 
\left(\frac{1}{\epsilon} + \log \left(\frac{\mu^2}{-p^2} \right) \right)
\langle Q_1 \rangle_{ tree} \, .
\label{eq:ME-Q2-0}
\end{align}\\
The $1/\epsilon$ poles in the square brackets of Eqs.~(\ref{eq:ME-Q1-0}), (\ref{eq:ME-Q2-0}), are removed again with the renormalisation of the quark field.
However, the results are still divergent and require in addition an operator renormalisation, i.e.\\ 
\begin{equation}
Q_i^{(0)} = Z_{ij} \, Q_j, \qquad i,j = 1,2\,,
\label{eq:Z-def}
\end{equation}\\
where the superscript $(0)$ refers to unrenormalised quantities,
and $\hat Z$ is a $2 \times 2$ renormalisation matrix. 
By taking into account also the field renormalisation $Z_q$, the relation between the unrenormalised and renormalised matrix elements, denoted by $\langle Q_i \rangle$, is given by\\
\begin{equation}
\langle Q_i \rangle^{(0)} = Z_q^{-2} Z_{ij} \, \langle Q_j \rangle,
\label{eq:operators-renormalisation}
\end{equation}\\ 
and in the MS scheme \cite{tHooft:1973mfk}, the explicit expression of $Z_q$ is \\
\begin{equation}
Z_q =  1 - \frac{1}{\epsilon}  \frac{C_F \,\alpha_s}{4 \pi} + {\cal O}(\alpha_s^2)\,.
\label{eq:Zq}
\end{equation}\\
Using Eq.~\eqref{eq:operators-renormalisation}, and Eqs.~\eqref{eq:ME-Q1-0}, 
\eqref{eq:ME-Q2-0}, \eqref{eq:Zq}, yields to the following result for the $\hat Z$ matrix in the MS scheme, namely \\
\begin{equation}
\hat Z = \mathds{1}_2 + \frac{\alpha_s}{4\pi} \frac{1}{\epsilon} 
\left(
\begin{array}{cc}
3/N_c & -3 \\
-3 & 3/N_c
\end{array} 
\right) + {\cal O}(\alpha_s^2) ,
\label{eq:Z-res}
\end{equation}\\
from which we obtain that the renormalised matrix elements of the local operators, respectively read\\
\begin{align}
\langle Q_1 \rangle
& = 
\left[
1 + 2 \, C_F \frac{\alpha_s}{4 \pi} 
\log \left(\frac{\mu^2}{-p^2} \right)
\right] \langle Q_1 \rangle_{ tree}
\nonumber \\[3mm]
& 
\, + \frac{3}{N_c} \frac{\alpha_s}{4 \pi} 
\log \left(\frac{\mu^2}{-p^2} \right)
\langle Q_1 \rangle_{ tree}
- 3 \frac{\alpha_s}{4 \pi} \log \left(\frac{\mu^2}{-p^2} \right)
\langle Q_2 \rangle_{ tree} \, ,
\label{eq:ME-Q1-ren}
\end{align}\\
and \\
\begin{align}
\langle Q_2 \rangle 
& =  
\left[
1 + 2 \, C_F \frac{\alpha_s}{4 \pi} 
\log \left(\frac{\mu^2}{-p^2} \right)
\right] \langle Q_2 \rangle_{ tree}
\nonumber \\[3mm]
& 
\, + \frac{3}{N_c} \frac{\alpha_s}{4 \pi} 
\log \left(\frac{\mu^2}{-p^2} \right)
\langle Q_2 \rangle_{ tree}
- 3 \frac{\alpha_s}{4 \pi} 
\log \left(\frac{\mu^2}{-p^2} \right)
\langle Q_1 \rangle_{ tree} \, .
\label{eq:ME-Q2-ren}
\end{align}\\
Finally, by substituting Eqs.~(\ref{eq:ME-Q1-ren}), (\ref{eq:ME-Q2-ren}), into Eq.~\eqref{eq:matching} and taking into account Eq.~\eqref{eq:A-full}, we can extract the corresponding expressions of the WCs, i.e.\\
\begin{equation}
C_1(\mu)  = 
1 + \frac{3}{N_c} \frac{\alpha_s}{4 \pi} \log \left(\frac{m_W^2}{\mu^2}  \right)\,,
\qquad
C_2(\mu)  = 
-3 \, \frac{\alpha_s}{4 \pi} \log \left(\frac{m_W^2}{\mu^2}  \right)\,.
\label{eq:C2-mW}
\end{equation}\\
Setting $\alpha_s$ to zero in Eq.~(\ref{eq:C2-mW}), we recover $C_1 = 1$, $C_2 = 0$.
Note that this same result is obtained also by setting $\mu = m_W$, corresponding to the matching scale.
\\[2mm]
From the above description it is evident that the main property of the construction of the OPE lies in the possibility to factorise the short and long distance contributions of the full amplitude, between the Wilson coefficients and the matrix element of local operators in the effective theory. In fact, up to terms of order $ {\cal O}(\alpha_s^2)$, we have\\
\begin{equation}
\left( 1 + \frac{3}{N_c}\frac{\alpha_s}{4 \pi} \log\left(\frac{m_W^2}{- p^2}\right) \right) = \left( 1 + \frac{3}{N_c}\frac{\alpha_s}{4 \pi} \log\left(\frac{m_W^2}{ \mu^2 }\right) \right) \left( 1 +\frac{3}{N_c}\frac{\alpha_s}{4 \pi} \log\left(\frac{\mu^2}{-p^2}\right) \right)  \,,
\end{equation}\\
and then \\
\begin{equation}
 \log\left(\frac{m_W^2}{- p^2}\right) =  \log\left(\frac{m_W^2}{ \mu^2}\right) +  \log\left(\frac{\mu^2}{ -p^2}\right) \,.
\end{equation}\\
By taking into account that the logarithms originate from the integration over a loop variable, it follows that we can schematically write\\
\begin{equation}
\int \limits_{-p^2}^{m_W^2} \frac{d k^2}{k^2} = \int \limits_{\mu^2}^{m_W^2} \frac{d k^2}{k^2} + \int \limits_{-p^2}^{\mu^2} \frac{d k^2}{k^2}\,,
\end{equation}\\
showing that the effect of large virtual momenta in the loop, e.g.\ from scales  $\mu \approx 1$ GeV to $m_W$ is absorbed in the expression of the Wilson coefficients, while the low energy contributions, depending on the off-shell momentum $p$, are encoded into the matrix elements of the local operators.\\[2mm]
However, it is easy to verify that at scales much smaller than $m_W$, 
the logarithms in Eq.~(\ref{eq:C2-mW}) become large, namely \\
\begin{equation}
\alpha_s \log \frac{m_W^2}{\mu^2} = {\cal O} (1), \qquad  {\rm with}\quad  \mu^2 \ll m_W^2\,,
\end{equation}\\
and therefore the series in powers of $\alpha_s \log (m_W^2/\mu^2)$ does not converge.  
 The solution is provided by employing the renormalisation group equations 
(RGEs), which allow to resum the leading logarithms of the type $\alpha_s^n \log(m_W^2/\mu^2)^n$ to all orders in perturbation theory.  Analogously to Eq.~(\ref{eq:Z-def}), we then introduce the unrenormalised Wilson coefficients \\
\begin{equation}
C_{i}^{(0)} = Z^c_{ij} C_{j}\,,
\end{equation}\\
where $Z^c_{ij}$ is the corresponding renormalisation matrix
\footnote{Because the effective Hamiltonian, proportional to $ \vec C \cdot  \langle \vec Q \rangle$, must be scale independent,  it follows that $Z_{ij}^c = Z_{ji}^{-1}$, 
where $Z_{ij}$ is given in Eq.~\eqref{eq:Z-res}. }.
From the definition of the anomalous dimension matrix $\hat \gamma$\\
\begin{equation}
\hat \gamma = \hat Z^{-1} \frac{d}{d \log \mu} \hat Z \,,
\label{eq:ADM-def}
\end{equation}\\
or explicitly, using Eq.~(\ref{eq:Z-res})
\begin{equation}
\gamma (\alpha_s) 
= \frac{\alpha_s}{4 \pi}
\left(
\begin{array}{cc}
- 6/N_c & 6 \\
6 & - 6/N_c
\end{array} 
\right)\,,
\end{equation}\\
it follows that, taking into account $\hat Z^{c\,T} = Z^{-1}$, the RGEs satisfied by the renormalised Wilson coefficients, read\\
\begin{equation}
\frac{d}{d \log \mu} C_i (\mu) = \gamma^T_{ij} (\alpha_s)  C_j (\mu)\,.
\label{eq:RGE-WC}
\end{equation}\\
The solution of Eq.~\eqref{eq:RGE-WC}, can be formally presented as\\
\begin{equation}
C_i (\mu) = U_{ij} (\mu, m_W) C_j (m_W)\,,
\label{eq:resummation}
\end{equation}\\
with $\hat U (\mu, m_W)$ being the evolution matrix describing the running of the Wilson coefficients from the matching scale $m_W$ to the lower scale $\mu$.
\\[2mm]
We conclude by emphasising that the presence of only two operators $Q_1(x)$ and $Q_2(x)$, in the effective Hamiltonian in Eq.~(\ref{eq:Heff}), follows from having considered the specific decay mode $c \to s \bar d u$. In fact, in the description of arbitrary $c$-quark decays,  additional operators are generated by including QCD corrections, these are the penguin operators\\
\begin{align}
Q_3(x)& = \bar u^i (x) \gamma^\mu(1-\gamma_5) c^i(x) \, \sum_q \bar q^j(x)\gamma_\mu (1 - \gamma_5) q^j(x)\,,
\\[3mm]
Q_4(x) &= \bar u^i (x) \gamma^\mu(1-\gamma_5) c^j(x) \, \sum_q \bar q^j(x)\gamma_\mu (1 - \gamma_5) q^i(x)\,,
\\[3mm]
Q_5(x)& = \bar u^i (x) \gamma^\mu(1-\gamma_5) c^i(x) \, \sum_q \bar q^j(x)\gamma_\mu (1 + \gamma_5) q^j(x)\,,
\\[3mm]
Q_6(x) &= \bar u^i (x) \gamma^\mu(1-\gamma_5) c^j(x) \, \sum_q \bar q^j(x)\gamma_\mu (1 + \gamma_5) q^i(x)\,.
\end{align}


\section{The heavy quark effective theory}
\label{sec:HQET}
The low-energy dynamics of hadrons is governed by the confining QCD interactions and the scale $\Lambda_{QCD}$, at which the strong coupling $\alpha_s(\mu)$ becomes non perturbative, provides a characteristic parameter for it. The inapplicability of standard perturbation theory poses a big challenge for the computation of any hadronic matrix element,
however, simplifications usually arise when considering special limiting cases. The description of hadrons containing a heavy quark $Q$, where by heavy it is meant that $m_Q \gg \Lambda_{QCD}$, leads to profound consequences, because under this condition, the hadronic system can be parametrised as an almost free heavy quark surrounded by a cloud of light degrees of freedom. In particular, in the limit $m_Q \to \infty$, it is only the four-velocity $v^\mu$ of the infinitely heavy quark, which coincides with the hadron velocity, 
that characterises the bound state dynamics. The QCD interaction with the light constituents, despite changing the heavy quark momentum $p_Q^\mu$, cannot affect its velocity, which is conserved because of $\Delta v^\mu = \Delta p^\mu_Q/m_Q$. In this limit, the heavy quark effectively acts as a static external colour source. The soft gluons and quarks are sensitive to the static colour field because of confinement but they are unable to resolve other quantum numbers of the heavy quark, like flavour and spin. These relativistic effects are suppressed by the heavy quark mass and can be systematically taken into account in a perturbative way, see e.g.\ the early review \cite{ Neubert:1993mb}. It follows that, in the heavy-quark limit, the QCD Lagrangian is  approximated by an effective theory, the heavy quark effective theory (HQET) \cite{Eichten:1979pu, Shuryak:1981fza,  Voloshin:1986dir, Shifman:1987rj, DAVIDPOLITZER1988504, Eichten:1989zv, ISGUR1989113, GEORGI1990447, Grinstein:1990mj, Falk:1990yz, Luke:1990eg}, where new symmetries, which are not present in the original theory, become manifest. Specifically, for a system with $f$ heavy flavours, there is a $SU(2f)$ symmetry group corresponding to rotations in spin and heavy flavour space. The possibility to exploit the existence of the heavy-quark symmetry in certain kinematical domains, leads to simplifications in the computation of hadronic matrix elements involving heavy quarks, in particular it allows to derive model-independent relations between hadronic form factors for weak decays, thus reducing significantly the number of independent input required, see e.g.\ the review \cite{Mannel:1997ky}. 

Far from being exhaustive, the rest of this section is mainly intended as a brief introduction to the HQET, in order to derive the basic properties that will be used in the sequent chapters.
The exposure closely follows the comprehensive monograph \cite{manohar_wise_2000} and the excellent reviews \cite{Neubert:1993mb, Mannel:1995dr, Mannel:1997ky, doi:10.1146/annurev.nucl.47.1.591}.
 
 The fundamental assumption for the construction of the HQET is that a heavy quark bounded in a QCD state with light constituents carries most of the four-momentum of the system and is quasi on-shell. Interactions with soft gluons and quarks can only change $p^\mu_Q$ by a fraction small compared to its large ``kinetic" component. According to this picture, the heavy quark momentum is parametrised as\\
\begin{equation}
p_Q^\mu = m_Q v^\mu + k^\mu\,,
\label{eq:HQET-momentum}
\end{equation}\\
where $v^\mu$ is the hadron velocity with $v^2 = 1$ and the ``residual" momentum $k^\mu$ determines by how much the heavy quark is off-shell because of the QCD interaction with the light degrees of freedom, so $k$ is of the order of $ \Lambda_{QCD}$. Substituting Eq.~(\ref{eq:HQET-momentum}) into the expression for the Feynman propagator for $Q$,
and expanding in the small quantity $k/m_Q$, yields
\\
\begin{align}
i \frac{\slashed p_Q + m_Q}{p_Q^2 - m_Q^2 + i \varepsilon} &= i \frac{m_Q \slashed v + \slashed k + m_Q}{\Big( m_Q^2 + 2 m_Q v \cdot k + k^2 - m_Q^2 + i \varepsilon\Big)}
\nonumber\\[3mm]
& = \frac{i}{  v \cdot k  + i \varepsilon}  \left( \frac{1 + \slashed v }{2} \right)  + {\cal O}\left(\frac{k}{m_Q}\right)\,.
\label{eq:HQ-propagator}
\end{align}\\
Eq.~(\ref{eq:HQ-propagator}) shows that the propagator of a heavy quark contains a velocity dependent operator which projects onto the positive energy components of the Dirac field. In fact it is immediate to verify that the operators \\
\begin{equation}
P_\pm = \frac{1 \pm \slashed v }{2} \,, 
\label{eq:projectors-Ppm}
\end{equation}\\
fulfil $P_\pm^2 = P_\pm$, $P_\pm P_\mp = 0$,
and are thus projectors. Their meaning becomes particularly transparent if we consider the rest frame of the heavy quark i.e.\ $v^\mu = (1, \bf 0)$, in fact in this case $ P_\pm = \big( 1 \pm \gamma_0\big)/2$,
or explicitly \\
\begin{equation}
P_+ = 
\begin{pmatrix}
\mathds 1_2 & 0 \\
0 & 0
\end{pmatrix} \,, \quad 
P_- = 
\begin{pmatrix}
0 & 0 \\
0 & \mathds 1_2  
\label{eq:projectors}
\end{pmatrix}
\,,
\end{equation}\\
indicating that $P_\pm$ respectively project onto the upper/lower two components of the Dirac spinor. 
Because in Eq.~(\ref{eq:HQ-propagator}) only the positive energy solutions of the Dirac equation are propagated, it appears appropriate to introduce the following parametrisation for the heavy quark field:\\
 \begin{equation}
 Q(x) = e^{-i m_Q v \cdot x} \, h_v(x) + {\cal O}\left(\frac{k}{m_Q}\right)\,,
 \label{eq:HQ-field}
 \end{equation}\\
 where the effective heavy quark $h_v(x)$ satisfies\\
 \begin{equation}
 h_v(x) = e^{i m_Q v \cdot x}\,  \frac{1 + \slashed v }{2}\,   Q(x)\,, 
  \label{eq:hv-def}
 \end{equation}\\ 
and hence 
\begin{equation}
P_+ h_v(x) = h_v(x)\,.
\label{eq:hv-proj}
\end{equation}\\
The presence of the exponential prefactor in Eq.~(\ref{eq:HQ-field}) removes the large ``kinetic'' part of the heavy-quark momentum, so that $h_v(x)$ contains only the small frequencies of the order of $k$. Notice also that due to Eq.~(\ref{eq:hv-def}), $h_v(x)$ is constrained to be effectively a two-component field.   
By expressing the QCD Lagrangian for $Q$, in terms of $h_v(x)$, gives
\begin{align}
{\cal L}_{QCD} &= \bar Q(x) \left( i \slashed D - m_Q \right) Q(x)
= \bar h_v(x)\, i \slashed D \, h_v(x) + {\cal O}\left(\frac{k}{m_Q}\right)
\nonumber \\[3mm]
& =  \bar h_v(x) P_+ \, i \slashed D \, P_+ \,h_v(x) + {\cal O}\left(\frac{k}{m_Q}\right) \,,
\label{eq:QCD-lagrangian}
\end{align}\\
where we have used Eq.~(\ref{eq:hv-proj}) and then $\slashed v h_v = h_v$. Taking into account the identity\\
\begin{equation}
P_+ \gamma^\mu P_+ = P_+ P_- \gamma^\mu + P_+ v^\mu
 = P_+ v^\mu P_+\,,
 \label{eq:P+gammaP+}
\end{equation}\\
we obtain that in the limit of a infinitely heavy quark, Eq.~(\ref{eq:QCD-lagrangian}) becomes\\
\begin{align}
{\cal L}_{HQET} = \bar h_v(x) (i v\cdot D) h_v(x)
 = \bar h_v^j(x) \left(i v\cdot \partial \, \delta_{jk} + g_s v \cdot  A^a \, t^a_{jk} \right) h^k_v(x)
\,,
\label{eq:L-HQET}
\end{align}\\
and for clarity the colour indices $j,k,$ have been explicitly indicated in the second equality. Eq.~(\ref{eq:L-HQET}) defines the Lagrangian of the HQET. The corresponding Feynman rules for the heavy quark propagator and for the coupling of a heavy quark to the gluon field can be easily read off Eq.~(\ref{eq:L-HQET}). They are respectively given by \footnote{The $+ i \varepsilon $ prescription is consistent with a heavy quark propagating forward in time, see Ref~\cite{Eichten:1989zv}. } \\
\begin{equation}
 \includegraphics[scale = 0.55]{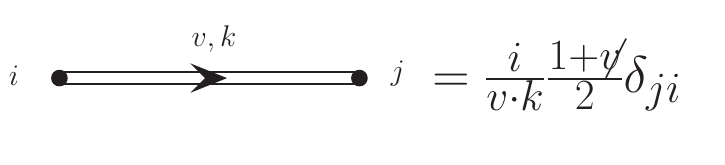}
 \label{eq:hv-prop}
 \end{equation}
 and
 \begin{equation}
 \includegraphics[scale = 0.55]{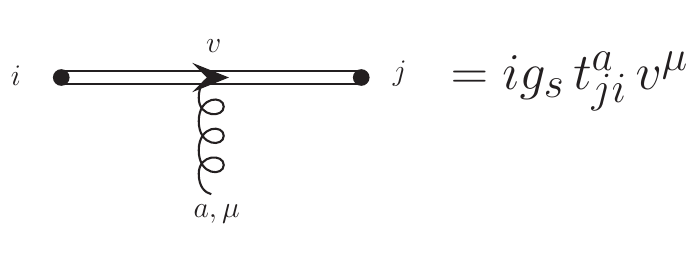} 
 \label{eq:hv-vertex}
 \end{equation}
Clearly Eq.~(\ref{eq:hv-prop}) reproduces the leading term in Eq.~(\ref{eq:HQ-propagator}). It is worthwhile to emphasise that the effective heavy field $h_v(x)$, by construction, annihilates a heavy particle with velocity $v^\mu$ but does not create a heavy antiparticle. The conjugate field $\bar h_v(x)$, on the other side, creates a heavy particle with velocity $v^\mu$ but does not annihilate a heavy antiparticle. Pair production is absent in the infinite heavy mass limit and the field-theoretic description becomes actually redundant, see e.g.\ Ref.~\cite{doi:10.1146/annurev.nucl.47.1.591}. Consistently, contrary to the full QCD propagator, Eq.~(\ref{eq:hv-prop}) has a single pole since only heavy particles are propagating in space and time, see e.g.\ Ref.~ \cite{Neubert:1993mb}. The contribution of heavy antiparticles is suppressed by the heavy quark mass and arises when power corrections are included, cf.\ Eq.~(\ref{eq:hv-def}). 
Note that Eq.~(\ref{eq:L-HQET}) does not depend on the heavy quark mass, so that the theoretical description stays unchanged if the heavy quark $Q$ is replaced by a different heavy quark $Q^\prime$ with the same velocity $v^\mu$, provided that the condition $m_{Q^\prime} \gg \Lambda_{QCD}$ is verified. Furthermore, since the vertex Eq.~(\ref{eq:hv-vertex}) does not contain any gamma matrix, which would act on the spin states of the heavy quark field, the interaction with the gluon is independent of the heavy quark spin. The effective theory exhibits a flavour-spin symmetry, broken by the inclusion of mass effects. 

Eq.~(\ref{eq:HQ-field}) describes only the contribution of the large component of the heavy quark field, $h_v(x)$. In order to derive power corrections to the HQET Lagrangian in Eq~(\ref{eq:L-HQET}), we introduce the small component $\mathfrak{h}_v(x)$, defined by \\
\begin{equation}
 \mathfrak{h}_v(x) = e^{i m_Q v \cdot x}\,  \frac{1 - \slashed v }{2}\,   Q(x)\,,
\label{eq:hv-frak-def}
\end{equation}\\ 
with
\begin{equation}
P_- \mathfrak{h}_v(x) = \mathfrak{h}_v(x) \,.
\label{eq:hv-frak-proj}
\end{equation}\\
In the rest frame of the heavy quark, see Eq.~(\ref{eq:projectors}), $\mathfrak{h}_v(x)$ corresponds to the lower two components of $Q(x)$ and creates a heavy antiquark with velocity $v^\mu$. Including also the effect of $\mathfrak{h}_v(x)$, Eq.~(\ref{eq:HQ-field}) reads \footnote{Note that the formalism introduced so far applies to the description of a bound state with a heavy quark. The case of a   hadron containing a heavy antiquark is obtained by replacing the sign of the velocity i.e. $v^\mu \to - v^\mu$.}\\
\begin{equation}
Q(x) = e^{- i m_Q v \cdot x} \Big[ h_v(x) + \mathfrak{h}_v(x)\Big]\,,
\label{eq:Qx-hqet}
\end{equation}\\
and correspondingly Eq.~(\ref{eq:QCD-lagrangian}) becomes\\
\vspace*{-5mm}
\begin{align}
{\cal L}_{QCD} &= \Big[ \bar h_v(x) +\bar {\mathfrak{h}}_v(x)\Big]  \Big( m_Q \slashed v + i \slashed D - m_Q\Big) \Big[ h_v(x) +{\mathfrak{h}}_v(x)\Big] 
\nonumber\\[3mm]
& = \bar h_v(x) (i v\cdot D) h_v(x) - \bar{\mathfrak{h}}_v(x) \big( i v \cdot D + 2 m_Q \big) \mathfrak{h}_v(x) 
\nonumber \\[3mm]
& +  \bar h_v(x) P_+ i\slashed D P_- \mathfrak{h}_v(x) +  \bar{\mathfrak{h}}_v(x) P_- i\slashed D P_+h_v(x)\,.
\label{eq:L-qcd-1}
\end{align}\\
Notice that in deriving the second equality in Eq.~(\ref{eq:L-qcd-1}), we have used Eqs.~(\ref{eq:hv-proj}), (\ref{eq:hv-frak-proj}), together with $\slashed v h_v(x) = h_v(x)$ and $\slashed v \mathfrak{h}_v(x) = - \mathfrak{h}_v(x)$. Moreover, the last line of Eq.~(\ref{eq:L-qcd-1}) can be further simplified. Because $\slashed D$ is squeezed between $P_+$ and $P_-$, the component of the covariant derivative parallel to $v^\mu$ vanishes and only the one orthogonal to the four-velocity actually contributes. Defining\\
\begin{equation}
D^\mu_\perp = D^\mu - v^\mu (v \cdot D)\,,
\end{equation} 
with $v \cdot D_\perp = 0$, gives
\begin{align}
{\cal L}_{QCD} 
& = \bar h_v(x) (i v\cdot D) h_v(x) - \bar{\mathfrak{h}}_v(x) \big( i v \cdot D + 2 m_Q \big) \mathfrak{h}_v(x) 
\nonumber \\[3mm]
& +  \bar h_v(x) i\slashed D_\perp \mathfrak{h}_v(x) +  \bar{\mathfrak{h}}_v(x) i\slashed D_\perp h_v(x)\,.
\label{eq:L-qcd-2}
\end{align}\\
\begin{figure}
\centering
\includegraphics[scale = 0.8]{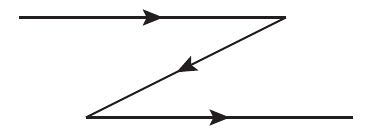}
\caption{Virtual fluctuation involving the creation and annihilation of a heavy antiquark. Time flows from left to right.}
\label{fig:fluct}
\end{figure}
Eq.~(\ref{eq:L-qcd-2}) shows that the Lagrangian of a heavy, but not infinitely heavy, quark $Q$ contains two independent fields $h_v(x)$ and $\mathfrak{h}_v(x)$, describing respectively massless degrees of freedom and massive excitations with mass twice as large as $m_Q$. These fields interact due to the presence of the two terms in the second line of Eq.~(\ref{eq:L-qcd-2}), so that the propagator of a heavy particle $h_v(x)$ receives virtual corrections from the coupling with the heavy antiparticle $\mathfrak{h}_v(x)$. Precisely, a heavy quark propagating forward in time can turn into a virtual heavy antiquark propagating backward in time and then turn back into a heavy quark as it is schematically shown in Figure~\ref{fig:fluct}, see also Ref.~\cite{Neubert:1993mb}. From Eq.~(\ref{eq:L-qcd-2}), it follows that the propagator of the virtual antiquark is suppressed by a factor of $2 m_Q$ and at energy scales of the order of $\Lambda_{QCD}$, the diagram in Figure~\ref{fig:fluct} can be effectively described by a local interaction of the form \\
\begin{equation}
\bar h_v (x) i \slashed D_\perp 1/(2 m_Q) i \slashed D_\perp h_v(x)\,,
\end{equation}\\
in which the heavy degrees of freedom corresponding to $\mathfrak{h}_v(x)$, appear decoupled. The process of integrating out the small component of the heavy quark field can be carried out in a systematic way by constructing an effective Lagrangian expressed only in terms of the large component $h_v(x)$. To this end, first we derive from Eq.~(\ref{eq:L-qcd-2}) the equations of motion for $h_v(x)$ and $\mathfrak{h}_v(x)$ by computing $\delta {\cal L}_{QCD}/\delta \bar h_v(x)$ and $\delta {\cal L}_{QCD}/\delta \bar {\mathfrak{h}}_v(x)$. This yields respectively\\
\begin{equation}
(i v \cdot D) h_v(x) = - i \slashed D_\perp \mathfrak{h}_v(x)\,,
\label{eq:EOM-hv}
\end{equation}\\
and
\begin{equation}
\big( i v \cdot D + 2 m_Q \big) \mathfrak{h}_v(x) =  i \slashed D_\perp h_v(x)\,.
\label{eq:EOM-hv-frak}
\end{equation}\\
Eq.~(\ref{eq:EOM-hv-frak}) can be inverted in order to find a relation between $\mathfrak{h}_v(x)$ and  $h_v(x)$, i.e. \\
\begin{equation}
\mathfrak{h}_v(x) = \big( i v \cdot D + 2 m_Q - i \varepsilon  \big)^{-1}  i \slashed D_\perp h_v(x)\,,
\label{eq:EOM-hv-frak-2}
\end{equation}\\
showing that $\mathfrak{h}_v(x)$ indeed represents the small component of the heavy field $Q(x)$, as it is suppressed with respect to $h_v(x)$, by the heavy quark mass $m_Q$. By substituting Eq.~(\ref{eq:EOM-hv-frak-2}) into the equation of motion for $h_v(x)$, Eq.~(\ref{eq:EOM-hv}), we readily arrive~at\\
\begin{equation}
( i v \cdot D ) h_v(x) + i\slashed D_\perp \big( i v \cdot D + 2 m_Q - i \varepsilon \big)^{-1}  i \slashed D_\perp  h_v(x)  = 0\,,
\label{eq:EOM-hv-2}
\end{equation}\\
which can be evidently traced back to the following Lagrangian \footnote{The Lagrangian in Eq.~(\ref{eq:Eff-L}) can be equivalently derived using the path integral formalism by integrating out the heavy degrees of freedom from the generating integral of the QCD Green functions with heavy quark fields, see Ref.~\cite{Mannel:1991mc}. }
\\
\begin{equation}
{\cal L}_{eff} = \bar h_v(x)( i v \cdot D ) h_v(x) + \bar h_v(x) i\slashed D_\perp \big( i v \cdot D + 2 m_Q - i \varepsilon \big)^{-1}  i \slashed D_\perp  h_v(x)  \,.
\label{eq:Eff-L}
\end{equation}\\
Eq.~(\ref{eq:Eff-L}) provides the appropriate theory to describe the strong interactions of a heavy quark at the energy scale of the order of $\Lambda_{QCD}$. It is expressed only in terms of the effective heavy field $h_v(x)$, as the dynamics of the massive degrees of freedom  becomes irrelevant at this scale. The information on $\mathfrak{h}_v(x)$ however, is contained in the second term of Eq.~(\ref{eq:Eff-L}), which represents a non local operator. Because the action of a derivative on $h_v(x)$, returns only the ``residual" momentum $k^\mu$, the non local contribution in Eq.~(\ref{eq:Eff-L}) can be consistently expanded in powers of $1/(2 m_Q)$ leading to higher dimensional operators built from covariant derivatives. Correspondingly, the equation of motion satisfied by $h_v(x)$, Eq.~(\ref{eq:EOM-hv-2}), explicitly depends on the heavy quark mass. In order to exploit the symmetries of the effective theory in the limit of a infinitely heavy quark, it appears convenient to regard Eq.~(\ref{eq:Eff-L}) in an alternative way, namely by  treating the tower of power suppressed operators arising from the expansion of the non local term in Eq.~(\ref{eq:Eff-L}), as perturbations to the HQET Lagrangian in Eq.~(\ref{eq:L-HQET}) \cite{Luke:1990eg, Georgi:1990ei, Falk:1992wt}. Accordingly, Eq.~(\ref{eq:Eff-L}) is recast in the form
\\
\begin{equation}
{\cal L}_{eff} = {\cal L}_{HQET} + {\cal L}_{power}\,,
\label{eq:L-eff-3}
\end{equation}
with 
\begin{align}
{\cal L}_{power} &= \bar h_v(x) i \slashed D_\perp \frac{1}{2 m_Q} i \slashed D_\perp h_v(x) + \bar h_v(x)\, i \slashed D_\perp \frac{(- i v \cdot D)}{(2 m_Q)^2}\, i \slashed D_\perp h_v(x) +\ldots \,.
\label{eq:L-power}
\end{align} \\
Here the ellipsis denote terms suppressed by higher powers of $m_Q$.
Now, the effective heavy quark field $h_v(x)$, satisfies the equation of motion following only from the leading term of Eq.~(\ref{eq:L-eff-3}) i.e.\\
\begin{equation}
(i  v \cdot D) h_v(x)=0 \,,
\label{eq:EOM-hv-free}
\end{equation}\\
and in the computation of hadronic matrix elements, the contribution of ${\cal L}_{power}$ in Eq.~(\ref{eq:L-power}) must be included in a standard perturbative way by taking the time order product with the respective leading order operators. Similarly, by substituting Eq.~(\ref{eq:EOM-hv-frak-2}) into the expression for the heavy quark field Eq.~(\ref{eq:Qx-hqet}), leads to the following expansion:\\
\begin{align}
Q(x) &= e^{- i m_Q v \cdot x} \left( 1 +  \big( i v \cdot D + 2 m_Q - i \epsilon  \big)^{-1} i \slashed D_\perp  \right) h_v(x)
\nonumber \\[3mm]
& = e^{- i m_Q v \cdot x} \left( 1 + \frac{1}{2 m_Q} i \slashed D_\perp + \frac{1}{(2 m_Q)^2} (-i v \cdot D) \, i \slashed D_\perp +  \ldots  \right) h_v(x)\,,
\label{eq:Qx-expansion-HQET}
\end{align}\\
which provides the prescription to consistently define in HQET any operator involving a heavy quark field $Q(x)$. Consider e.g.\ the heavy to light vector current ${\cal V}^\mu(x) = \bar q(x) \gamma^\mu Q(x)$, with $m_q \ll m_Q$. Up to leading power corrections, ${\cal V}^\mu(x)$ can be expressed as \\
\begin{equation}
{\cal V}^\mu(x) = e^{- i m_Q v \cdot x} \,  \bar q(x) \gamma^\mu  \left( 1 + \frac{i \slashed D_\perp}{2 m_Q} + \ldots \right) h_v(x)\,.
\label{eq:Vx-expansion-HQET}
\end{equation}\\
Due to the equation of motion Eq.~(\ref{eq:EOM-hv-free}), the effective heavy field $h_v(x)$ does not contain any information about the heavy quark mass. This has the advantage that the local hadronic matrix element $\langle 0 | {\cal V}^\mu(0) | M(v)\rangle$, where $M$ is the corresponding heavy meson, defined in full QCD, admits a systematic expansion in powers of $1/m_Q$, in which the dependence on the heavy quark mass results completely factored out. In fact, from Eqs.~(\ref{eq:Vx-expansion-HQET}), (\ref{eq:L-power}),  it follows that\\
\begin{align}
\langle 0 | {\cal V}^\mu | M(v)\rangle_{QCD} &= \langle 0| \bar q \gamma^\mu h_v|M(v)\rangle_{HQET}+ \frac{1}{2 m_Q} \langle 0| \bar q \gamma^\mu i \slashed D_\perp h_v | M(v) \rangle_{HQET} 
\nonumber \\[3mm]
& + \frac{1}{2 m_Q} \, \langle 0| i \int d^4z\, {\rm T} \left\{ \bar q \gamma^\mu h_v, {\cal L}_1(z) \right\} |M(v) \rangle_{HQET} +  {\cal O}\left(\frac{1}{m_Q^2}\right) \,,   
\label{eq:Exp-vector-current}
\end{align}\\
where we have introduced the notation\\
\begin{equation}
 {\cal L}_{power} = \frac{1}{2 m_Q} {\cal L}_1 + \frac{1}{4 m_Q^2} {\cal L}_2 + \ldots \,.
 \label{eq:Lpower-1}
\end{equation}\\
Contrary to the matrix element on the l.h.s.\ of Eq.~(\ref{eq:Exp-vector-current}), the ones on the r.h.s are independent on $m_Q$ and can be parametrised by universal form factors \cite{ISGUR1989113, Falk:1992wt}.  However, because of Eq.~(\ref{eq:EOM-hv-free}), the hadronic state $|M(v) \rangle_{HQET}$ differs from the original one $|M(v) \rangle_{QCD}$. This is encoded in the appearance of the time ordered product of the first term in ${\cal L}_{power}$, with the leading order part of ${\cal V}(0)$, which can be interpreted as a correction to the wave function of the heavy meson, see Ref.~\cite{Neubert:1993mb}. 
\\[2mm]
Finally, in order to identify the set of lowest dimensional operators generated by the Lagrangian in Eq.~(\ref{eq:L-power}), we can employ the identity
\begin{align}
P_+ \, i \slashed D_\perp i \slashed D_\perp P_+ &= P_+ \, i D_{\perp}^\mu\, i D_{\perp}^\nu \left( \frac{\{\gamma_\mu, \gamma_\nu \}}{2}  + \frac{[\gamma_\mu, \gamma_\nu]}{2}  \right) P_+
\nonumber \\[3mm]
& = P_+ (i D_{\perp})^2 P_+ + P_+\, i D^\mu\, i D^\nu \, (- i \sigma_{\mu \nu}) P_+
 \,,
 \label{eq:DslDsl}
\end{align}\\
where in the second line of Eq.~(\ref{eq:DslDsl}) we have replaced $D_\perp^\mu$ with the total derivative $D^\mu$ since the component of the covariant derivative parallel to the four velocity does not contribute, in fact\\
\begin{equation}
P_+ v^\mu \sigma_{\mu \nu} P_+ = \frac{i}{2}\, P_+( \slashed v \gamma_\nu - \gamma_\nu \slashed v) P_+  =  \frac{i}{2}\, P_+(  \gamma_\nu - \gamma_\nu ) P_+ = 0\,.
\label{eq:vsigmamunu}
\end{equation}\\ 
Recalling the definition of the gluon field strength tensor $G_{\mu \nu} = -i \,[i D_\mu, i D_\nu]$, see Eq.~(\ref{eq:G-munu}), it follows that at order $1/m_Q$, two operators appear in ${\cal L}_{power}$, namely\\
\begin{equation}
{\cal L}_{power} = \frac{1}{2 m_Q}\Big( {\cal O}_I(x) + {\cal O}_{II}(x) \Big) + \ldots \,, 
\label{eq:L-power-lead}
\end{equation}
with
\begin{align}
 {\cal O}_I(x) &= \bar h_v(x) (i \slashed D_\perp)^2 h_v(x) \,,
\label{eq:O-kin-HQET}
\end{align}
and
\begin{equation}
{\cal O}_{II}(x) = \frac{1}{2} \bar h_v(x)  G^{\mu \nu}  \sigma_{\mu \nu} h_v(x)  \,.
\label{eq:O-magn-HQET}
\end{equation}\\
The two contributions in Eq.~(\ref{eq:L-power-lead}), describe respectively the covariant extension of the kinetic energy of the heavy quark due to its off-shell motion inside the hadron and the chromo-magnetic interaction of the heavy quark spin with the  external gluon field. 


\section{Expansion of the quark-propagator in the Fock-Schwinger gauge}
\label{sec:FS}
In light of the primary role that it will play in the following chapters, we discuss the computation of the quark propagator \footnote{We now consider an arbitrary quark, without making any assumption on its mass.} in the presence of non perturbative QCD interactions, in a form suitable to describe the case in which the gluon field is soft, namely its characteristic momentum is much smaller than the one carried by the corresponding quark field. Under this assumption the dynamics reduces to that of a quark propagating in a weakly changing gluon background, see e.g.\ the lecture notes~\cite{Shifman:1995dn}, and the solution of the Green function equation can be build as an operator expansion in terms of the the external gauge field. The formulation is based on the Schwinger method which was introduced in the early 50's in the context of Electrodynamics in Ref.~\cite{Schwinger:1951nm}. Later it has been adapted to QCD where it has found a large number of applications, see e.g. Ref.~\cite{Novikov:1980uj}. A variation of the background field technique based on the Fock-Schwinger (FS) gauge \cite{Fock:1937dy, Schwinger:1951nm} has been first considered in Refs.~\cite{Dubovikov:1981bf, Shuryak:1981kj, Smilga:1982wx}. This alternative method results extremely convenient for calculations in gauge theories due to the remarkable property that only gauge covariant expressions appear in the intermediate steps of the computation of gauge invariant quantities. For details on the application of the Schwinger method and of the FS gauge in QCD we refer to the technical review Ref.~\cite{Novikov:1983gd} as well as to the references within. 

In the rest of the present section, after introducing the FS gauge and discussing its main features, we turn to the calculation of the quark propagator. Specifically, we use the FS gauge to compute the coefficients of the gluon operators that arise in the expansion of the quark propagator, up to terms proportional to one covariant derivative of the gluon field strength tensor $G_{\mu \nu}$. The corresponding expressions are derived both in momentum and in coordinate space.

Let us start by recalling that the vacuum expectation value of the time ordered product of two free-quark fields is defined as  \\
\begin{equation}
\langle 0 | {\rm T} \big\{\psi (x), \bar \psi (y) \big\} | 0 \rangle = i  S_0(x,y)\,.
\label{eq:corr-free-quark}
\end{equation}\\
For e.g.\ $x^0 > y^0$, the l.h.s.\ of Eq.~(\ref{eq:corr-free-quark}) describes a quark emitted at the space-time point $y^\mu$ and subsequently annihilated at point $x^\mu$, i.e.\ $S_0(x,y)$ denotes the propagator of a free-quark. Equivalently $S_0(x,y)$ constitutes the Green function for the Dirac equation, namely it satisfies the inhomogeneous differential equation\\
\begin{equation}
\big(i \slashed \partial_x  - m\big) S_0 (x, y) = \delta^{(4)} (x-y)\,.
\label{eq:free-prop-equation}
\end{equation} \\
It is worthwhile to emphasise how the translation invariance of Eq.~(\ref{eq:free-prop-equation})
implies that the free-quark propagator is also translation invariant, this is reflected by the condition $S_0(x,y) = S_0 (x-y)$. Eq.~(\ref{eq:free-prop-equation}) can be solved exactly. In momentum space $S_0(x-y)$ admits the well known Fourier representation, see e.g.\ the textbook~\cite{Itzykson:1980rh}\\
\begin{equation}
S_0(x-y) = \int \frac{d^4 p}{(2 \pi)^4} \, \, {\cal S}_0(p)\, e^{- i p \cdot (x - y)}\,,
\label{eq:Sxy-free-prop}
\end{equation} \\
with
\begin{equation}
{\cal S}_0(p) =  \frac{\slashed p + m}{p^2 - m^2 + i \varepsilon} \,.
\label{eq:S0p}
\end{equation}\\ 
A quark bounded in an hadronic state is subject to the long-distance interaction with the confining gluon field $A_\mu(x)$. Correspondingly, the quark propagator $S(x,y)$ defines the Green function of the coupled Dirac equation, see e.g.\ Ref~\cite{Novikov:1983gd} \footnote{Unless otherwise stated, in the following, the coupling constant $g_s$ is absorbed in the definition of the gauge field~$A^\mu(x)$.}\\
\begin{equation}
\big(i \slashed \partial_x + \slashed A (x) - m\big) S (x, y) = \delta^{(4)} (x-y)\,.
\label{eq:GreenFunEq}
\end{equation}\\
Eq.~(\ref{eq:GreenFunEq}) cannot be solved exactly, however in the kinematical region $k^2 \ll q^2$ where $k$ refers to the momentum of the gluon field and $q \gg \Lambda_{QCD}$, is a large perturbative scale saturated by the quark momentum $p$ i.e. $p^2 \sim q^2$, 
the quark propagates 
with a characteristic length scale that is much smaller than the one of the external gluon field, which effectively acts as a slowly changing background, see e.g.\ Ref.~\cite{Zakharov:1999jj}.
Under the assumption that the field $A_\mu(x)$ is weak and randomly orientated, it is possible to construct the solution of Eq.~(\ref{eq:GreenFunEq}) in the form of the series, see e.g.\ Ref.~\cite{Novikov:1983gd}\\
\begin{figure}
\centering
\includegraphics[scale = 0.7]{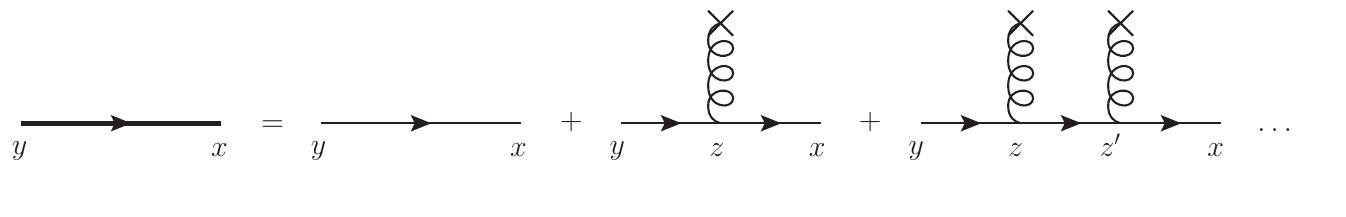}
\caption{ A quark propagating from $y$ to $x$ in the background gluon field, scatters off $0, 1, 2, \ldots$, soft gluons.}
\label{fig:propagator}
\end{figure}
\begin{equation}
i S (x, y) = i S_{0}(x-y) + i S_{1}(x,y) +  \ldots \,,
\label{eq.quark-propagator-definition1}
\end{equation}\\
where $S_{1}(x,y)$ denotes the first order correction, describing the interaction of the quark with one gluon field while the ellipsis stand for higher order terms with more than one gluon, explicitly\\
\begin{equation}
i S_{1}(x,y) =\int d^4 z \,\,  i S_{0} (x - z)\,\, i \slashed A (z)\,\, i S_{0} (z - y).
\label{eq.quark-propagator-definition}
\end{equation}\\
Eq.~(\ref{eq.quark-propagator-definition1}) is schematically represented in Figure~\ref{fig:propagator}.
It is straightforward to verify that Eq.~(\ref{eq.quark-propagator-definition1}) does indeed satisfy Eq.~(\ref{eq:GreenFunEq}) up to terms of first order in $A_\mu(x)$, by substituting Eq.~(\ref{eq.quark-propagator-definition}) and using Eq.~(\ref{eq:free-prop-equation}).
To fix the form of the gauge field in Eq.~(\ref{eq.quark-propagator-definition}), it is particularly convenient to employ the FS gauge. This is defined by\\ 
\begin{equation}
(x^\mu - x^\mu_0 ) A_\mu(x) = 0\,.
\label{eq:fs-gauge}
\end{equation}\\
In Eq.~(\ref{eq:fs-gauge}) the gauge fixing parameter $x^\mu_0$ is an arbitrary space-time point which we set for convenience to zero.  On one side, this will lead to simpler expressions, on the other, the possibility to use the independence of the final result on $x^\mu_0$ as a consistency check for the computation, is lost. The gauge condition then becomes\\
\begin{equation}
x^\mu  A_\mu (x) = 0\,.
\label{eq:FS-gauge}
\end{equation} \\
Let us immediately point out that the quark propagator $S(x,y)$ is not translation invariant anymore. First, the gauge field in Eq.~(\ref{eq.quark-propagator-definition}) depends on the space-time coordinate. This however only apparently breaks the translation symmetry, since after averaging, the background gluon field is actually translation invariant, see Ref.~\cite{Novikov:1983gd}. The true reason for the symmetry breaking lies in the choice of the FS gauge Eq.~(\ref{eq:FS-gauge}), which gives to the origin the special role of gauge fixing parameter. In general then\\
\begin{equation}
S(x,y) \neq S(x-y)\,.
\label{eq:trans-break-Sxy}
\end{equation}\\
Eq.~(\ref{eq:trans-break-Sxy}) can lead to differences in intermediate steps of the computation of a physical quantity although the translation invariance must be restored in any final meaningful expression. 

The FS gauge though, has many advantages. The first is that it allows for a simple relation between the gauge field $A_\mu(x)$ and the field strength tensor $G_{\mu \nu}(x)$.  This relation reads \footnote{In the literature this is known as inversion formula, in reference to the fact that it inverts the usual relation in which the field strength tensor is expressed in terms of the gauge field and of its derivative, for a comprehensive overview on the FS gauge see the PhD thesis \cite{1991fock}.} \\ 
\begin{equation}
A_\mu^a (x) = \int_0^1 \! \! d \alpha \, \alpha x^\rho \, G^a_{\rho\mu}(\alpha x).
\label{eq:Amu}
\end{equation}\\
The proof of Eq.~(\ref{eq:Amu}) proceeds as follows. We start with the identity \\
\begin{align}
A_\sigma(x) &=\frac{d}{d x^\sigma} \big( x \cdot  A(x)\big) -  x^\rho \frac{\partial}{\partial x^\sigma}A_\rho(x)
\nonumber
\\[3mm]
& = x^\rho G_{\rho \sigma}(x) - x^\rho \frac{\partial}{\partial x^\rho}A_\sigma(x) + i \, x^\rho \, \big[A_\rho(x), A_\sigma(x)\big]
\nonumber \\[3mm]
& =  x^\rho G_{\rho \sigma}(x) - x^\rho \frac{\partial}{\partial x^\rho}A_\sigma(x) \,,
\label{eq:Amu-der-1}
\end{align}\\
where the second and third equalities are consequence of the gauge condition Eq.~(\ref{eq:FS-gauge}). By performing the change of variable $x^\mu \rightarrow \alpha x^\mu$, it is easy to see that the dependence on $A_\sigma(\alpha x)$ is reduced to that of a total derivative i.e.\ Eq.~(\ref{eq:Amu-der-1}) becomes\\ 
\begin{equation}
\frac{d}{ d \alpha} \Big( \alpha A_\sigma(\alpha x) \Big)  = \alpha x^\rho G_{\rho \sigma}(\alpha x)\,,
\end{equation}\\
which reproduces Eq.~(\ref{eq:Amu}), after integrating both sides over $\alpha$ from $0$ to~$1$. In order to prove another property of the FS gauge, we expand $A_\mu (x)$ in Eq.~(\ref{eq:FS-gauge}) around $x = 0$, this yields  \footnote{Note that we often omit to explicitly write the space-time coordinate when this is zero.} \\
\begin{equation}
x^\mu \Big( A_\mu + x^{\nu_1} \partial_{\nu_1} A_\mu(x)\big|_{x=0} + \frac12 x^{\nu_1} x^{\nu_2}  \partial_{\nu_1} \partial_{\nu_2}  A_\mu(x)\big|_{x = 0} + \ldots \Big) = 0\,.
\label{eq:FS-gauge1}
\end{equation}\\
For arbitrary space-time coordinates, evidently Eq.~(\ref{eq:FS-gauge1}) requires that\\
\begin{equation}
x^\mu A_\mu = x^\mu x^{\nu_1} \partial_{\nu_1} A_\mu(x)\big|_{x=0} =  x^\mu x^{\nu_1} x^{\nu_2}  \partial_{\nu_1}\partial_{\nu_2}  A_\mu(x)\big|_{x=0}   = \ldots = 0 \,.
\label{eq:FS-gauge2}
\end{equation}\\
Eq.~(\ref{eq:FS-gauge2}) leads to the important result that in the expansion of an arbitrary function $f(x)$, the action of the partial derivative $\partial_\mu$ at the origin can be replaced with that of the covariant derivative $D_\mu$, in fact\\
\begin{align}
f(x) & = f+  x^\mu \partial_\mu f(x)\big|_{x=0}+ \frac12 x^\mu x^\nu \partial_\mu \partial_\nu f(x)\big|_{x=0}  + \ldots  
\nonumber \\[3mm]
& =  f + x^\mu \big( \partial_\mu  -  i A_\mu(x) \big) f(x)\big|_{x=0}
+  x^\mu  x^\nu \big( \partial_\mu  -  i A_\mu(x) \big)\big( \partial_\nu -  i A_\nu (x)\big) f(x)\big|_{x=0} + \ldots
\nonumber \\[3mm]
& =  f +  x^\mu D_\mu f(x)\big|_{x=0}  + \frac12 x^\mu x^\nu D_\mu D_\nu f(x)\big|_{x=0} + \ldots \,.
\label{eq:FS-gauge3}
\end{align}\\
We can now derive a convenient representation for the gauge field $A_\mu(x)$.
Expanding $G_{\rho \mu}(\alpha x)$ around $x = 0$ in Eq.~(\ref{eq:Amu}), and taking into account Eq.~(\ref{eq:FS-gauge3}), yields \\
\begin{equation}
A_\mu^a (x) = 
\frac{1}{2} x^\rho G_{\rho\mu}^a(0) + \frac13 x^\alpha x^\rho D_\alpha G^a_{\rho \mu} (0)+ \ldots \,,
\label{eq:A_mu}
\end{equation}\\
where the ellipsis denote terms with higher derivatives.\ Eq.~(\ref{eq:A_mu}) shows that the gauge field $A_\mu(x)$ can be  expanded directly in terms of the gluon field strength tensor and of its covariant derivatives evaluated at the origin and it constitutes the main result of the FS gauge. In particular it follows that $A_\mu (0) = 0$. This property will reveal to be very useful in practical calculations.  

To compute the first order correction to the free-quark propagator in Eq.~(\ref{eq.quark-propagator-definition1}), we substitute Eq.~(\ref{eq:A_mu}) into Eq.~(\ref{eq.quark-propagator-definition}), and choose for simplicity $y^\mu=0$. This gives\\
\begin{align}
S_1(x,0) & =  \int d^4 z \,\, \int \frac{d^4 p}{(2\pi)^4} \,  \left(  \frac{ i (\slashed p + m)}{p^2 - m^2 + i \varepsilon} \right) e^{-i p \cdot (x-z)}
\nonumber
\\[3mm]
& \times 
\left( \frac{1}{2}  \gamma^\mu z^\rho G_{\rho \mu}  + \frac13 \gamma^\mu z^\alpha z^\rho D_\alpha G^a_{\rho \mu} 
 \right) 
 \nonumber \\[3mm]
 & \times  \int \frac{d^4 k}{(2\pi)^4} \,  \left(  \frac{i (\slashed k + m)}{k^2 - m^2 + i \varepsilon} \right)  e^{-i k \cdot z}+ \ldots  \,.
 \label{eq:S1-x}
 \end{align}\\
 The functions $z^\rho$ and $z^\alpha z^\rho$ in the second line of Eq.~(\ref{eq:S1-x}) can be conveniently rewritten using the identity \footnote{Rewriting $z^\mu$ in terms of a derivative with respect to $k_\mu$ is the simplest choice. Equivalently one could write $z^\mu = -i \frac{\partial}{ \partial p_\mu} e^{i p\cdot z}$, in this case though, when integrating by parts one would have to differentiate also the function $e^{- i p \cdot x}$. }  $z^\mu = i \frac{\partial}{\partial k_{  \mu} }e^{-i k \cdot z}$, 
which leads to\\
 \begin{align}
S_1(x,0)  & =   \int d^4 z \,\, \int \frac{d^4 p}{(2\pi)^4} \, \left( \frac{i (\slashed p + m)}{p^2 - m^2 + i \varepsilon}\right) e^{-i p \cdot (x-z)} 
\nonumber
\\[3mm]
& \times 
\Bigg[ \frac12 \gamma^\mu G_{\rho \mu}
 \int \frac{d^4 k}{(2\pi)^4} \, \left(  \frac{i (\slashed k + m)}{k^2 - m^2 + i \varepsilon}  \right) \left(i \frac{\partial}{\partial k_{  \rho} }e^{-i k \cdot z}\right) 
 \nonumber \\[3mm]
 & +  
 \frac13 \gamma^\mu D_\alpha G_{\rho \mu}  \int \frac{d^4 k}{(2\pi)^4} \, \left(  \frac{i (\slashed k + m)}{k^2 - m^2 + i \varepsilon } \right)  \left(   - \frac{\partial}{\partial k_{  \alpha}}  \frac{ \partial}{\partial k_{  \rho}}e^{-i k \cdot z}\right) \Bigg] + \ldots \,.
 \label{eq:S1-x-0}
 \end{align}\\
Performing a single and double integration by parts, respectively in the second and third line of Eq.~(\ref{eq:S1-x-0}), we obtain \\
 \begin{align}
 S_1(x,0)
 &=  \int d^4 z \,\, \int \frac{d^4 p}{(2\pi)^4}  \,\, \int \frac{d^4 k}{(2\pi)^4}\,  e^{-i (k-p) \cdot z } e^{-i p \cdot x} 
 \nonumber \\[3mm]
 & \times \left(  \frac{i (\slashed p + m)}{p^2 - m^2 + i \varepsilon} \right)  \,  \Bigg[ \frac12 \gamma^\mu G_{\rho \mu} \, \left( -i \frac{\partial}{\partial k_\rho} \right) \left( \frac{i (\slashed k + m)}{k^2 -m^2 + i \varepsilon}  \right)
 \nonumber \\[3mm]
 & -   \frac13 \gamma^\mu D_\alpha G_{\rho \mu} \,  \frac{\partial}{\partial k_\rho} \frac{\partial}{\partial k_\alpha} \left( \frac{i (\slashed k + m)}{k^2 -m^2 + i \varepsilon} \right) \Bigg] + \ldots \ \,,
 \label{eq:S1-x-1}
\end{align}\\
also note that we have taken into account that all the boundary terms vanish, which can be easily verified by direct inspection.
The integral over the variable $z^\mu$ in Eq.~(\ref{eq:S1-x-1}), results in a delta function and enforces the momentum conservation $k^\mu = p^\mu$ when integrating over the variable $k^\mu$. Furthermore, the first and second order derivatives of the free-quark propagator in  the square brackets of Eq.~(\ref{eq:S1-x-1}), yield\\
\begin{align}
\frac{\partial}{\partial p_{\rho}}   \frac{(\slashed p + m)}{p^2 - m^2 + i \varepsilon} & =
 \frac{\gamma^\rho }{p^2 -m^2 + i \varepsilon} -  \frac{2 p^\rho (\slashed p + m)}{(p^2 -m^2 + i \varepsilon)^2} \, ,
 \label{eq:der1}
\end{align}\\
and\\
\begin{align}
\frac{\partial}{\partial p_\rho} \frac{\partial}{\partial p_\alpha}  \frac{ (\slashed p + m)}{p^2 -m^2 + i \varepsilon} & = - \frac{2 (p^\alpha \gamma^\rho + p^\rho \gamma^\alpha)}{(p^2 - m^2 + i \varepsilon)^2} - \frac{2 g^{\rho \alpha}  (\slashed p + m)}{(p^2 -m^2 + i \varepsilon)^2} 
+  \frac{8 p^\alpha p^\rho(\slashed p + m)}{(p^2-m^2 + i \varepsilon)^3}\,.
\label{eq:der2}
 \end{align}\\
By substituting Eqs.~(\ref{eq:der1}), (\ref{eq:der2}), the expression in Eq.~(\ref{eq:S1-x-1}) can be written as\\
\begin{align}
S_1(x,0) & =  \int \frac{d^4 p}{(2\pi)^4} \, e^{-i p \cdot x} \Big(  {g}_2^{\rho \mu}  G_{\rho \mu}  + g_3^{\alpha \rho \mu} D_\alpha G_{\rho \mu} + \ldots  \Big)  \,,
\label{eq:S1-gamma}
\end{align}\\
where for clarity we have introduced the compact notation\\
\begin{align}
g_2^{\rho \mu} &= \frac{i}{2}  \Bigg[ p_\eta \frac{\gamma^\eta \gamma^\mu \gamma^\rho}{(p^2 -m^2+ i \varepsilon)^2} + m \frac{\gamma^\mu \gamma^\rho}{(p^2 - m^2+ i \varepsilon)^2}
\nonumber \\[3mm]
& - 2 p_\eta p_\sigma p^\rho \frac{\gamma^\eta \gamma^\mu \gamma^\sigma}{(p^2 -m^2 + i \varepsilon)^3} - 2 m^2 p^\rho \frac{\gamma^\mu}{(p^2 -m^2 + i \varepsilon)^3} \Bigg] \,,
\label{eq:g2}
\end{align}\\
and
\begin{align}
g_3^{\alpha \rho \mu} &= - \frac23 \frac{1}{(p^2 -m^2 + i \varepsilon)^3} \,  \Bigg[ \Big( p_\eta p_\xi \gamma^\eta \gamma^\mu \gamma_\tau  + m\, p_\xi \gamma^\mu \gamma_\tau \Big)\Big(g^{\xi \alpha} g^{\tau \rho} + g^{\xi \rho} g^{\tau \alpha} \Big) \nonumber \\[3mm]
& + \left( g^{\rho \alpha}-  \frac{4 p^\rho p^\alpha}{(p^2 -m^2 + i \varepsilon)}  \right) \Big(p_\eta p_\sigma  \gamma^\eta \gamma^\mu \gamma^\sigma  
 + m^2 \gamma^\mu +2  m \, p^\mu  \Big) \Bigg] \,.
\nonumber \\[3mm]
\label{eq:g3}
\end{align}
We can simplify Eq.~(\ref{eq:S1-gamma}) by using the tensor decomposition of three gamma matrices, see Eq.~(\ref{eq:tensor-decomposition-gamma-mat}), together with the antisymmetry of the field strength tensor $G_{\mu \nu}$. A slightly lengthy yet simple algebraic manipulation leads to the final result for the quark propagator\\
\begin{equation}
S(x,0) =  \int \frac{d^4 p}{(2\pi)^4}\, {\cal S}(p)  \, e^{-i p \cdot x} \,,
\label{eq:Sx0}
\end{equation}\\
with
\begin{equation}
{\cal S}(p) = {\cal S}_0(p) + {\cal S}_1(p) + \ldots \,,
\label{eq:Sp-01}
\end{equation}\\
and
\begin{align}
{\cal S}_1(p) & = -\frac{m}{2} \frac{G_{\rho \mu }}{(p^2 -m^2 + i \varepsilon)^2} \sigma^{\rho \mu}  +  \frac{\tilde G_{\sigma \eta}}{(p^2 -m^2 + i \varepsilon)^2} p^\sigma \gamma^\eta \gamma^5 - \frac23 \frac{p^\alpha D_\alpha  G_{\rho \mu}}{(p^2 -m^2 + i \varepsilon)^3} \gamma^\mu p^\rho 
\nonumber \\[3mm]
&+ \frac23 \frac{D_\alpha G_{\alpha \mu}}{(p^2 - m^2 + i \varepsilon)^3} \Big[ \gamma^\mu  (p^2 -m^2) - p^\mu (\slashed p + 2 m) \Big]
 + 2 i \frac{D_\alpha \tilde G_{\tau \eta}}{(p^2 -m^2 + i \varepsilon)^3} p^\alpha p^\tau \gamma^\eta \gamma^5 
\nonumber \\[3mm]
& + \frac23 m \frac{D_\alpha G_{\rho \mu}}{(p^2 -m^2 + i \varepsilon)^3} \Big( p^\alpha \gamma^\rho \gamma^\mu - p^\rho \gamma^\mu \gamma^\alpha \Big)
+ \, \ldots \,.
\label{eq:S1p}
\end{align}\\
Here the ellipsis denote terms with higher order derivatives of the field strength tensor, while the dual field tensor is $\tilde G_{\mu \nu} = (1/2) \epsilon_{\mu \nu \rho \sigma} G^{\rho \sigma}$. Eq.~(\ref{eq:S1p}) has a transparent meaning. The interaction with the soft gluon field introduces corrections to the free-quark propagator parametrised by operators of higher dimensions built from the gluon field strength tensor and its covariant derivatives, evaluated at the origin. Each operator of dimension-$n$ is suppressed by $n$ powers of the quark momentum $p$, with the lowest order contribution being due to the dimension-two operator $G_{\mu \nu}$. 
In the limit of massless quark, Eq.~(\ref{eq:S1p}) reproduces the result of Ref.~\cite{Novikov:1983gd}, apart from the opposite sign in front of the two terms proportional to $\gamma_5$.
We trace this back to the different convention used in the Russian literature to define the fifth gamma matrix, namely $\gamma_5 = - i \, \gamma^0 \gamma^1 \gamma^2 \gamma^3$, cf.\ Eq.~(\ref{eq:gamma5-def}).
We stress that having fixed the notation and been consistent with it, the computation of any observable using Eq.~(\ref{eq:S1p}) or the expression in Ref.~\cite{Novikov:1983gd} must lead to the very same result. Moreover,
it is worthwhile to emphasise that the massless limit should be taken with care. Upon integration over $p^\mu$, it is only in the domain $p^2 \sim q^2$, where $q$ denotes a large perturbative scale, that the operator expansion in terms of the external gluon field is legitimate. However, as higher dimensional operators are considered and the power of the momentum variable in the denominator increases, the integral in Eq.~(\ref{eq:Sx0}) starts to be sensitive also to the long-distance region $p^2 = 0$ and to develop infrared (IR) divergences. In this case, the corresponding quark line becomes soft and the effect must be parametrised in terms of quark operators, see for details Ref.~\cite{Novikov:1983gd} or Chapter~\ref{ch:Darwin}.   
\\[2mm]
Because of Eq.~(\ref{eq:trans-break-Sxy}), the expression obtained in Eq.~(\ref{eq:S1p}) is valid only in the specific reference frame chosen, namely $y^\mu= 0$. In order to compute $S_1(0,y)$ we must repeat the calculation and set $x^\mu=0$. In this case substituting Eq.~(\ref{eq:A_mu}) into Eq.~(\ref{eq.quark-propagator-definition}) gives
\\
\begin{align}
S_1(0,y) & =  \int d^4 z \,\, \int \frac{d^4 p}{(2\pi)^4} \,  \left(  \frac{ i (\slashed p + m)}{p^2 - m^2 + i \varepsilon} \right) e^{i p \cdot z}
\nonumber
\\[3mm]
& \times 
\Bigg( \frac{1}{2}  \gamma^\mu z^\rho G_{\rho \mu}  + \frac13 \gamma^\mu z^\alpha z^\rho D_\alpha G^a_{\rho \mu} 
 \Bigg) 
 \nonumber \\[3mm]
 & \times  \int \frac{d^4 k}{(2\pi)^4} \,  \left(  \frac{i (\slashed k + m)}{k^2 - m^2 + i \varepsilon} \right)  e^{-i k \cdot (z-y)}+ \ldots  \,,
 \label{eq:S1-y}
 \end{align}\\
where now it is convenient to rewrite $z^\rho$ and $z^\alpha z^\rho$ in the second line of Eq.~(\ref{eq:S1-y}) using the identity $z^\mu = -i \frac{\partial}{\partial p_{  \mu} }e^{i p \cdot z}$.
The next intermediate steps proceed in analogy to the case of $S_1(x,0)$, for brevity  we omit them and state here only the final result, which reads\\
\begin{equation}
S(0, y) = \int \frac{d^4 p}{(2 \pi)^4} \, \tilde {\cal S}(p) \, e^{i p\cdot y} \,,
\label{eq:S0x}
\end{equation}\\
with\\
\begin{equation}
\tilde {\cal S}(p) = {\cal S}_0(p) + \tilde {\cal S}_1(p) + \ldots\,,
\end{equation}
and\\
\begin{align}
\tilde {\cal S}_1(p) & =  -\frac{m}{2} \frac{G_{\rho \mu }}{(p^2 -m^2 + i \varepsilon )^2} \sigma^{\rho \mu}  +  \frac{\tilde G_{\sigma \eta}}{(p^2 -m^2  + i \varepsilon)^2} p^\sigma \gamma^\eta \gamma^5 - \frac23 \frac{p^\alpha D_\alpha  G_{\rho \mu}}{(p^2 -m^2 + i \varepsilon )^3} \gamma^\mu p^\rho
\nonumber \\[3mm]
&+ \frac23 \frac{D_\alpha G_{\alpha \mu}}{(p^2 - m^2 + i \varepsilon )^3} \, \Big[ \gamma^\mu \, (p^2 -m^2) - p^\mu \slashed p \Big] 
 - 2 i \frac{D_\alpha \tilde G_{\tau \eta}}{(p^2 - m^2 + i \varepsilon )^3} p^\alpha p^\tau \gamma^\eta \gamma^5
\nonumber \\[3mm]
&  - \frac23 m \frac{D_\alpha G_{\rho \mu}}{(p^2 -m^2 + i \varepsilon )^3} \, \Big( p^\alpha \gamma^\rho \gamma^\mu - p^\rho \gamma^\mu \gamma^\alpha \Big) 
+ \ldots\,.
\label{eq:S1p-tilde}
\end{align}\\
The absence of translation symmetry Eq.~(\ref{eq:trans-break-Sxy}), is then reflected in momentum space by the condition $\tilde {\cal S}_1(p) \neq {\cal S}_1(p)$,
which actually holds true only starting from the operator of dimension-three $D_\rho G_{\mu \nu}$. In fact a comparison between  Eq.~(\ref{eq:S1p}) and Eq.~(\ref{eq:S1p-tilde}), shows that the translation invariance is still preserved in the coefficients of the dimension-two operator $G_{\mu \nu}$. 

For completeness, it is instructive to derive also an explicit representation of the quark propagator in coordinate space, by performing the anti-Fourier transform of the expressions in Eqs.~(\ref{eq:S1p}), (\ref{eq:S1p-tilde}). These contain both scalar and tensor functions of the variable $p^\mu$. However, it actually suffices to directly evaluate only the scalar integrals. In the case of $S(x,0)$ in Eq.~(\ref{eq:Sx0}), they have the following form\\
\begin{equation}
I_n (x) = \int \frac{d^4 p}{(2 \pi)^4} \frac{1 }{\big( p^2 - m^2 + i \varepsilon \big)^n} \, e^{-i p \cdot x}\,,  \qquad n \in  {\mathbb N}\,.
\label{eq:In}
\end{equation}\\
Given the analytic expression of $I_n(x)$, it is possible to obtain the tensor integrals appearing in Eq.~(\ref{eq:Sx0}), by differentiating Eq.~(\ref{eq:In}) with respect to $x_\mu$, namely\\
\begin{align}
I^{\mu_1 \ldots \mu_{n^\prime}}_n (x) &= \int \frac{d^4 p}{(2 \pi)^4} \frac{p^{\mu_1} \ldots p^{\mu_{n^\prime}}}{\big( p^2 - m^2 + i \varepsilon \big)^n} \, e^{-i p \cdot x} 
= \prod \limits_{j = 1}^{n^\prime}\left( i \frac{d}{d x_{\mu_j}}\right) I_n(x) \,,
\label{eq:In-tensor}
\end{align} \\
with $n^\prime \leq n$ \footnote{Note that for $n\geq 2$ only $n-1$ powers of the four-momentum can appear in the numerator. This follows from the fact that ${\cal S}(p)$ must have mass dimension of $-1$ and that for each gluon operator of dimension $n$ there are $2n$ powers of the momentum in the denominator.}. The integral in Eq.~(\ref{eq:In}) can be conveniently computed in Euclidean space by performing the Wick rotations $p_0 \to -i p_4$ and $x_0 \to -i x_4$, see e.g.\ the textbook~\cite{Peskin:1995ev} or the lecture notes~\cite{Soldati:QFT1}. This gives \\
\begin{equation}
I_n(x_E) =i \, (-1)^n  \int \frac{d^4 p_E}{(2 \pi)^4} \frac{1 }{\big( p_E^2 + m^2 \big)^n} \, e^{i p_E \cdot x_E}\,.
\label{eq:In-1}
\end{equation}\\
The Euclidean four-vectors are defined as $p_E^\mu = ( {\bf p}, p_4)$ and $x_E^\mu = ({\bf x}, x_4)$,
while the Euclidean metric reads $diag (1, 1, 1, 1)$. The factor of $(-1)^n$ in Eq.~(\ref{eq:In-1}), follows from the fact that $(p^2 - m^2) \to - (p_E^2 + m^2)$, also note that we have dropped the $+ i \varepsilon$ prescription since the denominator is now positive definite. The latter can be suitably expressed in an integral form. To this end, we start by writing $\Gamma(s) =\{ {\cal M} e^{-t} \} (s),$ where $\{ {\cal M } g(t)\}$ denotes the Mellin transform of $g(t)$ and $\Gamma(s)$ is the gamma function, see Ref.~\cite{gradshteyn2007} for exhaustive tables with definitions and useful properties, i.e.\\
\begin{equation}
\Gamma(s) = \int_0^\infty dt \, t^{s-1} e^{-t}\,, \qquad {\rm Re } \, s >0\,.
\end{equation}\\
Performing the change of variable $t \to \lambda t$ with $\lambda > 0$, yields $\lambda^{-s} \Gamma(s) = \{ {\cal M } e^{-\lambda t}\}(s)$, and after setting $\lambda = (p^2_E + m^2)$ and $s = n$, we readily obtain that\\
\begin{equation}
\frac{1}{(p_E^2 + m^2)^n} =\frac{1}{\Gamma(n)} \int_0^\infty dt\, t^{n-1} e^{-t (p_E^2 + m^2)}\,.
\label{eq:Mellin}
\end{equation}\\
Equivalently, Eq.~(\ref{eq:Mellin}) can be derived by taking into account the identity\\
\begin{equation}
\frac{1}{(p_E^2 + m^2)} = \int_0^\infty dt\, e^{-t (p_E^2 + m^2)}\,,
\label{eq:Mellin-2}
\end{equation}\\
easily proved by directly computing the integral on the r.h.s. The result for $n >1$ follows from differentiating $n-1$ times both sides of Eq.~(\ref{eq:Mellin-2}) with respect to the parameter $m^2$ and using that $\Gamma(n) = (n-1)!$.
Substituting Eq.~(\ref{eq:Mellin}) into Eq.~(\ref{eq:In-1}), we then arrive at\\
\begin{equation}
I_n(x_E) = i \, (-1)^n \int_0^\infty d t  \, \, t^{n-1} e^{- t m^2 }  \int \frac{d^4 p_E}{(2 \pi)^4}  e^{-t p_E^2 + i p_E \cdot x_E} \,.
\label{eq:In-2}
\end{equation}\\
The second integral on the r.h.s.\ of Eq.~(\ref{eq:In-2}) reduces to a standard four-dimensional Gaussian integral, see e.g.\ Ref~\cite{abramowitz+stegun}, after shifting the integration variable by a constant Euclidean four-vector i.e. $p_E^\mu \to p_E^\mu -  i x_E^\mu/ (2t)$. The solution reads   \\
\begin{equation}
\int d^4 p_E \, e^{-t p_E^2 + i p_E \cdot x_E} = \left( \frac{\pi}{t} \right)^2 e^{- \frac{x_E^2}{4 t}}\,,
\end{equation}\\
which we insert into Eq.~(\ref{eq:In-2}) to obtain \\
\begin{equation}
I_n(x_E) = i \,\frac{ (-1)^n}{16 \pi^2} \int_0^\infty d t  \, \,  t^{n-3}  e^{- t m^2 - \frac{x_E^2}{4 t}}\,.
\label{eq:In-3}
\end{equation}\\
It is easy to show that Eq.~(\ref{eq:In-3}) can be expressed in terms of the modified Bessel functions of the second type $K_\nu (z)$. Starting with the integral representation, see e.g.\ Ref.~\cite{gradshteyn2007}\\
\begin{equation}
K_\nu(z) = \frac12 \left( \frac{z}{2}\right)^\nu \int_0^\infty dt\,  t^{-\nu -1} e^{-t -\frac{z^2}{4 t } }  \,, \qquad |{\rm arg}\, z| < \frac{\pi}{2}\,, \,\,  {\rm Re} \, z^2 >0\,,
\end{equation}\\
and performing the change of variables $t \to \alpha t$, $z^2 \to \alpha z^2$, with $\alpha > 0$, gives\\ 
\begin{equation}
\int_0^\infty d t \, \, t^{-\nu -1 } e^{- \alpha t -\frac{ z^2}{4t}} = 2 \left(\frac{4 a }{z^2}\right)^\frac{\nu}{2}   K_{\nu} \left( \sqrt{a} z \right) \,.
\label{eq:Int-Knu}
\end{equation}\\
From Eq.~(\ref{eq:Int-Knu}), it then follows that  \\
 \begin{equation}
I_n(x_E) = i \,\frac{ (-1)^n}{8 \pi^2}  \left(\frac{4 m^2}{x_E^2}\right)^{\frac{2-n}{2}} K_{2-n} \left( m \sqrt{x_E^2} \right)\,,
\label{eq:In-4}
\end{equation}\\
and the condition ${\rm Re} (m  x_E^2) > 0$ translates into $x_E^2 >0$ for positive and real values of $m$ as well as real Euclidean distances. We can now rotate back to the Minkowski space-time and arrive at the final expression\\
 \begin{equation}
I_n(x) = i \,\frac{ (-1)^n}{8 \pi^2}  \left(\frac{4 m^2}{- x^2}\right)^{\frac{2-n}{2}} K_{2-n} \left( m \sqrt{- x^2} \right)\,,
\label{eq:In-5}
\end{equation}\\
which is valid for space-like separations $x^2 < 0$. The solution corresponding to the space-time region $x^2 > 0$ can be derived from Eq.~(\ref{eq:In-5}) by analytic continuation, taking into account, see e.g.\ Ref.~\cite{gradshteyn2007}, that\\
\begin{equation}
K_\nu (i z ) = \frac{-\pi i }{2} e^{ -  \frac{\pi}{2} \nu i} \, H^{(2)}_{-\nu}(z)\,,
\end{equation}\\
where $H^{(2)}_\nu(z)$ denotes the Hankel's function of the second kind.\ Finally, recall that the Bessel functions $K_\nu (z)$ satisfy the following recursive relation\\
\begin{equation}
\frac{d}{d z} K_\nu (z)= \frac{\nu}{z} K_\nu(z) - K_{\nu+1}(z)\,,
\end{equation}\\
that allows to compute the tensor integrals appearing in Eq.~(\ref{eq:Sx0}) according to Eq.~(\ref{eq:In-tensor}).
Let us consider explicitly the case $n= 1$. Eqs.~(\ref{eq:In-5}), (\ref{eq:In-tensor}) then read \\
\begin{equation}
I_1(x) = - \frac{i }{4 \pi^2} \frac{m}{\sqrt{- x^2}} K_1 \left( m \sqrt{-x^2} \right)\,,
\end{equation}\\
and
\begin{align}
I^\mu_1(x) 
& = - \frac{1}{4 \pi^2} \frac{m^2}{x^2} K_2 \left( m \sqrt{-x^2} \right) x^\mu\,,
\end{align}\\
from which we can readily derive the expression of the anti-Fourier transform of the free-quark propagator ${\cal S}_0(p)$, namely \\
\begin{equation}
S_0(x) = -\frac{1}{4 \pi^2} \frac{m^2}{x^2} K_2 \left( m \sqrt{-x^2} \right) \slashed x - \frac{i }{4 \pi^2} \frac{m^2}{\sqrt{- x^2}} K_1 \left( m \sqrt{-x^2} \right)\,.
\label{eq:Sx0-free}
\end{equation}\\
Proceeding in a similar way for the remaining cases $ n= 2, 3 $, it is straightforward  to verify that the coordinate representation of the first order correction in Eq.~(\ref{eq:S1p}), has the following form\\
\begin{align}
S_1(x,0) & = - \frac{\tilde G_{\alpha \beta}}{8 \pi^2} x^\alpha \gamma^\beta \gamma_5
    \frac{m K_1 (m \sqrt{-x^2})}{\sqrt{-x^2}} 
 - i  \frac{G_{\alpha \beta}}{16 \pi^2} \sigma^{\alpha \beta} m K_0 (m \sqrt{-x^2}) 
\nonumber \\[2mm]
&+ i \, \frac{D_\alpha G^{\alpha \beta}}{24 \pi^2}  \gamma_\beta K_0 (m \sqrt{-x^2})
  - i\, \frac{D_\alpha G^{\alpha \beta}}{48 \pi^2}  x_\beta \, \slashed x \,
    \frac{m K_1 (m \sqrt{-x^2})}{\sqrt{-x^2}}  
\nonumber \\[2mm]
&- \frac{D_{\alpha} G^{\alpha \beta} }{24 \pi^2}  x_\beta \, m K_0 (m \sqrt{-x^2})
  -i\, \frac{ D^{\alpha} G^{\beta \rho}}{48 \pi^2} \gamma_\rho x_\alpha x_\beta \, 
    \frac{m K_1 (m \sqrt{-x^2})}{\sqrt{-x^2}}
\nonumber \\[2mm]
&- \frac{ D^{\alpha} \tilde G^{\beta \rho} }{16 \pi^2} \gamma_\rho \gamma_5 x_\alpha x_\beta \, 
    \frac{m K_1 (m \sqrt{-x^2})}{\sqrt{-x^2}}
    - \frac{D^{\alpha} G^{\beta \rho} }{48 \pi^2} \gamma_\rho \gamma_\alpha x_\beta \, 
    m K_0 (m \sqrt{-x^2}) 
\nonumber \\[2mm]
& -  \frac{D^{\alpha} G^{\beta \rho}}{48 \pi^2}  \gamma_\rho \gamma_\beta 
    x_\alpha \, m K_0 (m \sqrt{-x^2}) +\ldots\,.
\label{eq:prop-coordinate-space}
\end{align}\\
The result in Eq.~(\ref{eq:prop-coordinate-space}) was first derived in Ref.~\cite{Belyaev:1985wza} up to terms proportional to $G_{\mu \nu}$, while the contribution of $D_\rho G_{\mu \nu}$ can be found in Ref.~\cite{Blok:1994cd} \footnote{Note that it is presented only the expression relevant for the computation described in the paper, namely with an odd number of gamma-matrices.}. In order to compute the corresponding expression for $S(0,y)$ we can make the replacement $x^\mu \to - y^\mu$ in Eqs.~(\ref{eq:In}), (\ref{eq:In-tensor}). Notice that because of Eq.~(\ref{eq:In-5}), the function $I_n(x)$ is even i.e.\ $I_n(-x) = I_n(x)$ and from this we obtain that\\
\begin{align}
I^{\mu_1 \ldots \mu_n}_n(-y) = (-1)^{n} I^{\mu_1 \ldots \mu_n}_n(y)\,.
\label{eq:In-y-tensor}
\end{align}\\
The above relation allows to immediately write \\
\begin{equation}
S_0(-y) = \frac{1}{4 \pi^2} \frac{m^2}{y^2} K_2 \left( m \sqrt{-y^2} \right) \slashed y - \frac{i }{4 \pi^2} \frac{m^2}{\sqrt{- y^2}} K_1 \left( m \sqrt{-y^2} \right)\,,
\label{eq:S0y-free}
\end{equation}\\
consistently with the fact that free-quark propagator is translation invariant. Moreover it is easy to check that the first order correction now reads \\
\begin{align}
S_1(0,y) & = \frac{\tilde G_{\alpha \beta}}{8 \pi^2} y^\alpha \gamma^\beta \gamma_5
    \frac{m K_1 (m \sqrt{-y^2})}{\sqrt{-y^2}} 
 - i  \frac{G_{\alpha \beta}}{16 \pi^2} \sigma^{\alpha \beta} m K_0 (m \sqrt{-y^2}) 
\nonumber \\[2mm]
&+i\,  \frac{D_\alpha G^{\alpha \beta}}{24 \pi^2}  \gamma_\beta K_0 (m \sqrt{-y^2})
  -i\,  \frac{D_\alpha G^{\alpha \beta} }{48 \pi^2} y_\beta \, \slashed y \,
    \frac{m K_1 (m \sqrt{-y^2})}{\sqrt{-y^2}}  
\nonumber \\[2mm]
&
  -i\, \frac{D^{\alpha} G^{\beta \rho} }{48 \pi^2} \gamma_\rho y_\alpha y_\beta \, 
    \frac{m K_1 (m \sqrt{-y^2})}{\sqrt{-y^2}} 
    + \frac{D^{\alpha} G^{\beta \rho} }{48 \pi^2} \gamma_\rho \gamma_\alpha y_\beta \, 
    m K_0 (m \sqrt{-y^2})
\nonumber \\[2mm]
&
 + \frac{D^{\alpha} G^{\beta \rho}}{48 \pi^2}  \gamma_\rho \gamma_\beta 
    y_\alpha \, m K_0 (m \sqrt{-y^2})
    + \frac{D^{\alpha} \tilde G^{\beta \rho}}{16 \pi^2}  \gamma_\rho \gamma_5 y_\alpha y_\beta \, 
    \frac{m K_1 (m \sqrt{-y^2})}{\sqrt{-y^2}}
 +\ldots\,,
\label{eq:prop-coordinate-space-y}
\end{align}\\
which clearly indicates that the translation symmetry is firstly broken in the coefficients of the dimension-three operator $D_\rho G_{\mu \nu}$, due to the fact that some of the quadratic functions in Eqs.~(\ref{eq:prop-coordinate-space}), (\ref{eq:prop-coordinate-space-y}), appear with a negative relative sign. 

We conclude this section with a final remark about the dependence of the quark propagator on the mass parameter $m$.\ Following the comment made on the massless limit of Eq.~(\ref{eq:S1p}), the expressions in Eqs.~(\ref{eq:prop-coordinate-space}), (\ref{eq:prop-coordinate-space-y}), become divergent for $m \to 0$. In coordinate space the singularity derives from  the asymptotic behaviour of the Bessel functions in the limit of small argument, see e.g.\ Ref.~\cite{gradshteyn2007}, namely\\
\begin{equation}
K_\nu(z) \sim 
\left\{
\begin{matrix}
- \log\left( \frac{z}{2} \right) - \gamma_E\,, &\quad \nu = 0\,,
\\[3mm]
\frac{\Gamma(\nu)}{2} \left( \frac{2}{z} \right)^\nu\,, & \quad \nu >0 \,.
\end{matrix}
\right.
\label{eq:massless-lim-prop}
\end{equation}\\
Notice that in the specific case of Eqs.~(\ref{eq:prop-coordinate-space}), (\ref{eq:prop-coordinate-space-y}), the divergence originates from the dimension-three contribution $D_\alpha G^{\alpha \beta}\gamma_\beta K_0 (m \sqrt{-x^2})$.


\section{The heavy quark expansion}
\label{sec:HQE}
The total decay width $\Gamma$ or equivalently its inverse, the total lifetime $\tau = \Gamma^{-1}$, defines one of the  fundamental properties of elementary and composite particles and hence constitutes an observable of phenomenological primary importance.
In the study of lifetimes a special role is occupied by heavy hadrons, due to the interplay that strong and weak interactions have in determining their decay, see e.g.\ Ref.~\cite{Shifman:1984wx}. 
As discussed already in Section~\ref{sec:HQET}, a heavy flavour hadron $H_Q$ is a QCD bound state that can be conveniently  represented as heavy quark $Q$ surrounded by a cloud of light quarks, antiquarks and gluons, where the distinction between heavy and light degrees of freedom is meant with respect to the typical hadronic scale $\Lambda_{QCD}$. 
The description of these systems considerably simplifies by considering the limit of an infinitely heavy quark $Q$, i.e.\ $m_Q \to \infty$. In this case it is possible to neglect the effect of the non perturbative QCD interactions and the hadron dynamics results entirely determined by that of a free quark $Q$, with the light constituents reducing to passive spectators, see e.g.\ Ref.~\cite{Shifman:1994mg}. This approximation leads to the result
$\Gamma(H_Q) = \Gamma_Q$, and then to the theoretical prediction that the lifetimes of hadrons containing the same heavy flavour but different spectator quarks should be equal. While among bottom hadrons this statement can be experimentally accommodated within deviations of few percent \cite{HFLAV:2019otj}, the pattern in the charm family is far less monotonous and lifetime ratios of charmed hadrons can be as large as $7$ \cite{LHCb:2018nfa, LHCb:2019ldj}~\footnote{More precisely, the lifetime ratio between the longest and the shortest living $b$-hadrons, decaying weakly and containing only one heavy quark, is $\tau(\Omega_b)/\tau(\Lambda_b) = 1.11$, to be compared with $\tau(\Xi_c^0)/\tau(D^+) = 6.8$, in the charm sector \cite{HFLAV:2019otj, LHCb:2018nfa, LHCb:2019ldj}. }. The infinite mass limit is clearly not sufficient for a proper interpretation of the experimental data, particularly in the charm sector where deviations from the free quark decay approximation are expected to give the dominant contribution. Before discussing how the corrections to this limit can be systematically taken into account, in what is the current theoretical framework for the study of inclusive decays of heavy hadrons, it is instructive to briefly retrace the main developments that have brought to its construction. In this respect we refer to the comprehensive review~\cite{Lenz:2014jha}.

The possibility to describe the decay of a heavy hadron in terms of an asymptotically free constituent heavy quark was first exploited in 1973 by Nikolaev in Ref.~\cite{Nikolaev:1973uu}, where constraints on the decay properties of a 'supercharged' (charmed) hadron, which at the time was only a theoretical particle, were proposed as an indirect test of its existence. Furthermore, after their discovery, the description of charmed hadrons decays was initially performed by considering only the dominant partonic contribution see e.g.\ Refs.~\cite{Gaillard:1974mw, Kingsley:1975fe, Ellis:1975hr, Altarelli:1975zz}. 
A pioneering work for the study of the lifetime of heavy hadrons is the one of Shifman and Voloshin, Ref.~\cite{Shifman:1984wx}, where many of the ingredients that contributed to the formulation of the heavy quark expansion (HQE) were originally presented. The HQE is a theoretical framework in which inclusive decays of heavy hadrons can be computed in terms of an operator product expansion (OPE) \cite{Wilson:1969zs, Wilson:1973jj, Novikov:1984rf}, by exploiting the large scale hierarchy $m_Q \gg \Lambda_{QCD}$. This method was firstly applied in a systematic way by Chay, Georgi and Grinstein for the analysis of inclusive semileptonic decays of heavy hadrons in Ref.~\cite{Chay:1990da}, and briefly after employed in the work of Bigi, Shifman, Uraltsev and Vainshtein \cite{Bigi:1992su, Bigi:1993fe} and of Blok, Koyrakh, Shifman and Vainshtein \cite{Blok:1993va}, for the description of inclusive non-leptonic as well as semileptonic decays, see e.g.\ Ref.~\cite{Mannel:1993su}. 

In order to discuss the construction of the HQE, it is convenient to review some important results in scattering theory, focusing in particular on the case of particle decay. 
We recall that the scattering process is described by introducing a unitary operator $S$, which governs the evolution of an asymptotically free state $|i\rangle$ into an asymptotically free state $| f \rangle$ for a given interaction theory, where the information about the non trivial part of the dynamics is encoded in the action of a transition operator $T$. The $S$ matrix is then decomposed into\\
\begin{equation}
S_{fi} \equiv \langle f | S | i \rangle = \delta_{f i} + i\, T_{f i} \,.
\label{eq:S}
\end{equation}\\
By taking into account that the interaction must conserve four-momentum, the transition amplitude $T_{fi}$ can be further parametrised as \\
\begin{equation} 
T_{fi} =  (2 \pi)^4 \delta^{(4)}(p_f - p_i)  {\cal M}_{fi}\,,
\label{eq:Tfi}
\end{equation}\\
where $p_i^\mu$, $p_f^\mu$ label respectively the momentum of the initial and final states and ${\cal M}_{fi}$ denotes the invariant  scattering amplitude. 
In the case that $|i\rangle$ contains only one particle, say $A$, the total decay width $\Gamma(A)$, is obtained by computing the amplitude squared for the process $A \to n$, where $|n \rangle$ represents an allowed $n$-particle final state, by summing over all the possible values of $n$ and finally by accounting for the flux factor $2 m_A$, namely, see e.g.\ the textbook~\cite{Itzykson:1980rh}\\
\begin{align}
\Gamma (A) = \frac{1}{2m_A} \sum \limits_n\,  \int_{n}\, 
(2 \pi)^4 \delta^{(4)} \left(  \sum \limits_{j=1}^n p_j - p_A\right)
\left| \langle n | {\cal M} | A \rangle\right|^2
\, ,
\label{eq:decay_rate_4.1}
\end{align}\\
here 
\begin{equation}
\int_n \equiv  \int \prod \limits_{j = 1}^n 
 \frac{d^3p_j}{(2 \pi)^3 2 E_j} \,,
 \label{eq:integral-n}
 \end{equation}\\
denotes the integration over the Lorentz invariant $n$-particle phase space while the presence of
the delta function ensures that
four-momentum is conserved in each decay. 
An equivalent and, for practical calculations, more advantageous representation of Eq.~(\ref{eq:decay_rate_4.1}), 
can be obtained by employing the optical theorem, see e.g.\ the textbook \cite{Peskin:1995ev}, which states that in a given interaction theory, the imaginary part of the forward scattering amplitude is proportional to the total cross
section for the production of all final states. 
The optical theorem follows from the unitarity of the scattering operator $S$ and hence from the mathematical requirement for the conservation of probability in Quantum Field Theory. The unitarity condition reads\\
\begin{equation}
S^\dagger S = \mathbb{I} = S S^\dagger\,.
\label{eq:unitarity-S}
\end{equation}
\begin{figure}
\centering
\includegraphics[scale = 0.5]{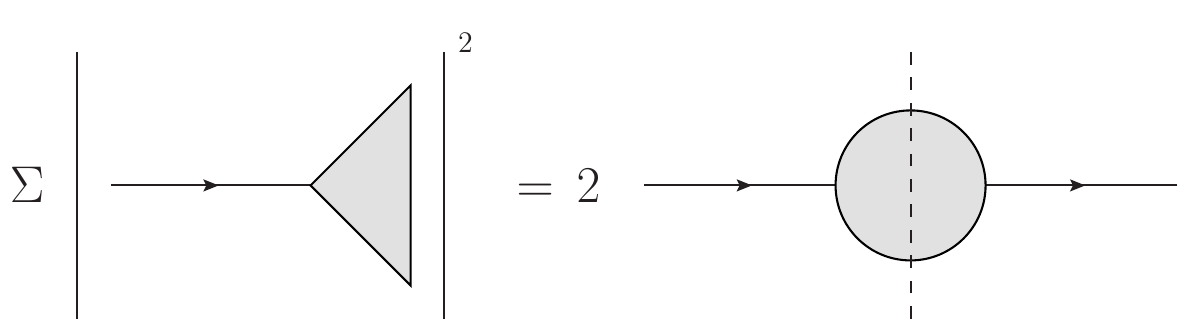}
\caption{Schematic representation of the optical theorem: the amplitude squared for the production of all final states is proportional to the absorptive part of the forward scattering amplitude.}
\label{fig:opt-th}
\end{figure}
Considering the matrix element between the states $|i\rangle$, $|f\rangle$, and using the completeness relation $\sum_n^\prime |n \rangle \langle n| = \mathbb{I}$, where $\sum_n^\prime \equiv \sum_n \int_n$ implies the sum over all the particles in $|n\rangle$ as well as the integration over their momenta, cf.\ Eq.~(\ref{eq:integral-n}), the first equality of Eq.~(\ref{eq:unitarity-S}) can be recast as\\
\begin{equation}
{\sum_n}^\prime \, \langle f | S^\dagger | n \rangle \langle n | S | i \rangle = \delta_{f i} \, .
\label{eq:unitarity_S}
\end{equation}\\
In the special case of forward scattering i.e.\ $ | f \rangle = | i \rangle$, taking into account that $\langle i | S^\dagger | n \rangle = \langle n | S | i \rangle^\dagger =  \langle n | S | i \rangle^*$, from Eqs.~(\ref{eq:S}), (\ref{eq:unitarity_S}), it follows that \\
\begin{equation}
 {\sum_n}^\prime \, (\delta_{ni} - i \, T^*_{ni}) (\delta_{ni} + i \, T_{ni} )  = \delta_{ii} \,,
 \label{eq:opt-th-1}
\end{equation}\\
and expanding the l.h.s.\ of Eq.~(\ref{eq:opt-th-1}), readily yields \\
\begin{equation}
i \, (  T_{ii} -  T_{ii}^* ) +  {\sum_n}^\prime \, T^*_{ni} T_{ni} = 0  \,,
\end{equation}\\
or equivalently  \\
\begin{equation}
2 \, \mbox{Im} \, T_{ii}  =   {\sum_n}^\prime\, |T_{ni}|^2  \,.
\label{eq:opt_th}
\end{equation}\\
Finally we can rewrite Eq.~(\ref{eq:opt_th}) by substituting Eq.~(\ref{eq:Tfi}) on both sides and by using $(\delta^{(4)} (z))^2 =\delta^{(4)}(0) \delta^{(4)} (z)$ to evaluate the square of the delta function, namely\\
\begin{equation}
2 \, {\rm Im} \,{\cal M}_{ii} = \sum_n \, \int_n \, (2 \pi)^4 \delta^{(4)} \left(\sum_{j=1}^n p_j - p_i\right) |{\cal M }_{ni}|^2\,,
\label{eq:Opt-th-2}
\end{equation}\\
which corresponds to the standard formulation of the optical theorem, schematically sketched in Figure~\ref{fig:opt-th}. Setting $|i \rangle = |A \rangle$, a comparison between  Eq.~(\ref{eq:Opt-th-2}) and Eq.~(\ref{eq:decay_rate_4.1}), evidently gives\\
\begin{equation}
\Gamma (A) = \frac{1}{ m_A} \,{\rm Im} \, {\cal M}_{AA}  \, ,
\label{eq:Gamma-A-opt-th}
\end{equation}\\
showing that the total decay width $\Gamma(A)$ can be obtained by computing the imaginary part of the forward scattering amplitude $A \to A$. 

We now apply the result in Eq.~(\ref{eq:Gamma-A-opt-th}) to the decay of a heavy quark $Q$ \footnote{The choice of a heavy quark is just for future convenience, the same description applies, taking into account the proper replacements, to the weak decay of any elementary fermion.}. We assume that at the renormalisation scale $\mu = m_Q$ the weak interaction is described by an effective Hamiltonian ${\cal H}_{eff}(x)$, governing the transition of the heavy quark into all possible lighter fermions, see Section~\ref{sec:Heff}, so that the scattering operator $S$ can be written as \footnote{It is worthwhile to emphasise that ${\cal H}_{eff}$ must be intended as supplemented with the QCD as well as with the QED Lagrangian, responsible for higher order corrections to the leading weak decay.} \\
\begin{equation}
S = {\rm T}\exp  \left\{ {- i \int d^4 x\,  {\cal H}_{eff}(x) }\right\}\,,
\end{equation}\\
where T is the time-ordering operator. The first non vanishing contribution to the forward scattering amplitude ${\cal M}_{QQ}$, is obtained by expanding $S$ to second order in the weak effective coupling. Up to terms of higher order this gives\\
\begin{align}
T_{QQ} &=  \frac{1}{2}  \,  \langle Q|\,  i\, \int d^4 x \int d^4 y  \,{\rm T} \Big \{  {\cal H}_{eff}(x), {\cal H}_{eff}(y) \Big \} | Q \rangle\,.
\label{eq:TQQ-1}
\end{align}\\
Using that the translation invariance of the Hamiltonian operator implies \\
\begin{equation}
{\cal H}_{eff}(x) = e^{i \hat P \cdot x} \, {\cal H}_{eff}(0)\, e^{-i \hat P \cdot x} \,,
\end{equation}\\
with $\hat P_\mu = i\, \partial/\partial x_\mu $, and that $|Q \rangle$ corresponds to a state with definite momentum $p_Q^\mu$, namely \\
\begin{equation}
e^{-i \hat P \cdot x} |Q \rangle = e^{- i p_Q \cdot x} |Q\rangle \,,
\end{equation}\\
Eq.~(\ref{eq:TQQ-1}) can be simplified as, see for a similar derivation the textbook~\cite{bookAK} \\
\begin{align} 
T_{QQ} &= \frac12  \,  \langle Q|\,  i\, \int d^4 x \int d^4 y  \,{\rm T} \Big \{ \underbrace{e^{-i \hat P \cdot y}\, {\cal H}_{eff}(x) \,e^{i \hat P \cdot y}}_{{\cal H}_{eff}(x-y)}, {\cal H}_{eff}(0) \Big \} | Q \rangle
\nonumber \\[3mm]
& = \frac{1}{2}  (2 \pi)^4  \delta^{(4)} (0) \,  \langle Q|\,  i\, \int d^4 x \,{\rm T}\, \Big \{  {\cal H}_{eff}(x), {\cal H}_{eff}(0) \Big \} | Q \rangle\,,
\end{align}\\
where in the last step we have performed the change of variable $x^\mu-y^\mu \to x^\mu$ under the integration over $x^\mu$ and used that $\int d^4 y = (2 \pi)^4 \delta^{(4)}(0)$. From Eqs.~(\ref{eq:Tfi}), (\ref{eq:Gamma-A-opt-th}), we finally obtain that\\
\begin{equation}
\Gamma(Q) = \frac{1}{2 m_Q}\, {\rm Im} \,\langle Q| {\cal T}| Q\rangle\,,
\label{eq:Gamma-Q-opt-th}
\end{equation}\\
with\\
\begin{equation}
{\cal T} = i \int d^4x 
\,  {\rm T} \Big \{ {\cal H} _{eff} (x) \, ,
 {\cal H} _{eff} (0)  \Big \} \,.
\label{eq:cal-T-opt-th}
\end{equation}\\
We can interpret Eqs.~(\ref{eq:Gamma-Q-opt-th}), (\ref{eq:cal-T-opt-th}), as the statement that, due to the optical theorem, the total decay width of $Q$ 
is proportional to
the amplitude for the process $Q \to X \to Q$, 
describing
the forward scattering of $Q$ via the production and annihilation of all the possible intermediate states $X$. This corresponds to computing the imaginary part of the time ordered product of the Hamiltonian operator evaluated at two different space-time points, namely the non local operator ${\cal T}$, and to determining its expectation value between external $|Q\rangle$ states.

An analogous description for the decay of a hadronic state is plagued by the presence of the non perturbative QCD effects  responsible for the confinement dynamics. However, it was first proposed by Shifman and Voloshin in Ref.~\cite{Shifman:1984wx}, that the inclusive decay width of a heavy hadron, in their specific case a charmed meson, in the assumption of an infinitely heavy constituent quark, could be obtained using the partonic description, and hence by computing the probability for the free heavy quark to decay into all the lighter fermions. Following their formulation, the total decay width of a heavy hadron $H_Q$ can be expressed as the imaginary part of the non local operator ${\cal T}$ in Eq.~(\ref{eq:cal-T-opt-th}), evaluated between external hadronic states, namely  \\
\begin{equation}
\Gamma(H_Q) = \frac{1}{2 m_{H_Q}} {\rm Im} \, \langle H_Q| {\cal T}| H_Q \rangle\,.
\label{eq:Gamma-HQ}
\end{equation}\\
When $m_Q \to \infty$, the hadronic state and mass coincide with those of the heavy quark and Eq.~(\ref{eq:Gamma-HQ}) becomes $\Gamma(H_Q) = \Gamma(Q)$. However, this approximation is not sufficient for phenomenological applications and corrections to this limit must be systematically included. The HQE provides a theoretical framework to compute $\Gamma(H_Q)$ in Eq.~(\ref{eq:cal-T-opt-th}), in the case of large, but finite, heavy quark mass $m_Q$. The fundamental assumption is that, inside a heavy hadron, the heavy quark, propagating in the soft background generated by the non perturbative gluon field, interacts with the light degrees of freedom exchanging momenta of the order of $\Lambda_{QCD}$, much smaller than $m_Q$, meaning that there is a large part in the heavy quark momentum, which is proportional to the heavy quark mass and that can be extracted by means of a  field redefinition, see e.g.\ Ref.~\cite{Shifman:1994yf},  i.e.\\
\begin{equation}
Q(x) = e^{- i m_Q v \cdot x} \, Q_v(x)\,,
\label{eq:Qx-HQE}
\end{equation}\\
where $v^\mu$ denotes the hadron velocity. It is worth emphasising that despite the strong analogy, $Q_v(x)$ in Eq.~(\ref{eq:Qx-HQE}) constitutes a rescaled four-component QCD field and not the two-component non relativistic field introduced in the context of the HQET, cf.\ Section~\ref{sec:HQET}, more details on the difference between the two methods can be found e.g.\ in Ref.~\cite{Shifman:1995dn}. From Eq.~(\ref{eq:Qx-HQE}) it then follows that\\
\begin{equation}
i D_\mu \, Q(x) = e^{-i m_Q v \cdot x} \Big( m_Q v_\mu + i D_\mu \Big) Q_v(x)\,,
\label{eq:DmuQ}
\end{equation} \\
which combined with the equation of motion $(i \slashed D - m_Q) Q(x) = 0$, gives\\
\begin{equation}
P_+ Q_v(x) = Q_v(x) - \frac{i \slashed D}{2 m_Q} Q_v(x)\,,
\label{eq:p+Qv}
\end{equation}\\
and
\begin{equation}
P_- Q_v(x) = \frac{i \slashed D}{2 m_Q} Q_v(x)\,,
\label{eq:p-Qv}
\end{equation}\\
with the projector operators $P_\pm$ defined as in Eq.~(\ref{eq:projectors-Ppm}). Moreover, acting with $P_+$ on both sides of Eq.~(\ref{eq:p-Qv}) and using that $P_+ i \slashed D = i \slashed D P_- +( i v \cdot D) $, yields\\
\begin{equation}
(i v\cdot D) Q_v(x) = - \frac{1}{2 m_Q} i \slashed D i \slashed D Q_v(x)\,.
\label{eq:vdotD-Qv}
\end{equation}\\
The relations in Eqs.~(\ref{eq:Qx-HQE})-(\ref{eq:vdotD-Qv}), allow to construct a systematic procedure to compute the inclusive decay width $\Gamma(H_Q)$. Specifically, Eq.~(\ref{eq:Gamma-HQ}) can be evaluated in two steps, see e.g.\ Ref.~\cite{Bigi:1992ne}. First, by taking into account the soft interaction with the background gluon field as well as with the light spectator quarks, the imaginary part of the non local second order operator ${\cal T}$ is expanded in a series of local operators ${\cal O}_d$ with increasing dimension $d$, where the corresponding coefficients $c_d$ are suppressed by $d-3$ powers of the heavy quark mass $m_Q$, namely\\
\begin{equation}
{\rm Im} \, {\cal T} = \sum_d \, c_d \, \frac{{\cal O}_d}{m_Q^{d-3}}\,.
\label{eq:OPE-Im-T}
\end{equation}\\
In general all possible Lorentz and gauge invariant operators, bilinear in the heavy quark field, can appear on the r.h.s\ of Eq.~(\ref{eq:OPE-Im-T}) and for large values of $m_Q$, it is sufficient to consider only those of lowest dimension. These respectively are $\bar Q Q$, $(1/2) \bar Q \sigma_{\mu \nu} G^{\mu \nu}Q $, $\bar Q \Gamma q \bar q \Gamma Q$, etc., where $\Gamma$ denotes a combination of gamma matrices and colour matrices. The corresponding coefficients in Eq.~(\ref{eq:OPE-Im-T}) are extracted by taking the matrix element of both sides of Eq.~(\ref{eq:OPE-Im-T}) between external quark and gluon states, see e.g.\  Ref.~\cite{Shifman:1995dn}. Notice that there is no dimension-four operator, since $\bar Q \slashed D Q$ can be reduced to $\bar Q Q$ by means of the equation of motion for $Q$ \cite{Bigi:1992ne}.
The series in Eq.~(\ref{eq:OPE-Im-T}) starts at dimension-three with the operator $ \bar Q Q = \bar Q_v Q_v $. This is not suppressed by the heavy quark mass and at leading order in $1/m_Q$ it reproduces the partonic result in Eq.~(\ref{eq:Gamma-Q-opt-th}). In fact, $\bar Q Q$ receives non perturbative corrections from higher order operators, see Ref.~\cite{Bigi:1992su}. The proof starts with the following identity\\
\begin{equation}
\bar Q Q = \bar Q_v \slashed v Q_v + 2  \bar Q_v P_- Q_v  =  \bar Q_v \slashed v Q_v + 2  \bar Q_v P_- P_-  Q_v\,,
\end{equation}\\
which, using Eq.~(\ref{eq:p-Qv}) together with $  \bar Q_v P_- = \bar Q_v (- i \overset{\leftarrow}{\slashed D} )/2 m_Q$, leads to\\
\begin{equation}
\bar Q Q=  \bar Q_v \slashed v Q_v - 2 \bar Q_v \frac{i \overset{\leftarrow}{\slashed D}}{2 m_Q} \frac{i \overset{\rightarrow}{ \slashed D}}{2 m_Q} Q_v = \bar Q_v \slashed v Q_v + \bar Q_v  \frac{(i \slashed D)^2}{2 m_Q^2} Q_v + {\rm total \, derivative}\,,
\label{eq:barQvQv-exp-BUV}
\end{equation}\\
where the contribution of the total derivative can be neglected since, in forward matrix elements with zero momentum transfer, it vanishes, see e.g.\ Ref.~\cite{Bigi:1994ga}. The first operator on the r.h.s.\ of Eq.~(\ref{eq:barQvQv-exp-BUV}) is the generator of the conserved charge associated to the heavy flavour $Q$, its matrix element between external hadronic states is one, up to a normalisation factor \cite{Bigi:1992su}. Note that in Eq.~(\ref{eq:barQvQv-exp-BUV}) there are no linear terms in $1/m_Q$. These would be generated by operators of dimension-four, however, containing only one covariant derivative, they would either correspond to a total derivate, which does not contribute, as stated above, or to a derivative acting on the heavy quark field, which, by means of the equation of motion Eq.~(\ref{eq:vdotD-Qv}), is proportional to operators of higher order. The absence in the HQE of linear terms in $1/m_Q$ was first discussed by Chay, Georgi and Grinstein in Ref.~\cite{Chay:1990da}, and subsequently by Bigi, Uraltsev, and Vainshtein in Ref.~\cite{Bigi:1992su} and is known as CGG/BUV theorem, see Ref.~\cite{Shifman:1995dn} \footnote{In the framework of the HQET, the absence of linear terms in $1/m_Q$ to the forward matrix element of a heavy quark current is known as Luke's theorem \cite{Luke:1990eg}, see e.g.\ Ref.~\cite{Mannel:1995dr}.}. 
First corrections to the infinite mass limit arise at dimension-five, and correspond to operators with two covariant derivatives acting on the heavy quark field. We can identify them from Eq.~(\ref{eq:barQvQv-exp-BUV}), i.e.\\
\begin{equation}
 \bar Q_v  \frac{(i \slashed D)^2}{2 m_Q^2} Q_v = \frac{1}{2 m_Q^2} \bar Q_v (i D_\mu) (i D^\mu) Q_v + \frac{1}{2 m_Q^2} \bar Q_v (i D_\mu) (i D_\nu) (- i \sigma^{\mu \nu}) Q_v\,,
\end{equation}\\
where we have used that $\gamma^\mu \gamma^\nu = \{ \gamma^\mu, \gamma^\nu\}/2 + [\gamma^\mu, \gamma^\nu]/2$. The kinetic and chromo-magnetic operators are then defined respectively as\\
\begin{align}
{\cal O}_{kin} &= \bar Q_v (i D_\mu) (i D^\mu) Q_v \,,  
\label{eq:O-kin}
\\[3mm]
{\cal O}_{mag}& = \bar Q_v (i D_\mu) (i D_\nu) (- i \sigma^{\mu \nu}) Q_v\,.
\label{eq:O-magn}
\end{align}\\
At dimension-six, operators generated from the action of three covariant derivatives, but also four-quark operators, contribute. The former correspond to the spin-orbit and Darwin operators, defined respectively as, see e.g.\ Ref.~\cite{Dassinger:2006md}\\
\begin{align}
{\cal O}_{LS} &= \bar Q_v (i D_\mu)(i v \cdot D) (i D_\nu) (- i \sigma^{\mu \nu}) Q_v  \,,
\label{eq:O-LS}
\\[3mm]
{\cal O}_{\rho_D} &= \bar Q_v (i D_\mu) (i v \cdot D)(i D_\nu) Q_v \,.
\label{eq:O-Darwin}
\end{align}\\
Four-quark operators have the schematic form $\bar Q \Gamma q \bar q \Gamma Q$, where $\Gamma$ refers to a combination of gamma matrices as well as colour matrices, compatible with the $V-A$ structure of the effective Hamiltonian, and $q$ denotes a light spectator quark. It is worth mentioning that, using the equation of motion for the gluon field $D_\mu G^{\mu \nu  a} = - g_s \sum_q \bar q \gamma^\nu t^a q $, the Darwin operator can be expressed, at leading order in $1/m_Q$, in terms of four-quark operators, see e.g.\ Ref.~\cite{Bigi:1992su} and also Chapter~\ref{ch:pheno}. Finally, operators of higher dimension are built by further expanding in the number of covariant derivatives and of light quark fields. 

Having constructed the series in Eq.~(\ref{eq:OPE-Im-T}) up to the desired order in $1/m_Q$, the second step in the calculation of Eq.~(\ref{eq:Gamma-HQ}) is to evaluate the matrix element of the local operators obtained, between external hadronic states, see Ref.~\cite{Bigi:1992ne}. These encode the large distance dynamics responsible for the hadronic structure and require non perturbative methods like Lattice QCD \cite{Wilson:1974sk} or QCD Sum Rules \cite{Shifman:1978bx, Shifman:1978by} to be determined. In some cases they can also be extracted performing fits to the experimental data, see e.g.\ Ref.~\cite{Alberti_2015}. Alternatively, using the framework of the HQET, the dependence of the heavy quark field and of the hadronic state on the heavy quark mass, see Section~\ref{sec:HQET}, can be further factored out, the corresponding matrix elements are then expanded in inverse powers of $m_Q$ and expressed in terms of a minimal set of elementary parameters, which must again be determined by means of the non perturbative methods mentioned above, or in some cases via spectroscopy relations \cite{Neubert:1993mb}. For the matrix element of the dimension-three operator in Eq.~(\ref{eq:barQvQv-exp-BUV}), the HQET expansion has the following form \cite{Bigi:1992su, Dassinger:2006md}  \\
\begin{equation}
\frac{ \langle H_Q| \bar Q Q |H_Q \rangle}{2 m_{H_Q}} = 1 - \frac{\mu_{\pi}^2(H_Q) - \mu_{G}^2(H_Q)}{2 m_Q^2} + {\cal O}\left( \frac{1}{m_Q^5} \right)\,,
\label{eq:barQ-Q-expansion-ME}
\end{equation}\\
where the non perturbative parameters $\mu_\pi^2$, $\mu_G^2$ are related to the expectation value of the operators in Eqs.~(\ref{eq:O-kin}), (\ref{eq:O-magn}), as\\
\begin{equation}
2 m_{H_Q}\, \mu_\pi^2(H_Q)  = - \langle H_Q|{\cal O}_{kin}|H_Q\rangle\,, \quad
2 m_{H_Q} \, \mu_G^2 (H_Q) = \langle H_Q|{\cal O}_{mag}|H_Q\rangle\,.
\label{eq:dim-5-ME-parameters}
\end{equation}\\
Similarly, the matrix elements of the spin-orbit and Darwin operators, are expressed in terms of the two non perturbative parameters $\rho_{LS}^3$, $\rho_D^3$, i.e.\\
\begin{equation}
2 m_{H_Q}\, \rho_{LS}^3(H_Q)  = - \langle H_Q|{\cal O}_{LS}|H_Q\rangle\,, \quad
2 m_{H_Q} \, \rho_D^3 (H_Q) = \langle H_Q|{\cal O}_{D}|H_Q\rangle\,.
\label{eq:dim-6-ME-parameters}
\end{equation}\\
In the case of four-quark operators, a simple way to estimate the matrix elements between external mesons states, is the so called `vacuum insertion approximation' (VIA), corresponding to the assumption that the matrix elements can be saturated by the vacuum intermediate state, see e.g.\ Ref.~\cite{Shifman:1995dn}, namely \\
\begin{equation}
\langle H_Q | \bar Q \Gamma q \bar q \Gamma Q | H_Q \rangle \overset{\rm VIA}{=} \langle H_Q| \Bar Q  \Gamma q| 0 \rangle \langle 0| \bar q \Gamma Q |  H_Q \rangle \,.
\label{eq:VIA}
\end{equation}\\
Set for definiteness $\Gamma = \gamma^\mu \gamma_5$ and consider $H_Q$ to be e.g.\ a pseudoscalar $B$ meson. It follows that the matrix element on the l.h.s.\ of Eq.~(\ref{eq:VIA}) is parametrised in terms of the $B$ meson mass $m_B$ and decay constant $f_{B}$, where the latter is defined as 
\begin{equation}
\langle 0 | \bar q \gamma^\mu \gamma_5 b | B \rangle =  i f_{B} p^\mu_{B}\,,
\end{equation}
here $p^\mu_{B}$ denotes the meson four-momentum with $p_B^2 = m_B^2$. By taking into account that the matrix element of the corresponding vector current vanishes due to parity conservation in QCD, see e.g.\ the textbook \cite{ellis_stirling_webber_1996}, we obtain that \\
\begin{equation}
\langle B | \bar b \gamma_\mu (1- \gamma_5) q|0\rangle \langle 0 | \bar q \gamma^\mu(1- \gamma_5) b | B \rangle =  f_B^2\, m_B^2\,.
\end{equation}\\
\begin{figure}
\centering
\includegraphics[scale=0.34]{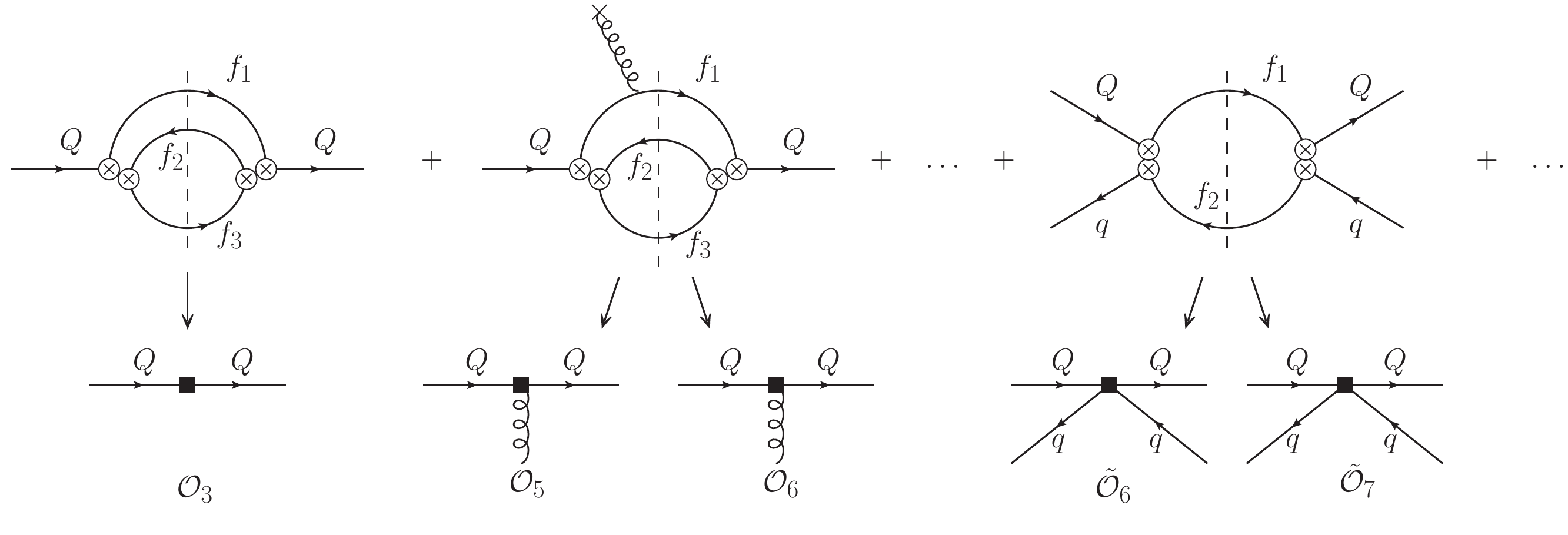}
\caption{Schematic representation of the HQE in Eq.~(\ref{eq:HQE}). The imaginary part of the double insertion of the effective Hamiltonian (top line), is matched into a series of local operators (bottom line).}
\label{fig:HQE-exp}
\end{figure}
Finally, the construction of the HQE leads to the following expansion for $\Gamma(H_Q)$ \\
\begin{equation}
\Gamma(H_Q) = 
\Gamma_3  +
\Gamma_5 \frac{\langle {\cal O}_5 \rangle}{m_Q^2} + 
\Gamma_6 \frac{\langle {\cal O}_6 \rangle}{m_Q^3} + \ldots
 + 16 \pi^2 
\left[ 
  \tilde{\Gamma}_6 \frac{\langle \tilde{\mathcal{O}}_6 \rangle}{m_Q^3} 
+ \tilde{\Gamma}_7 \frac{\langle \tilde{\mathcal{O}}_7 \rangle}{m_Q^4} + \dots 
\right]\,,
\label{eq:HQE}
\end{equation}\\
schematically sketched in Figure~\ref{fig:HQE-exp}. Eq.~(\ref{eq:HQE}) shows that, by exploiting the large hierarchy $m_Q \gg \Lambda_{QCD}$, the total decay width of a heavy hadron can be systematically computed as a series in inverse powers of the heavy quark mass. 
The lowest order contributions describe the effect of two- and four-quark operators and, in Eq.~(\ref{eq:HQE}), the latter are labelled  with a tilde. Moreover, from the diagrammatic representation in Figure~\ref{fig:HQE-exp}, we see that while the contribution of four-quark operators corresponds to one-loop diagrams at LO-QCD, two-quark operators are generated only at two-loop, again at LO-QCD, and this mismatch is reflected in the presence of the enhancement factor of $16 \pi^2$ in front of the square brackets in Eq.~(\ref{eq:HQE}) \cite{Khoze:1983yp, Shifman:1986mx, Neubert:1996we, Uraltsev:1996ta}. 
As already stressed in Section~\ref{sec:Heff}, the essential feature of the OPE is the separation between 
short- and long-distance effects. Namely, the non perturbative dynamics is
absorbed in the matrix element of local operators, whereas
the short distance contribution is encoded in the corresponding coefficients. The latter, in fact, obey the perturbation expansion\\
\begin{equation}
    \Gamma_d =  \Gamma_d^{(0)} + \left( \frac{\alpha_s}{4 \pi} \right) \Gamma_d^{(1)} + \left( \frac{\alpha_s}{4 \pi} \right)^2 \Gamma_d^{(2)} + \ldots  \, ,
\label{eq:pert-Gamma}
\end{equation}\\
and can be computed within standard perturbation theory. Extensive work has been put in this direction, here a brief summary of the current status. The complete calculation of $\Gamma_3$ up to NLO-QCD corrections has been obtained in Refs.~\cite{Hokim:1983yt,Altarelli:1991dx,Voloshin:1994sn,Bagan:1994zd,Bagan:1995yf,Lenz:1997aa,Lenz:1998qp,Krinner:2013cja}. Currently, also NNLO-QCD corrections are known for semileptonic decays
\cite{Czarnecki:1997hc,Czarnecki:1998kt,vanRitbergen:1999gs,Melnikov:2008qs,Pak:2008cp,Pak:2008qt,Dowling:2008ap,Bonciani:2008wf,Biswas:2009rb,Brucherseifer:2013cu}, while for non-leptonic decays, these have only been determined, for massless final quarks and in full QCD, i.e.\ without using the effective Hamiltonian, in Ref.~\cite{Czarnecki:2005vr}.
$\Gamma_5$ has been computed at LO-QCD for both non-leptonic and semileptonic decays
\cite{Bigi:1992su,Blok:1992hw,Blok:1992he,Bigi:1992ne}, for the latter even NLO-QCD corrections
are available \cite{Alberti:2013kxa,Mannel:2014xza,Mannel:2015jka}. For semileptonic decays, 
$\Gamma_6$~was first computed at LO-QCD in Ref.~\cite{Gremm:1996df} 
and recently the NLO-QCD corrections were determined in Ref.~\cite{Mannel:2019qel}, while the LO-QCD computation 
for non-leptonic decays has been performed for the first time in Refs.~\cite{Lenz:2020oce, Mannel:2020fts, Moreno:2020rmk } for the $b$-system and in Ref.~\cite{King:2021xqp} for $c$-quark decays, see also Section~\ref{sec:pheno1}. Finally 
$\tilde{\Gamma}_6$ is known at NLO-QCD 
\cite{Beneke:2002rj,Franco:2002fc, Lenz:2013aua}, while $\tilde \Gamma_7$ only
at LO-QCD \cite{Gabbiani:2004tp}.
\\[3mm]
We conclude by emphasising that the construction of the HQE is based on the validity of the so called quark-hadron duality (QHD). This refers to the assumption that the inclusive rate determined by summing over all the exclusive hadronic decay channels, and the one predicted by the HQE, are dual to each other, in the sense that they provide two valid representations of the same quantity, using respectively the hadron-level and the quark-level description. However, violations of QHD constitute a systematic uncertainty of the HQE, and one simple argument is the fact that, by computing the total decay width in terms of a series expansion in powers of $\Lambda_{QCD}/m_Q$, any term of the type $\exp(-m_Q/\Lambda_{QCD}) \sin(m_Q/\Lambda_{QCD})$, would be systematically  neglected, since $\exp(-1/x)$ is non-analytic and its expansion around $x = 0$, yields identically zero. Despite deviations of HQD cannot be excluded, there is no experimental evidence so far for sizeable violations that might compromise the applicably of the HQE. For a detailed discussion of QHD see e.g.\  Refs.~\cite{Blok:1997hs, Chibisov:1996wf, Shifman:2000jv} .


\chapter{Practical Calculations within the HQE}
\label{ch:HQE-ex}
With the theoretical background discussed in Chapter~\ref{ch:theory-bg}, we can now show the explicit computation of the lowest-order contributions to the total decay width of a heavy hadron, Eq.~(\ref{eq:Gamma-HQ}). For definiteness we assume $H_Q$ to be a $B$ meson with $B = \{ \bar B_d, B^- ,\bar B_s \}$, i.e.\ we limit ourselves to systems containing a heavy $b$ quark and a light antiquark $\bar q = \{ \bar d, \bar u, \bar s \}$, without discussing the case of the $B_c$ meson \footnote{In this case the HQE must be properly generalised in order to include a double expansion in inverse powers of the bottom as well as the charm quark mass, see e.g.\  Refs.~\cite{Beneke:1996xe, Aebischer:2021ilm, Aebischer:2021eio }.}.  We stress, however, that the expressions obtained, taking into account the appropriate replacements, e.g.\ of the CKM factors and masses, can be also applied to the study of the charm system and of the $b$-baryons with one heavy quark. Furthermore, we emphasise that all the calculations presented are only at LO-QCD. 

The total decay width in Eq.~(\ref{eq:HQE}) can be decomposed in the sum of semileptonic and non-leptonic  widths, namely\\
\begin{equation}
\Gamma(B) = \Gamma^{\rm  (SL)}(B) + \Gamma^{\rm  (NL)} (B) \,.
\end{equation}\\
For simplicity, in the following, we consider only the computation of $\Gamma^{\rm NL}(B)$, again, the corresponding results for the semileptonic case can be easily derived by setting $N_c = 1$, $C_1 =1$ and $C_2 = 0$. 
According to Eq.~(\ref{eq:Gamma-HQ}), the total non-leptonic decay width of a $B$ meson is induced at the quark level by the flavour-changing transition $b \to q_1 \bar q_2 q_3$, with $q_1,q_2 =\{ u, c \}$ and $q_3 = \{ d, s \}$, described, at the renormalisation scale $\mu_1 \sim m_b$,  by the effective weak Hamiltonian ${\cal H}_{eff}(x)$, see Section~\ref{sec:Heff}, i.e. \\
\begin{equation}
{\cal H}_{\rm eff} (x) =  \frac{G_F}{\sqrt 2} V_{q_1 b}^* V_{q_2  q_3}  
\Big[ C_1 \,  Q_1 (x) + C_2 \, Q_2 (x) \Big] + {\rm h.c.}\,.
\label{eq:H-eff-b}
\end{equation}\\
The colour-singlet and colour-rearranged local four-quark operators $Q_1,(x)$, $ Q_2(x)$, in Eq.~(\ref{eq:H-eff-b}), are respectively given by \\
\begin{align}
Q_1(x) =  \Big(\bar q_1^i (x) \Gamma_\mu b^i(x) \Big) \Big(\bar q_3^{j}(x) \Gamma^\mu q_2^j(x) \Big)\, ,
\label{eq:O1-operator}
\\[3mm]
Q_2(x) =  \Big(\bar q_1^i (x) \Gamma_\mu b^{j}(x)\Big) \Big(\bar q_3^{j}(x) \Gamma^\mu q_2^i(x)\Big)\,,
\label{eq:O2-operator}
\end{align}\\
where $\Gamma_\mu = \gamma_\mu (1-\gamma_5)$. In Eq.~(\ref{eq:H-eff-b}), $C_1(\mu_1)$, $ C_2(\mu_1)$, define the corresponding Wilson coefficients, their scale dependence is often omitted in order to simplify the notation. Note also that, being interested in discussing only the general structure of the computation, in Eq.~(\ref{eq:H-eff-b}) we have neglected the contribution of the penguin operators, however the expressions can be easily generalised to include them. 
Substituting Eq.~(\ref{eq:H-eff-b}) into Eq.~(\ref{eq:cal-T-opt-th}), leads to the following decomposition for the non local second-order operator ${\cal T}$ i.e.\\\
\begin{equation}
{\cal T}^{(q_1 \bar q_2 q_3)}   = C_1^2 \,  {\cal T}_{11}^{(q_1 \bar q_2 q_3)} + \,  2 \,C_1 C_2 \,  {\cal T}_{12}^{(q_1 \bar q_2 q_3)}  +C_2^2 \, {\cal T}_{22}^{(q_1 \bar q_2 q_3)}
\, , 
\label{eq:Tsecond}
\end{equation}\\
here the superscript $(q_1 \bar q_2 q_3)$ refers to the specific decay mode of the $b$ quark, which for the sake of a more compact notation will be sometimes dropped, and \\
\begin{equation}
{\cal T}_{mn}^{(q_1 \bar q_2 q_3)}   =
\frac{G_F^2 |V_{q_1 b} |^2 |V_{q_3 q_2}|^2}{2}
\,  i \int d^4x 
\, 
{\rm T} 
\Big\{ Q_{m}  (x) \, ,  Q^\dagger_{n} (0)  \Big\} + (x \leftrightarrow 0)
\, .
\label{eq:T_mn}
\end{equation}\\
In Eq.~(\ref{eq:T_mn}), the corresponding term due to the exchange of coordinates $(x \leftrightarrow 0)$ must be considered separately, and cannot be in general reduced to a symmetry factor of 2, since the computation of power corrections will be performed in the FS gauge, which, it is worth remarking, explicitly breaks the translation invariance of the propagator, see Section~\ref{sec:FS}. 
\\[3mm]
The time-ordered product in Eq.~(\ref{eq:T_mn}) is written, by means of the Wick's theorem \cite{Wick:1950ee}, as a linear combination of terms where only normal products, normal products and contractions and only contractions of fields appear, see e.g.\ the textbook~\cite{Bogolyubov:1959bfo}. The lowest-order contributions in the HQE correspond to two- and four-quark operators and are generated respectively from the contraction of three- and two-pairs of light quark fields while leaving the $b$-quark fields uncontracted. The first case is discussed in Section~\ref{sec:2q-contr}, the second in Section~\ref{sec:4q-contr}. 
\section{Contribution of two-quark operators}
\label{sec:2q-contr}
Following Refs.~\cite{Blok:1992hw, Blok:1992he}, for a straightforward treatment of colour in the computation of power corrections due to the expansion of the quark propagator, it is convenient to perform in ${\cal T}_{12}$, Eq.~(\ref{eq:Tsecond}), the change of basis\\
\begin{equation}
\Big\{Q_1(x), Q_2(x) \Big\} \to \Big\{Q_1(x), Q_3(x) \Big\}\,,
\label{eq:basis-change}
\end{equation}\\
where $Q_3(x)$ denotes the colour-octet operator \footnote{Note that in Refs.~\cite{Blok:1992hw, Blok:1992he} the colour octet-operator is denoted by $\tilde Q_1$. }\\
\begin{equation}
 Q_3(x) = \Big(\bar q_1^i(x) \Gamma_\mu \, t^a_{ij} \, b^j(x)\Big)\Big(\bar q_3^{l}(x) \Gamma^\mu \, t^a_{lm}\, q_2^m(x)\Big)\,.
 \label{eq:Q3-def}
\end{equation} \\
The relation between $Q_2(x)$, and $Q_3(x)$, in Eq.~(\ref{eq:basis-change}) is obtained by taking into account the completeness property of the $SU(3)_c$ generators, i.e. \\
\begin{equation}
t^a_{ij} t^a_{lm} = \frac12 \left( \delta_{im}\delta_{jl} - \frac{1}{N_c} \delta_{ij}\delta_{lm} \right)\,.
\label{eq:Fiez-id-colour-matr}
\end{equation}\\
Substituting $Q_2(x) = (1/N_c) Q_1(x) + 2 Q_3(x)$ in ${\cal T}_{12}$ in Eq.~(\ref{eq:Tsecond}), and considering only the contribution of two-quark operators, leads to the general decomposition\\
\begin{align}
{\cal T}^{(2q)} = C_1^2 \,  {\cal T}^{ (2q)}_{11}+ \,  2 \,C_1 C_2 \, \left(\frac{1}{N_c} {\cal T}^{ (2q)}_{11} +  2 \, {\cal T}^{ (2q)}_{13} \right)  +C_2^2 \, {\cal T}^{ (2q)}_{22}\,,
\label{eq:T2q}
\end{align}\\
where the superscript ${(2q)}$ indicates that all pairs of light quarks fields in the time-ordered product in Eq.~(\ref{eq:T_mn}) have been contracted and replaced with the corresponding propagators, namely, without specifying the colour structure \footnote{ We recall that we often omit to explicitly indicate the dependence on the space-time coordinate for fields evaluated at the origin, so unless otherwise stated, we assume $q= q(0)$, for generic $q(x)$.}\\
\begin{align}
{\rm T}  \Big\{ & \bar q_1(x)  \Gamma_\mu b(x) \bar q_3(x)\Gamma^\mu  q_2(x),  \bar b  \Gamma_\nu q_1\bar q_2 \Gamma^\nu  q_3 \Big\}  
 = : 
 \wick{
 \c1{\bar q}_1(x) \Gamma_\mu b(x) \c3{\bar q_3}(x)  \Gamma^\mu  \c2{q}_2(x) \bar b \Gamma_\nu \c1{q}_1 \c2{\bar q}_2 \Gamma^\nu \c3{q_3}:
  }
\,,
 \label{eq:Wick-2q}
\end{align}\\
here the two colons denote the normal product. Eq.~(\ref{eq:Wick-2q}) can be schematically visualised in Figure~\ref{fig:dim3}. Taking into account Eq.~(\ref{eq:Wick-2q}), it follows that the expressions of the non local operators ${\cal T}^{\rm (2q)}_{mn}$ in Eq.~(\ref{eq:T2q}), are respectively given by\\
\begin{align}
{\cal T}_{11}^{ (2q)} & =  - \,  \frac{G_F^2}{2}  |V_{q_1b}|^2 |V_{q_3q_2}|^2\,  i \int d^4 x \,\, \bar b^j \gamma_\nu (1-\gamma_5) \, i S^{( q_1)}_{jk} (0, x) \gamma_\mu (1-\gamma_5) b^k (x)
\nonumber \\[3mm]
& \times   {\rm Tr} \Biggl[\gamma^\nu (1-\gamma_5) i S^{( q_3)}_{lm} (0, x) \gamma^\mu (1-\gamma_5) i S^{( q_2)}_{ml} (x,0) \Biggr]  + (x \leftrightarrow 0)
\, ,
\label{eq:T2q-11}
\end{align}\\
\begin{align}
{\cal T}_{13}^{ (2q)} & =  - \,  \frac{G_F^2}{2}  |V_{q_1b}|^2 |V_{q_3q_2}|^2 \, 
 i \int d^4 x \, \,  \bar b^j \gamma_\nu (1-\gamma_5) t^a_{jl} \, i S^{( q_1)}_{lk} (0, x) \gamma_\mu (1-\gamma_5) b^k (x) 
\nonumber \\[3mm]
& \times    {\rm Tr} \Biggl[\gamma^\nu (1-\gamma_5) t^a_{mn} \, i S^{( q_3)}_{nr} (0, x) \gamma^\mu (1-\gamma_5) i S^{( q_2)}_{rm} (x,0) \Biggr]  + (x \leftrightarrow 0)
\, ,
\label{eq:T2q-13}
\end{align}\\
and\\
\begin{align}
{\cal T}_{22}^{ (2q)}& =  - \,  \frac{G_F^2}{2}  |V_{q_1b}|^2 |V_{q_3q_2}|^2 \, 
  i \int d^4 x \, \,  \bar b^j  \gamma_\nu (1-\gamma_5) \, i S^{( q_1)}_{lm} (0, x) \gamma_\mu (1-\gamma_5) b^k (x)
\nonumber \\[3mm]
& \times   {\rm Tr} \Biggl[\gamma^\nu (1-\gamma_5) i S^{( q_3)}_{jk} (0, x) \gamma^\mu (1-\gamma_5) i S^{( q_2)}_{lm} (x,0) \Biggr]  + (x \leftrightarrow 0)
\, ,
\label{eq:T2q-22}
\end{align}\\
where the minus sign and the trace over spinor indices, derive from the fermion loop.
Note that applying the Fierz identity \\
\begin{equation}
\Big(\bar q_1^i \Gamma_\mu q_2^i \Big) \Big(\bar q_3^j \Gamma^\mu q_4^j \Big) = \Big(\bar q_3^j \Gamma_\mu q_2^i \Big) \Big(\bar q_1^i \Gamma^\mu  q_4^j \Big) \,, 
\label{eq:Fierz-identity}
\end{equation}\\
to the four-quark operators in Eqs.~(\ref{eq:O1-operator}), (\ref{eq:O2-operator}), gives\\ 
\begin{equation}
Q_{1, 2} = Q_{2,1}^{(q_1 \leftrightarrow q_3)}\,.
\label{eq:Fierz-identity-2}
\end{equation} \\
meaning that the four-quark operator $Q_{1}$ is equal to $Q_2$ with the exchange $q_1 \leftrightarrow q_3$ and vice versa.
This important relation implies that in Eq.~(\ref{eq:T2q-22}) we can write\\
\begin{equation}
S^{(q_1)}_{lm, \, \alpha \beta} \, S^{(q_3)}_{jk, \, \gamma \delta} \to  S^{(q_1)}_{lm, \, \gamma \delta} \,  S^{(q_3)}_{jk, \, \alpha \beta} \,,
\end{equation} 
where the Greek letters denote spinor indices, and then that\\
\begin{equation}
{\cal T}^{(2q)}_{22} = {\cal T}^{{( 2q)} \, (q_1\leftrightarrow q_3 )}_{11} \,.
\label{eq:Fierz-13}
\end{equation}\\
By taking into account Eq.~(\ref{eq:Fierz-13}), the contribution of $Q_2 \otimes Q_2$ in Eq.~(\ref{eq:T_mn}) can be obtained from that of $Q_1 \otimes Q_1$ after performing the replacement $q_1 \leftrightarrow q_3$. 

The coefficients of the two-quark operators up to order $1/m_b^3$, are computed in detail in Chapter~\ref{ch:Darwin}, for generic non-leptonic decays of the $b$-quark and using the representation of the quark-propagator in momentum space, Eqs.~(\ref{eq:S1p}), (\ref{eq:S1p-tilde}). However, it is instructive to perform the same calculation also using the expression of the quark-propagator in coordinate space given in Eqs.~(\ref{eq:prop-coordinate-space}), (\ref{eq:prop-coordinate-space-y}). This is discussed in the next two sections, respectively for the case of dimension-three and dimension-five contributions, and for the single mode $b \to c \bar u d$.


\subsection{Computation of $\Gamma_3^{(c \bar u d)}$}
\begin{figure}
\centering
\includegraphics[scale = 0.5]{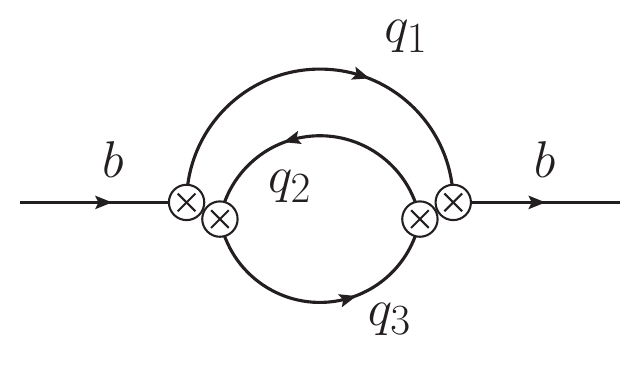}
\caption{Diagram describing the leading order contribution to the free $b$-quark decay.}
\label{fig:dim3}
\end{figure} 
The leading term in Eq.~(\ref{eq:HQE}) corresponds to the decay of a free $b$ quark, as shown in Figure~\ref{fig:dim3}. Neglecting the interaction with the background gluon field, see Section~\ref{sec:FS}, all the propagators in Eqs.~(\ref{eq:T2q-11})-(\ref{eq:T2q-22}), reduce to\\
\begin{equation}
S^{(q)}_{jk}(x,y) = S_0^{(q)}(x-y) \, \delta_{jk}\,, \qquad q = c,u,d\,,
\label{eq:S-jk-dim3}
\end{equation}\\
where $S_0^{(q)}(x-y)$ is the free-quark propagator defined in Eq.~(\ref{eq:free-prop-equation}), and a superscript has been introduced in order to distinguish between the different quarks consistently with the description of the $b \to c \bar u d$ decay. The presence of the Kronecker delta in Eq.~(\ref{eq:S-jk-dim3}) leads to immediate simplifications. In fact, by enforcing the trace of the $SU(3)_c$ generators $t^a$ in the square brackets of Eq.~(\ref{eq:T2q-13}), we readily obtain that ${\cal T}^{(2q)}_{13}$ must vanish at this order i.e.~\footnote{Note that the first correction to the free-quark propagator arises at order ${\cal O}(1/m_b^2)$.}\
\begin{equation}
{\cal T}^{(2q)}_{13} = {\cal O}\left(\frac{1}{m_b^2} \right) \,.
\label{eq:T13-d3}
\end{equation}\\
Moreover, it follows that Eqs.~(\ref{eq:T2q-11}), (\ref{eq:T2q-22}), exactly coincide, so that ${\cal T}^{(2q)}_{11}$ and ${\cal T}^{(2q)}_{22}$ are equal up to higher order corrections, namely\\
\begin{align}
{\cal T}^{(2q)}_{22}\Big|_{d=3} = {\cal T}^{(2q)}_{11}\Big|_{d=3} \, .
\label{eq:T22-equal-T11-d3}
\end{align}\\
Because of Eqs.~(\ref{eq:T13-d3}), (\ref{eq:T22-equal-T11-d3}), we need to compute only one expression, i.e. \\\
\begin{align}
{\cal T}^{(2q)}_{11}& =  -  \frac{  N_c G_F^2}{2}  |V_{q_1b}|^2 |V_{q_3q_2}|^2  \, \int d^4 x \, \, \bar b \, \gamma_\nu (1-\gamma_5)  S^{(c)}_0 (- x) \gamma_\mu (1-\gamma_5) b(x) 
\nonumber \\[3mm]
& \times    {\rm Tr} \Biggl[\gamma^\nu (1-\gamma_5) S^{(d)}_0 (- x) \gamma^\mu (1-\gamma_5) S^{(u)}_0 (x) \Biggr]  +  (x \leftrightarrow 0) + \ldots \,.
\label{eq:T2q-11-2}
\end{align}\\
In Eq.~(\ref{eq:T2q-11-2}), the colour factor is $\delta_{ii} = N_c$, and the ellipsis denote power suppressed contributions due to higher order terms in the quark-propagator. Notice also that the result in Eq.~(\ref{eq:T22-equal-T11-d3}), combined with Eq.~(\ref{eq:Fierz-13}), shows that Eq.~(\ref{eq:T2q-11-2}) must be a symmetric function under the exchange $c \leftrightarrow d$. 
A further simplification in the computation of the dimension-three contribution, derives from the translation invariance of the free-quark propagator, meaning that the integral in Eq.~(\ref{eq:T2q-11-2}) is also translation invariant, and that the second term on the r.h.s.\ of Eq.~(\ref{eq:T2q-11-2}), at this order, reduces to a symmetry factor of $2$. 

In Eq.~(\ref{eq:T2q-11-2}), the corresponding expressions for the three propagators follow from Eq.~(\ref{eq:S0y-free}). For the charm quark, it is \\
\begin{equation}
S_0^{(c)}(-x) = -\frac{i }{4 \pi^2} \frac{m_c^2 K_1(m_c \sqrt{-x^2})}{\sqrt{-x^2}} + \frac{ \slashed x}{4 \pi^2} \frac{ m_c^2  K_2(m_c \sqrt{- x^2})}{x^2} \,,
\label{eq:S0-c}
\end{equation}\\
and, due to the chiral structure of Eq.~(\ref{eq:T2q-11-2}), only the term in Eq.~(\ref{eq:S0-c}) proportional to an odd number of gamma matrices contributes. Furthermore, neglecting the mass of the up- and down-quarks, we need to consider the following limits of the Bessel functions \\
\begin{equation}
\lim_{m \to 0} m^2 K_1(m \sqrt{- x^2})  =  0 \, , \qquad
\lim_{m \to 0} m^2 K_2(m \sqrt{- x^2})  =  - \frac{2}{x^2} \, ,
\label{eq:lim_Bessel}
\end{equation}\\
from which we obtain that \\
\begin{equation}
S_0^{(u)}(x) =  \frac{\slashed x}{2 \pi^2 x^4} = - S_0^{(d)}(-x) \,.
\label{eq:S0-ud}
\end{equation}\\
Moreover, we recall that the coefficient of the dimension-three operator $\bar b b$ in the OPE in Eq.~(\ref{eq:OPE-Im-T}), is obtained by evaluating Im ${\cal T}^{(\rm 2q)}$ between external $b$ states with momentum $p_b$, hence we can make the following replacement in Eq.~(\ref{eq:T2q-11-2}) \footnote{This follows from $b(x) |b\rangle = \exp(- i p_b \cdot x) u_b(p_b)$ where $u_b(p_b) = b(0) |b\rangle$, is the $b$-quark spinor.}, see also Refs.~\cite{Georgi:1990ei, Shifman:1984wx} \\
\begin{equation}
b(x) \to e^{-i p_b \cdot x}\,  b(0)\,.
\label{eq:bx-d3}
\end{equation}\\
By substituting Eqs.~(\ref{eq:S0-c}), (\ref{eq:S0-ud}) and (\ref{eq:bx-d3}) into Eq.~(\ref{eq:T2q-11-2}), it is then straightforward to arrive at \\
\begin{align}
{\cal T}^{ (2q)} &  =  4\,  \frac{G_F^2}{ \pi^6} \, |V_{cb}|^2 |V_{ud}|^2 \, \Bigl( N_c\,  \big( C_1^2 +  C_2^2 \big) + 2 \, C_1 C_2 \Bigr) 
\nonumber 
\\[3mm]
&  \times    \bar b\, \Bigg\{
 \int d^4 x \, e^{- i p_b \cdot x} \,  \frac{m_c^2 K_2(m_c \sqrt{-x^2})}{x^{8}} \, \slashed x \Bigg\}   (1-\gamma_5) \, b  + \ldots \, .
\label{eq:T3}
\end{align}\\
The next step is to compute the imaginary part of the integral in the curly brackets of Eq.~(\ref{eq:T3}). This is easily obtained using the formalism presented in Ref.~\cite{Blok:1992hw}, based on the technique developed by Belyaev and Blok in Ref.~\cite{Belyaev:1985wza} for the computation of the spectral representation of integrals appearing in the Fourier transform of the product of several massless and one massive quark propagator, expressed in coordinate space, see Ref.~\cite{Blok:1992hw}. The result reads~\footnote{Note that the result in Ref.~\cite{Blok:1992hw} must be multiplied by $-\pi$. This might be due to a different convention used to define the discontinuity of a complex function.} \\
\begin{equation}
{\rm Im} \int d^4 x\, e^{- i p \cdot x}  \, \frac{K_\nu (m \sqrt{-x^2})}{(\sqrt{-x^2})^n} \, x^\mu = G_{\nu, n}(p^2) \, p^\mu\,,
\label{eq:Im-Bessel}
\end{equation}\\
with\\
\begin{equation}
G_{\nu, n}(p^2)  =  \frac{\pi^3}{2^{(n-4)} m^\nu} \frac{1}{\Gamma \left( \frac{n-\nu}{2}\right)}
 \sum\limits_{k=0}^{\frac{n-\nu}{2} -1 }  \, (-1)^{\frac{n-\nu}{2}-1-k}  \, C^k_{\frac{n-\nu}{2}-1} \, U^{n+\nu-4}_{\frac{n+\nu}{2}-k} (p^2)\,.
\label{eq:Bessel-integral}
\end{equation}\\
In Eq.~(\ref{eq:Bessel-integral}), the coefficients $C^m_n$ are given by\\
\begin{equation}
C^m_n =\binom{n}{m} =  \frac{n!}{m! \,(n-m)!}\,,
\end{equation} \\
while the functions $U_i^{2j}(p)$ are \\
\begin{equation}
U_i^{2 j} (p^2) = \frac{m^{2(i-1)}}{(j-1)!} \, \int \limits_{m^2}^{p^2} dz\, \frac{(p^2 - z)^{j-1}}{z^i} \,.
\end{equation}\\
Setting $\nu = 2$ and $n=8$ in Eq.~(\ref{eq:Im-Bessel}) yields\\
\begin{equation}
 {\rm Im } \, \Bigg\{
  \int \, d^4 x \, e^{- i p_b \cdot x} \,  \frac{m_c^2 K_2(m_c \sqrt{-x^2})}{x^{8}} \, \slashed x \Bigg\}  
 =   \frac{ \pi^3}{768}\,  p_b^4 \, \slashed p_b \left( 1 - 8 r - 12 r^2 \log r + 8 r^3 - r^4 \right) \, ,
 \label{eq:Im-int-d3}
\end{equation}\\
where we have introduced the dimensionless parameter $r = m_c^2/p_b^2$. Substituting Eq.~(\ref{eq:Im-int-d3}) into Eq.~(\ref{eq:T3}),  we then obtain \\
\begin{align}
{\rm Im} {\cal T}^{( 2q)} & = \frac{ G_F^2}{ 192 \pi^3} |V_{cb}|^2 |V_{ud}|^2  \Bigl( N_c  \, \Big( C_1^2 +  C_2^2 \Big) + 2 \, C_1 C_2 \Bigr) 
\nonumber \\[3mm]
&\times  p_b^4 \,  p_b^\mu  \left( 1 - 8 r - 12 r^2 \log r + 8 r^3 - r^4 \right) \bar b \gamma_\mu (1-\gamma_5) b + \ldots\,.
\label{eq:T3-1}
\end{align}\\
Finally, Eq.~(\ref{eq:T3-1}) depends on the heavy quark momentum $p_b$, which, it is worth remarking, admits the general parametrisation $p_b^\mu = m_b v^\mu + k^\mu$, here $v^\mu$ is the hadron velocity and the `residual' momentum $k^\mu$ describes the interaction of the heavy quark with the light degrees of freedom, see Chapter~\ref{ch:theory-bg}. As stated above, the dimension-three contribution to ${\rm Im}{\cal T}^{(\rm 2q)}$ corresponds to the decay of a free $b$-quark, meaning that all the interactions with the soft gluons and quarks can be neglected at this order. In this case the heavy quark momentum reduces to $p_b^\mu = m_b v^\mu + {\cal O}(1/m_b)$, and recalling the definition of the rescaled heavy quark field $b(x) = \exp(- i m_b v \cdot x) b_v(x)$, see Eq.~(\ref{eq:Qx-HQE}), satisfying $\slashed v b_v = b_v + {\cal O}(1/m_b)$, cf.\ Eq.~(\ref{eq:p-Qv}), from Eq.~(\ref{eq:Gamma-HQ}) and Eq.~(\ref{eq:barQ-Q-expansion-ME}), we arrive at the well known expression  
\begin{align}
\Gamma_3^{(c  \bar u d)} & =  \frac{ G_F^2 m_b^5}{ 192 \pi^3} |V_{cb}|^2 |V_{ud}|^2  \Bigl( N_c  \, \big( C_1^2 +  C_2^2 \big) + 2 \, C_1 C_2 \Bigr)  
  \left( 1 - 8 \rho - 12 \rho^2 \log \rho + 8 \rho^3 - \rho^4 \right) 
 \, ,
 \label{eq:G3-cud-final}
\end{align}\\
where $\rho = m_c^2/m_b^2$ is a dimensionless mass parameter and we have taken into account that the contribution of the axial current to the matrix element between $B$ mesons states vanishes due to conservation of parity in QCD \cite{ellis_stirling_webber_1996}. 



\subsection{Computation of $\Gamma_5^{(c \bar u d)}$ }
\label{sec:Gamma-5}
\begin{figure}
\centering
\includegraphics[scale = 0.5]{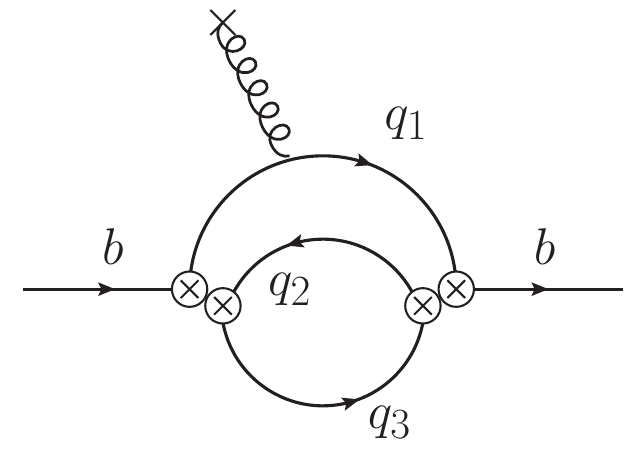}
\caption{Soft gluon corrections to the free $b$-quark decay.}
\label{fig:dim5}
\end{figure} 

In order to compute power corrections to the free $b$-quark decay in Eq.~(\ref{eq:G3-cud-final}), we must include the effect of the QCD interaction of the heavy quark field with the soft degrees of freedom inside the heavy hadron. This results in three contributions, generated by expanding respectively, the propagator of the quarks inside the two-loop diagram, because of the large $b$-quark momentum flowing into it, the heavy-quark momentum and the relevant matrix elements up to the order in $1/m_b$ considered. 
\\
In the case of two-quark operators, the starting point is represented by Eqs.~(\ref{eq:T2q-11})-(\ref{eq:T2q-22}), where now, being interested in computing dimension-five contributions, we must take into account, in the expression of the quark propagator, also terms proportional to the gluon field strength tensor $G_{\rho \sigma}$, see Section~\ref{sec:FS} and Figure~\ref{fig:dim5}, namely\\
\begin{equation}
S_{jk}^{(q)}(x,y) = S^{(q)}_0 (x-y) \delta_{jk} + S^{(q)}_1(x-y)^{a}\, t^{a}_{jk}\,, \quad q = c,u,d\,,
\label{eq:d5-prop}
\end{equation}\\
here $S_1^{(q)}(x-y)$ contains only corrections to the free-quark propagator due to operators of dimension-two, cf.\ Eq.~(\ref{eq:prop-coordinate-space}) and we have explicitly indicated that up to this order the translation invariance is still preserved. The presence of the Kronecker delta and of the $SU(3)_c$ generators in Eq.~(\ref{eq:d5-prop}), significantly simplifies the colour structure of Eqs.~(\ref{eq:T2q-11})-(\ref{eq:T2q-22}), in fact, it is easy to verify that, since ${\rm Tr}[t^a ]= 0$ and terms proportional to $G_{\rho \sigma} G_{\mu \nu}$ correspond to corrections of at least dimension-seven, a soft gluon cannot be radiated off every propagator for each of the expressions in Eqs.~(\ref{eq:T2q-11})-(\ref{eq:T2q-22}). Namely, only $S_1^{(c)}(-x)$ can contribute to ${\cal T}^{(2q)}_{11}$ and as consequence of Eq.~(\ref{eq:Fierz-13}) only $S_1^{(d)}(-x)$ can contribute to ${\cal T}^{(2q)}_{22}$. Finally, in Eq.~(\ref{eq:T2q-13}) only the expansion of the two propagators inside the trace is non vanishing and both $S_1^{(u)}(x)$ and $S_1^{(d)}(-x)$ must be independently taken into account in ${\cal T}^{(2q)}_{13}$. However, it is clear that by substituting Eq.~(\ref{eq:d5-prop}) into Eqs.~(\ref{eq:T2q-11})-(\ref{eq:T2q-22}), we also recover the leading order result already discussed in the previous section. This generates the remaining dimension-five contributions, once the corrections to the heavy quark momentum in Eq.~(\ref{eq:T3-1}) and to the matrix element of the dimension-three and dimension-four operators are respectively included.
To avoid confusion we introduce the notation\\
\begin{equation}
\Gamma_5^{(c \bar u d)} = \Gamma_5^{(c \bar u d)}\Big|_{\rm I} +  \Gamma_5^{(c \bar u d)}\Big|_{\rm II}\,,
\label{eq:calT-dim5-I-II}
\end{equation}\\
corresponding to the sum of the dimension-five contributions arising from the expansion of the quark-propagator on one side (I) and of the heavy-quark momentum as well as of the matrix elements of dimension lower than five, on the other (II). 

Let us start by considering (I). In the case of ${\cal T}_{11}^{( 2q)}$, as commented above, only the contribution of the gluon emitted from the charm-quark line is non vanishing and the three propagators in Eq.~(\ref{eq:T2q-11}) are respectively given by \footnote{We already take into account that in the propagator of the charm quark only terms proportional to an odd number of gamma matrices contribute.}\\ 
\begin{equation}
S_1^{(c)} (- x )  =  \frac{\tilde G^{\rho \eta} x_\rho \gamma_\eta \gamma_5}{8 \pi^2} 
    \frac{m_c K_1 (m_c \sqrt{-x^2})}{\sqrt{-x^2}}  \, ,
\end{equation}\\
and\\ 
\begin{equation}
S_0^{(u)} (x)  = 
\frac{1}{2 \pi^2} \frac{\slashed x }{x^4} 
= - S_0^{(d)} (-x) 
\, .
\end{equation}\\
Substituting these expressions as well as including a symmetry factor of 2 due to the translation invariance \footnote{The same factor appear also in the remaining contributions.}, Eq.~(\ref{eq:T2q-11}) becomes \footnote{Note that the replacement in Eq.~(\ref{eq:bx-d3}) is again introduced, since at this order the contribution from the dimension-five operator with one gluon field is obtained taking the matrix element $\langle b g |{\rm Im} {\cal T}^{( 2q)}| b\rangle$ between external $b$-quark and gluon states.}\\
\begin{align}
{\cal T}^{( 2q)}_{11}\Big|_{d=5} &=  \frac{G^2_F}{ 64 \pi^6} \, |V_{c b}|^2 |V_{u d}|^2  \, \bar b\,  \Bigg\{  \int d^4 x \,  e^{-i p_b \cdot x}
 \, \frac{\tilde G^{\rho \eta} \, x_\rho x_\xi  x_\tau }{x^8}  \frac{m_c\, K_1 (m_c \sqrt{-x^2})}{\sqrt{-x^2}} 
\nonumber \\[3mm]
& \times  \Big( \gamma_\mu (1-\gamma_5)
 \, \gamma_\eta \gamma_5 \, \gamma_\nu (1-\gamma_5) \Big) {\rm Tr} \Big[\, \gamma^\mu (1-\gamma_5) \, \gamma^\xi
\, \gamma^\nu \, (1-\gamma_5) \, \gamma^\tau\, \Big]  \Bigg\} \, b \,.
\label{eq:T2q-11-d5-I}
\end{align}\\
From \\
\begin{equation}
 x_\xi x_\tau \,\big(  \gamma_\mu \gamma_\eta \gamma_\nu \big)  \ {\rm Tr} \Big[\, \gamma^\mu (1-\gamma_5) \, \gamma^\xi
\, \gamma^\nu \, (1-\gamma_5) \, \gamma^\tau\, \Big]  =  32 \, x_\eta \, \slashed x \, ,
\end{equation}\\
it is clear that the contribution in Eq.~(\ref{eq:T2q-11-d5-I}) must vanish since the dual field strength tensor $\tilde G^{\rho \eta}$ is contracted with the symmetric tensor $x_\rho x_\eta$ i.e.\\
\begin{equation}
{\cal T}^{( 2q)}_{11}\Big|_{d=5}  = 0\,.
\label{eq:T2q-11-d5-I-0}
\end{equation}\\
In the case of ${\cal T}^{( 2q)}_{22}$, in Eq.~(\ref{eq:T2q-22}) we can only expand the propagator of the down quark while the $c$- and $u$-quark propagators are free. Using the following limits\\
\begin{equation}
 \lim_{m \to 0 } m \, K_0 (m \sqrt{-x^2}) = 0 \, , \qquad
\lim_{m \to 0 } m \, K_1 (m \sqrt{-x^2}) = \frac{1}{\sqrt{-x^2}} \, ,
\end{equation}\\
we evidently obtain that\\
\begin{equation}
S_0^{(u)} (x)  =  \frac{1}{2 \pi^2} \frac{\slashed x }{x^4} \, , \qquad
S_1^{(d)} (- x)  =  - \frac{1}{8 \pi^2} \frac{ x^\rho }{x^2}  \tilde G_{\rho \xi} \gamma^\xi \gamma_5\, ,
\end{equation}\\
and\\
\begin{equation}
S_0^{(c)} (-x )  =  \frac{ \slashed x}{4 \pi^2} \frac{ m_c^2  K_2(m_c \sqrt{- x^2})}{x^2}   \, .
\end{equation}\\
Substituting these expressions into Eq.~(\ref{eq:T2q-22}) gives\\
\begin{align}
{\cal T}^{( 2q)}_{22}\Big|_{d=5}
 &  =    \frac{G^2_F}{ 128 \pi^6} |V_{c b}|^2 \, |V_{u d}|^2  \, \bar b\, \,  \Bigg\{   \int d^4 x \,  e^{-i p_b \cdot x} 
 \, \frac{\tilde G_{\rho \xi}   x^\rho x^\eta  x_\tau }{x^8}   m_c^2 \, K_2(m_c \sqrt{- x^2})  
\nonumber \\[3mm]
& \times \Big(   \gamma_\mu (1-\gamma_5)
 \, \gamma_\eta\, \gamma_\nu (1-\gamma_5)   \Big)\,
 {\rm Tr} \Big[\, \gamma^\mu (1-\gamma_5) \, \gamma^\xi\, \gamma_5
\, \gamma^\nu \, (1-\gamma_5) \, \gamma^\tau\, \Big] \Bigg\}\,  b \,,
\label{eq:T2q-22-d5-I-0-1}
\end{align}\\
where now the corresponding gamma structure can be simplified as \\
\begin{align}
x^\eta x_\tau \Big(  \gamma_\mu \gamma_\eta \gamma_\nu (1-\gamma_5) \Big) \,  {\rm Tr} \Big[\, \gamma^\mu (1+\gamma_5) \, \gamma^\xi\,
\, \gamma^\nu \, (1-\gamma_5) \, \gamma^\tau\, \Big] 
=  32 \, x^\xi \, \slashed x (1-\gamma_5) \,,
\end{align}\\
showing that also this contribution must vanish, since again the dual field tensor $\tilde G_{\rho \xi }$ in Eq.~(\ref{eq:T2q-22-d5-I-0-1}) is contracted with the symmetric tensor $x^\rho x^\xi$, namely \\
\begin{equation}
{\cal T}^{( 2q)}_{22}\Big|_{d=5} = 0\,.
\label{eq:T2q-22-d5-I-0}
\end{equation}\\
We see that the emission of a soft gluon from both the charm and down quark propagators does not contribute at dimension-five, therefore in Eq.~(\ref{eq:T2q-13}) it is sufficient to consider the expansion of the $u$-quark propagator only. However, we notice that the result in Eqs.~(\ref{eq:T2q-11-d5-I-0}), (\ref{eq:T2q-22-d5-I-0}) is general and that the fact that contributions to the propagator of the $q_1$- and $q_3$-quarks vanish at order $1/m_b^2$, holds independently of the specific mode considered see Refs.~\cite{Blok:1992hw, Blok:1992he}. We will discuss this again in Chapter~\ref{ch:Darwin}.
\\[2mm]
Finally in the case of ${\cal T}_{13}^{( 2q)}$, considering only corrections to the propagator of the up quark, we must
substitute the following expressions into Eq.~(\ref{eq:T2q-13}), i.e.\\
\begin{equation}
S_1^{(u)} (x)  =  \frac{1}{8 \pi^2} \frac{ x^\rho }{x^2}  \tilde G_{\rho \tau} \gamma^\tau \gamma_5 \, , \qquad 
S_0^{(d)} (-x) =   -   \frac{1}{2 \pi^2} \frac{\slashed x }{x^4} \, ,
\end{equation}\\
and
\begin{equation}
S_0^{(c)} (- x )  =  \frac{ \slashed x}{4 \pi^2} \frac{ m_c^2  K_2(m_c \sqrt{- x^2})}{x^2} \,,
\end{equation}\\
which, taking into account that Tr$[t^a t^b]= (1/2) \delta^{ab}$,  leads to\\
\begin{align}
{\cal T}^{(2q)}_{13}\Big|_{d=5} & =    \frac{G^2_F}{ 128 \pi^6}\,  |V_{c b}|^2 |V_{u d}|^2 \,\,  \bar b\,  \Bigg\{   \int d^4 x \,  e^{-i p_b \cdot x}
 \, \frac{\tilde G_{\rho \tau}   x^\rho x^\eta  x_\xi}{x^8}  m_c^2  K_2(m_c \sqrt{- x^2})
\nonumber \\[3mm]
& \times   \Big(  \gamma_\mu (1-\gamma_5)
 \, \gamma_\eta\, \gamma_\nu (1-\gamma_5) \Big)
  {\rm Tr} \Big[\, \gamma^\mu (1-\gamma_5) \, \gamma^\xi
\, \gamma^\nu \, (1-\gamma_5) \, \gamma^\tau\, \gamma_5 \Big] \Bigg\} \, b \,.
\end{align}\\
The gamma structure simplifies as\\
\begin{equation}
x^\eta x_\xi \big(  \gamma_\mu \gamma_\eta \gamma_\nu (1-\gamma_5) \big) \,   {\rm Tr} \Big[ \gamma^\mu (1+\gamma_5) \, \gamma^\xi\,
\, \gamma^\nu \, (1-\gamma_5) \, \gamma^\tau \Big] 
=  32 \, \gamma^\tau \, x^2 (1-\gamma_5) \,,
\end{equation}\\
from which it follows that, since the dual field tensor $\tilde G_{\rho \tau}$ is now contracted with $x^\rho \gamma^\tau$, this contribution is non vanishing and equal to\\
\begin{align}
{\cal T}^{( 2q)}_{13}\Big|_{d=5}   =    \frac{G^2_F}{ 2 \pi^6} |V_{c b}|^2 |V_{u d}|^2  \, \bar b\, \tilde G_{\rho \tau}  \gamma^\tau 
 \Bigg\{  \int d^4 x \,  e^{-i p_b \cdot x}
   \,\frac{  m_c^2 \,K_2(m_c \sqrt{- x^2})}{x^6} \, x^\rho \, \Bigg\} \,(1-\gamma_5) \, b \,.
   \label{eq:T13-2q-dim5}
\end{align}\\
The imaginary part of the integral in the curly brackets of Eq.~(\ref{eq:T13-2q-dim5}) can be again obtained using the general result   derived in Ref.~\cite{Blok:1992hw}. Setting $\nu = 2$ and $n = 6$ in Eq.~(\ref{eq:Im-Bessel}), yields \\
\begin{align}
 {\rm Im } \, \Bigg\{
 \,  \int \, d^4 x \, e^{- i p_b \cdot x} \,  \frac{m_c^2 K_2(m_c \sqrt{-x^2})}{x^{6}} \,  x^\rho \Bigg\}  
 =  - \frac{ \pi^3}{24}\,  p_b^2 \, p_b^\rho \left(1 - r \right)^3 \, ,
 \label{eq:Im-integral-dim5}
\end{align}\\
where we recall that the dimensionless parameter $r$ is defined as $r = m_c^2/p_b^2$. In order to single out the operator appearing in Eq.~(\ref{eq:T13-2q-dim5}), we use that $\tilde G_{\rho \tau} = (1/2) \epsilon_{\rho \tau \mu \nu} G^{\mu \nu}$, then, from the the tensor decomposition of three gamma matrices Eq.~(\ref{eq:tensor-decomposition-gamma-mat}) we have \\
\begin{align}
\epsilon_{\rho \mu \nu \tau} \gamma^\tau = -i \gamma_\rho \gamma_\mu \gamma_\nu \gamma_5 + i g_{\rho \mu} \gamma_\nu \gamma_5 - i g_{\rho \nu} \gamma_\mu \gamma_5 + i g_{\mu \nu} \gamma_\rho \gamma_5\,,
\label{eq:epsilon-gamma}
\end{align}\\
which leads to \\
\begin{align}
\bar b \, \tilde G_{\rho \tau} \gamma^\tau p_b^\rho (1- \gamma_5) \, b  &=  \frac12 \, \bar b  \,  G^{\mu \nu} \epsilon_{\rho \mu \nu \tau}  \gamma^\tau p_b^\rho (1- \gamma_5) \, b
\nonumber \\[3mm]
&=\bar b \, \Bigg\{ \frac{i}{2}\, \, G^{\mu \nu} \gamma_\rho \gamma_\mu \gamma_\nu \, p_b^\rho (1-  \gamma_5  \, ) 
  - i \,  \, G^{\mu \nu}  \gamma_\mu \, p_{b \nu} (1-  \gamma_5)  \Bigg\} b \,.
\label{eq:G-mug-exp}
\end{align}\\
Note that, when contracted with the antisymmetric tensor $G^{\mu \nu}$, the fourth term on the r.h.s\ of Eq.~(\ref{eq:epsilon-gamma}) vanishes while the second and third terms give the same contribution.
Taking into account that at this order $p_b^\mu = m_b v^\mu + {\cal O}(1/m_b)$ and introducing the rescaled heavy quark field $b(x) = \exp(- i m_b v \cdot x) b_v(x)$, see Eq.~(\ref{eq:Qx-HQE}), satisfying $b_v(x) = \slashed v b_v(x) + {\cal O}(1/m_b) $, cf.\ Eq.~(\ref{eq:p-Qv}), it is easy to show that the second term on the r.h.s.\ of the second line in Eq.~(\ref{eq:G-mug-exp}) is zero, in fact \\
\begin{align}
  - i \,  \bar b \, G^{\mu \nu}  \gamma_\mu \, p_{b \nu}  \, b & =   - i  m_b \,  \bar b_v  \, G^{\mu \nu}  \gamma_\mu v_ \nu \, b_v + {\cal O}(1/m_b)
  \nonumber \\[3mm]
& =   - i  m_b \,  \bar b_v  \, G^{\mu \nu}  v_\mu v_ \nu  \,  b_v  + {\cal O}(1/m_b) 
\nonumber \\[3mm]
& =   {\cal O}(1/m_b) \,,
\end{align}\\
where the contributions proportional to $\gamma_5$ in Eq.~(\ref{eq:G-mug-exp}) can be neglected, since they vanish in matrix elements between $B$ meson states due to parity conservation, and we have used the following identity\\
\begin{equation}
 \bar b_v  \gamma_\mu b_v = \bar b_v \slashed v \gamma_\mu \slashed v b_v  = -\bar b_v  \gamma_\mu b_v + 2 \bar b_v v_\mu b_v\,,
 \end{equation}\\
valid up to $1/m_b$ corrections, to write $\bar b_v \gamma_\mu b_v = \bar b_v v_\mu b_v$, see e.g.\ Ref.~\cite{Georgi:1991mr} or cf.\ Eq.~(\ref{eq:P+gammaP+}). Conversely, the first term on the r.h.s.\ of the second line in Eq.~(\ref{eq:G-mug-exp}) gives \\
\begin{align}
(i/2) \, \bar b \, G^{\mu \nu} \gamma_\rho \gamma_\mu \gamma_\nu \, p_b^\rho   b &  = (1/2) m_b \, \bar b_v \, G^{\mu \nu} \slashed v  \sigma_{\mu \nu}   b_v +  {\cal O}(1/m_b) 
\nonumber \\[3mm]
& =  (1/2)\, m_b \, \bar b_v \, G^{\mu \nu}  \sigma_{\mu \nu}  b_v +  {\cal O}(1/m_b) 
\nonumber \\[3mm]
& =   {\cal O}_{mag} +  {\cal O}(1/m_b)  \,,
\label{eq:O-magn-dim5}
\end{align}
here the first equality follows from taking the antisymmetric part of $\gamma_\mu \gamma_\nu$ upon contraction with $G^{\mu \nu}$, while the chromo-magnetic operator ${\cal O}_{mag}$ is defined in Eq.~(\ref{eq:O-magn}). From Eqs.~(\ref{eq:Im-integral-dim5}), (\ref{eq:G-mug-exp}), (\ref{eq:O-magn-dim5}), and (\ref{eq:T2q}), we finally obtain, in agreement with Ref.~\cite{Blok:1992hw}, that\\
\begin{align}
\Gamma_5^{(c \bar u d)}\Big|_{\rm I} =  -  \frac{G^2_F m_b^3}{ 192 \pi^3} \, |V_{c b}|^2 |V_{u d}|^2  \, 2 \, C_1 C_2    \, 8 (1-\rho)^3 \,\mu_G^2(B)\,,
\end{align}\\
%
where the dimensionless mass parameter $\rho$ is again defined as $\rho = m_c^2/m_b^2$ and the non perturbative parameter $\mu_G^2$ is given in Eq.~(\ref{eq:dim-5-ME-parameters}).  

The second type of contribution (II) in Eq.~(\ref{eq:calT-dim5-I-II}) is obtained by expanding the expression in Eq.~(\ref{eq:T3-1}) up to order $1/m_b^2$. Using the results in Appendix~\ref{app:1}, a slightly long yet straightforward computation, which can be easily performed with e.g.\ Mathematica \cite{Mathematica}, leads to \\
\begin{align}
\Gamma_5^{(c \bar u d)}\Big|_{\rm II} & = \frac{ G_F^2 m_b^3}{ 192 \pi^3}\, |V_{cb}|^2 |V_{ud}|^2  \, \Bigl( N_c  \, \big( C_1^2 +  C_2^2 \big) + 2 \, C_1 C_2 \Bigr) 
\nonumber 
\\[3mm]
& \times \Bigg[ \left( 1 - 8 \rho - 12 \rho^2 \log \rho + 8 \rho^3 - \rho^4 \right) \left(- \frac{\mu_\pi^2(B)}{2} \right)
\nonumber \\[3mm]
& +  \frac{1}{2} \left(-3 + 8 \rho -12 \rho^2 \log \rho - 24 \rho ^2   
+ 24 \rho^3 - 5 \rho^4 \right) {\mu_G^2(B)}\Bigg] \,,
\end{align} \\
with the non perturbative parameter $\mu_\pi^2$ given in Eq.~(\ref{eq:dim-5-ME-parameters}).



\section{Contribution of four-quark operators}
\label{sec:4q-contr}
\begin{figure}
\centering
\includegraphics[scale = 0.45]{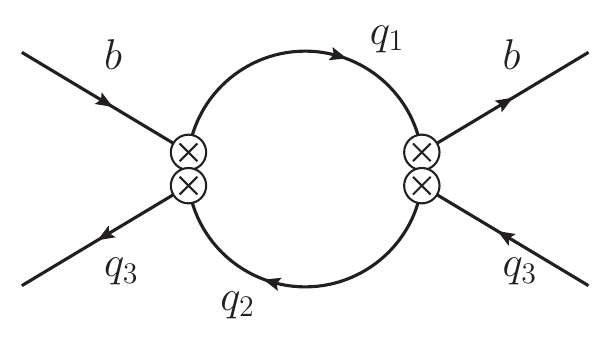}\,\,
\includegraphics[scale = 0.45]{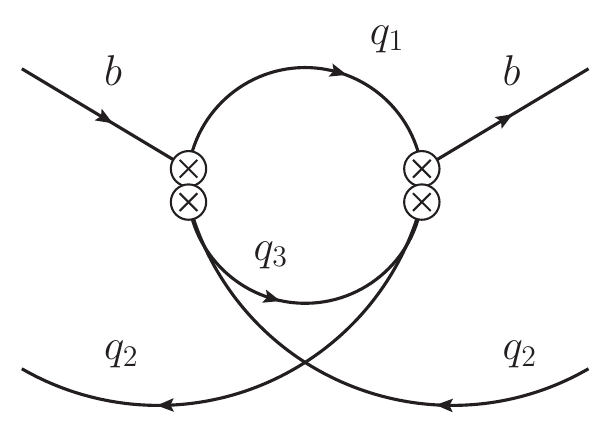}\,\,
\includegraphics[scale = 0.45]{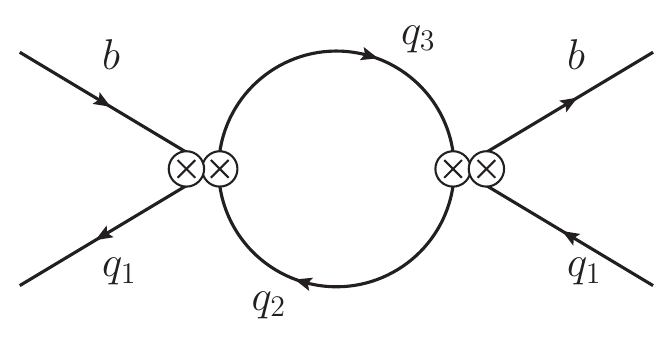}
\caption{Diagrams describing respectively, from left to right, the WE, PI and WA  topologies.}
\label{fig:WE-WA-PI}
\end{figure}
We now turn to discuss the contribution of four-quark operators. By applying the Wick's theorem in Eq.~(\ref{eq:T_mn}), we consider all possible contractions of two pairs of light-quark fields in the time ordered product, which leave the $b$-quark as well as a pair of light-quark fields uncontracted, namely 
\begin{align}
{\rm T}  \Bigg\{ & \Big(  \bar q_1(x)\,  \Gamma_\mu \, b(x) \, \bar q_3(x) \, \Gamma^\mu \,  q_2(x) \Big)  ,  \Big( \bar b\,  \Gamma_\nu \, q_1 \, \bar q_2 \, \Gamma^\nu \,  q_3 \Big) \Bigg\}  
 \nonumber 
 \\
 & = \, : 
 \wick{
 \c1{\bar q}_1(x)\,  \Gamma_\mu \, b(x) \, \bar q_3(x) \, \Gamma^\mu \,  \c2{q}_2(x) \, \bar b\,  \Gamma_\nu \, \c1{q}_1 \,\c2{\bar q}_2 \, \Gamma^\nu \,  q_3 :
  }
 \nonumber \\[3mm]
 & +  : \wick{
 \c1{\bar q}_1(x)\,  \Gamma_\mu \, b(x) \, \c2{\bar q}_3(x) \, \Gamma^\mu \,  q_2(x) \, \bar b\,  \Gamma_\nu \, \c1{q}_1 \,\bar q_2 \, \Gamma^\nu \,  \c2 q_3 :
 }
 \nonumber \\[3mm]
 & +  : \wick{
 \bar q_1(x)\,  \Gamma_\mu \, b(x) \, \c2{\bar q}_3(x) \, \Gamma^\mu \, \c1 q_2(x) \, \bar b\,  \Gamma_\nu \, q_1 \, \c1{\bar q}_2 \, \Gamma^\nu \,  \c2 q_3 :
 }
 \,.
 \label{eq:Wick-4q}
\end{align}\\
Note that for simplicity we have omitted to specify the colour structure. 
The three terms on the r.h.s.\ of Eq.~(\ref{eq:Wick-4q}) generate different topologies, usually referred to, respectively, as Weak Exchange (WE), Pauli Interference (PI) and Weak Annihilation (WA) \footnote{It is worth mentioning that for the description of baryons, the WE and PI topologies are interchanged.}, schematically shown in Figure~\ref{fig:WE-WA-PI}. In analogy to Eq.~(\ref{eq:T2q}) we then introduce the notation\\
\begin{equation}
{\cal T}^{(4q)}  = {\cal T}^{\rm WE}  +  {\cal T}^{\rm PI}  +  {\cal T}^{\rm WA} 
\, , 
\label{eq:T4q}
\end{equation}\\
where the superscript $(4q)$ indicates that four-quark fields are not contracted and \\
\begin{equation}
{\cal T}^{X}  =  \Bigg( C_1^2 \,  {\cal T}_{11}^{X} + 2 \, C_1 \, C_2\, {\cal T}^{X}_{12}  + C_2^2 \, {\cal T}_{22}^{X} \Bigg)
\, ,
\label{eq:T4q-X}
\end{equation}\\
with $X$ labelling the specific topology. We emphasise that the three contributions in Eq.~(\ref{eq:T4q-X}) correspond to the same Dirac structure and differ only by the contraction of the colour indices, namely, starting with the case of Weak Exchange, we can conveniently define the following tensor in colour space  \\
\begin{align}
{\cal A}^{ijmn }_{klrs}  &=   \frac{G_F^2}{2}  |V_{q_1b}|^2 |V_{q_3q_2}|^2\  i \int d^4 x \, \Bigl( \bar b^i \, \gamma_\nu (1-\gamma_5) \, i S^{( q_1)}_{kl} (0, x) \gamma_\mu (1-\gamma_5) b^j (x)
\nonumber \\[3mm]
& \times   \, \bar q_3^m(x) \, \gamma^\mu (1-\gamma_5) \,i S^{( q_2)}_{rs} (x,0)\, \gamma^\nu (1-\gamma_5) \, q_3^n   \Bigr) 
 + (x \leftrightarrow 0)
\, ,
\label{eq:WE-dirac-str}
\end{align}\\
so that the three terms in Eq.~(\ref{eq:T4q-X}) are respectively given by \\
\begin{align}
{\cal T}_{11}^{\rm WE }  = {\cal A}^{ijmn }_{klrs}  \, \delta^i_k  \delta^j_l  \delta^m_r  \delta^n_s\,,
\quad 
{\cal T}_{12}^{\rm WE }  = {\cal A}^{ijmn }_{klrs}  \, \delta^i_s  \delta^j_l  \delta^m_r \delta^n_k\,,
\quad
{\cal T}_{22}^{\rm WE }  = {\cal A}^{ijmn }_{klrs}  \, \delta^{in} \delta^{jm} \delta_{ks}  \delta_{lr}\,.
\label{eq:T-WE-22}
\end{align}\\
Similarly, for the Pauli Interference topology, we introduce the tensor\\
\begin{align}
{\cal B}^{ijmn}_{klrs} &= \frac{G_F^2}{2}  |V_{q_1b}|^2 |V_{q_3q_2}|^2\ i \int d^4 x \, \Bigl( \bar b^i \, \gamma_\nu (1-\gamma_5) \, i S^{( q_1)}_{kl} (0, x) \gamma_\mu (1-\gamma_5) b^j (x)
\nonumber \\[3mm]
& \times   \, \bar q_2^m \, \gamma^\nu (1-\gamma_5) \,i S^{(q_3)}_{rs} (0,x)\, \gamma^\mu (1-\gamma_5) \, q_2^n(x)   \Bigr) 
 + (x \leftrightarrow 0)
\, ,
\label{eq:PI-dirac-str}
\end{align}\\
and the three contributions in Eq.~(\ref{eq:T4q-X}) read \\
\begin{align}
{\cal T}_{11}^{\rm PI }  = {\cal B}^{ijmn }_{klrs}  \, \delta^i_k  \delta^j_l  \delta^m_r  \delta^n_s\,,
\quad 
{\cal T}_{12}^{\rm PI }  = {\cal B}^{ijmn }_{klrs}  \, \delta^i_k \delta^j_r  \delta^m_l \delta^n_s\,,
\quad
{\cal T}_{22}^{\rm PI }  =  {\cal T}_{11}^{\rm PI }\,.
\label{eq:T-PI-22}
\end{align}\\
Finally, in the case of Weak Annihilation, we define \\
\begin{align}
{\cal C}^{ijmn}_{klrs} & =  - \frac{G_F^2}{2}  |V_{q_1b}|^2 |V_{q_3q_2}|^2\  \,  i \int d^4 x \, \Bigl( \bar b^i \, \gamma_\nu (1-\gamma_5) \,  q^j_1 \, \bar q^m_1(x) \gamma_\mu (1-\gamma_5)  b^n(x) \Bigr)
\nonumber \\[3mm]
& \times   \,{\rm Tr} \Big[ \gamma^\nu (1-\gamma_5) \,  i S^{( q_2)}_{kl} (0, x) \, \gamma^\mu (1-\gamma_5) \,i S^{( q_3)}_{rs} (x,0) \Big] 
+ (x \leftrightarrow 0)
\, ,
\label{eq:WA-dirac-str}
\end{align}\\
where the minus sign and the trace over spinor indices follow from the fermion-loop and \\
\begin{align}
{\cal T}_{11}^{\rm WA }  = {\cal C}^{ijmn }_{klrs}  \, \delta^{ij} \delta^{mn}  \delta_{ks}  \delta_{lr}\,,
\quad
{\cal T}_{12}^{\rm WA }  = {\cal C}^{ijmn }_{klrs}  \, \delta^i_k  \delta^j_s  \delta_{lr}  \delta^{mn}  \,,
\quad
{\cal T}_{22}^{\rm WA }  = {\cal C}^{ijmn }_{klrs}  \, \delta^i_k  \delta^j_s  \delta^m_r \delta^n_l\,.
\label{eq:T-WA-22}
\end{align}


\subsection{Computation of Im${\cal T}^{ (4q)}_6$}
\label{sec:T-4q-6}
To compute the leading power corrections to the WE, PI and WA topologies, namely the dimension-six contribution to the four-quark operators, we can ignore the QCD interaction with the background field and set in Eqs.~(\ref{eq:WE-dirac-str}), (\ref{eq:PI-dirac-str}), (\ref{eq:WA-dirac-str})\\
\begin{equation}
S^{(q_i)}_{jk}(x,y) = S^{(q_i)}_0(x-y)\,  \delta_{jk}\,, \qquad i=1,2,3\,,
\label{eq:Free-quark-prop-4q}
\end{equation}\\
where $S_0^{(q_i)}(x-y)$ is the free-quark propagator associated to $q_i$, and note that we consider now the general case of $b \to q_1 \bar q_2 q_3$ transition.
Higher order terms in Eq.~(\ref{eq:Free-quark-prop-4q}), cf. Eqs.~(\ref{eq.quark-propagator-definition1}), (\ref{eq:prop-coordinate-space}), describe soft gluon corrections to the four-quark operators and lead to contributions to the HQE of at least dimension-eight. These will not be discussed in the present work. However now, the loop-computation is more easily performed in momentum space and we will use, contrary to what we have done in the previous section, the Fourier representation of the propagator in Eq.~(\ref{eq:Free-quark-prop-4q}).
Substituting Eq.~(\ref{eq:Sxy-free-prop}) into the first of the three expressions in Eq.~(\ref{eq:T-WE-22}) yields \footnote{Note that we take into account the replacement in Eq.~(\ref{eq:bx-d3}), supplemented with the corresponding one for the light spectator antiquark.}\\
\begin{align}
 {\cal T}_{11}^{\rm WE }& =  -   \,G_F^2 |V_{q_1b}|^2 |V_{q_3q_2}|^2 \,  i \int d^4 x \, \int \frac{d^4 l}{(2 \pi)^4} \,   \int \frac{d^4 k }{(2 \pi)^4} e^{- i (p + k -l) \cdot x}\, \Bigl( \bar b^i \gamma_\nu (1-\gamma_5) 
 \nonumber \\[3mm]
&\times \left(\frac{\slashed l }{l^2 - m_1^2+ i \varepsilon} \right)  \gamma_\mu (1-\gamma_5) b^i 
 \bar q_3^j  \gamma^\mu (1-\gamma_5) \left(\frac{\slashed k }{k^2 - m_2^2 + i \varepsilon} \right)  \gamma^\nu (1-\gamma_5) q_3^j   \Bigr)\,,
\label{eq:T-WE-11-1}
\end{align}\\
where $p^\mu = p^\mu_b + p^\mu_{q_3}$.
In deriving Eq.~(\ref{eq:T-WE-11-1}) we have used that $b(x)$ and $\bar q_3(x)$ describe respectively, an incoming quark with momentum $p^\mu_b$ and an incoming antiquark with momentum $p^\mu_{q_3}$ \footnote{For baryons, being both $b$ and $q_3$ quarks, it would be $p^\mu = p^\mu_b - p^\mu_{q_3}$.}. We notice that a symmetry factor of $2$ due to translation invariance has been already included and that the chiral structure of Eq.~(\ref{eq:T-WE-11-1}) implies that all terms in the propagator proportional to an even number of gamma matrices must vanish. Performing the integration over the variables $x^\mu$ and $k^\mu$, leads to\\
\begin{align}
 {\cal T}_{11}^{\rm WE }& =  -  4  \,G_F^2 |V_{q_1b}|^2 |V_{q_3q_2}|^2 \, i\, \int \frac{d^4 l}{(2 \pi)^4} \, \frac{\big( l^\rho l^\sigma - p^\rho l^\sigma \big)}{ (l^2 - m_1^2+ i \varepsilon) ((l-p)^2 - m_2^2+ i \varepsilon) }
  \nonumber \\[3mm]
 & \times \, g_{\sigma \xi} \, \Big( \bar b^i  \gamma_\nu  (1-\gamma_5)  \gamma_\rho  \gamma_\mu b^i \Big) \Big( \bar q_3^j \gamma^\mu  \gamma^\xi  \gamma^\nu (1-\gamma_5) q_3^j \Big)  \,,
 \label{eq:T-WE-11-2}
\end{align}\\
where the structure of the four-quark operator in Eq.~(\ref{eq:T-WE-11-2}) can be simplified taking into account the tensor decomposition of three gamma matrices in Eq.~(\ref{eq:tensor-decomposition-gamma-mat}), and the Fierz identity in Eq.~(\ref{eq:Fierz-identity}), i.e. \\
\begin{align}
& g_{\sigma \xi} \Big( \bar b^i   \gamma_\nu (1-\gamma_5)    \gamma_\rho \gamma_\mu b^i \Big)  \Big( \bar q_3^j \gamma^\mu  \gamma^\xi   \gamma^\nu (1-\gamma_5) q_3^j \Big)
 \nonumber \\[2mm]
= -2 \, \Big( \bar b^i  & \gamma^\nu (1-\gamma_5) q_3^j\Big) \Big( \bar q_3^j (g_{\sigma \nu} \gamma_\rho + g_{\nu \rho} \gamma_\sigma - g_{\rho \sigma} \gamma_{\nu} ) (1-\gamma_5) b^i \Big)\,,
   \label{eq:gamma-structure-WE}
 \end{align}\\
 note that we have dropped the contribution of the Levi-Civita tensor since the integral in Eq.~(\ref{eq:Im-WE}) is a symmetric function in the indices $\{\rho, \sigma\}$.
%
Using the Passarino-Veltman reduction algorithm \cite{Passarino:1978jh},  the one-loop tensor integral in Eq.~(\ref{eq:T-WE-11-2}) can be decomposed in terms of one- and two-point one-loop scalar integrals. Note that the latter are ultraviolet (UV) divergent and must be regularised. Performing the computation in dimensional regularisation \cite{tHooft:1972tcz, Bollini:1972ui, Cicuta:1972jf, Ashmore:1972uj } with $D = 4 - 2 \epsilon$ space-time dimensions, one obtains the results listed in Appendix~\ref{app:2}, in which the singularity appears as a single pole in $\epsilon$, see Eqs.~(\ref{eq:A0}), (\ref{eq:B0}). However, being interested only in the imaginary part of ${\cal T}_{11}^{\rm WE}$ which, cf.~Eqs.~(\ref{eq:iA0}), (\ref{eq:iB0}), is finite for $\epsilon \to 0$, we can use the expressions in Eqs.~(\ref{eq:r1-bubble})-(\ref{eq:B22}) setting $D = 4$. This gives \\
\begin{align}
{\rm Im}\,  \Bigg( i \, & \int  \frac{d^4 l}{(2 \pi)^4} \,   \frac{\big( l^\rho l^\sigma - p^\rho l^\sigma \big) }{(l^2 - m_1^2+ i \varepsilon)((l-p)^2 - m_2^2 + i \varepsilon) } \Bigg) =
\nonumber \\[3mm]
 \frac{p^2  \sqrt{\lambda(1, r_1,r_2)}}{192 \pi} &  \Bigg[g^{\rho \sigma}  \lambda(1, r_1,r_2)
- 2 \frac{p^\rho p^\sigma}{p^2} \Big( 2 (r_1 -r_2)^2  - (1+r_1 + r_2)  \Big)    \Bigg]\,,
\label{eq:Im-WE}
\end{align}\\
where $r_i = m_i^2/p^2$ 
and $\lambda(a,b,c) \equiv (a -b -c)^2 - 4 bc$ is the K\"allen function. Notice that Eq.~(\ref{eq:Im-WE}) is a symmetric function under the exchange $m_1 \leftrightarrow m_2$. Substituting Eqs.~(\ref{eq:gamma-structure-WE}), (\ref{eq:Im-WE}) into Eq.~(\ref{eq:T-WE-11-2}), we then obtain\\
\begin{align}
{\rm Im}\,  {\cal T}_{11}^{\rm WE} &  =    \frac{G_F^2 }{12 \pi} |V_{q_1b}|^2 |V_{q_3q_2}|^2\,  p^2   \sqrt{\lambda(1, r_1,r_2)}  \, 
\nonumber \\[3mm]
&\times \Bigg[ \Big( (r_1 -r_2)^2 + r_1 + r_2 -2 \Big)  \Big( \bar b^i \Gamma_\mu  q_3^j  \Big)\Big( \bar q_3^j  \Gamma^\mu b^i \Big)
\nonumber \\[3mm]
& - 2\, \frac{ p^\mu p^\nu}{p^2} \Big( 2 (r_1 -r_2)^2 - (1+ r_1 +r_2) \Big) \Big( \bar b^i \Gamma_\mu q_3^j  \Big) \, \Big( \bar q_3^j  \Gamma_\nu b^i \Big) \Bigg]\,.
\label{eq:T-WE-11-3}
\end{align}\\
Finally, we can express the colour-rearranged operators in Eq.~(\ref{eq:T-WE-11-3}) in terms of colour-singlet and colour-octet ones using Eq.~(\ref{eq:Fiez-id-colour-matr}), this yields\\
\begin{align}
{\rm Im}\,  {\cal T}_{11}^{\rm WE} &  =    \frac{G_F^2}{12 \pi}  |V_{q_1b}|^2 |V_{q_3q_2}|^2 \, p^2   \sqrt{\lambda(1, r_1,r_2)} \, \Bigg\{ \Big( (r_1 -r_2)^2 + r_1 + r_2 -2 \Big) 
\nonumber \\[3mm]
&\times \, \Bigg[ \frac{1}{N_c} \Big( \bar b^i \Gamma_\nu  q_3^i  \Big) \Big( \bar q_3^j  \Gamma^\nu b^j \Big) 
+ 2 \,  \Big( \bar b^i  \Gamma_\nu  t^a_{ij} q_3^j  \Big) \Big( \bar q_3^l  \Gamma^\nu  t^a_{lm} b^m \Big)  \Bigg] 
\nonumber \\[3mm]
&   - \,2  \frac{p^\mu p^\nu}{p^2} \Big( 2 (r_1 -r_2)^2 - (1+ r_1 + r_2) \Big)  \,  \Bigg[ \frac{1}{N_c} \Big( \bar b^i \Gamma_\mu q_3^i  \Big)  \Big( \bar q_3^j  \Gamma_\nu b^j \Big)  
\nonumber \\[3mm]
& + 2 \Big( \bar b^i \, \Gamma_\mu \, q_3^i  \Big) \, \Big( \bar q_3^j \, \Gamma_\nu \, b^j \Big) \Bigg] \Bigg\}\,.
\label{eq:T-WE-11-4}
\end{align}\\
To compute the remaining colour structures we substitute Eq.~(\ref{eq:Sxy-free-prop}) into the second and third term of Eq.~(\ref{eq:T-WE-22}),  obtaining respectively\\
\begin{align}
 {\cal T}_{12}^{\rm WE }& =  -   \,G_F^2 |V_{q_1b}|^2 |V_{q_3q_2}|^2 \,  i \int d^4 x \, \int \frac{d^4 l}{(2 \pi)^4} \,   \int \frac{d^4 k }{(2 \pi)^4} e^{- i (p + k -l) \cdot x} \, \Biggl[ \bar b^i \gamma_\nu (1-\gamma_5) 
 \nonumber \\[3mm]
 & \times  \left(\frac{\slashed l }{l^2 - m_1^2 + i \varepsilon} \right) \gamma_\mu (1-\gamma_5) b^j 
\bar q_3^j  \gamma^\mu (1-\gamma_5)  \left(\frac{\slashed k }{k^2 - m_2^2+ i \varepsilon} \right)  \gamma^\nu (1-\gamma_5) q_3^i   \Biggr]\,,
\label{eq:T-WE-12-1}
\end{align}\\
and
\begin{equation}
{\cal T}_{22}^{\rm WE} = N_c \, {\cal T}_{12}^{\rm WE } \,,
\label{eq:T-22-WE}
\end{equation}\\
with the colour factor in Eq.~(\ref{eq:T-22-WE}) following from $\delta_{ii}= N_c$.
The calculation of Eq.~(\ref{eq:T-WE-12-1}) proceeds in the very same way as for Eq.~(\ref{eq:T-WE-11-1}). For brevity we only state here the final expression which reads\\
\begin{align}
{\rm Im} \,{\cal T}_{12}^{\rm WE }& =    \frac{G_F^2}{12 \pi}  |V_{q_1b}|^2 |V_{q_3q_2}|^2 \, p^2   \sqrt{\lambda(1, r_1,r_2)} 
\, \Bigg[ \Big( (r_1 -r_2)^2 + r_1 + r_2 -2 \Big) 
\nonumber \\[3mm]
&\times  \Big( \bar b^i \Gamma_\nu  q_3^i  \Big) \Big( \bar q_3^j \Gamma^\nu b^j \Big) 
 - 2  \frac{p^\mu p^\nu}{p^2}  \Big( 2 (r_1 - r_2)^2 - (1+ r_1 + r_2) \Big) \Big( \bar b^i \Gamma_\mu  q_3^i  \Big)  \Big( \bar q_3^j  \Gamma_\nu b^j \Big) \Bigg] \,,
\label{eq:T-WE-12-2}
\end{align}\\
note that in this case, the result contains already colour-singlet and colour-rearranged operators and we do not need to further use Eq.~(\ref{eq:Fiez-id-colour-matr}). 
%
Substituting Eqs.~(\ref{eq:T-WE-11-4}), (\ref{eq:T-WE-12-2}), and (\ref{eq:T-22-WE}), into Eq.~(\ref{eq:T4q-X}), we readily arrive at\\
\begin{align}
{\rm Im} \,{\cal T}^{\rm WE }&  = \frac{G_F^2}{12 \pi}  |V_{q_1b}|^2 |V_{q_3q_2}|^2 \, p^2   \sqrt{\lambda(1, r_1,r_2)}  \, \Bigg\{ k_1 \Bigg[\, \omega_1(r_1,r_2)  \,  \Big( \bar b^i \Gamma_\mu q_3^i  \Big) \Big( \bar q_3^j \Gamma^\mu b^j \Big)  
\nonumber \\[3mm]
&  - 2 \frac{p^\mu p^\nu}{p^2} \omega_2(r_1, r_2) \Big( \bar b^i \Gamma_\mu q_3^i  \Big) \Big( \bar q_3^j \Gamma_\nu b^j \Big) \Bigg]
+ k_2 
  \Bigg[ \omega_1(r_1, r_2) \Big( \bar b^i \Gamma_\mu  t^a_{ij} q_3^j  \Big) \Big( \bar q_3^l \Gamma^\mu  t^a_{lm} b^m \Big)
  \nonumber \\[3mm]
& - 2   \frac{p^\mu p^\nu}{p^2} \omega_2(r_1, r_2) \Big( \bar b^i \Gamma_\mu t^a_{ij} q_3^j  \Big) \Big( \bar q_3^l \Gamma_\nu   t^a_{lm} b^m \Big)\Bigg] \Bigg\} + \ldots \,,
\label{eq:ImT-WE}
\end{align}\\
where the ellipsis refer to power suppressed contributions arising from corrections to the propagator of order ${\cal O}(1/m_b^2)$ and for the sake of a more compact notation we have introduced the following combinations of Wilson coefficients\\
\begin{align}
k_1 =  \frac{1}{N_c} \, C_1^2 + 2 \, C_1 C_2 + N_c \, C_2^2 \,, \qquad 
k_2 = 2 \, C_1^2\,,
\label{eq:WC-WE}
\end{align}\\
while
\begin{align}
\omega_1(a,b)  = (a - b)^2 + a + b - 2\,,
\qquad
\omega_2(a,b) = 2 \, (a -b)^2 - (1 + a + b)\,.
\end{align}\\
We now turn to consider the contribution of Pauli Interference. Inserting Eq.~(\ref{eq:Sxy-free-prop}) into the first expression in Eq.~(\ref{eq:T-PI-22}), gives\\\
\begin{align}
 {\cal T}_{11}^{\rm PI }& =  -   \,G_F^2 |V_{q_1b}|^2 |V_{q_3q_2}|^2 \,  i \int d^4 x \, \int \frac{d^4 l}{(2 \pi)^4} \,   \int \frac{d^4 k }{(2 \pi)^4}  e^{- i (p - k - l) \cdot x} \, \Biggl( \bar b^i \gamma_\nu (1-\gamma_5)
 \nonumber \\[3mm]
 & \times \left(\frac{\slashed k }{k^2 - m_1^2 + i \varepsilon } \right) \gamma_\mu (1-\gamma_5) b^i 
\bar q_2^j  \gamma^\nu (1-\gamma_5)  \left(\frac{\slashed l }{l^2 - m_3^2+ i \varepsilon} \right) \gamma^\mu (1-\gamma_5) q_2^j   \Biggr)\,,
\label{eq:T-PI-11-1}
\end{align}\\
here $p^\mu = p_b^\mu - p_{q_2}^\mu $, which follows from the fact that $q_2(x)$ describes an outgoing antiquark with momentum $p^\mu_{q_2}$ \footnote{For baryons, being both $b$ and $q_2$ quarks, it would be $p^\mu = p^\mu_b + p^\mu_{q_2}$.}.
The integration over the variables $x^\mu$ and $k^\mu$ yields\\
\begin{align}
 {\cal T}_{11}^{\rm PI }& =  4  \,G_F^2 |V_{q_1b}|^2 |V_{q_3q_2}|^2 \, i\, \int \frac{d^4 l}{(2 \pi)^4} \, \frac{ (l^\rho l^\sigma - p^\rho l^\sigma)}{ (l^2 - m_3^2  + i \varepsilon) ((l-p)^2 - m_1^2 + i \varepsilon) }
  \nonumber \\[3mm]
 & \times \, g_{\sigma \xi} \, \Big( \bar b^i \gamma_\nu  (1-\gamma_5) \gamma_\rho \gamma_\mu b^i \Big) \Big( \bar q_3^j  \gamma^\nu  \gamma^\xi  \gamma^\mu (1-\gamma_5)  q_3^j \Big)  \,,
 \label{eq:T-PI-11-2}
\end{align}\\
where the four-quark operator can be simplified similarly to Eq.~(\ref{eq:gamma-structure-WE}) and using again Eq.~(\ref{eq:Fiez-id-colour-matr}),
while the imaginary part of the one-loop integral in Eq.~(\ref{eq:T-PI-11-2}) is obtained, taking into account the symmetry under the exchange $m_1 \leftrightarrow m_3$, by replacing $r_2 \to  r_3$ into Eq.~(\ref{eq:Im-WE}). A straightforward calculation leads to\\
\begin{align}
{\rm Im} \, {\cal T}_{11}^{\rm PI }& =  \frac{G_F^2}{2 \pi}  |V_{q_1b}|^2 |V_{q_3q_2}|^2 p^2 \, \sqrt{\lambda(1, r_1, r_3)}\, (1- r_1 - r_3 ) 
  \nonumber \\[3mm]
 & \times \, \Bigg( \frac{1}{N_c}   \Big( \bar b^i  \Gamma_\mu q_2^i  \Big)  \Big( \bar q_2^j  \Gamma^\mu b^j \Big)   + 2   \Big( \bar b^i  \Gamma_\mu t^a_{ij}  q_2^j  \Big)  \Big( \bar q_2^l  \Gamma^\mu\, t^a_{lm}  b^m \Big)     \Bigg)\,.
 \label{eq:T-PI-11-4}
\end{align}\\
The remaining colour structure is obtained by substituting Eq.~(\ref{eq:Sxy-free-prop}) into the second term of Eq.~(\ref{eq:T-PI-22}), i.e.\\
\begin{align}
 {\cal T}_{12}^{\rm PI }& =  -   \,G_F^2 |V_{q_1b}|^2 |V_{q_3q_2}|^2 \,  i \int d^4 x \, \int \frac{d^4 l}{(2 \pi)^4} \,   \int \frac{d^4 k }{(2 \pi)^4}  e^{- i (p - k - l) \cdot x} \, \Biggl( \bar b^i  \gamma_\nu (1-\gamma_5)
 \nonumber \\[3mm]
 & \times  \left(\frac{\slashed k }{k^2 - m_1^2 + i \varepsilon} \right) \gamma_\mu (1-\gamma_5) b^j
 \bar q_2^j \gamma^\nu (1-\gamma_5) \left(\frac{\slashed l }{l^2 - m_3^2 + i \varepsilon} \right)  \gamma^\mu (1-\gamma_5) q_2^i   \Biggr)\,.
\label{eq:T-PI-12-1}
\end{align}\\
All the steps proceed analogously to the case of ${\cal T}_{11}^{\rm PI}$, the only difference being that the result is already expressed in terms of colour-singlet operators and we do not need to use Eq.~(\ref{eq:Fiez-id-colour-matr}). It is in fact easy to show that\\
\begin{align}
{\rm Im} \, {\cal T}_{12}^{\rm PI }& =  \frac{G_F^2}{2 \pi}  |V_{q_1b}|^2 |V_{q_3q_2}|^2 \, p^2 \, \sqrt{\lambda(1, r_1, r_3)}
(1- r_1 - r_3 ) \, \Big( \bar b^i \Gamma_\mu q_2^i  \Big) \Big( \bar q_2^j \Gamma^\mu  b^j \Big)  \,,
 \label{eq:T-PI-12-2}
\end{align}\\
and from Eqs.~(\ref{eq:T-PI-11-4}), (\ref{eq:T-PI-12-2}), (\ref{eq:T4q-X}) we readily obtain that\\
\begin{align}
{\rm Im} \,{\cal T}^{\rm PI }& =  \frac{G_F^2 }{2 \pi}|V_{q_1b}|^2 |V_{q_3q_2}|^2 p^2   \, \sqrt{\lambda(1, r_1, r_3)} \, (1 - r_1 - r_3)
\nonumber \\[3mm]
& \times \Bigg[ k_3 \Big( \bar b^i \Gamma_\mu q_2^i  \Big) \Big( \bar q_2^j \Gamma^\mu b^j \Big)
+ k_4  \Big( \bar b^i \Gamma_\mu  t^a_{ij} q_2^j  \Big) \Big( \bar q_2^l \Gamma^\mu t^a_{lm} b^m \Big)  \Bigg] + \ldots \,,
\label{eq:ImT-PI}
\end{align}\\
with\\
\begin{equation}
k_3 = \frac{1}{N_c} \Big( C_1^2 + C_2^2 \Big) + 2 \, C_1 C_2 \,, \qquad
k_4 = 2 \, \Big( C_1^2 + C_2^2 \Big) \,.
\label{eq:WC-PI}
\end{equation}\\
Finally we discuss the Weak Annihilation topology. Substituting Eq.~(\ref{eq:Sxy-free-prop}) into the first expression of Eq.~(\ref{eq:T-WA-22}) yields\\
\begin{align}
 {\cal T}_{11}^{\rm WA }& = N_c\,  G_F^2 |V_{q_1b}|^2 |V_{q_3q_2}|^2  i \int d^4 x \int \frac{d^4 l}{(2 \pi)^4}  \int \frac{d^4 k }{(2 \pi)^4} e^{- i (p + k -l) \cdot x} \,  \Big( \bar b^i \gamma_\nu (1-\gamma_5) q_1^i \Big) 
\nonumber \\[3mm]
& \times  \Big(  \bar q_1^j \gamma_\mu (1-\gamma_5)  b^j \Big)  {\rm Tr} \left[ \gamma^\nu (1-\gamma_5) \left(\frac{\slashed l }{l^2 - m_2^2 + i \varepsilon} \right) \gamma^\mu (1-\gamma_5) \left(\frac{\slashed k }{k^2 - m_3^2 + i \varepsilon} \right)   \right]\,,
\label{eq:T-WA-11-1}
\end{align}\\
where $p^\mu = p_b^\mu + p_{q_1}^\mu$, due to the fact that $\bar q_1(x)$ describes an incoming antiparticle with momentum $p^\mu_{q_1}$ \footnote{For baryons, being both $b$ and $q_1$ quarks, it would be $p^\mu = p^\mu_b - p^\mu_{q_1}$.}, and we have taken into account that only terms in the propagator proportional to an odd number of gamma matrices contribute to the trace.\\
Performing the integration over the variables $x^\mu$ and $k^\mu$ and evaluating the trace in Eq.~(\ref{eq:T-WA-11-1}), leads to\\
\begin{align}
 {\cal T}_{11}^{\rm WA }& = 8N_c G_F^2 |V_{q_1b}|^2 |V_{q_3q_2}|^2 \, i\, \int \frac{d^4 l}{(2 \pi)^4} \, \frac{ l_\rho l_\sigma - p_\rho l_\sigma}{ (l^2 - m_2^2 + i \varepsilon) ((l-p)^2 - m_3^2 + i \varepsilon) }
  \nonumber \\[3mm]
 & \times  \Big( \bar b^i \gamma_\nu (1-\gamma_5) q_1^i \Big) \Big(  \bar q_1^j  \gamma_\mu (1-\gamma_5)  b^j \Big) 
 \Big( g^{\nu \rho} g^{\mu \sigma} + g^{\nu \sigma} g^{\mu \rho} - g^{\nu \mu} g^{\rho \sigma} + i \epsilon^{\nu \rho \sigma \mu} \Big) 
  \,.
 \label{eq:T-WA-11-2}
\end{align}\\
Again, since the one-loop integral in Eq.~(\ref{eq:T-WA-11-2}) is symmetric under the exchange $m_2 \leftrightarrow m_3$, its imaginary part can be obtained from Eq.~(\ref{eq:Im-WE}) making the replacement $r_1 \to r_3$, namely\\
\begin{align}
{\rm Im}\,  {\cal T}_{11}^{\rm WA} &  = \frac{N_c}{12 \pi} G_F^2 |V_{q_1b}|^2 |V_{q_3q_2}|^2 p^2   \sqrt{\lambda(1, r_3, r_2)} 
 \Bigg[ \Big( (r_3 - r_2)^2 + r_3 + r_2 -2 \Big) 
\nonumber \\[2mm]
& \times \Big( \bar b^i \Gamma_\nu q_1^i  \Big) \Big( \bar q_1^j \Gamma^\nu b^j \Big) - 2 \frac{ p^\mu p^\nu}{p^2} \Big( 2 (r_3 - r_2)^2 - (1 + r_3 + r_2) \Big) \Big( \bar b^i  \Gamma_\mu q_1^i  \Big) \Big( \bar q_1^j \Gamma_\nu b^j \Big) \Bigg]\,.
\label{eq:T-WA-11-3}
\end{align}\\
Note that Eq.~(\ref{eq:T-WA-11-3}) reproduces ${\rm Im}{\cal T}_{22}^{\rm WE}$ in Eq.~(\ref{eq:T-22-WE}) with the exchange $q_1 \leftrightarrow q_3$. This result is consequence of the Fierz identity in Eq.~(\ref{eq:Fierz-identity}), and by taking into account Eq.~(\ref{eq:Fierz-identity-2}) it follows that  \\
\begin{equation}
 {\cal T}_{mn}^{\rm WA} =  {\cal T}_{nm}^{{\rm WE}\,  (q_1 \leftrightarrow q_3)}\,.
 \label{eq:WE-WA-Fierz}
 \end{equation}\\
From Eq.~(\ref{eq:WE-WA-Fierz}), we see that the expression for Im${\cal T}^{\rm WA}$ can be immediately obtained from Eq.~(\ref{eq:ImT-WE}) by replacing $C_1 \leftrightarrow C_2$, and $q_1 \leftrightarrow q_3$, namely\\
\begin{align}
{\rm Im} \,{\cal T}^{\rm WA }&  = \frac{G_F^2}{12 \pi}  |V_{q_1b}|^2 |V_{q_3q_2}|^2 \, p^2   \sqrt{\lambda(1, r_3,r_2)}  \, \Bigg\{ k_5 \Bigg[\, \omega_1(r_3,r_2)  \,  \Big( \bar b^i \Gamma_\mu q_1^i  \Big) \Big( \bar q_1^j \Gamma^\mu b^j \Big)  
\nonumber \\[3mm]
&  - 2 \frac{p^\mu p^\nu}{p^2} \omega_2(r_3, r_2) \Big( \bar b^i \Gamma_\mu q_1^i  \Big) \Big( \bar q_1^j \Gamma_\nu b^j \Big) \Bigg]
+ k_6 
  \Bigg[ \omega_1(r_3, r_2) \Big( \bar b^i \Gamma_\mu  t^a_{ij} q_1^j  \Big) \Big( \bar q_1^l \Gamma^\mu  t^a_{lm} b^m \Big)
  \nonumber \\[3mm]
& - 2   \frac{p^\mu p^\nu}{p^2} \omega_2(r_3, r_2) \Big( \bar b^i \Gamma_\mu t^a_{ij} q_1^j  \Big) \Big( \bar q_1^l \Gamma_\nu   t^a_{lm} b^m \Big)\Bigg] \Bigg\} + \ldots \,,
\label{eq:ImT-WA}
\end{align}\\
with
\begin{align}
k_5 =  N_c  \, C_1^2 + 2 \, C_1 C_2 + \frac{1}{N_c}  \, C_2^2 \,, \qquad 
k_6 = 2 \, C_2^2\,.
\label{eq:WC-WA}
\end{align}\\
In deriving Eqs.~(\ref{eq:ImT-WE}), (\ref{eq:ImT-PI}) and (\ref{eq:ImT-WA}) we have only neglected power corrections due to the expansion of the quark propagator, however, because $p^\mu$ depends on the residual component of the heavy quark momentum $k^\mu$ as well as on the soft momentum of the light spectator quark $p^\mu_{q_i}$, the expressions above contain also the information about the interaction of the heavy quark with the light degrees of freedom. In order to single out the dimension-six result, we set $p^\mu = m_b v^\mu + {\cal O}(1/m_b)$ and introduce the rescaled field $b(x) = \exp(- i m_b v \cdot x) b_v(x)$,  see  Eq.~(\ref{eq:Qx-HQE}), satisfying $b_v(x) = \slashed v b_v(x) + {\cal O}(1/m_b) $, cf.\ Eq.~(\ref{eq:p-Qv}). We then obtain that\\
\begin{equation}
\frac{p^\mu p^\mu }{p^2} \Big( \bar b \gamma_\mu (1- \gamma_5) q_i  \Big)  \Big( \bar q_i \gamma_\nu (1-\gamma_5) b \Big)  = \Big( \bar b_v (1- \gamma_5) q_i  \Big)  \Big( \bar q_i  (1+\gamma_5) b_v\Big)  + {\cal O}\left( \frac{1}{m_b} \right)\,,
\end{equation}\\
where for simplicity we have not specified the colour structure, and that the dimension-six contributions to ${\rm Im}{\cal T}^{\rm WE}$, ${\rm Im}{\cal T}^{\rm PI}$ and ${\rm Im}{\cal T}^{\rm WA}$ are given respectively by\\
\begin{align}
{\rm Im} \,{\cal T}^{\rm WE }_6 &  = \frac{G_F^2}{12 \pi} |V_{q_1b}|^2 |V_{q_3q_2}|^2 m_b^2 \sqrt{\lambda(1, x_1, x_2)} \,
\Bigg\{ k_1 \Big[ \, \omega_1(x_1,x_2)\, { O}^{(q_3)}_{1} 
 - 2  \omega_2(x_1, x_2) { O}^{(q_3)}_{2}  \Big]
 \nonumber \\[3mm]
& + k_2 \Big[
 \omega_1(x_1, x_2) \tilde { O}^{(q_3)}_{1} 
- 2  \omega_2(x_1, x_2) \tilde {O}^{(q_3)}_{2} \Big] \Bigg\}\,,
\label{eq:ImT-WE-6}
\end{align}\\
\begin{align}
{\rm Im} \,{\cal T}^{\rm PI }_6 & =   \frac{G_F^2}{2 \pi}  |V_{q_1b}|^2 |V_{q_3q_2}|^2 m_b^2  \sqrt{\lambda(1, x_1, x_3)}  \, (1 - x_1 - x_3) \, \Big[ k_3  {O}^{(q_2)}_{1} + k_4 \tilde {O}^{(q_2)}_{1} \Big]
\,,
\label{eq:ImT-PI-6}
\end{align}\\
and\\
\begin{align}
{\rm Im} \,{\cal T}^{\rm WA }_6 &  = \frac{G_F^2}{12 \pi} |V_{q_1b}|^2 |V_{q_3 q_2}|^2 m_b^2 \sqrt{\lambda(1, x_3, x_2)} \,
\Bigg\{ k_5 \Big[ \, \omega_1(x_3,x_2)\, { O}^{(q_1)}_{1} 
 - 2  \omega_2(x_3, x_2) { O}^{(q_1)}_{2}  \Big]
 \nonumber \\[3mm]
& + k_6 \Big[
 \omega_1(x_3, x_2) \tilde { O}^{(q_1)}_{1} 
- 2  \omega_2(x_3, x_2) \tilde {O}^{(q_1)}_{2} \Big] \Bigg\}\,.
\label{eq:ImT-WA-6}
\end{align}\\
In Eqs.~(\ref{eq:ImT-WE-6})-(\ref{eq:ImT-WA-6}), $x_i$ denotes the dimensionless mass parameters $x_i = m_i^2/m_b^2 $, and the following basis for the dimension-six four-quark operators has been introduced \\
\begin{align}
& {O}^{(q)}_{1} = \Big( \bar b^i_v \gamma_\nu  (1-\gamma_5) q^i  \Big)  \Big( \bar q^j  \gamma^\nu (1-\gamma_5) b^j_v \Big)\,,
\label{eq:Ova-4q}
\\[3mm]
&  { O}^{(q)}_{2} = \Big( \bar b^i_v (1-\gamma_5) q^i  \Big) \Big( \bar q^j  (1+\gamma_5)  b^j_v \Big)\,, 
\label{eq:Osp-4q}
\\[3mm]
& \tilde {O}^{(q)}_{1} = \Big( \bar b^i_v \gamma_\nu  (1-\gamma_5)  t^a_{ij}  q^j \Big) \Big( \bar q^l \gamma^\nu (1-\gamma_5) t^a_{lm} b^m_v \Big)\,,
\label{eq:Tva-4q}
\\[3mm]
&\tilde {O}^{(q)}_{2} = \Big( \bar b^i_v  (1-\gamma_5) t^a_{ij} q^j \Big) \Big( \bar q^l (1+\gamma_5) t^a_{lm} b^m_v \Big)\,.
\label{eq:Tsp-4q}
\end{align}\\


\subsection{Computation of Im${\cal T}^{ (4q)}_7$}
\label{sec:dim-7-four-q}
By including $1/m_b$ corrections to the incoming four-momentum $p^\mu$ as well as to the heavy quark field $b(x)$, in Eqs.~(\ref{eq:ImT-WE}), (\ref{eq:ImT-PI}), and (\ref{eq:ImT-WA}), leads to dimension-seven contributions to the WE, PI, and WA topologies. In the following, we discuss in detail only the first two cases since the corresponding expression for WA can be immediately obtained using Eq.~(\ref{eq:WE-WA-Fierz}). However, 
before considering the specific contributions separately, it is convenient to derive some general results which will facilitate the computation. 
We recall that the incoming momentum $p^\mu$ is the sum (for WE and WA), or the difference (for PI), of the $b$-quark and of the light-quark momentum, i.e.\  $p^\mu = p_b^\mu \pm p_q^\mu$, here $p^\mu_b = m_b v^\mu + k^\mu$, while $q = q_1, q_2, q_3$, respectively for the case of WA, PI and WE. Taking into account that $k \sim p_q \ll m_b$, the square of the incoming momentum can be written, up to terms of order ${\cal O}(1/m_b^2)$, as $p^2 = m_b^2 \, (1 + z)$, with the small parameter $z$ given by\\
\begin{equation}
z =  2 \,  \frac{v \cdot k}{m_b} \pm \,2 \, \frac{v \cdot p_{q}}{m_b} \ll1\,.
\label{eq:z}
\end{equation}\\
Correspondingly, the heavy quark field $b(x)$, using the framework of the HQET, see Section~\ref{sec:HQET}, can be expanded in powers of $1/(2 m_b)$, as\\
\begin{equation}
b(x) = e^{- i m_b v \cdot x} \left( h_v(x) + \frac{i \slashed D_\perp}{2 m_b} h_v(x) +{\cal O}\left(\frac{1}{m_b^2}\right) \right)\,,
\label{eq:bv-exp-HQET-d7}
\end{equation}\\
where $h_v(x)$ denotes the effective heavy quark field, which coincides with $b_v(x)$ at leading order in $1/m_b$, and which obeys to the equation of motion $(i v \cdot D) h_v(x) = 0$, following from the HQET Lagrangian in Eq.~(\ref{eq:L-HQET}). However, in the original QCD Lagrangian in Eq.~(\ref{eq:Eff-L}), there are also subleading contributions suppressed by the heavy quark mass, which are treated as perturbations to ${\cal L}_{HQET}$, cf.\ ${\cal L}_{power}$ in Eq.~(\ref{eq:L-eff-3}), so that, in the expansion of the matrix element of an operator containing the heavy quark field, their effect must be included in a standard perturbative way by taking the time ordered product of ${\cal L}_{power}$ with the corresponding leading order operators, see Eq.~(\ref{eq:Exp-vector-current}). It then follows that \\ 
\begin{align}
\bar b \Gamma_\mu q \bar q \Gamma_\nu b
& \simeq \bar h_v  \Gamma_\mu q  \bar q \Gamma_\nu h_v + \frac{1}{2 m_b} \bigg[  \bar h_v  \Gamma_\mu q  \bar q  \Gamma_\nu i \slashed D h_v  + \bar h_v (- i \overset{\leftarrow}{\slashed D}) \Gamma_\mu q \bar q \Gamma_\nu h_v  \bigg] 
\nonumber \\[3mm]
 & + \frac{1}{m_b}  i \int d^4y\,   {\rm T} \Big\{ \bar h_v  \Gamma_\mu q  \bar q \Gamma_\nu  h_v , {\cal L}_1 (y) \Big\} +{\cal O}\left(\frac{1}{m_b^2}\right)  \,,
\label{eq:exp-four-quark-dim7}
\end{align}\\
and for brevity we have not specified the colour structure.
Note that in Eq.~(\ref{eq:exp-four-quark-dim7}), the equation of motion for $h_v(x)$ has been used to replace the action of $D_\perp^\mu $ with that of $D_\mu$ and that the symbol $\simeq$ refers to the fact that left and right hand side of the equation must be evaluated respectively between QCD and HQET states, cf.\ Eq.~(\ref{eq:Exp-vector-current}) and see for more details Ref.~\cite{Neubert:1996we} \footnote{ However, for simplicity, in the following we will just use an equal sign.}.
Moreover, 
the non local operator in the second line of Eq.~(\ref{eq:exp-four-quark-dim7}), parametrises the contribution of the first power  correction to the QCD Lagrangian, i.e.\ ${\cal L}_1$ in Eq.~(\ref{eq:Lpower-1}). 
\\[2mm]
Taking into account that the action of a derivative on $h_v(x)$, returns only small frequencies of the order of $k$ and that $q(x)$ describes an outgoing antiquark with momentum $p_q^\mu$, we respectively have \footnote{Recall that in the FS gauge $A_\mu (0) =0 $, and that the action of a partial derivative at the origin can be replaced with that of a covariant derivative, see Section~\ref{sec:FS} and also Appendix~\ref{app:1}.} \\
\begin{equation}
v \cdot k \, \Big( \bar h_v  \Gamma_\mu q  \bar q \Gamma_\nu  h_v \Big) =\lim_{x \to 0} \Big( \bar h_v(x) \Gamma_\mu  q(x)  \bar q(x)  \Gamma_\nu i v \cdot D h_v(x) \Big) = 0\,,
\label{eq:z-vk}
\end{equation}\\
while\\
\begin{equation}
v \cdot p_{q}\, \Big( \bar h_v  \Gamma_\mu q  \bar q \Gamma_\nu h_v  \Big) 
 = \lim_{x \to 0}  \Big( \bar h_v (x)  \Gamma_\mu (- i v \cdot D) q(x) \bar q(x) \Gamma_\nu h_v(x) \Big) 
 = - \bar h_v \Gamma_\mu  i v \cdot D q  \bar q \Gamma_\nu h_v 
 \,,
 \label{eq:z-vpq}
\end{equation}\\
where Eq.~(\ref{eq:z-vk}) vanishes due to the equation of motion for $h_v$, and in Eq.~(\ref{eq:z-vpq}), the small momentum $p_q$ has resulted in a dimension-seven operator with a covariant derivative acting on $q$. 
Consider now the expansion of the product of two momenta, namely\\
\begin{equation}
p^\mu p^\nu = m_b^2 \, \left( v^\mu v^\nu + \frac{v^\mu  k^\nu}{m_b} + \frac{v^\nu  k^\mu}{m_b} +  \frac{v^\mu p_{q}^\nu}{m_b} +  \frac{v^\nu p_{q}^\mu}{m_b} +{\cal O}\left(\frac{1}{m_b^2}\right)  \right)\,,
\end{equation}\\
which, combined with the result in Eq.~(\ref{eq:exp-four-quark-dim7}), yields\\
\begin{align}
& \hspace*{6.5cm} p^\mu p^\nu  \Big(\bar b\Gamma_\mu q \bar q \Gamma_\nu  b  \Big)
\nonumber \\[3mm]
& \hspace*{1cm }= m_b^2 \, \Bigg(\bar h_v (1- \gamma_5) q   \bar q  (1 + \gamma_5)  h_v 
 + \frac{1}{m_b}  
 \bar h_v (1- \gamma_5) q  \bar q (1 + \gamma_5) i \slashed D h_v 
 \nonumber \\[3mm]
& 
+ \frac{1}{m_b} i \int d^4 y {\rm T}\Big\{ \bar h_v (1- \gamma_5) q   \bar q  (1 + \gamma_5)  h_v , {\cal L}_1(y)\Big\}  
-  2 \frac{m_q}{m_b}   \bar h_v (1- \gamma_5)q  \bar q (1 - \gamma_5) h_v + \ldots
\Bigg) \,,
\label{eq:pmu-pnu}
\end{align}\\
where the ellipsis stand for terms of order ${\cal O}(1/m_b^2)$. Notice that in Eq.~(\ref{eq:pmu-pnu}), the anti-commutation relation $\slashed v \slashed D = - \slashed D \slashed v + 2 v\cdot D$, 
and the corresponding equations of motion for $h_v(x)$ and for the light field, i.e.\ $( i \slashed D - m_q ) q(x) = 0$,
have been used. Moreover, we have taken into account that operators related by Hermitian conjugation lead to the same matrix element, hence a factor of 2 has been included. Specifically these are\\
 \begin{equation}
\bar h_v (- i \overset{\leftarrow}{\slashed D}) (1 - \gamma_5)  \bar q ( 1+ \gamma_5) h_v = \Big[ \bar h_v (1 - \gamma_5)  q \bar q (1+ \gamma_5) (i \slashed D) h_v \Big]^\dagger\,,
\end{equation}\\
and
\begin{equation}
 \bar h_v (1+ \gamma_5)q  \bar q (1+  \gamma_5) h_v = \Big[  \bar h_v (1- \gamma_5)q  \bar q (1 - \gamma_5) h_v
 \Big]^\dagger\,.
\end{equation}\\
Finally, we introduce the dimension-seven four-quark operator basis, which for clarity, we split into three categories, namely \\
\begin{align}
{\cal P}^{(q)}_1 &= m_q \, \big( \bar h_v^i (1-\gamma_5) q^i \big) \big( \bar q^j (1- \gamma_5) h^j_v \big)\,,
\label{eq:P1-d7}
\\[3mm]
{\cal P}^{(q)}_2 &=  \big( \bar h^i_v \gamma_\mu (1-\gamma_5) i v \cdot D q^i \big) \big( \bar q^j \gamma^\mu (1- \gamma_5) h^j_v \big)\,,
\\[3mm]
{\cal P}^{(q)}_3&=  \big( \bar h^i_v (1-\gamma_5) i v \cdot D q^i \big) \big( \bar q^j (1+ \gamma_5) h^j_v \big)\,,
\end{align}\\
parametrising the effect of the light-quark momentum $p_q$,\\
\begin{align}
{\cal R}^{(q)}_1 &=  \big( \bar h^i_v \gamma_\mu (1-\gamma_5)  q^i \big) \big( \bar q^j  \gamma^\mu (1- \gamma_5) i \slashed D   h_v^j \big)\,,
\\[3mm]
{\cal R}^{(q)}_2 &=  \big( \bar h^i_v (1-\gamma_5)  q^i \big) \big( \bar q^j  (1+ \gamma_5) i \slashed D h_v^j \big)\,,
\end{align}\\
describing local contributions due to the expansion of the heavy-quark field,
and \\
\begin{align}
{\cal M}_1^{(q)} &= i \int d^4 y {\rm T}\Big\{  {\cal O}^{(q)}_{1} , {\cal O}_I(y) \Big\}\,,
\\[3mm]
{\cal M}_2^{(q)} &= i \int d^4 y {\rm T}\Big\{  {\cal O}^{(q)}_{1}  , {\cal O}_{II}(y) \Big\}\,,
\\[3mm]
{\cal M}_3^{(q)} &= i \int d^4 y {\rm T}\Big\{ {\cal O}^{(q)}_{2}  , {\cal O}_I(y) \Big\}\,,
\\[3mm]
{\cal M}_4^{(q)} &= i \int d^4 y {\rm T}\Big\{  {\cal O}^{(q)}_{2}  , {\cal O}_{II}(y) \Big\}\,,
\label{eq:M4-d7}
\end{align}\\
with ${\cal O}_{I, II}$ defined in Eqs.~(\ref{eq:O-kin-HQET}), (\ref{eq:O-magn-HQET}),
corresponding to the non-local contributions generated by taking the time ordered product of the $1/m_b$ correction in the Lagrangian in Eq.~(\ref{eq:L-eff-3}), with the dimension-six local HQET operators\\
\begin{align}
& {\cal O}^{(q)}_{1} = \Big( \bar h^i_v \gamma_\nu  (1-\gamma_5) q^i  \Big)  \Big( \bar q^j  \gamma^\nu (1-\gamma_5) h^j_v \Big)\,,
\label{eq:Ova-4q-hqet}
\\[3mm]
&  { \cal O}^{(q)}_{2} = \Big( \bar h^i_v (1-\gamma_5) q^i  \Big) \Big( \bar q^j  (1+\gamma_5)  h^j_v \Big)\,, 
\label{eq:Osp-4q-hqet}
\\[3mm]
& \tilde {\cal O}^{(q)}_{1} = \Big( \bar h^i_v \gamma_\nu  (1-\gamma_5)  t^a_{ij}  q^j \Big) \Big( \bar q^l \gamma^\nu (1-\gamma_5) t^a_{lm} h^m_v \Big)\,,
\label{eq:Tva-4q-hqet}
\\[3mm]
&\tilde {\cal O}^{(q)}_{2} = \Big( \bar h^i_v  (1-\gamma_5) t^a_{ij} q^j \Big) \Big( \bar q^l (1+\gamma_5) t^a_{lm} h^m_v \Big)\,.
\label{eq:Tsp-4q-hqet}
\end{align}\\
Note that the $1/m_b$ contributions arising from the expansion of the heavy quark momentum vanish due to the equation of motion for $h_v(x)$.
Apart from the operators in Eqs.~(\ref{eq:P1-d7})-(\ref{eq:M4-d7}), the basis includes also the corresponding colour-octet operators $\tilde {\cal P}_1^{(q)}, \tilde {\cal P}_2^{(q)}, \tilde {\cal P}_3^{(q)},$ $\tilde {\cal R}_1^{(q)}, \tilde {\cal R}_2^{(q)}$, containing respectively the colour matrices $t^a$, and $\tilde {\cal M}_1^{(q)}, \tilde {\cal M}_2^{(q)}, \tilde {\cal M}_3^{(q)} \tilde {\cal M}_4^{(q)}$, in which we must replace ${\cal O}_i \to \tilde {\cal O}_i$. For brevity however, we omit to explicitly show them.
\\[2mm]
With the above results, it is straightforward to obtain the expansion of Eq.~(\ref{eq:ImT-WE}), up to order ${\cal O}(1/m_b^2)$, namely
\begin{align}
{\rm Im} \,{\cal T}^{\rm WE } &= \frac{G_F^2}{12 \pi}   |V_{q_1b}|^2 |V_{q_3q_2}|^2  m_b^2  \sqrt{\lambda(1, x_1, x_2)}   \,\left( 1 + \frac{(1- x_1 - x_2)}{\lambda(1, x_1, x_2)} z  \right) 
\nonumber \\[3mm]
& \times  \Bigg\{ k_1\,  \Bigg[\big( \omega_1(x_1, x_2) - y_1(x_1, x_2) \, z \big) 
\Big( {\cal O}_{1}^{(q_3)}  + \frac{ {\cal R}_1^{(q_3)} }{m_b} + \frac{ {\cal M}_1^{(q_3)} }{m_b} +  \frac{{\cal M}_2^{(q_3)}}{m_b}   \Big) 
\nonumber \\[3mm]
&- 2 \, \big(\omega_2 (x_1, x_2) - y_2 (x_1, x_2) \, z \big) \left(1- z \right)  \Big(  {\cal O}_{2}^{(q_3)}  + \frac{ {\cal R}_2^{(q_3)}}{m_b} + \frac{ {\cal M}_3^{(q_3)} }{m_b} + \frac{ {\cal M}_4^{(q_3)} }{m_b} - 2 \frac{ {\cal P}_1^{(q_3)}}{m_b} \Big) 
 \Bigg] 
\nonumber \\[2mm]
&+ \Big( k_1 \to k_2 , \, {\rm singlet} \to {\rm octet} \Big) +{\cal O}\left(\frac{1}{m_b^2}\right)   \Bigg\}  \,,
\label{eq:ImT-WE-expanded}
\end{align}\\ 
where
\begin{align}
y_1(a,b) = 2 \, (a -b)^2 + (a+b) \,, \qquad 
y_2(a,b) =  4 \, (a - b)^2 - (a+b) \,,
\end{align}\\
and the small parameter $z$ is defined in Eq.~(\ref{eq:z}) (with the plus sign). 
The leading order result in Eq.~(\ref{eq:ImT-WE-expanded}) reproduces the dimension-six expression obtained in Eq.~(\ref{eq:ImT-WE-6}), but with the QCD operators, i.e.\ $Q_i^{(q_3)}, \tilde Q_i^{(q_3)}$, see Eqs.~(\ref{eq:Ova-4q})-(\ref{eq:Tsp-4q}), replaced by the corresponding HQET ones i.e.\ ${\cal O}_i^{(q_3)}, \tilde {\cal O}_i^{(q_3)}$, see Eqs.~(\ref{eq:Ova-4q-hqet})-(\ref{eq:Tsp-4q-hqet}). Moreover, using Eqs.~(\ref{eq:z-vk}), (\ref{eq:z-vpq}), to rewrite $z$ in terms of derivatives acting on light-quark field, we arrive at the final dimension-seven contribution, which reads\\
\begin{align}
{\rm Im} \,{\cal T}^{\rm WE}_7 &= \frac{G_F^2}{12 \pi}  |V_{q_1b}|^2 |V_{q_3q_2}|^2 m_b  \sqrt{\lambda(1, x_1,x_2)} \, \Bigg\{   
 k_1  \Bigg[  \omega_1(x_1, x_2) \Big( {\cal R}_1^{(q_3)} +   {\cal M}_1^{(q_3)}  +  {\cal M}_2^{(q_3)} \Big)
\nonumber \\[3mm]
& +2\,   \frac{(1- x_1 - x_2)}{\lambda(1, x_1, x_2)}\, \Big( 2 \, \omega_2(x_1, x_2) \, {\cal P}_3^{(q_3)} - \omega_1(x_1, x_2) \, {\cal P}_2^{(q_3)} \Big)
\nonumber \\[3mm]
& + 2\, \omega_2(x_1, x_2) \Bigg( 2\, \Big( {\cal P}_1^{(q_3)}  -   {\cal P}_3^{(q_3)} \Big) - \Big(  {\cal R}_2^{(q_3)} +  {\cal M}_3^{(q_3)}  +  {\cal M}_4^{(q_3)} \Big) \Bigg) 
\nonumber \\[3mm]
& + 2 \, \Big( y_1(x_1, x_2) {\cal P}_2^{(q_3)} - 2 \, y_2(x_1, x_2)  \, {\cal P}_3^{(q_3)} \Big) \Bigg]
\nonumber \\[2mm]
&+ \Big( k_1 \to k_2 , \, {\rm singlet} \to {\rm octet} \Big)   \Bigg\} \,.
\label{eq:ImT-WE-7}
\end{align}\\ 
Similarly, 
it follows that the expansion of Eq.~(\ref{eq:ImT-PI}) up order ${\cal O}\left(1/m_b^2\right)$, yields\\
\begin{align}
{\rm Im} \,{\cal T}^{\rm PI}& =  \frac{G_F^2}{2 \pi}  |V_{q_1b}|^2 |V_{q_3q_2}|^2 m_b^2  \sqrt{\lambda(1, x_1, x_3)} \, (1 - x_1 -x_3) 
\nonumber \\[3mm]
& \times   \Bigg[1+ \Bigg(\frac{ (x_1 + x_3) }{(1 - x_1 -x_3) } + \frac{(1 - x_1 -x_3) }{\lambda(1, x_1,x_3)} \Bigg) \, z \Bigg] 
\nonumber \\[3mm]
& \times \Bigg\{ k_3\,  \Big[  {\cal O}_{1}^{(q_2)}+ \frac{  {\cal R}_1^{(q_2)} }{m_b} +  \frac{ {\cal M}_1^{(q_2)} }{m_b} + \frac{ {\cal M}_2^{(q_2)} }{m_b}
\Big]
\nonumber \\[3mm]
& 
+ \Big( k_3 \to k_4 \,, \, {\rm singlet} \to {\rm octet} \Big) +{\cal O}\left(\frac{1}{m_b^2}\right)   \Bigg\} \,,
\label{eq:ImT-PI-exp}
\end{align}\\
with the small parameter $z$ defined in Eq.~(\ref{eq:z}) (with the minus sign).
Again, the leading order contribution in Eq.~(\ref{eq:ImT-PI-exp}) reproduces to the dimension-six expression obtained in Eq.~(\ref{eq:ImT-PI-6}), but with the QCD operators, i.e.\ $Q_i^{(q_2)}, \tilde Q_i^{(q_2)}$, see Eqs.~(\ref{eq:Ova-4q})-(\ref{eq:Tsp-4q}) replaced by the corresponding HQET ones i.e.\ ${\cal O}_i^{(q_2)}, \tilde {\cal O}_i^{(q_2)}$, see Eqs.~(\ref{eq:Ova-4q-hqet})-(\ref{eq:Tsp-4q-hqet}). Moreover, using Eqs.~(\ref{eq:z-vk}), (\ref{eq:z-vpq}), to rewrite $z$ in terms of derivatives acting on light-quark field we arrive at final dimension-seven result, which reads\\\
\begin{align}
{\rm Im} \,{\cal T}^{\rm PI}_7& =  \frac{G_F^2}{2 \pi}  |V_{q_1b}|^2 |V_{q_3q_2}|^2 m_b \sqrt{\lambda(1, x_1, x_3)} \, (1 - x_1 -x_3)  \,
\nonumber \\[3mm]
& \times \Bigg\{ k_3 \, \Bigg[   {\cal R}_1^{(q_2)} +   {\cal M}_1^{(q_2)}  +  {\cal M}_2^{(q_2)}   + 2\, \Bigg(\frac{ (x_1 + x_3) }{(1 - x_1 -x_3) } + \frac{(1 - x_1 -x_3) }{\lambda(1, x_1,x_3)} \Bigg)  \, {\cal P}^{(q_2)}_2 \Bigg] 
\nonumber \\[3mm]
& + \, \Bigg( k_3 \to k_4 , \, {\rm singlet} \to {\rm octet} \Bigg) \Bigg\} \,.
\label{eq:ImT-PI-7}
\end{align}\\
Finally, the corresponding dimension-seven contribution to the WA topology is simply obtained by replacing $C_1 \leftrightarrow C_2$, and $q_1 \leftrightarrow q_3$, into Eq.~(\ref{eq:ImT-WE-7}), namely\\
\begin{align}
{\rm Im} \,{\cal T}^{\rm WA}_7 &= \frac{G_F^2}{12 \pi}  |V_{q_1b}|^2 |V_{q_3q_2}|^2 m_b  \sqrt{\lambda(1, x_3,x_2)} \, \Bigg\{   
 k_5  \Bigg[  \omega_1(x_3, x_2) \Big( {\cal R}_1^{(q_1)} +   {\cal M}_1^{(q_1)}  +  {\cal M}_2^{(q_1)} \Big)
\nonumber \\[3mm]
& +2\,   \frac{(1- x_3 - x_2)}{\lambda(1, x_3, x_2)}\, \Big( 2 \, \omega_2(x_3, x_2) \, {\cal P}_3^{(q_1)} - \omega_1(x_3, x_2) \, {\cal P}_2^{(q_1)} \Big)
\nonumber \\[3mm]
& + 2\, \omega_2(x_3, x_2) \Bigg( 2\, \Big( {\cal P}_1^{(q_1)}  -   {\cal P}_3^{(q_1)} \Big) - \Big(  {\cal R}_2^{(q_1)} +  {\cal M}_3^{(q_1)}  +  {\cal M}_4^{(q_1)} \Big) \Bigg) 
\nonumber \\[3mm]
& + 2 \, \Big( y_1(x_3, x_2) {\cal P}_2^{(q_1)} - 2 \, y_2(x_3, x_2)  \, {\cal P}_3^{(q_1)} \Big) \Bigg]
\nonumber \\[2mm]
&+ \Big( k_5 \to k_6 , \, {\rm singlet} \to {\rm octet} \Big)   \Bigg\} \,.
\label{eq:ImT-WA-7}
\end{align}\\ 
We emphasise that the results in Eqs.~(\ref{eq:ImT-WE-7}), (\ref{eq:ImT-PI-7}), and (\ref{eq:ImT-WA-7}), can be applied without any difference, to the description of baryons as well. In fact, in this case, the sign change in front of the light-quark momentum i.e.\ $p^\mu = p_b^\mu \mp p_q^\mu$, is compensated by the same sign change in front of the operators ${\cal P}_i^{(q)}$, $i = 1, 2,3$, since now $q(x)$ describes an incoming quark with momentum $p_q^\mu$. \\
To conclude, we note that we have derived the dimension-seven contribution in terms of operators defined in HQET, the corresponding expressions in terms of QCD fields are obtained by expanding Eqs.~(\ref{eq:ImT-WE}), (\ref{eq:ImT-PI}), and (\ref{eq:ImT-WA}), only in the small momentum $p_q$ of the light spectator quark, but not in the HQET field. This results in the following basis\\
\begin{align}
{ P}^{(q)}_1 &= m_q \, \big( \bar b^i (1-\gamma_5) q^i \big) \big( \bar q^j (1- \gamma_5) b^j \big)\,,
\label{eq:P1}
\\[3mm]
{ P}^{(q)}_2 &= \frac{1}{m_b} \big( \bar b^i \overset{\leftarrow}{D_\nu} \gamma_\mu (1-\gamma_5)  D^\nu q^i \big) \big( \bar q^j \gamma^\mu (1- \gamma_5) b^j \big)\,,
\label{eq:P2}
\\[3mm]
{ P}^{(q)}_3&= \frac{1}{m_b} \big( \bar b^i \overset{\leftarrow}{D_\nu} (1-\gamma_5) D^\nu q^i \big) \big( \bar q^j (1+ \gamma_5) b^j \big)\,,
\label{eq:P3}
\end{align}\\
together with the colour-octet operators $\tilde P^{(q)}_i$. Due to the presence, in Eqs.~(\ref{eq:P2}), (\ref{eq:P3}), of a covariant derivative acting on the $b$-quark field, which scales as $m_b$ at this order, there is no explicit power counting, differently to the HQET basis. The corresponding QCD result for Eqs.~(\ref{eq:ImT-WE-7}), (\ref{eq:ImT-PI-7}), (\ref{eq:ImT-WA-7}), can be immediately obtained by setting in these expressions, 
 ${\cal R}^{(q)}_i, {\cal R}^{(q)}_i, {\cal M}^{(q)}_i, {\cal M}^{(q)}_i$ to zero and by replacing ${\cal P}^{(q)}_i \to  P^{(q)}_i$ and $\tilde{\cal  P}^{(q)}_i \to \tilde P^{(q)}_i$. We stress that, in this case, the difference in the operator basis is compensated by the different parametrisation of the corresponding matrix elements in QCD and in HQET, cf.\ Chapter~\ref{ch:pheno} and Appendix~\ref{app:7}.


\chapter{Contribution of the Darwin Operator}
\label{ch:Darwin}
In Section~\ref{sec:2q-contr} we have discussed the calculation of the dimension-three and dimension-five contributions to the HQE of a $B$ meson due to the single mode $b \to c \bar u d $ and using the representation of the quark-propagator in coordinate space given in Eqs.~(\ref{eq:prop-coordinate-space}), (\ref{eq:prop-coordinate-space-y}). In the present chapter, we generalise the above results by computing the coefficients of the two-quark operators in the HQE, for the generic $b \to q_1 \bar q_2 q_3$ transition and up to terms of order $1/m_b^3$, where the latter describe the contribution of the Darwin operator, see Eq.~(\ref{eq:O-Darwin}) \footnote{We stress that again, the computation is performed only at LO-QCD.}. The representation derived in Ref.~\cite{Blok:1992hw} for the calculation of the imaginary part of integrals containing Bessel functions, is not sufficient for the case in which more than one propagator is massive, hence now the whole computation is performed in momentum space, using the expression of the quark-propagator given in Eqs.~(\ref{eq:S1p}), (\ref{eq:S1p-tilde}). Specifically, we start  in Section~\ref{sec:Dar-sec-1}, by computing the expansion of ${\rm Im} {\cal T}^{(2q)}$ up to order $1/m_b^3$. We will find that the coefficients of the Darwin operator develop IR divergences in correspondence of the emission of a soft gluon from a light quark propagator. In particular, the singularities originate from the expansion of the propagator of the up, down, and strange quarks, which we consider massless, and are logarithmic, namely the corresponding coefficients have the asymptotic form $\sim \log(m_q^2/m_b^2)$ in the limit $m_q \to 0$, with $q = u,d,s$. As described e.g.\ in Ref.~\cite{Novikov:1980uj}, these logarithmic infrared divergences are due to the mixing between operators of the same dimension under renormalisation. This is discussed in Section~\ref{sec:Dar-sec-2}, where we compute the one-loop diagram describing the mixing of the four-quark and the Darwin operators and perform their renormalisation in order to ensure the cancellation of the IR divergences. Finally, the complete expressions for the coefficients of the Darwin operator are presented in Section~\ref{sec:Dar-sec-3} together with a discussion of the results. The structure closely follows the one of Ref.~\cite{Lenz:2020oce}. 


\section{Computation of $\Gamma^{(2q)}(B)$ up to order $1/m_b^3$}
\label{sec:Dar-sec-1}
The starting point for the computation of the contribution of two-quark operators is Eq.~(\ref{eq:T2q}), which we state here again for practicality, i.e.  \footnote{Note that the superscript $(q_1 \bar q_2 q_3)$ is often omitted for the sake of a cleaner notation, however it must be always understood.}\\
\begin{align}
{\cal T}^{ (2q)} = C_1^2 \,  {\cal T}^{ (2q)}_{11}+ \,  2 \,C_1 C_2 \, \left(\frac{1}{N_c} {\cal T}^{ (2q)}_{11} +  2 \, {\cal T}^{ (2q)}_{13} \right)  +C_2^2 \, {\cal T}^{ (2q)}_{22}\,,
\label{eq:T2q-D}
\end{align}\\
with Eqs.~(\ref{eq:T2q-11}), (\ref{eq:T2q-13}), compactly written as\\
\begin{align}
{\cal T}_{11 \{13\}}^{ (2q)}& =  - \frac{G_F^2}{2}  |V_{q_1b}|^2 |V_{q_3q_2}|^2 \,  i  \int  d^4 x \, \, \bar b\, \gamma_\nu (1-\gamma_5) \{t^a\} i S^{( q_1)} (0, x) \gamma_\mu (1-\gamma_5) b(x) 
\nonumber \\[3mm]
& \times    {\rm Tr} \Biggl[ \gamma^\nu (1-\gamma_5) \{t^a\} i S^{( q_3)}(0, x) \gamma^\mu (1-\gamma_5) i S^{( q_2)} (x,0) \Biggr]  + (x \leftrightarrow 0)
\, .
\label{eq:T2q-11-D}
\end{align}\\
Note that only in the case of ${\cal T}_{13}^{ (2q)}$ the $SU(3)_c$ generators $t^a$ appear on the r.h.s.\ of Eq.~(\ref{eq:T2q-11-D}), as indicated by the curly brackets, and that a summation over colour indices is understood. Furthermore, the corresponding expression for ${\cal T}_{22}^{ (2q)}$ has been omitted, since it can be derived from that of ${\cal T}_{11}^{(2q)}$ by taking into account the result in Eq.~(\ref{eq:Fierz-13}).

\begin{figure}
\centering
\includegraphics[scale=0.5]{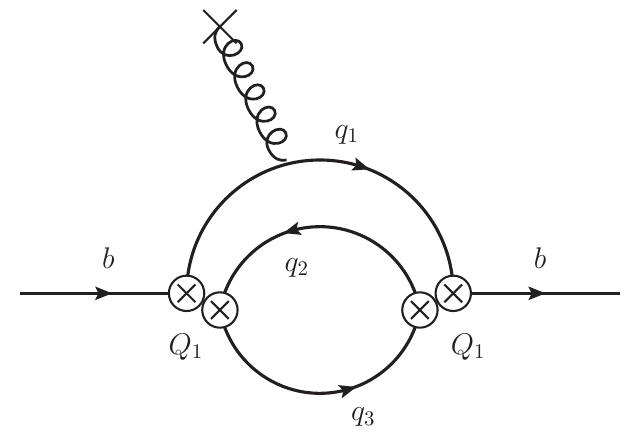}
\includegraphics[scale= 0.5]{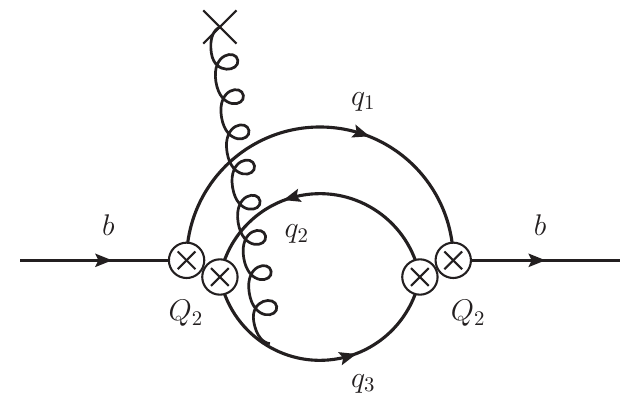}
\caption{Two-loop diagrams describing power corrections due to the expansion of the quark propagator up to order $1/m_b^3$, from the $Q_1 \otimes Q_1$ (left) and $Q_2 \otimes Q_2$ (right) contributions.}
\label{fig:O1O1}
\end{figure}

\begin{figure}
\centering
\includegraphics[scale = 0.5]{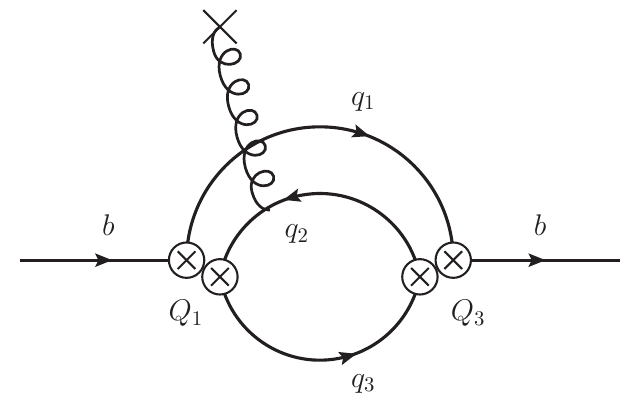}
\includegraphics[scale = 0.5]{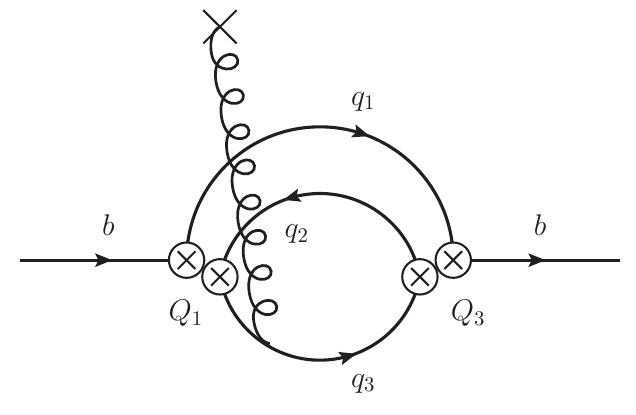}
\caption{Two-loop diagrams describing power corrections due to the expansion of the quark propagator up to order $1/m_b^3$, from the $Q_1 \otimes Q_3$ contribution.}
\label{fig:O1O3}
\end{figure}

Being interested in the expansion of ${\cal T}^{(2q)}$ up to order $1/m_b^3$, we must now substitute in Eq.~(\ref{eq:T2q-11-D}), the complete expression of the quark propagator derived in Section~\ref{sec:FS}, namely including also terms proportional to one covariant derivative of the gluon field strength tensor $D_\rho G_{\mu \nu}$. It is worth remarking that the colour structure of Eq.~(\ref{eq:T2q-11-D}) allows for a straightforward treatment of colour, in complete analogy to what has already been discussed  in Section~\ref{sec:Gamma-5}. To this end, it is convenient to single out the colour structure of the propagator in Eq.~(\ref{eq:Sp-01}), which schematically reads \\
 \begin{equation}
{\cal S}_{jk}^{(q_i)}(p) ={\cal S}_{0}^{(q_i)} (p) \, \delta_{jk} + {\cal S}_{1}^{(q_i)\, a} (p) \,  t^a_{jk} + {\cal O} \big( t^a t^b \big)\,, \qquad i = 1,2,3\,,
\label{eq:prop-colour-struct}
\end{equation}\\
where ${\cal S}_0^{(q_i)}(p)$ denotes the Fourier transform of the free-quark propagator given in Eq.~(\ref{eq:S0p}), while ${\cal S}_1^{(q_i)}(p)$ includes higher order corrections due to operators of dimension-two and dimension-three describing the emission of one soft gluon field, see Eq.~(\ref{eq:S1p}), and again the superscript $(q_i)$ has been introduced in order to distinguish between different quarks. As consequence of fact that ${\rm Tr}[t^a] = 0$ and that terms quadratic in the gluon field strength tensor correspond to operators in the HQE of at least dimension-seven, a soft gluon cannot be emitted from all the propagators in Eq.~(\ref{eq:T2q-11-D}). Specifically, in the case of ${\cal T}_{11}^{(2q)}$ only the contribution of $S_1^{(q_1)}$ is non zero and hence, because of Eq.~(\ref{eq:Fierz-13}), only $S_1^{(q_3)}$ contributes to ${\cal T}_{22}^{(2q)}$, see Figure~\ref{fig:O1O1}. Finally, in the case of ${\cal T}_{13}^{ (2q)}$, we can independently expand the two propagators inside the trace and both $S_1^{(q_2)}$ and $S_1^{(q_3)}$ give a non vanishing contribution, see Figure~\ref{fig:O1O3}. 
Substituting Eqs.~(\ref{eq:Sx0}), (\ref{eq:S0x}) into Eq.~(\ref{eq:T2q-11-D}), and taking into account Eq.~(\ref{eq:Fierz-13}), we respectively obtain \footnote{Note that the replacement in Eq.~(\ref{eq:bx-d3}) is used.}  \\
\begin{align}
{\cal T}_{11}^{ (2q)}& =  -  \frac{G_F^2}{2}  |V_{q_1b}|^2 |V_{q_3q_2}|^2  \int d^4 x\,  \int \frac{d^4 l_1}{(2 \pi)^4} \int  \frac{d^4 l_2}{(2 \pi)^4}   \int  \frac{d^4 l_3}{(2 \pi)^4} \, \Bigg\{ e^{-i (p_b + l_2  -l_1 -l_3 )  \cdot x}
\nonumber \\[3mm]
& \times  \bar b \gamma_\nu (1-\gamma_5) \left({\cal S}_0^{(q_1)}(l_1) + \tilde {\cal S}_1^{(q_1)} (l_1) \right) \gamma_\mu (1-\gamma_5) b 
\nonumber \\[3mm]
& \times    {\rm Tr} \Biggl[\gamma^\nu (1-\gamma_5) {\cal S}_0^{(q_3)}(l_3) \gamma^\mu (1-\gamma_5) {\cal S}_0^{(q_2)} (l_2) \Biggr]  
\nonumber \\[3mm]
& + \bar b \gamma_\nu (1-\gamma_5)  \left( {\cal S}_0^{(q_1)}(l_1) + {\cal S}_1^{(q_1)} (l_1) \right) \gamma_\mu (1-\gamma_5) b 
\nonumber \\[3mm]
& \times    {\rm Tr} \Biggl[\gamma^\nu (1-\gamma_5) \, {\cal S}_0^{(q_3)}(l_3) \gamma^\mu (1-\gamma_5) {\cal S}_0^{(q_2)} (l_2) \Biggr] \,   e^{i (p_b + l_2  -l_1 -l_3 )  \cdot x}  \Bigg\}  + {\cal O}\left(\frac{1}{m_b^4}\right) 
\, ,
\label{eq:T2q-11-D-2}
\end{align}\\
\begin{align}
{\cal T}_{22}^{ (2q)}& =  -   \frac{ G_F^2}{2}  |V_{q_1b}|^2 |V_{q_3q_2}|^2  \int d^4 x \,  \int \frac{d^4 l_1}{(2 \pi)^4} \int  \frac{d^4 l_2}{(2 \pi)^4}   \int  \frac{d^4 l_3}{(2 \pi)^4} \,\Bigg\{  e^{-i (p_b + l_2  -l_1 -l_3 )  \cdot x}
\nonumber \\[3mm]
& \times   \bar b \gamma_\nu (1-\gamma_5) \,  \, {\cal S}_0^{(q_1)} (l_1)\,  \gamma_\mu (1-\gamma_5) b
\nonumber \\[3mm]
& \times    {\rm Tr} \Biggl[\gamma^\nu (1-\gamma_5)  \left( {\cal S}_0^{(q_3)}(l_3) + \tilde {\cal S}_1^{(q_3)} (l_3) \right) \gamma^\mu (1-\gamma_5) {\cal S}_0^{(q_2)} (l_2) \Biggr] 
\nonumber \\[3mm]
& + e^{i (p_b + l_2  -l_1 -l_3)  \cdot x} \,\,\, \bar b \gamma_\nu (1-\gamma_5) {\cal S}_0^{(q_1)} (l_1)  \gamma_\mu (1-\gamma_5) b  
\nonumber \\[3mm]
& \times    {\rm Tr} \Biggl[\gamma^\nu (1-\gamma_5) \left( {\cal S}_0^{(q_3)}(l_3) + {\cal S}_1^{(q_3)} (l_3) \right) \gamma^\mu (1-\gamma_5) {\cal S}_0^{(q_2)} (l_2) \Biggr]  \Bigg\}
+ {\cal O}\left(\frac{1}{m_b^4}\right)  
\, ,
\label{eq:T2q-22-D-2}
\end{align}\\
and
\begin{align}
{\cal T}_{13}^{ (2q)}& =  -  \frac{ G_F^2}{2}  |V_{q_1b}|^2 |V_{q_3q_2}|^2  \int d^4 x  \int \frac{d^4 l_1}{(2 \pi)^4} \int  \frac{d^4 l_2}{(2 \pi)^4}   \int  \frac{d^4 l_3}{(2 \pi)^4} \, \Bigg\{ \, e^{-i (p_b + l_2  -l_1 -l_3 )  \cdot x}
\nonumber \\[3mm]
& \times  \bar b \gamma_\nu (1-\gamma_5) {\cal S}_0^{(q_1)} (l_1) \gamma_\mu (1-\gamma_5) b 
\nonumber \\[3mm]
& \times  \Bigg\{  {\rm Tr} \Biggl[\gamma^\nu (1-\gamma_5)  \tilde {\cal S}_1^{(q_3)} (l_3)  \gamma^\mu (1-\gamma_5) {\cal S}_0^{(q_2)} (l_2) \Biggr] 
 \nonumber \\[3mm]
& +    {\rm Tr} \Biggl[\gamma^\nu (1-\gamma_5)  {\cal S}_0^{(q_3)}(l_3) \gamma^\mu (1-\gamma_5) {\cal S}_1^{(q_2)} (l_2)  \Biggr] \Bigg\} 
\nonumber \\[3mm]
& + e^{i (p_b + l_2  -l_1 -l_3 )  \cdot x} \,\, \bar b \gamma_\nu (1-\gamma_5)   {\cal S}_0^{(q_1)} (l_1) \gamma_\mu (1-\gamma_5) b 
\nonumber \\[3mm]
& \times  \Bigg\{  {\rm Tr} \Biggl[\gamma^\nu (1-\gamma_5)  {\cal S}_1^{(q_3)} (l_3)  \gamma^\mu (1-\gamma_5) {\cal S}_0^{(q_2)} (l_2) \Biggr]
 \nonumber \\[3mm]
& +    {\rm Tr} \Biggl[\gamma^\nu (1-\gamma_5) {\cal S}_0^{(q_3)}(l_3) \gamma^\mu (1-\gamma_5)  \tilde {\cal S}_1^{(q_2)} (l_2)  \Biggr] \Bigg\}  \Bigg\}
 + {\cal O}\left(\frac{1}{m_b^4}\right) 
\, ,
\label{eq:T2q-13-D-2}
\end{align}\\
where, owing to the fact that starting with the dimension-three operator $D_\rho G_{\mu \nu}$, the translation invariance of the quark-propagator is broken, see Section~\ref{sec:FS}, the second term in Eq.~(\ref{eq:T2q-11-D}) must be explicitly computed and cannot be derived from symmetry arguments. 
Note also that in writing Eq.~(\ref{eq:T2q-13-D-2}) we have already taken into account that the contribution of the free-quark propagator alone, vanishes because of the traceless property of the colour matrices, cf.\ Eq.~(\ref{eq:T13-d3}).

It is worth remarking that the expression of the propagator in Eqs.~(\ref{eq:prop-coordinate-space}), (\ref{eq:prop-coordinate-space-y}) is infrared divergent in the limit of massless quark, cf.\ Eq.~(\ref{eq:massless-lim-prop}). This point will be further discussed later on, for the moment it is important to stress that by setting the mass of the up, down and strange quarks to zero, the logarithmic divergences would appear in Eqs.~(\ref{eq:T2q-11-D-2})-(\ref{eq:T2q-13-D-2}) from the contribution of $S_1^{(u)},  S_1^{(d)}$ and $S_1^{(s)}$, starting at order $1/m_b^3$. In the following, in order to regularise the integrals, we keep the light-quark masses finite and only in the final expressions, once the infrared divergences have been subtracted, we take the limit $m_q \to 0 $, where $q = u,d,s$. Moreover, the computation is performed in $D = 4$ space-time dimensions since the imaginary part of the integrals Eqs.~(\ref{eq:T2q-11-D-2})-(\ref{eq:T2q-13-D-2}) is ultraviolet (UV) finite at LO-QCD. Alternatively, Eqs.~(\ref{eq:T2q-11-D-2})-(\ref{eq:T2q-13-D-2}) could be calculated in dimensional regularisation setting in this case $m_q = 0$ from the beginning \cite{Mannel:2020fts, Moreno:2020rmk}. 

The manipulation of Eqs.~(\ref{eq:T2q-11-D-2})-(\ref{eq:T2q-13-D-2}) proceeds in a similar way and can be conveniently performed using e.g.\ the Mathematica package {\it FeynCalc} \cite{Mertig:1990an}. After integrating over the variables $x^\mu$ and $l_3^\mu$, Eqs.~(\ref{eq:T2q-11-D-2})-(\ref{eq:T2q-13-D-2}) reduce to a linear combination of two-point two-loop tensor integrals with one external momentum $p^\mu_b$ and of possible rank $r = 1, \ldots, 4$, of the type\\
\begin{equation}
  \int \frac{d^4 l_1}{(2 \pi)^4} \, \int \frac{d^4 l_2}{(2 \pi)^4}   \frac{ \big\{l_1^\mu, l_2^\mu, l_1^\mu l_2^\nu, \ldots, l_1^\mu l_1^\nu l_2^\rho, \ldots, l_1^\mu l_1^\nu l_1^\rho l_2^\sigma, \ldots \big\} }{ \big[ l_1^2 - m_1^2 + i \varepsilon\big]^{a_1} \big[ l_2^2 - m_2^2 + i \varepsilon \big]^{a_2} \big[ (l_1 + l_2 -p_b)^2 - m_3^2 + i \varepsilon \big]^{a_3}} \, ,
  \label{eq:tensor-integrals-D}
\end{equation}\\
with $a_i = 1,2,3$. 
The tensor structure of the integrals in Eq.~(\ref{eq:tensor-integrals-D}) can be simplified using the procedure discussed in Appendix~\ref{app:3} for $D=4$. As a result, each integral of rank-$r$ in Eq.~(\ref{eq:tensor-integrals-D}), is expressed in terms of a linear combination of tensors of the same rank built from the metric tensor $g^{\mu \nu}$ and the external momentum $p_b^\mu$ where the corresponding coefficients represent scalar integrals of the form \\
\begin{equation}
  \int \frac{d^4 l_1}{(2 \pi)^4} \, \int \frac{d^4 l_2}{(2 \pi)^4} \frac{ \Big\{ l_1^2, l_1 \cdot l_2, l_1 \cdot p_b, \ldots \Big\}  }{ \big[ l_1^2 - m_1^2 + i \varepsilon \big]^{a_1} \big[ l_2^2 - m_2^2 + i \varepsilon \big]^{a_2} \big[ (l_1 + l_2 -p_b)^2 - m_3^2 + i \varepsilon  \big]^{a_3}} \, .
\label{eq:scalar-int-D}
\end{equation}\\
Note that all the possible scalar products of the three momenta $ l_1^\mu, l_2^\mu, p_b^\mu $ appear in the numerator of Eq.~(\ref{eq:scalar-int-D}). In the next step, we use the Mathematica package {\it LiteRed} \cite{Lee:2012cn,Lee:2013mka } to perform the reduction of the set of scalar integrals obtained, to a liner combination of master integrals (MIs). 
Let us introduce the notation\\
\begin{equation}
\hspace*{-3.5mm}
 {\cal I}_{n_1 n_2 n_3}
\equiv 
\int \frac{d^4 l_1}{(2 \pi)^4} \, \int \frac{d^4 l_2}{(2 \pi)^4}  \frac{ 1 }{ \big[ l_1^2 - m_1^2 + i \varepsilon \big]^{n_1} \big[ l_2^2 - m_2^2 + i \varepsilon \big]^{n_2} \big[ (l_1 + l_2 -p_b)^2 - m_3^2 + i \varepsilon \big]^{n_3}} \,,
\label{eq:MI-def}
\end{equation}\\
with $n_i \in {\mathbb N}_0$, then the set of master integrals reads\\
\begin{equation}
\Bigg\{ {\cal I}_{111}, \, {\cal I}_{211}, \, {\cal I}_{121}, \, {\cal I}_{112}, \,  {\cal I}_{011}, \, {\cal I}_{101}, \, {\cal I}_{110}\Bigg\}  \,.
\label{eq:MIs}
\end{equation}\\
The first four integrals in Eq.~(\ref{eq:MIs}) correspond to the MIs of the sunrise graph with three different masses \cite{Tarasov:1997kx, Caffo:1998du}. From the definition in Eq.~(\ref{eq:MI-def}), it evidently follows that once the expression of ${\cal I}_{111}$ is known, the solution for the remaining three MIs can be obtained by differentiating ${\cal I}_{111}$ with respect to the appropriate mass parameter, i.e.\
\begin{align}
 {\cal I}_{211} = \frac{\partial}{\partial m_1^2} \, {\cal I}_{111}\,, \quad
{\cal I}_{121} = \frac{\partial}{\partial m_2^2}\,  {\cal I}_{111}\,, \quad
 {\cal I}_{112} = \frac{\partial}{\partial m_3^2}\,  {\cal I}_{111}\,.
\label{eq:Im_MI112}
\end{align}\\
Furthermore, the last three integrals in Eq.~(\ref{eq:MIs}) factorise into the product of two scalar tadpoles, in fact, e.g.\ $ {\cal I}_{101}$, can be rewritten as \\
\begin{equation}
{\cal I}_{1 0 1} = \int \frac{d^4 l_1}{(2 \pi)^4} \,  \frac{ 1 }{  l_1^2 - m_1^2 + i \varepsilon}  \, \int \frac{d^4 l_2}{(2 \pi)^4}  \frac{ 1 }{ l_2^2 - m_3^2 + i \varepsilon }  \,,
\label{eq:prod-scal-tadopole}
\end{equation}\\
by performing the change of variable $ l^\mu_1 + l^\mu_2 - p_b^\mu \to  l_2^\mu$, and similarly for $ {\cal I}_{011}$ and $ {\cal I}_{110}$. However, we immediately point out that these do not contribute to ${\rm Im} {\cal T}^{(2q)}$, since the imaginary part of the product of two tadpoles vanishes, cf.\ Eq.~(\ref{eq:iA0}), namely\\
\begin{equation}
{\rm Im} \, {\cal I}_{1 0 1} = {\rm Im} \,  {\cal I}_{1 1 0} = {\rm Im} \, {\cal I}_{0 1 1}  = 0\,.
\label{eq:Im-prod-tadpole}
\end{equation}\\
From Eq.~(\ref{eq:Im_MI112}), (\ref{eq:Im-prod-tadpole}), it then follows that in order to compute the imaginary part of the set of integrals in Eq.~(\ref{eq:scalar-int-D}), it suffices to know the solution of the master integral ${\cal I}_{111}$ in the physical decay region $p_b^2 \geq (m_1 + m_2 + m_3)^2$, in correspondence of which its integrand develops a discontinuity. Using the result presented in 
Ref.~\cite{Remiddi:2016gno}, we obtain that \footnote{Note that we set $d=4$ in the result of Ref.~\cite{Remiddi:2016gno}. }\\
\begin{equation} 
{\rm Im} \, {\cal I}_{111} = \frac{1}{256 \pi^3}\!  \int\limits_{(m_2 + m_3) ^2}^{(\sqrt{s} - m_1)^2} 
\! \! d t \, \frac{\sqrt{\lambda(t, m_2^2, m_3^2)\, \lambda(s, t, m_1^2)}}{t\, s}\, ,
\label{eq:Im_sunset}
\end{equation}\\
where $s = p_b^2$. 
However, the integral in Eq.~(\ref{eq:Im_sunset}) admits a simple analytical expression if at most two masses are non zero. For three non vanishing masses, its complexity highly increases and the solution involves elliptic functions, see e.g.\ Refs.~\cite{Pivovarov:1984ij,Groote:2000kz,Broedel:2017kkb,Broedel:2017siw}. We emphasise that in the approximation of massless $u, d$ and $s$ quarks, it is always possible to set at least one mass to zero and to compute all the corresponding master integrals analytically, except in the case of $b \to c \bar c s$ transition where, as discussed above, we need to keep the mass of the $s$-quark finite in order to regularise the IR divergence originating from
$S_1^{(s)}$. It follows that for this specific mode, we do not provide an analytical expression for all the corresponding MIs, and our results still require a numerical integration.

The integrals in Eqs.~(\ref{eq:Im_sunset})-(\ref{eq:Im_MI112}), are scalar functions of the external momentum $p_b^\mu$ and depend on the dimensionless  parameters $r_i = m_i^2/p_b^2$. At this point of the computation then, taking into account that $G_{\mu \nu}  = -i \left[i D_\mu, i D_\nu \right]$, see Eq.~(\ref{eq:G-munu}), together with $D_\rho \,G_{\mu \nu}  = - \left[i D_\rho, \left[i D_\mu, i D_\nu \right] \right]$, see Eq.~(\ref{eq:G-munu-DG}), the imaginary part of Eqs.~(\ref{eq:T2q-11-D-2})-(\ref{eq:T2q-22-D-2}) can be schematically written in the following form
 \footnote{Note that for brevity spinor and colour indices are not shown.}\\
\begin{align}
{\rm Im}  {\cal T}_{mn}^{(2q)}  & = {\cal F}_{mn}(p_b, r_i) \, \bar b  b  +  {\cal G}_{mn}^{\mu \nu} (p_b, r_i)  \, \bar b( i D_\mu)( i D_\nu) b
\nonumber \\[3mm]
& + {\cal D}_{mn}^{\mu \nu \rho} (p_b, r_i)  \, \bar b (i D_\mu) (i D_\nu) (i D_\rho) b + {\cal O}\left(\frac{1}{m_b^4}\right)\,,
\label{eq:Im-Tmn}
\end{align}\\
corresponding, respectively, to the sum of the contributions due to the free quark propagator and to the two lowest dimensional corrections to this, namely the ones proportional to the gluon field strength tensor and to its first covariant derivative. 
We notice that ${\cal F}_{13}(p_b, r_i) = 0$,  
since the free-quark propagator is colour singlet and the contribution of the colour octet operator $Q_3$, cf.\ Eq.~(\ref{eq:Q3-def}), vanishes. Moreover, another feature of Eq.~(\ref{eq:Im-Tmn}) is that\\ 
\begin{equation}
{\cal F}_{22}(p_b, r_i) = {\cal F}_{11}(p_b, r_i) \,,
\label{eq:F22-equal-F11}
\end{equation}\\
which is the generalisation of the result given in Eq.~(\ref{eq:T22-equal-T11-d3}), to the case of an arbitrary decay mode of the $b$ quark. Singling out the colour factor due to $\delta_{ii} = N_c$, we can write
\begin{equation}
{\cal F}_{11}(p_b, r_i) = N_c \, \tilde {\cal F}_{11}(p_b, r_i)\,,
\end{equation}\\
and then from Eq.~(\ref{eq:T2q-D}) we readily obtain that\\
\begin{equation}
{\rm Im } {\cal T}^{(2q)} \Big|^{(q_1 \bar q_2 q_3)}_{d=3} =  \Big( N_c \,\big( C_1^2 + C_2^2 \big) + 2 \, C_1 C_2  \Big) \tilde {\cal F}_{11}^{(q_1 \bar q_2 q_3)}(p_b, r_i) \, \bar b b \,,
\end{equation}\\
which again generalises the result in Eq.~(\ref{eq:T3-1}).
Furthermore, Eq.~(\ref{eq:Im-Tmn}) presents also the important property, already encountered in Section~\ref{sec:Gamma-5} for the specific case of $b \to c \bar u d$ transition, that at order $1/m_b^2$ only the expansion of the $q_2$ quark propagator gives a non vanishing contribution, namely\\
\begin{equation}
{\cal G}^{\mu \nu}_{11}(p_b, r_i) = {\cal G}^{\mu \nu}_{22}(p_b,r_i) = 0\,, \quad  \, {\cal G}^{\mu \nu}_{13}(p_b, r_i) \neq 0\,.
\label{eq:zero-contr-prop-d5}
\end{equation}\\
This result has relevant numerical consequences, since the coefficient of $C_1^2$ at order $1/m_b^2$, is strongly suppressed.
So far only power corrections deriving from the expansion of the quark-propagator have been taken into account. The coefficient functions in Eq.~(\ref{eq:Im-Tmn}) depend on the heavy quark momentum $p_b^\mu$ explicitly and implicitly through the variable $r_i$. Introducing the standard parametrisation $p_b^\mu = m_b v^\mu + k^\mu$, see Section~\ref{sec:HQET}, and recalling the definition of the rescaled heavy quark field $b(x)= \exp (- i m_b v\cdot x) b_v(x)$, see Eq.~(\ref{eq:Qx-HQE}), each term in Eq.~(\ref{eq:Im-Tmn}) can be further expanded in powers of $1/m_b$, resulting in higher dimensional operators with additional covariant derivatives acting on the $b_v$ field. Specifically, in our case, the expansion must be performed up to order $1/m_b^3$, hence  power corrections to the dimension-six coefficients ${\cal D}_{ab}^{\mu \nu \rho} (p_b, r_i)$ in Eq.~(\ref{eq:Im-Tmn}), can be neglected as these would lead to contributions of order ${\cal O}(1/m_b^4)$.
Using the procedure described in Appendix~\ref{app:1}, we then obtain \\
\begin{align}
  {\cal F}_{mn}(p_b, r_i)  \,  \bar b  b  
&  = {\cal F}_{mn}(m_b v, \rho_i)  \, \bar b_v  b_v 
 + {\cal K}_{mn}^\mu (m_b v, \rho_i)  \, \bar b_v  i D_\mu b_v 
\nonumber \\[3mm]
& +  {\cal G}_{mn}^{\prime \mu \nu}(m_b v, \rho_i)  \,  \bar b_v i D_\mu i D_\nu b_v 
+  {\cal D}_{mn}^{\prime \mu \nu \rho}(m_b v, \rho_i) \,   \bar b_v  i D_\mu i D_\nu i D_\rho  b_v + \ldots
\,,
\label{eq:exp-Fmn}
\end{align}\\
\begin{equation}
 {\cal G}^{\mu \nu}_{mn}(p_b, r_i)   \,  \bar b i D_\mu i D_\nu  b
  = {\cal G}_{mn}^{\mu \nu}(m_b v, \rho_i)  \,  \bar b_v i D_\mu i D_\nu  b_v  
 +   {\cal D}_{mn}^{\prime \prime \mu \nu \rho}(m_b v, \rho_i) \,   \bar b_v i D_\mu i D_\nu i D_\rho  b_v + \ldots\,, 
\label{eq:exp-Gmn}
\end{equation}\\
and
\begin{equation}
 {\cal D}^{\mu \nu \rho}_{mn}(p_b, r_i) \, \bar b i D_\mu i D_\nu i D_\rho  b = {\cal D}^{\mu \nu \rho}_{mn}(m_b v, \rho_i) \, 
 \bar b_v  i D_\mu i D_\nu i D_\rho b_v + \ldots \,,
\label{eq:exp-Dmn}
\end{equation}\\
where the ellipsis denote power suppressed contributions of order ${\cal O}(1/m_b^4)$ and we have introduced the dimensionless mass parameters $\rho_i = m_i^2/m_b^2$.
Finally, in order to compute $\Gamma^{(2q)}(B)$, we must evaluate the matrix element of ${\rm Im} {\cal T}^{(2q)}$ between external $B$ states. This can be conveniently done in the framework of the HQET, in which, the residual mass dependence of the $b_v$ field and of the $B$ meson state, can be systematically extracted, leading to a further expansion in $1/m_b$. 
A consistent procedure to determine the forward matrix element of operators containing multiple covariant derivatives acting on the heavy quark field and to express them in terms of a minimal set of non perturbative parameters, has been presented in Ref.~\cite{Dassinger:2006md}. Using their results, we readily arrive at the final expression for 
the contribution of two-quark operators to the total decay width of a $B$ meson, namely\\
\begin{align}
\Gamma^{ (2q)} (B) & =  \Gamma_0 \,
\Biggl[\Big(N_c C_1^2  + 2 \, C_1  C_2  + N_c C_2^2 \Big) {\cal C}_0^{(q_1 \bar q_2 q_3)}
\left(1 - \frac{\mu_\pi^2 (B)} {2 m_b^2} \right) 
\nonumber \\[3mm]
&   + \Big(N_c C_1^2 \,\, {\cal C}_{G, 11}^{(q_1 \bar q_2 q_3)} 
+ 2 \, C_1  C_2 \,\, {\cal C}_{G, 12}^{(q_1 \bar q_2 q_3)} 
+ N_c C_2^2 \,\, {\cal C}_{G, 22}^{(q_1 \bar q_2 q_3) }\Big) \frac{\mu_G^2 (B)} {m_b^2}  
\nonumber \\[2mm]
&  +\Big(N_c C_1^2 \,\, {\cal C}_{D, 11}^{(q_1 \bar q_2 q_3)}
+ 2 \, C_1  C_2 \,\, {\cal C}_{D,12}^{(q_1 \bar q_2 q_3)} 
+ N_c  C_2^2 \,\, {\cal C}_{D, 22}^{(q_1 \bar q_2 q_3)}   \Big) \frac{\rho_D^3 (B)} {m_b^3}  
\Biggr] + {\cal O}\left( \frac{1}{m_b^4}\right)\,,
\label{eq:Gamma-NL-res-scheme}
\end{align}\\
where we have introduced\\
\begin{equation}
\Gamma_0 = \frac{G_F^2 m_b^5}{192 \, \pi^3} \, |V_{q_1 b}|^2 |V_{q_2  q_3}|^2\,,
\label{eq:Gamma0}
\end{equation}\\
and the non perturbative parameters $\mu_\pi^2(B)$, $\mu_G^2(B)$ and $\rho_D^3(B)$ are defined as in Eqs.~(\ref{eq:dim-5-ME-parameters}), (\ref{eq:dim-6-ME-parameters}).
In Eq.~(\ref{eq:Gamma-NL-res-scheme}), ${\cal C}_0^{(q_1 \bar q_2 q_3)}$ refers to the partonic-level 
coefficient, which coincides, up to a factor of $(-1/2)$, with that of the kinetic operator ${\cal O}_{kin}/m_b^2$~\footnote{This follows from the reparametrisation invariance of the HQE, see e.g.\ Ref.~\cite{Mannel:2018mqv}. }, cf.\ Eq.~(\ref{eq:O-kin}), while ${\cal C}_{G, mn}^{(q_1 \bar q_2 q_3)}$ and ${\cal C}_{D,  mn}^{(q_1 \bar q_2 q_3)}$, respectively describe the contribution of the chromo-magnetic and of the Darwin operators. Note that having adopted a covariant definition for these operators, the coefficient of the spin-orbit operator ${\cal O}_{LS}$, cf.\ Eq.~(\ref{eq:O-LS}), is identically zero, see for more details e.g.\ Refs.~\cite{Dassinger:2006md, Mannel:2020fts, Mannel:2018mqv}. 
Moreover, we stress that while ${\cal C}_0^{(q_1 \bar q_2 q_3)}$ and ${\cal C}_{G, nm}^{(q_1 \bar q_2 q_3)}$ are finite functions of at most one dimensionless mass parameter $\rho = m_c^2/m_b^2$, and 
we list their complete expressions in Appendix~\ref{app:4}, as previously discussed, the coefficients of the Darwin operator still depend on the infrared regulator $m_q$, with $q = u,d,s$. These in fact, have the following schematic form\\
\begin{equation}
{\cal C}_{D, mn}^{(q_1 \bar q_2 q_3)} = {\cal R}_{mn}^{(q_1 \bar q_2 q_3)} + {\cal D}_{mn}^{(q_1 \bar q_2 q_3)}\,,
\label{eq:CDnm}
\end{equation}\\
where $ {\cal R}_{mn}^{(q_1 \bar q_2 q_3)} $ are finite functions of at most the dimensionless parameter $\rho$, while ${\cal D}_{nm}^{(q_1 \bar q_2 q_3)}$ absorb the remaining divergent contributions, namely \\
\begin{equation}
{\cal D}_{mn}^{(q_1 \bar q_2 q_3)}  \sim \log \left(\frac{m_q^2}{m_b^2}\right)\,,
\label{eq:lim-Dnm}
\end{equation}\\
and their explicit expression can be found in Appendix~\ref{app:5}.
Eq.~(\ref{eq:lim-Dnm}) shows that the functions ${\cal C}_{D, nm}^{(q_1 \bar q_2 q_3)}$ are logarithmically sensitive to the light quark mass $m_q \ll m_b $. However, the advantage of constructing the operator product expansion lies in the introduction of a factorisation scale $\mu$, in terms of which the dependence on the hard, i.e.\ $\mu_h$ and soft i.e.\ $\mu_s$ scales with $ m_b \geq \mu_h \geq \mu$, and $\mu_s \leq \mu$, is respectively factorised between short distance coefficients and matrix elements of local operators, see e.g.\ Ref.~\cite{Shifman:1995dn}. Eq.~(\ref{eq:Gamma-NL-res-scheme}) alone, including the effect of two-quark operators only, does not correspond to the complete OPE up to dimension-six because starting at this order also four-quark operators contribute. In fact, the logarithmic infrared divergence in Eq.~(\ref{eq:lim-Dnm}) reflects the mixing between operators at order $1/m_b^3$ under renormalisation. It follows that by solving the corresponding RGEs, the dependence on the light quark mass in the coefficients in Eq.~(\ref{eq:CDnm}) can be correctly absorbed in the matrix element of local operators, making then manifest the factorisation between hard and soft scales. This is discussed in detail in the next section. 


\section{Operator-mixing at order $1/m_b^3$}
\label{sec:Dar-sec-2}
\begin{figure}
\centering
\includegraphics[scale = 0.35]{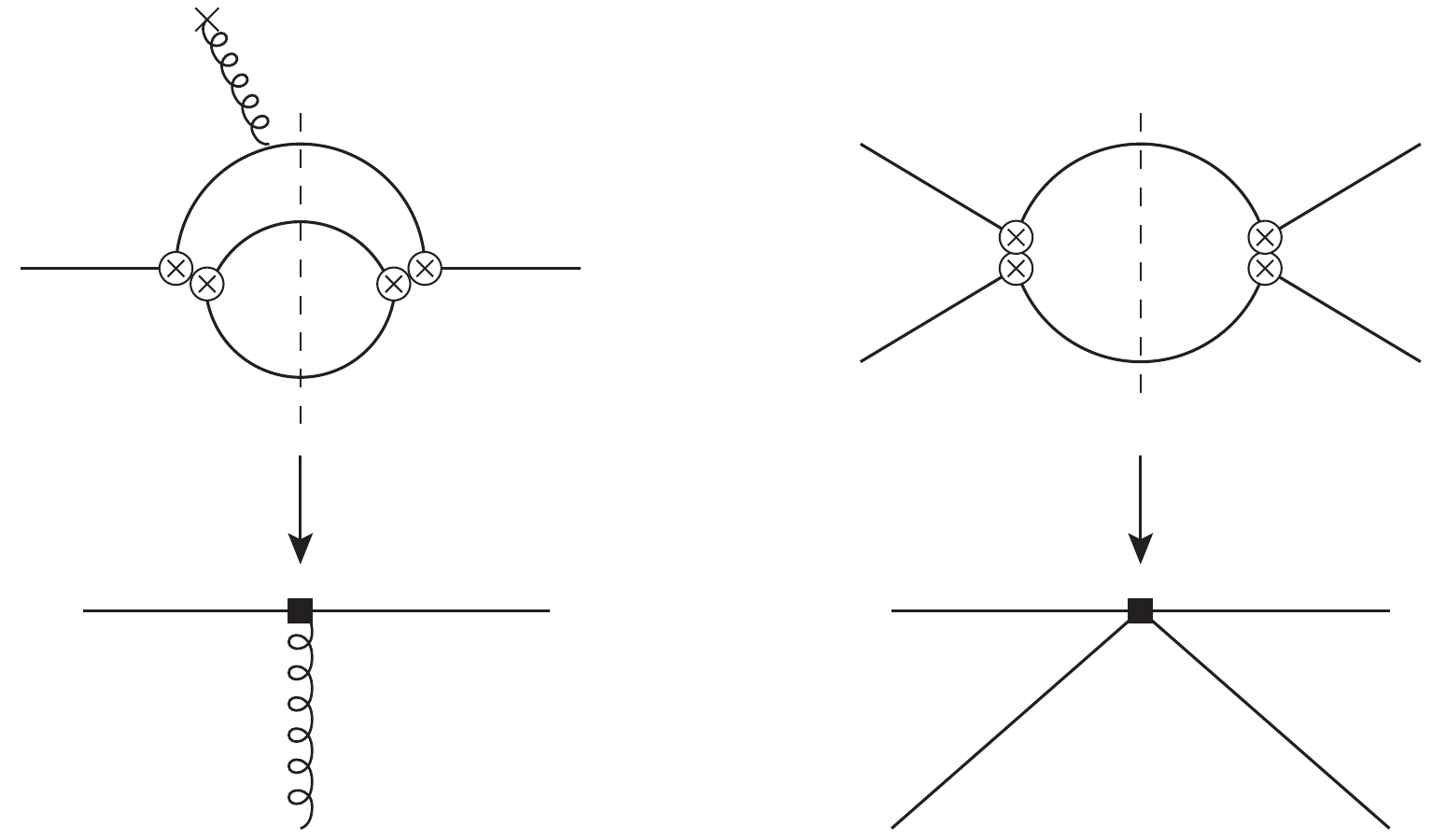}
\caption{Schematic representation of the OPE at order $1/m_b^3$. At LO-QCD, the Darwin operator is generated at two-loop, whereas the four-quark operators arise already at one-loop.}
\label{fig:OPE}
\end{figure}
\begin{figure}
\centering
\includegraphics[scale =0.6]{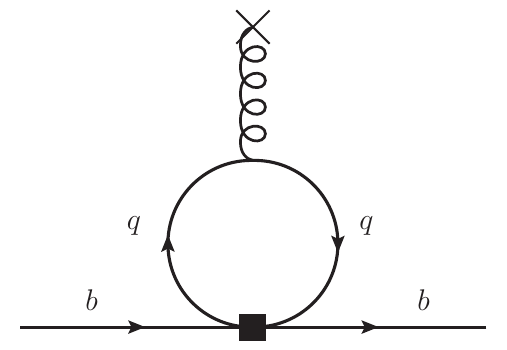}
\caption{One-loop diagram describing the mixing of four-quark operators with the Darwin operator.}
\label{fig:Mixing}
\end{figure}

To understand the origin of the IR divergences in Eq.~(\ref{eq:lim-Dnm}), and how these are properly subtracted, we study the structure of Eq.~(\ref{eq:Gamma-HQ}) at order $1/m_b^3$.  Within the HQE, the imaginary part of time ordered product of the double insertion of the effective Hamiltonian is expanded in a series of local operators with new effective couplings. This can be schematically written as  \footnote{Note that the superscript $(q_1 \bar q_2 q_3)$ and the subscript $mn$ are omitted for the sake of a cleaner notation, however they must be always understood.}
\begin{equation}
\langle {\rm Im} {\cal T} \rangle \, \Big|_{d=6} =  c_{\rho_D}(\mu_0) \, \langle {\cal O}_{\rho_D} \rangle(\mu_0) + \sum \limits_{q = q_1, q_2, q_3}  \, \vec c_{4q}^{\, (q)}(\mu_0) \cdot  \langle {\vec {\cal O}_{4q}}^{\, (q)} \rangle(\mu_0) \, ,
\label{eq:OPE-d6}
\end{equation}\\
where the shorthand  $\langle \ldots \rangle$ denotes a matrix element between external $B$ meson states and the dependence on the scale $\mu_0$ at which the matrix element of the local dimension-six operators are renormalised, is now explicitly indicated. In Eq.~(\ref{eq:OPE-d6}), the Darwin operator ${\cal O}_{\rho_D}$ is defined as in Eq.~(\ref{eq:O-Darwin}) and we have introduced a compact notation for the four-quark operators listed in Eqs.~(\ref{eq:Ova-4q})-(\ref{eq:Tsp-4q}), namely \\
\begin{equation}
\vec {\cal O}_{4q}^{(q)} = \Big( { O}_{1}^{(q)}, \,   {O}_{2}^{(q)},  \, \tilde { O}_{1}^{(q)}, \, \tilde {O}_{2}^{(q)}   \Big)\,.
\label{eq:O4q}
\end{equation}\\
In the following we want to determine the short distance coefficients $c_{\rho_D}(\mu_0)$ and $\vec c_{4q}^{\, (q)}(\mu_0)$.
The OPE leading to Eq.~(\ref{eq:OPE-d6}) can be schematically visualised as in Figure~\ref{fig:OPE}. It has the peculiarity that the order of the Darwin and of the four-quark operators, in terms of the loop- and $\alpha_s$-expansion, does not coincide. Specifically, while the four-quark operators are generated at one-loop at order $\alpha_s^{0}$, the Darwin operator arises only at two-loop again at order $\alpha_s^0$. It follows that the one-loop correction to the four-quark operators, shown in Figure~\ref{fig:Mixing}, is of the same order in terms of loop- and $\alpha_s$-expansion as the  coefficient of the Darwin operator and must be included to obtain the complete contribution to $c_{\rho_D}(\mu_0)$ at LO-QCD, see e.g.\ Refs.~\cite{Gambino:2005tp, Bigi:2005bh, Breidenbach:2008ua}. We note that operator mixing at zeroth-order in $\alpha_s$ has been extensively discussed for the $b \to s \gamma $ effective Hamiltonian, see e.g.\ Refs.~\cite{Grigjanis:1989py, Misiak:1991dj, Ciuchini:1993fk}.
In order to evaluate the diagram in Figure~\ref{fig:Mixing} for all four operators in Eq.~(\ref{eq:O4q}), we start by considering 
the one-loop matrix element of the colour-singlet operators in the presence of a soft gluon field $A_\mu (x)$, which is obtained by computing
the time ordered product of ${\cal O}_{4q, j}^{(q)}$, $j = 1,2,$ with the interaction part of the QCD Lagrangian ${\cal L}_{QCD}^{int}(x) = \sum_{q} \bar q \slashed A(x) q$. This reads \footnote{Note that summation over colour indices is understood.}\
\begin{align}
\langle {\cal O}_{4q, j}^{(q)} \rangle_{1-loop} & = \langle  i  \int d^4 z \, {\rm T} \Bigg\{\vec {\cal O}_{4q, j}^{(q)}, \sum_{q^\prime}\,  \bar q^\prime(z) \slashed A(z) q^\prime (z) \Bigg\}  \rangle + \ldots\,
\nonumber \\[3mm]
&  = 
 \langle \bar b_v   \bar \Gamma_j   \Bigg( \int d^4 z \,  i S_0^{(q)}(- z) i \slashed A(z) iS_0^{(q)}(z) \Bigg) \Gamma_j  b_v \rangle + \ldots \,,
\label{eq:Q1-Darwin-mixing}
\end{align} \\ 
where the ellipsis denote terms with more than one gluon field, $S_0^{(q)}(x-y)$ is the free-quark propagator 
and for convenience we have introduced the notation\\
\begin{equation}
\Gamma_1 = \gamma_\mu (1-  \, \gamma_5)  \qquad \Gamma_2 = (1+ \gamma_5)\,,
\qquad  \bar \Gamma_j = \gamma_0  \Gamma_j^\dagger  \gamma_0 \,.
\end{equation}\\
We recognise in the term in round brackets of Eq.~(\ref{eq:Q1-Darwin-mixing}) the first order correction to the free-quark propagator for vanishing space-time separation, cf.\ Eq.~(\ref{eq.quark-propagator-definition}), namely\\
\begin{equation}
\int d^4 z \, i S_0^{(q)}(- z) i \slashed A(z) i S_0^{(q)}(z) = \lim_{x \to 0} \, i S^{(q)}_1 (x,0)\,,
\label{eq:limS1}
\end{equation}\\
which, in the FS gauge, admits the Fourier representation \\
\begin{equation}
\lim_{x \to 0} S^{(q)}_1(x,0) = \int \frac{d^4 l}{(2 \pi)^4} \, {\cal S}^{(q)}_1(l) \,,
\label{eq:limS1-1}
\end{equation}\\
with ${\cal S}^{(q)}_1(l)$ given in Eq.~(\ref{eq:S1p}).
The integral in Eq.~(\ref{eq:limS1-1}) has both UV- and IR-divergences in the limit of massless quark $q$. To regularise the former we use  dimensional regularisation, setting the number of space-time dimension to $D = 4 - 2 \epsilon$, for the latter we must choose the same regularisation scheme applied to the computation of the coefficients in Eq.~(\ref{eq:CDnm}), hence we keep the mass of the light-quark running into the loop finite. The algebra of gamma matrices in $D$ dimensions is computed using the naive dimensional regularisation (NDR) scheme, with\\
\begin{equation}
\{\gamma_\mu , \gamma_\nu \} = 2  g_{\mu \nu}\,, \qquad g_{\mu \nu} g^{\mu \nu} = D\,, \qquad
\{\gamma_\mu, \gamma_5\} = 0\,.
\end{equation}\\
Taking into account in the expression for the quark-propagator ${\cal S}^{(q)}_1(l)$ that $G_{\mu \nu}  = -i \left[i D_\mu, i D_\nu \right]$ and that $D_\rho \,G_{\mu \nu}  = - \left[i D_\rho, \left[i D_\mu, i D_\nu \right] \right]$, Eq.~(\ref{eq:Q1-Darwin-mixing}) becomes\\
\begin{align}
\langle {\cal O}_{4q, j}^{(q)} \rangle_{1-loop}&=  \langle \bar b_v  \bar \Gamma_j  i \int \frac{d^D l}{(2 \pi)^D}  \,  \Bigg\{
 - \frac{i}{2}\frac{ [D^\sigma, D^\tau]}{(l^2 -m_q^2+ i \varepsilon)^2}  l^\rho \gamma^\mu  \epsilon_{\rho \mu \sigma \tau} 
\nonumber \\[3mm]
& +  i \frac23 \frac{[D_\rho , [D^\rho, D^\mu]]}{(l^2 - m_q^2 + i \varepsilon)^2} \left( \gamma_\mu - l_\mu \frac{\slashed l}{(l^2 - m_q^2 + i \varepsilon )} \right) 
\nonumber \\[3mm]
& - i \frac23 \,  [D_\nu,[D_\rho, D_\mu]] \frac{l^\nu l^\rho \gamma^\mu}{(l^2 - m_q^2 + i \varepsilon)^3} 
\nonumber \\[3mm]
& +  i  \frac{[D_\nu,[ D^\sigma, D^\tau]]}{(l^2 -m_q^2 + i \varepsilon)^3}  l^\nu l^\rho \gamma^\mu   \epsilon_{\rho \mu \sigma \tau}\Bigg\} 
 \Gamma_j  b_v \rangle + \ldots \,.
\label{eq:unren-one-loop-me}
 \end{align}\\
We stress that due to the chiral structure of Eq.~(\ref{eq:unren-one-loop-me}), terms in the propagator proportional to an even number of gamma matrices do not contribute and that we have used that $\gamma_5  \Gamma_j = \Gamma_j$.
Moreover, the first integral on the r.h.s of Eq.~(\ref{eq:unren-one-loop-me}) vanishes, being the integrand an odd function of~$l^\mu$ and the integration domain even. 
This result is independent of the colour structure of the operator inserted in the vertex of the diagram in Figure~\ref{fig:Mixing} and applies also to the matrix element of the remaining colour-octet operators in Eq.~(\ref{eq:O4q}). Hence, 
the four-quark operators ${\vec {\cal O}}^{(q)}_{4q}$ do not mix with the chromo-magnetic operator at order $\alpha_s^0$ \footnote{Note that mixing with lower-dimensional operators can arise at NLO-QCD, see e.g.\ Ref.~\cite{Bauer:1996ma}.}. 
Eq.~(\ref{eq:unren-one-loop-me}) then simplifies to\\
\begin{align}
\langle {\cal O}_{4q, j}^{(q)} \rangle_{1-loop} &=   - \langle \bar b_v \bar \Gamma_j  \int \frac{d^D l}{(2 \pi)^D}  \, \Bigg\{ 
  \frac{[D_\nu,[ D^\sigma, D^\tau]]}{(l^2 -m_q^2 + i \varepsilon)^3}  l^\nu l^\rho \gamma^\mu   \epsilon_{\rho \mu \sigma \tau}
\nonumber \\[3mm]
& + \frac23  \frac{[D_\rho , [D^\rho, D^\mu]]}{(l^2 - m_q^2 + i \varepsilon)^2} \left( \gamma_\mu - l_\mu \frac{\slashed l}{(l^2 - m_q^2 + i \varepsilon )} \right) 
\nonumber \\[3mm]
& -  \frac23   [D_\nu,[D_\rho, D_\mu]]  \frac{l^\nu l^\rho \gamma^\mu}{(l^2 - m_q^2 + i \varepsilon)^3}  \Bigg\} 
 \Gamma_j  b_v \rangle
+ \ldots\,.
\label{eq:unren-one-loop-me-2}
 \end{align}\\
The dimensional analysis of Eq.~(\ref{eq:unren-one-loop-me-2}) reveals an interesting subtlety. From   \\
\begin{equation}
[\psi] = \frac{D-1}{2}\,, \qquad [D_\mu] = 1\,, \qquad [l_\mu] = 1 \,,
\end{equation}\\
where the square brackets denote the dimension in units of mass i.e.\ $[m] = 1$ and $\psi$ an arbitrary fermion field, see e.g.\ Ref.~\cite{Itzykson:1980rh}, we obtain that while Eq.~(\ref{eq:unren-one-loop-me-2}) is dimensionally correct, since both sides have the same mass dimension, the integrals in Eq.~(\ref{eq:unren-one-loop-me-2}) have now become dimensionful, namely  \\
\begin{equation}
\left[ \int \frac{d^4 l}{(l^2 - m_q^2 + i \varepsilon)^2}  \right] = 0 \quad \to \quad  \left[ \int \frac{d^D l}{(l^2 - m_q^2 + i \varepsilon)^2}  \right] =  - 2 \epsilon\,.
\end{equation}\\
This fictitious dimension is an artefact of the regularisation scheme and does not correspond to any physical parameter. In fact, in this case, the integrals in Eq.~(\ref{eq:unren-one-loop-me-2}) would be expressed in terms of ill defined logarithms with a dimensionful argument, cf.\ Eqs.~(\ref{eq:in-l}), (\ref{eq:I0-tadpole}). The origin of this mismatch lies in the different scaling between the Darwin and the four-quark operators with the number of dimensions, see a similar discussion, for the case of $b \to s \gamma$ and $b \to s g$,  in Ref.~\cite{Ciuchini:1993fk}, i.e.\\
\begin{equation}
\Big[{\vec {\cal O}}^{(q)}_{4q}\Big] = 6 - 4  \epsilon \quad  \neq \quad  6 - 2 \epsilon =   \Big[{\cal O}_{\rho_D}\Big] \,.
\end{equation}\\
In order to keep the integrals in Eq.~(\ref{eq:unren-one-loop-me-2})  dimensionless, we must introduce a scale factor of $\mu^{2 \epsilon}$ in the integration measure. Eq.~(\ref{eq:unren-one-loop-me-2}) then becomes\\
\begin{align}
\langle {\cal O}_{4q, j}^{(q)} \rangle_{1-loop} &=-  \mu^{- 2\epsilon}  \langle \bar b_v  \bar \Gamma_j \mu^{2 \epsilon} \int \frac{d^D l}{(2 \pi)^D}   \Bigg\{ 
  \frac{[D_\nu,[ D^\sigma, D^\tau]]}{(l^2 -m_q^2 + i \varepsilon)^3} l^\nu l^\rho \gamma^\mu   \epsilon_{\rho \mu \sigma \tau}
\nonumber \\[3mm]
& +  \frac23  \frac{[D_\rho , [D^\rho, D^\mu]]}{(l^2 - m_q^2 + i \varepsilon)^2}  \left( \gamma_\mu - l_\mu \frac{\slashed l}{(l^2 - m_q^2 + i \varepsilon )} \right) 
\nonumber \\[3mm]
& -  \frac23  [D_\nu,[D_\rho, D_\mu]]  \frac{l^\nu l^\rho \gamma^\mu}{(l^2 - m_q^2 + i \varepsilon)^3}  \Bigg\} \,
 \Gamma_j  b_v \rangle + \ldots  \,.
\label{eq:unren-one-loop-me-3}
 \end{align}\\
Using the results in Eqs.~(\ref{eq:tadpole-order2}), (\ref{eq:r2-tensor-integral-tadpole}), to evaluate the scalar and tensor integrals in Eq.~(\ref{eq:unren-one-loop-me-3}), we  obtain that\\
\begin{align}
\langle {\cal O}_{4q, j}^{(q)} \rangle_{1-loop} &=  -  \mu^{- 2\epsilon}  \langle \bar b_v  \bar \Gamma_j  \, \Bigg\{ 
  \frac14   {\cal I}_0(m_q^2)  [D_\nu,[ D_\sigma, D_\tau]]   \epsilon^{\nu \mu \sigma \tau}
\nonumber \\[3mm]
& +  \frac13  {\cal I}_0(m_q^2) [D_\rho , [D^\rho, D^\mu]]  \,
\Bigg\}   \gamma_\mu \,
 \Gamma_j  \, b_v \rangle + \ldots \,,
\label{eq:unren-one-loop-me-4}
\end{align}\\
where ${\cal I}_0(m_q^2)$ is given in Eq.~(\ref{eq:I0-tadpole}).
The gamma structure on the r.h.s.\ of Eq.~(\ref{eq:unren-one-loop-me-4}) reduces to\\
\begin{align}
\gamma_\nu \, (1- \gamma_5)\, \gamma_\mu \, \gamma^\nu \, (1- \gamma_5) &= 2 \,  (2 - D) \, \gamma_\mu \, (1- \gamma_5)\,, \quad j=1 \,,
\label{eq:gamma-str-tadpole}
\\[3mm]
 (1- \gamma_5)\, \gamma_\mu \, (1+ \gamma_5) & = 2 \,  \gamma_\mu \, (1 + \gamma_5)\,, \qquad  \,\, \, \qquad  j=2\,,
\label{eq:gamma-str-tadpole1}
\end{align}\\
and by substituting the expression in Eq.~(\ref{eq:I0-tadpole}), it follows that\\
\begin{align}
\langle {\cal O}_{4q, j}^{(q)} \rangle_{1-loop} &=  \frac{a_j}{24 \pi^2} \Bigg[  \mu^{- 2\epsilon} \left( \frac{1}{\epsilon} - \gamma_E + \log(4 \pi) \right) + \log\left( \frac{\mu^2}{m_q^2} \right) + b_j \Bigg]
\nonumber \\[3mm]
& \times  \Bigg[ \frac34\langle \bar b_v   [D_\nu, [D_\sigma, D_\tau]]  \gamma^\nu \gamma^\sigma \gamma^\tau \big( \gamma_5  + (-1)^j \big)  b_v \rangle 
\nonumber \\[3mm]
&  + i \langle \bar b_v  [D_\rho , [D^\rho,  D^\mu]] \gamma_\mu \big( 1  + (-1)^j  \gamma_5 \big)  b_v \rangle   \,
\Bigg] + \ldots \,,
\label{eq:unren-one-loop-me-5}
\end{align}\\
with
\begin{equation}
a_1 = 2 \,, \quad b_1 = -1\,, \qquad a_2 = - 1 \,, \quad b_2 = 0\,.
\label{eq:as-bs}
\end{equation}\\
Note that the tensor decomposition of three gamma matrices Eq.~(\ref{eq:tensor-decomposition-gamma-mat}) has been used to rewrite $i\, \epsilon^{\nu \sigma \tau \mu} \, \gamma_\mu$ in the second line of Eq.~(\ref{eq:unren-one-loop-me-5}), see also Eq.~(\ref{eq:epsilon-gamma}), and that terms ${\cal O}(\epsilon)$ have been already neglected. We emphasise that the presence of the constant terms $b_j$ in Eq.~(\ref{eq:unren-one-loop-me-5}) depends on the choice of the four-quark operators basis in Eq.~(\ref{eq:O4q}). This will be discussed further at the end of this chapter. 
\\[2mm]
In order to simplify the structure of Eq.~(\ref{eq:unren-one-loop-me-5}) and identity the relevant operators, first we take into account that due to parity conservation, matrix elements with an odd number of $\gamma_5$ vanish, then 
we recall that the rescaled heavy quark field $b_v(x)$ satisfies $(i v \cdot D) b_v(x) =  {\cal O}(1/m_b)$, cf.\ Eq.~(\ref{eq:vdotD-Qv}), together with $\slashed v b_v(x) = b_v(x) +  {\cal O}(1/m_b)$, cf.\ Eq.~(\ref{eq:p-Qv}). It follows that the two operators in the second and third line of Eq.~(\ref{eq:unren-one-loop-me-5}), respectively give\\
\begin{align}
&\langle \bar b_v   [D_\nu, [D_\sigma, D_\tau]] \gamma^\nu \gamma^\sigma \gamma^\tau \, b_v \rangle 
\nonumber \\[3mm]
=  \frac12  \Bigg( \langle \bar b_v \slashed v   [D_\nu,   [D_\sigma, D_\tau]] & \gamma^\nu \gamma^\sigma \gamma^\tau   b_v \rangle  +  \langle \bar b_v  [D_\nu, [D_\sigma, D_\tau]]  \gamma^\nu \gamma^\sigma \gamma^\tau \slashed v  b_v \rangle \Bigg) + \ldots 
\nonumber \\[3mm]
=  \frac12  \langle \bar b_v  [D_\nu, [D_\sigma, D_\tau]]  \Big (\slashed v & \gamma^\nu  \gamma^\sigma \gamma^\tau   + \gamma^\nu \slashed v \gamma^\sigma \gamma^\tau + 2  v^\tau \gamma^\nu \gamma^\sigma - 2  v^\sigma \gamma^\nu \gamma^\tau \Big) b_v \rangle  + \ldots 
\nonumber \\[3mm]
= - i \langle \bar b_v  [D_\nu, & [D_\sigma,  D_\tau]]  \Big(  v^\nu  \sigma^{\sigma \tau}  +  v^\tau \sigma^{\nu \sigma} -  v^\sigma \sigma^{\nu \tau} \Big) \, b_v \rangle + \ldots 
\nonumber \\[3mm] 
&\qquad= {\cal O}\left(\frac{1}{m_b}\right)  \,,
\label{eq:Orls-me}
\end{align}\\
where in the last step we have taken into account the result in Eq.~(\ref{eq:vsigmamunu}), which applies up to order ${\cal O}(1/m_b)$, moreover \\
\begin{align}
&\quad i \langle \bar b_v  [D_\rho , [D^\rho,  D^\mu]] \gamma_\mu  b_v \rangle  
 \nonumber \\[3mm]
 =  \frac{i}{2} \Big(  \langle \bar b_v \slashed v [ D_\rho ,  [ D^\rho,  & D^\mu]]
\gamma_\mu b_v \rangle +  \langle \bar b_v  [ D_\rho , [ D^\rho,  D^\mu]] \gamma_\mu  \slashed v
 b_v \rangle  \Big) + \ldots 
 \nonumber \\[3mm]
= \frac{i}{2} & \langle \bar b_v  [ D_\rho ,  [ D^\rho,   D^\mu]]
\{ \slashed v, \gamma_\mu \}    b_v \rangle  + \ldots 
\nonumber \\[3mm]
= &\, \langle \bar b_v  [ i D_\rho ,  [ i D^\mu,   i D^\rho]]
  v_\mu b_v \rangle + \ldots
  \nonumber \\[3mm]
& = 2 \, \langle {\cal O}_{\rho_D}\rangle_{tree}  + {\cal O}\left(\frac{1}{m_b}\right)   \,,
 \label{eq:Ord-ME}
 \end{align}\\
with
\begin{equation}
 \langle {\cal O}_{\rho_D}\rangle_{tree} = \frac{1 }{2} \, \langle B|  \bar b _v \, [i D_\rho, [i v \cdot  D, i D^\rho] ] \, b_v | B \rangle \,.
 \end{equation}\\
Using Eqs.~(\ref{eq:Orls-me}), (\ref{eq:Ord-ME}), from Eq.~(\ref{eq:unren-one-loop-me-5}) we finally obtain\\
\begin{align}
\langle {\cal O}_{4q, j}^{(q)} \rangle_{1-loop} &= \frac{a_j }{12 \pi^2}\, \Bigg[  \mu^{- 2 \epsilon} \,  \left( \frac{1}{\epsilon} - \gamma_E + \log(4 \pi) \right) + \log\left( \frac{\mu^2}{m_q^2} \right) + b_j \Bigg] \langle {\cal O}_{\rho_D}\rangle_{tree}
  +  \ldots  \,,
\label{eq:unren-O1}
\end{align}\\
and the ellipsis refer to terms of higher order in $\alpha_s$ and $1/m_b$.
We can now consider the one-loop matrix element of the colour octet operators $ {\cal O}_{4q, j+2}^{(q)}$, $ j = 1,2$, namely \\
\begin{align}
\langle {\cal O}_{4q, j+2}^{(q)} \rangle_{1-loop} & =  \langle i  \int d^4 z \, {\rm T} \Bigg\{\vec {\cal O}_{4q, j+2}^{(q)},  \sum_{q^\prime} \bar q^\prime(z)  \slashed A(z) q^\prime (z) \Bigg\} \rangle + \ldots
\nonumber \\[3mm]
&  = 
 \langle \bar b_v^l \bar \Gamma_j t^a_{lm} \Bigg( \int d^4 z \, i S_0^{(q)}(- z)  i \slashed A^b(z)  t^b_{mr} \, iS_0^{(q)}(z)  \Bigg)  \Gamma_j  t^a_{rs} b^s_v \rangle + \ldots  \,.
\label{eq:Q-octet-Darwin-mixing}
\end{align} \\
where for clarity the colour indices have been explicitly indicated. Taking into account the relations in Eqs.~(\ref{eq:commut-t}), (\ref{eq:ta-properties}), it is straightforward to simplify the product of three colour matrices on the r.h.s.\  of Eq.~(\ref{eq:Q-octet-Darwin-mixing}), i.e.\\
\begin{align}
\big( t^a \cdot t^b \cdot t^a \big)_{ls} &= \big( t^b \cdot t^a \cdot t^a \big)_{ls} + i f^{abc}  \big( t^c \cdot t^a \big)_{ls}
\nonumber \\[3mm]
&= C_F   t^b_{ls} - \frac12  f^{bac}  f^{dac}  t^d_{ls}
\nonumber \\[3mm]
&= t^b_{ls}  \left( C_F - \frac{C_A}{2}\right)
 = - \frac{1}{2 N_c}  t^b_{ls}\,.
\label{eq:3-colour-matrices}
\end{align}\\
Substituting Eq.~(\ref{eq:3-colour-matrices}) into Eq.~(\ref{eq:Q-octet-Darwin-mixing}) and comparing with Eq.~(\ref{eq:Q1-Darwin-mixing}) we can readily obtain that\\
\begin{align}
\langle {\cal O}_{4q, j+2}^{(q)} \rangle_{1-loop}  & = 
 - \frac{1}{2 N_c} \,
  \langle \bar b_v \bar \Gamma_j  \Bigg( \int d^4 z  \, i S_0^{(q)}(- z) i \slashed A(z)  iS_0^{(q)}(z)  \Bigg)  \Gamma_j  b_v\rangle 
  \nonumber \\[3mm]
& =   - \frac{1}{2 N_c} \,  \langle {\cal O}_{4q, j}^{(q)} \rangle_{1-loop} + \ldots \,,
\label{eq:octet-1loop-darwin}
\end{align}\\
and the ellipsis stand for terms of order ${\cal O}(\alpha_s)$ and ${\cal O}(1/m_b)$.
Eq.~(\ref{eq:octet-1loop-darwin}) shows that up to higher order corrections, to compute the mixing between the four-quark and the Darwin operators, it is sufficient to know only the contribution of the colour singlet operators $\langle {\cal O}_{4q, j}^{(q)} \rangle_{1-loop}$. The basis in Eq.~(\ref{eq:O4q}) is then redundant and the computation would appear simpler if we would had chosen the equivalent basis \footnote{We stress that the computation in Ref.~\cite{Lenz:2020oce} has been performed in this basis. }\\
\begin{equation}
\vec {\cal O}_{4q}^{(q) \, \prime} = \Big( {\cal O}_{1}^{(q)}, \,   {\cal O}_{2}^{(q)}, {\cal O}_{3}^{(q) }, \,   {\cal O}_{4}^{(q) }  \Big)\,,
\label{eq:O4q-p}
\end{equation}\\
in terms of the colour-rearranged four-quark operators \\
\begin{align}
{\cal O}_{3}^{(q) } &=  \big( \bar b_v^l  \gamma_\nu  (1-\gamma_5) q^m \big)  \big( \bar q^m  \gamma^\nu (1-\gamma_5) b_v^l \big) \,,
\\[3mm]
{\cal O}_{4}^{(q) } &=   \big( \bar b_v^l   (1-\gamma_5) q^m  \big)  \big( \bar q^m (1+ \gamma_5) b_v^l \big)\,.
\end{align}\\
In fact, from \\
\begin{equation}
{\cal O}_{4q, j+2}^{(q) \, \prime} =\left( 2\,  {\cal O}_{4q,  j+2}^{(q)} + \frac{1}{N_c} \, {\cal O}_{4q, j}^{(q)} \right)\,,
\end{equation}\\
it immediately follows that\\
\begin{equation}
\langle {\cal O}_{4q, j+2}^{(q)\, \prime} \rangle_{1-loop}   =   {\cal O}\left(\frac{1}{m_b}\right) + \,  {\cal O}(\alpha_s)\,.
\label{eq:octet-1loop-darwin-final}
\end{equation}\\
However, we will continue the computation using the original basis in Eq.~(\ref{eq:O4q}). 

The one-loop matrix elements in Eq.~(\ref{eq:unren-O1}), and then also in Eq.~(\ref{eq:octet-1loop-darwin}), are divergent in the limit $\epsilon \to 0$ and need to be renormalised. For the sake of clarity, let us introduce the compact notation for the dimension-six operators in Eq.~(\ref{eq:OPE-d6}) and their corresponding coefficients i.e. \\
\begin{align} 
\vec {\cal O} = \Big(  {\cal O}_{\rho_D}, \sum_q \vec {\cal O}_{4q}^{(q)} \Big)\,,
\qquad 
\vec c = \Big( c_{\rho_D}, \sum_q \, \vec c_{4q}^{\,(q)} \Big)\,
\end{align}\\
so that the operator renormalisation reads\\
\begin{equation}
\langle \vec {\cal O} \rangle^{(0)} = \hat Z \, \langle \vec {\cal O} \rangle\,,
\label{eq:O-bare}
\end{equation}\\
where $\langle \vec {\cal O} \rangle^{(0)}$ and $\langle \vec {\cal O} \rangle$ denote respectively the bare and the renormalised matrix elements and the $\hat Z$ matrix is constructed, by definition, to absorb the divergences of $\langle \vec {\cal O} \rangle^{(0)}$. From\\
\begin{equation}
\langle \vec {\cal O} \rangle^{(0)} = \langle \vec {\cal O} \rangle_{tree} + \langle  \vec {\cal O} \rangle_{1-loop} + \ldots\,,
\end{equation}\\
and Eqs.~(\ref{eq:unren-O1}), (\ref{eq:octet-1loop-darwin}), we can read the expression of the $\hat Z$ matrix and of the renormalised matrix elements $\langle \vec {\cal O} \rangle$, in the $\overline{\rm MS}$ scheme \cite{Bardeen:1978yd} and at order $\alpha_s^0$, namely \\
\begin{equation}
\hat Z = 
\begin{pmatrix}
1 & 0 & 0 & 0 & 0 \\[2mm]
Z_{21} & 1 & 0 & 0 & 0 \\[2mm]
Z_{31}  & 0 & 1 & 0 & 0 \\[2mm]
Z_{41} & 0 & 0 & 1 & 0 \\[2mm]
Z_{51} & 0 & 0 & 0 & 1 \\[2mm]
\end{pmatrix}
+ \, {\cal O}(\alpha_s)\,,
\label{eq:Z-matrix}
\end{equation}\\
with
\begin{align}
Z_{21} = \frac{a_1}{12 \pi^2}  \mu^{- 2\epsilon}   \left( \frac{1}{\epsilon} - \gamma_E  + \log(4\pi)  \right) \,,
\\[3mm]
Z_{31} = \frac{a_2}{12 \pi^2}   \mu^{- 2\epsilon}  \left( \frac{1}{\epsilon} - \gamma_E  + \log(4\pi)  \right) \,,
\\[3mm]
Z_{41} = - \frac{1}{2 N_c} \, Z_{21} \,, \quad Z_{51} = - \frac{1}{2 N_c} \, Z_{31}\,,
\end{align}\\
and
\begin{align}
&\langle {\cal O}_{\rho_D}\rangle (\mu) =  \langle {\cal O}_{\rho_D}\rangle_{tree} +  {\cal O}(\alpha_s)\,,
\label{eq:Ord-ren}
\\[3mm]
&\langle {\cal O}^{(q)}_{4q, j}\rangle (\mu) = \langle {\cal O}_{4q, j}\rangle_{tree} + \frac{a_j}{12 \pi^2} \left( \log \left( \frac{\mu^2}{m_q^2}\right) + b_j \right) \langle {\cal O}_{\rho_D}\rangle_{tree}  +  {\cal O}(\alpha_s)\,,
\label{eq:O-sin-ren}
\\[3mm]
&\langle {\cal O}^{(q)}_{4q, j+2} \rangle (\mu) = \langle {\cal O}_{4q, j+2 } \rangle_{tree} - \frac{1}{2 N_c}\, \frac{a_j}{12 \pi^2} \left( \log \left( \frac{\mu^2}{m_q^2} \right) + b_j \right) \langle {\cal O}_{\rho_D}\rangle_{tree}  +  {\cal O}(\alpha_s)\,,
\label{eq:O-oct-ren}
\end{align}\\
where $j= 1,2$ and $a_j, b_j$ are given in Eq.~(\ref{eq:as-bs}). While Eqs.~(\ref{eq:Ord-ren})-(\ref{eq:O-oct-ren}) describe how the renormalised matrix elements depend on the renormalisation scale $\mu$, the behaviour under a variation of $\mu$ is obtained by requiring that the bare matrix elements $\langle \vec {\cal O} \rangle^{(0)}$ in Eq.~(\ref{eq:O-bare}), are scale independent, i.e.\\
\begin{equation}
\frac{d } {d \log \mu^2}\, \langle \vec {\cal O }\rangle^{(0)}  = 0\,, \quad \Rightarrow \quad  \frac{d\, \langle \vec {\cal O} \rangle} {d \log \mu^2}  = -  \hat \gamma \, \langle \vec {\cal O} \rangle\,,
\label{eq:RG-O}
\end{equation}\\
with the anomalous dimension matrix (ADM) $ \hat \gamma$ defined as \\
\begin{equation}
\hat \gamma = \hat Z^{-1}\, \frac{d}{d \log \mu^2} \, \hat Z\,,
\end{equation}\\
or explicitly, taking into account Eq.~(\ref{eq:Z-matrix})\\
\begin{equation}
\hat \gamma = 
\begin{pmatrix}
0 & 0 & 0 & 0 & 0 \\[2mm]
\gamma_{21} & 0 & 0 & 0 & 0 \\[2mm]
\gamma_{31} & 0 & 0 & 0 & 0 \\[2mm]
\gamma_{41} & 0 & 0 & 0 & 0 \\[2mm]
\gamma_{51} & 0 & 0 & 0 & 0 \\[2mm]
\end{pmatrix}  
+ \, {\cal O}(\alpha_s)\,,
\end{equation}\\
where
\begin{equation}
\gamma_{21} =- \frac{a_1}{12 \pi^2}\,,  \quad  \gamma_{31} = - \frac{a_2}{12 \pi^2}\,,
\quad 
\gamma_{41} = - \frac{\gamma_{21}}{2 N_c} \,, \quad  \gamma_{51} = -\frac{\gamma_{31}}{2 N_c} \,.
\end{equation}\\
The RGEs for the renormalised matrix elements in Eq.~(\ref{eq:RG-O}) lead to the corresponding ones for the renormalised coefficients $\vec c$, since the product \\
\begin{equation}
\langle {\rm Im} {\cal T}\rangle \Big|_{d = 6}= \vec c \cdot \, \langle \vec {\cal O}\rangle\,,
\end{equation}\\
must be scale independent. We then obtain the following system of equations\\
\begin{equation}
\left\{
\begin{array}{cl}
\displaystyle{\frac{d \, \langle \vec {\cal O} \rangle}{d \log \mu^2}}  & = -\hat \gamma \, \langle \vec {\cal O} \rangle\,,
\label{eq:RGE-Q}
\\[6mm]
\displaystyle{\frac{d\, \vec c}{d \log \mu^2}}  & =  \hat \gamma^T \,  \vec c\,, 
\end{array}
\right.
\end{equation}\\
which can be easily solved since $\hat \gamma$ is constant at this order. Integrating from the matching scale $\mu = m_b$ to $\mu = \mu_0$, respectively yields\\
\begin{align}
\langle {\cal O}_{\rho_D} \rangle (\mu_0) &=  \langle {\cal O}_{\rho_D} \rangle (m_b) +  {\cal O}(\alpha_s)\,,
\label{eq:O-rD-mu-var}
\\[3mm]
\langle  \vec{ \cal O}^{(q)}_{4q} \rangle (\mu_0) &= \langle \vec {\cal O}^{(q) }_{4q} \rangle (m_b) - \vec  \gamma \, \log \left( \frac{\mu_0^2}{m_b^2} \right) \langle {\cal O}_{\rho_D} \rangle (m_b) +  {\cal O}(\alpha_s)\,,
\label{eq:Oi-mu-var}
\end{align}
together with
\begin{align}
 \vec c^{\, (q)}_{4q}(\mu_0) &= \vec c^{\, (q)}_{4q}(m_b) + {\cal O}(\alpha_s)\,,
 \label{eq:ci-mu-var}
 \\[3mm]
c_{\rho_D} (\mu_0) &=  c_{\rho_D}(m_b) + \sum_q  \vec \gamma \cdot \vec{c}^{\, (q)}_{4q}(m_b)   \log \left( \frac{\mu_0^2}{m_b^2} \right)  + {\cal O}(\alpha_s)\,,
\label{eq:c-rD-mu-var}
\end{align}\\
and we have defined $\vec \gamma = \big( \gamma_{21},  \gamma_{31} , \gamma_{41}, \gamma_{51} \big)$.
Substituting Eqs.~(\ref{eq:O-rD-mu-var})-(\ref{eq:c-rD-mu-var}) into Eq.~(\ref{eq:OPE-d6}) and taking into account Eqs.~(\ref{eq:Ord-ren})-(\ref{eq:O-oct-ren}), we arrive at\\
\begin{align}
\langle {\rm Im} {\cal T}  \rangle \Big|_{d=6} &=  \left[ c_{\rho_D}(m_b) + \sum \limits_{q } \vec \gamma \cdot  \vec{c}^{\, (q)}_{4q}(m_b) \log \left( \frac{\mu_0^2}{m_b^2} \right) \right] \, \langle {\cal O}_{\rho_D} \rangle_{tree} 
\nonumber \\[3mm]
& + \sum \limits_{q} \sum \limits_{j = 1}^2 \left( c_{4q, j}^{\, (q)}(m_b) - \frac{ c_{4q, j+2}^{\, (q)}(m_b)  }{2 N_c}  \right)  \frac{a_j}{12 \pi^2} \left[ \log \left( \frac{m_b^2}{m_q^2}\right) + b_j \right]  \langle {\cal O}_{\rho_D}\rangle_{tree} 
\nonumber \\[3mm]
&  + \sum \limits_{q }  \vec{c}_{4q}^{\, (q)}(m_b) \cdot \, \left[ \langle { \vec {\cal O}_{4q }}^{\, (q)} \rangle_{tree} - \vec \gamma \, \log \left( \frac{\mu_0^2}{m_b^2} \right) \langle {\cal O}_{\rho_D} \rangle_{tree} \right]
+ {\cal O}(\alpha_s)\,,
\end{align}\\
in which the dependence on the renormalisation scale $\mu_0$ cancels, consistently with a calculation of order $\alpha_s^0$. Dropping for simplicity the the suffix $tree$, we then obtain\\
\begin{align}
\langle {\rm Im} {\cal T}  \rangle \Big|_{d=6} &=   c_{\rho_D} (m_b) \langle {\cal O}_{\rho_D}\rangle+ \sum \limits_{q }  \vec c_{4q}^{\, (q)} (m_b)  \cdot  \langle { \vec {\cal O}_{4q}}^{\, (q)} \rangle 
\nonumber \\[3mm]
& + \sum \limits_{q }  \sum \limits_{j = 1}^2 \left( c_{4q, j}^{\, (q)}(m_b) - \frac{ c_{4q, j+2}^{\, (q)}(m_b) }{2 N_c} \right)  \frac{a_j}{12 \pi^2}  \left[ \log \left( \frac{m_b^2}{m_q^2}\right) + b_j \right] \langle {\cal O}_{\rho_D}\rangle
 + {\cal O}(\alpha_s)
  \,.
\label{eq:Im-T-d6}
\end{align}\\
The l.h.s.\ of Eq.~(\ref{eq:Im-T-d6}) has been computed in Section~\ref{sec:T-4q-6} and in Section~\ref{sec:Dar-sec-1} and schematically reads\\
\begin{equation}
\langle {\rm Im} {\cal T}  \rangle \Big|_{d=6}  = {\cal C}_D \langle {\cal O}_{\rho_D} \rangle  +  \vec{\cal C}_{ \rm WE} \cdot \langle \vec{\cal O}^{(q_3)}_{4q} \rangle 
 +  \vec{\cal C}_{ \rm WA} \cdot \langle \vec{\cal O}^{(q_1)}_{4q} \rangle  +  \vec{\cal C}_{ \rm PI} \cdot \langle \vec{\cal O}^{(q_2)}_{4q} \rangle\,,
\label{eq:Im-T-d6-2}
\end{equation}\\
where the coefficients ${\cal C}_D$ have the divergent behaviour shown in Eq.~(\ref{eq:lim-Dnm}). Equating the respective r.h.s.\ of Eqs.~(\ref{eq:Im-T-d6}), (\ref{eq:Im-T-d6-2}), we can finally read the expression of $\vec{c}^{\, (q)}_{4q}$ and $c_{\rho_D}$ at the matching scale $m_b$, namely\\
\begin{align}
\vec{c}^{\,(q)}_{4q}(m_b) & = \vec{\cal C}_{ \rm WE}\,  \delta^{q q_3} + \vec{\cal C}_{ \rm WA} \,\delta^{q q_1} + \vec{\cal C}_{ \rm PI} \, \delta^{q q_2}\,,
\end{align}\\
and 
\begin{equation}
c_{\rho_D}(m_b)  = {\cal C}_D \,  -  \sum \limits_{q }  \sum \limits_{j = 1}^2 \left( c_{4q, j}^{\, (q)}(m_b) - \frac{1}{2 N_c}  c_{4q, j+2}^{\, (q)}(m_b)  \right) 
 \frac{a_j}{12 \pi^2}  \left[ \log \left( \frac{m_b^2}{m_q^2}\right) + b_j \right] \,.
\label{eq:crD-finite}
\end{equation}\\
It is straightforward to check, by taking into account the results for the coefficients of the four-quark operators computed in Section~\ref{sec:T-4q-6}, together with the expressions of the divergent functions listed in Appendix~\ref{app:5}, that all the dependence on $\log \left(m_q^2/ m_b^2 \right)$ exactly cancels on the r.h.s.\ of Eq.~(\ref{eq:crD-finite}), leaving the final coefficient of the Darwin operator $c_{\rho_D}$, free of any IR divergences. Notice  though that as consequence of the operator mixing, ${\cal C}_{D}$ and $c_{\rho_D}$ differ also by a finite contribution due to the presence of the constants $b_j$. 
The complete expressions for the coefficients $c_{\rho_D}$ are presented in the next section.


\section{Analytical expressions for the coefficients of the Darwin operator}
\label{sec:Dar-sec-3}
\begin{figure}
\centering
\includegraphics[scale=1]{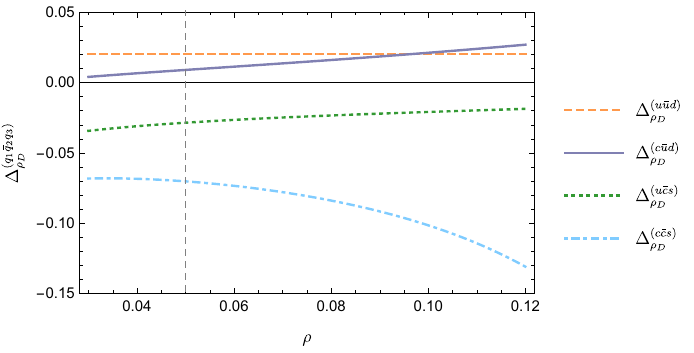} 
\caption{Relative size of the Darwin term compared to the partonic-level contribution respectively for the $b \to u \bar u d$ (dashed orange), $b \to c \bar u d$ (solid purple), $b \to u \bar c s$ (dotted green)
  and $b \to c \bar c s$ (dot-dashed cyan) modes. The reference values $m_b=4.5$ GeV and $\rho_D^3 = 0.2$~GeV$^3$ have been used and the dashed vertical line indicates the approximate value $\rho = 0.05$ 
  in the $\overline{ {\rm MS}}$ scheme.
}
\label{fig:Deltas-sum}
\end{figure}

The final expression for the contribution of the Darwin operator to the inclusive decay width of a $B$ meson, induced by the flavour-changing transition
$b \to q_1 \bar q_2 q_3$, with $q_1, q_2 = \{u,c\}$ and $q_3 = \{d, s\}$, is presented in the following form   \footnote{Note that the coefficients originally presented in Ref.~\cite{Lenz:2020oce}, still depend on the renormalisation scale $\mu_0$ since the running of the matrix element of the corresponding four-quark operators was not explicitly included.}\\
\begin{align}
\Gamma_{\rho_D}^{(q_1 \bar q_2 q_3)}(B) & = 
\Gamma_0 \, \Bigl( N_c C_1^2 \, {c}_{\rho_D, \, 11}^{(q_1 \bar q_2 q_3)} 
+ 2 \, C_1 C_2 \, {c}_{\rho_D,\, 12}^{(q_1 \bar q_2 q_3)} 
+ N_c C_2^2  \, {c}_{\rho_D,\, 22}^{(q_1 \bar q_2 q_3) }  \Bigr) \, \frac{\rho_D^3(B)}{m_b^3},
\label{eq:Gamma-NL-final-res}
\end{align}\\
where the non perturbative parameter $\rho_D(B)$ is defined as in Eq.~(\ref{eq:dim-6-ME-parameters}), $\Gamma_0$ is given in Eq.~(\ref{eq:Gamma0}) and the coefficients $c_{\rho_D, mn}^{(q_1 \bar q_2 q_3)}$ respectively read \\
\begin{align}
c_{\rho_D, 11}^{(u \bar u d)} = 6 \,, 
\qquad 
c_{\rho_D, 12}^{(u \bar u d)}  = 
-\frac{34}{3}\,, 
\qquad 
{c}_{\rho_D, 22}^{(u \bar u d)}  = 6\,,
\label{eq:Ruud}
\end{align}\\
for the $b \to u \bar u d$ mode, \\
\begin{align}
c_{\rho_D, 11}^{(u \bar c s)} 
& = 
\frac{2}{3} (1 - \rho) \biggl[ 9  + 11 \rho - 12 \rho ^2 \log (\rho ) 
 - \, 24 \left(1 - \rho^2 \right) \log (1-\rho )- 25 \rho ^2  + 5 \rho^3 \biggl]\,,
\label{eq:R11ucs} 
\\[3mm]
c_{\rho_D, 12}^{(u \bar c s)} 
& = 
\frac{2}{3} \biggl[ - 41 - 12 \left(2 + 5 \rho + 2 \rho ^2 - 2  \rho ^3 \right) \log( \rho)   - \, 48 (1 - \rho)^2 (1 + \rho ) \log (1-\rho) 
\nonumber \\[3mm]
&
\ + 26  \rho - 18 \rho^2 + 38 \rho^3 - 5 \rho ^4 \biggl]\,,
\label{eq:R12ucs} 
\\[3mm]
c_{\rho_D, 22}^{(u \bar c s)} 
& = 
\frac{2}{3} (1 - \rho) \biggl[ 9  + 11 \rho - 12 \rho ^2 \log (\rho )  - \, 24 \left(1 - \rho^2 \right) \log (1-\rho )- 25 \rho ^2  + 5 \rho^3 \biggr]\,,
\label{eq:R22ucs}
\end{align}\\
in the case of $b \to u \bar c s$ transition,\\
\begin{align}
c_{\rho_D, 11}^{(c \bar u d)} 
& =  
\frac{2}{3} \biggl[17 +12 \log (\rho)-16 \rho -12 \rho ^2 + 16 \rho ^3 - 5 \rho ^4 \biggr], 
\label{eq:R11cud} 
\\[4mm]
c_{\rho_D, 12}^{(c \bar u d)} 
& =  
\frac{2}{3} \biggl[- 9 + 12 \left(1 - 3 \rho ^2 + \rho ^3 \right) \log( \rho) 
\nonumber \\[4mm]
&
 + \, 24 (1 - \rho)^3 \log (1-\rho) +  50 \rho - 90 \rho ^2 + 
54 \rho ^3 - 5 \rho ^4 \biggr]\,, 
\label{eq:R12cud} 
\\[3mm]
c_{\rho_D, 22}^{(c \bar u d)} 
& = 
\frac{2}{3} (1 - \rho) \biggl[ 9  + 11 \rho - 12 \rho ^2 \log (\rho ) 
\nonumber \\[4mm]
& 
 - \, 24 \left(1 - \rho^2 \right) \log (1-\rho )- 25 \rho ^2  + 5 \rho^3 \biggr]\,,
\label{eq:R22cud}
\end{align}\\
for the $b \to u \bar c s$ mode and finally\\
\begin{align}
c_{\rho_D, 11}^{(c \bar c s)} & = 
\frac{2}{3} \Biggl[ \sqrt{1 - 4 \rho} \left(17 + 8 \rho - 22 \rho^2 - 60 \rho^3 \right)
\nonumber \\
&  \qquad- \, 12 \left(1 - \rho - 2 \rho^2 + 2 \rho^3 + 10 \rho^4 \right) 
\log \left(\frac{1 + \sqrt{1 - 4 \rho^{\phantom{\! 1}}}}
{1 - \sqrt{1 - 4 \rho^{\phantom{\! 1}}}} \right) \Biggr],
 \label{eq:R11ccs} \\[5mm]
c_{\rho_D, 12}^{(c \bar c s)} & = 
\frac{2}{3} \Biggl[ \sqrt{1 - 4 \rho} \left(-45 + 46 \rho - 106 \rho^2 - 60 \rho^3 \right)
\nonumber \\
& \quad \quad + \, 12 \left(1 + 4 \rho^2 - 16 \rho^3 - 10 \rho^4 \right) 
\log \left(\frac{1 + \sqrt{1 - 4 \rho^{\phantom{\! 1}}}}
{1 - \sqrt{1 - 4 \rho^{\phantom{\! 1}}}} \right) \Biggr] 
\nonumber \\[2mm]
& + \, 8 \, \left[ {\cal M}_{112} (\rho, \eta) - \sqrt{1 - 4 \rho} \, \log( \eta) \right]\Bigr|_{\eta \to 0} \,,
\label{eq:R12ccs} \\[5mm]
c_{\rho_D, 22}^{(c \bar c s)} & = 
\frac{2}{3} \Biggl[ \sqrt{1 - 4 \rho} \left(-3 + 22 \rho - 34 \rho^2 - 60 \rho^3 \right)
\nonumber \\
& \quad \quad  - \, 24 \rho \left(1 + \rho + 2 \rho^2 + 5 \rho^3 \right) 
\log \left(\frac{1 + \sqrt{1 - 4 \rho^{\phantom{\! 1}}}}
{1 - \sqrt{1 - 4 \rho^{\phantom{\! 1}}}} \right) \Biggr] 
\nonumber \\[2mm]
& + \, 8 \, \left[ {\cal M}_{112} (\rho, \eta) - \sqrt{1 - 4 \rho} \, \log (\eta) \right]\Bigr|_{\eta \to 0}\,,
\label{eq:R22ccs}
\end{align}\\
in the case of $b \to c \bar c s$ decay.
\begin{figure}[t]
\hspace*{-0.5cm}
\includegraphics[scale=0.65]{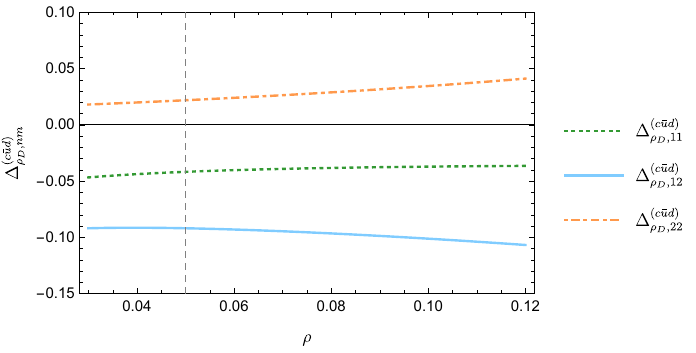} 
\includegraphics[scale=0.65]{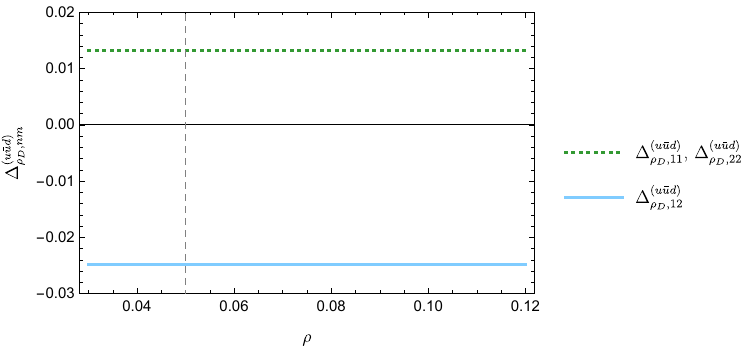}
\hspace*{-0.5cm}
\includegraphics[scale=0.65]{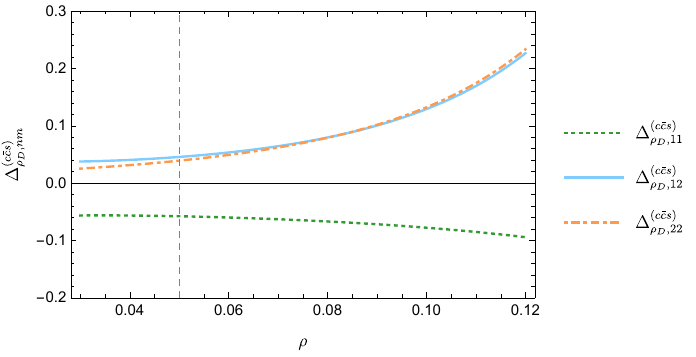} 
\includegraphics[scale=0.65]{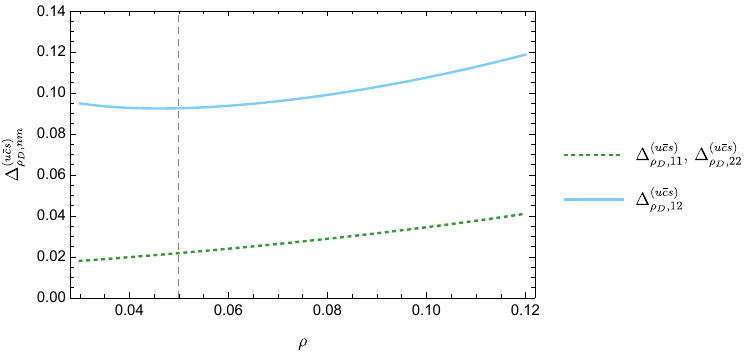}
\caption{Relative effect of the Darwin operator with respect to the dimension-three term for the $b \to c \bar u d$ (top left),
  $b \to u \bar u d$ (top right), $b \to c \bar c s$ (bottom left), and $b \to u \bar c s$ (bottom right) transitions. For each mode, the green dotted, the solid cyan and the dotted-dashed orange lines correspond respectively to the $Q_1 \otimes Q_1$, $Q_1 \otimes Q_2$ and $Q_2 \otimes Q_2$ contributions. The reference values $m_b=4.5$ GeV and $\rho_D^3 = 0.2$ GeV$^3$ have been used and the dashed vertical line indicates the approximate
  value $\rho = 0.05$ in the $\overline{ {\rm MS}}$ scheme.
}
\label{fig:Deltas}
\end{figure}
In Eqs.~(\ref{eq:R11ucs})-(\ref{eq:R22ccs}) the dimensionless mass parameter is $\rho = m_c^2/m_b^2$, moreover the master integral ${\cal M}_{112}$ in Eqs.~(\ref{eq:R12ccs}), (\ref{eq:R22ccs}) is defined~as\\
\begin{equation}
{\cal M}_{112} (\rho, \eta) = 
- \! \! \! \int\limits_{(\sqrt \rho + \sqrt \eta)^2}^{(1 - \sqrt{\rho})^2} \! \! \! d t \,
\frac{\left(t^2 - 2 (1 + \rho) t + (1 - \rho)^2 \right) (t - \eta + \rho)}{t 
\sqrt{\left(t^2 - 2 (1 + \rho) t + (1 - \rho)^2 \right) 
\left(t^2 - 2 t (\eta +\rho) + (\eta -\rho )^2\right)}} \,,
\end{equation}\\
with $\eta = m_q^2/m_b^2$. We emphasise however, that the analytical expression for the limits in Eqs.~(\ref{eq:R12ccs}), (\ref{eq:R22ccs}), has been derived in Ref.~\cite{Mannel:2020fts}, namely\\
\begin{align}
\left[ {\cal M}_{112} (\rho, \eta) - \sqrt{1^{\! \! \! \phantom 1} - 4 \rho} \, \log (\eta) \right]\Bigr|_{\eta \to 0} &= 2 (1- \rho) \log \left( \frac{1 + \sqrt{1- 4 \rho}}{1 - \sqrt{1 - 4 \rho}}\right) 
\nonumber \\[3mm]
& +  \sqrt{1 - 4 \rho}\, \left[ 1 + 2 \log (\rho) - 4 \log \left( \sqrt{1 - 4 \rho}\right)\right] \,,
\end{align}
in terms of which Eqs.~(\ref{eq:R12ccs}), (\ref{eq:R22ccs}), can be simplified as\\
\begin{align}
c_{\rho_D, 12}^{(c \bar c s)} & = 
\frac{2}{3} \Biggl[ \sqrt{1 - 4 \rho} \left(-33 +  24 \log(\rho) - 24 \log(1 - 4 \rho) + 46 \rho - 106 \rho^2 - 60 \rho^3 \right)
\nonumber \\
&  \qquad + \, 12 \left(3 - 2 \rho + 4 \rho^2 - 16 \rho^3 - 10 \rho^4 \right) 
\log \left(\frac{1 + \sqrt{1 - 4 \rho^{\phantom{\! 1}}}}
{1 - \sqrt{1 - 4 \rho^{\phantom{\! 1}}}} \right) \Biggr],
 \label{eq:R12ccs-new} \\[5mm]
 c_{\rho_D, 22}^{(c \bar c s)} & = 
\frac{2}{3} \Biggl[ \sqrt{1 - 4 \rho} \left(9 + 24 \log(\rho) - 24 \log(1 - 4 \rho) + 22  \rho - 34 \rho^2 - 60 \rho^3 \right)
\nonumber \\
&  \qquad + \, 24 \left(1- 2 \rho - \rho^2 - 2 \rho^3 - 5 \rho^4 \right) 
\log \left(\frac{1 + \sqrt{1 - 4 \rho^{\phantom{\! 1}}}}
{1 - \sqrt{1 - 4 \rho^{\phantom{\! 1}}}} \right) \Biggr].
 \label{eq:R22ccs-new} 
 \end{align}\\
Finally, we stress that for $m_d = m_s = 0$, the following relations hold, i.e.\\
\begin{equation}
c_{\rho_D, mn}^{(c \bar u d)} = c_{\rho_D, mn}^{(c \bar u s)}\,,
 \qquad
c_{\rho_D, mn}^{(c \bar c s)} = c_{\rho_D, mn}^{(c \bar c d)}\,,
\qquad
c_{\rho_D, mn}^{(u \bar u d)} = c_{\rho_D, mn}^{(u \bar u s)}\,,
 \qquad
c_{\rho_D, mn}^{(u \bar c s)} = c_{\rho_D, mn}^{(u \bar c d)}\,.
\end{equation}
The relative effect of the Darwin operator with respect to the corresponding partonic-level contribution ${\cal C}_0^{(q_1 \bar q_2 q_3)}$ is given by\\ 
\begin{equation}
\Delta_{\rho_D,mn}^{(q_1 \bar q_2 q_3)} = 
\frac{{c}_{\rho_D,mn}^{(q_1 \bar q_2 q_3)}}{{\cal C}_0^{(q_1 \bar q_2 q_3)}} \, \frac{\rho_D^3 }{m_b^3} \, .
\label{eq:ratios-Delta}
\end{equation}\\
In Figure~\ref{fig:Deltas}, the dependence of the functions in Eq.~(\ref{eq:ratios-Delta}) on the dimensionless mass parameter $\rho$, is plotted for all the three colour structures and the four modes,
using for reference the values $m_b=4.5$ GeV and $\rho_D^3 = 0.2$ GeV$^3$. Furthermore,  Figure~\ref{fig:Deltas-sum}
 shows the total relative contribution for each mode, namely\\
\begin{equation}
\Delta_{\rho_D}^{(q_1 \bar q_2 q_3)} =  \frac{ 3 \, C_1^2 \, {c}_{\rho_D, \, 11}^{(q_1 \bar q_2 q_3)} 
+ 2 \, C_1  C_2 \, {c}_{\rho_D,\, 12}^{(q_1 \bar q_2 q_3)} 
+ 3 \, C_2^2 \, {c}_{\rho_D,\, 22}^{(q_1 \bar q_2 q_3) } }
      {\Big( 3 \, C_1^2  + 2 \,  C_1 C_2 +3 \, C_2^2 \Big) \,  {\cal C}_0^{(q_1 \bar q_2 q_3)}  }\,  \frac{\rho_D^3}{m_b^3} \, ,
\end{equation}\\
indicating that the Darwin operator can lead to sizeable corrections to the $b \to q_1 \bar q_2 q_3$ decay width, of the order of $1-7\, \% $ (for $\rho = 0.05$).

A final comment about the numerical effect of the constant terms $b_j$ in Eq.~(\ref{eq:crD-finite})
for the coefficients ${c}_{\rho_D, mm}^{(q_1 \bar q_2 q_3)}$. 
We have already noticed that their values depend on the choice of the four-quark operators basis. 
Consider as an example the coefficient ${c}_{\rho_D ,12}^{(c \bar u d)}$. In our basis given by Eq.~(\ref{eq:O4q}), 
${c}_{\rho_D ,12}^{(c \bar u d)} = - 29.0$, for the reference value $\rho = 0.05$. 
Had we chosen the same basis as done in Ref.~\cite{Mannel:2020fts}, namely with ${O}_1^{(u)}$ and $O_2^{(u)}$, replaced by \\
\begin{equation}
{\cal O}_{1}^{(u)} = 
(\bar b_v^{\, i} \Gamma^\sigma \gamma^\mu \Gamma^\rho u^i)
(\bar u^j \Gamma_\sigma \gamma_\mu \Gamma_\rho b_v^{\, j}),
\qquad
{\cal O}_{2}^{(u)} = 
(\bar b_v^{\, i} \Gamma^\sigma \slashed v \Gamma^\rho u^i)
(\bar u^j \Gamma_\sigma \slashed v \Gamma_\rho b_v^{\,j}),
\end{equation} \\
we would have obtained in Eq.~(\ref{eq:as-bs}), that 
$a_1 = 8, \, $
$b_1 = -5/4$ and 
$a_2 = 2, \,$ 
$ b_2 = -3/2$, 
leading to 
${c}_{\rho_D ,12}^{(c \bar u d)} = - 24.0$, for the same value of $\rho $. This shift of $\sim 17 \%$ must be compensated, up to corrections of higher orders, by the different value of the matrix element of the operators defined in these two bases.


\chapter{Phenomenology of Lifetime and Mixing}
\label{ch:pheno}
In this last chapter, we consider two phenomenological applications of the HQE in the charm sector, specifically, the study of the lifetime of charmed mesons and of neutral $D$-meson mixing. In light of the large amount of current and future charm data collected
by LHCb \cite{Bediaga:2018lhg}, BESIII \cite{Ablikim:2019hff}, and Belle-II \cite{Kou:2018nap}, an improvement of the theoretical understanding of charm physics, see Refs.~\cite{Bellini:1996ra, Bianco:2003vb, Artuso:2008vf, Gersabeck:2012rp, Lenz:2020awd } for a comprehensive introduction to the subject, is crucial to fully exploit
the significant experimental progress in this field.
The recent discovery by the LHCb collaboration
\cite{Aaij:2019kcg} of direct CP violation in the charm sector, specifically in the non-leptonic decays $D^0 \to \pi^+ + \pi^-$ and  $D^0 \to K^+ + K^-$, provides one example of this necessity, since after its announcement,
both SM and BSM interpretations of the measurement have been proposed, see Refs.~\cite{Li:2019hho,Grossman:2019xcj,Cheng:2019ggx,Soni:2020kse} for the former, and Refs.~\cite{Chala:2019fdb, Dery:2019ysp}, partly based on the calculation of Ref.~\cite{Khodjamirian:2017zdu}, for the latter \footnote{A summary of references investigating a previous claim for evidence of CP violation can be found in Ref.~\cite{Lenz:2013pwa}.}. 
Exclusive non-leptonic decays of charm hadrons and even of $b$-hadrons are among the 
most challenging observables in quark flavour physics from a theoretical point of view. In the following, we will instead focus on the study of inclusive quantities like the total decay width, for which the HQE provides a systematic theoretical framework, see Section~\ref{sec:HQE}. 
However,
due to the size of its mass, the charm quark sits at the boundary between the heavy- and light-quark region, making the applicability of the HQE a priory questionable. This is clearly signalled by the fact that contrary to the $b$-sector, lifetime ratios of charmed hadrons can significantly differ from one, which represents the naive expectation in the heavy quark limit. Specifically,  lifetimes of charmed hadrons are experimentally determined very precisely  
\cite{Zyla:2020zbs} and also inclusive semileptonic branching fractions have been measured~\cite{Zyla:2020zbs}, with a recent update for the $D_s$-meson released by the BESIII Collaboration \cite{Ablikim:2021qvs}
 \footnote{New results from Belle II have recently been published \cite{Belle-II:2021cxx}: $\tau(D^0)= 410.5 \pm 1.1 \pm 0.8$~fs, \, $\tau (D^+) = 1030.4 \pm 4.7 \pm 3.1$~fs.
}. 
A summary of the current experimental status for the lightest $D$-mesons, is shown in Table~\ref{tab:exp-data}.
\begin{table}[t]
\centering
\renewcommand{\arraystretch}{1.5}
    \begin{tabular}{|c||c|c|c|}
    \hline
         & $D^0$ & $D^+$ & $D_s^+$ 
         \\
         \hline
         \hline
    $\tau \, [{\rm ps}]$ & $0.4101(15)$ & $1.040(7)$ & $0.504 (4)$
    \\
    \hline
     $\Gamma \, [{\rm ps}^{-1}]$ & $2.44(1)$ & $0.96(1)$ & $1.98 (2)$
    \\
    \hline
    $\tau (D_X)/\tau (D^0)$ & $1$ & $2.54(2)$ & $1.20 (1)$
    \\
    \hline
    \hline
     ${\rm Br}(D_X \to X e^+ \nu_e)  [\%]$ & 
     $6.49(11)$  & $16.07(30)$ & $6.30(16)$
     \\
    \hline
    $\displaystyle\frac{\Gamma (D_X \to X e^+ \nu_e)}{\Gamma (D^0 \to X e^+ \nu_e)}$ & $1$ & $0.977(26)$ & $0.790 (26)$    \\
    \hline
    \end{tabular}
    \caption{Status of the experimental determinations of the lifetime and the semileptonic branching fractions of the lightest charmed mesons. All values are taken from the 
    PDG~\cite{Zyla:2020zbs}, apart from the ones for the semileptonic 
    $D_s$-meson decays, which were recently measured by the BESIII Collaboration~\cite{Ablikim:2021qvs}. }
    \label{tab:exp-data}
\end{table}
Finally, a long-standing puzzle in charm physics, is the theoretical description of mixing of neutral
$D$ mesons, see e.g.\ the excellent reviews \cite{Proceedings:2001rdi, Nierste:2009wg, Silvestrini:2019sey}. Charm-mixing is experimentally well established and the HFLAV \cite{Amhis:2019ckw} average 
of Refs. \cite{Aitala:1996vz,Cawlfield:2005ze,Aubert:2007aa,Bitenc:2008bk,Aitala:1996fg,Godang:1999yd,Link:2004vk,Zhang:2006dp,Aubert:2007wf,Aaltonen:2013pja,Ko:2014qvu,Aaij:2017urz,Aubert:2008zh,Aaij:2016rhq,Aitala:1999dt,Link:2000cu,Csorna:2001ww,Lees:2012qh,Aaltonen:2014efa,Ablikim:2015hih,Aaij:2015yda,Staric:2015sta,Aaij:2017idz,Aaij:2018qiw,Aubert:2007if,Canto:2013fza,Aaij:2019kcg,delAmoSanchez:2010xz,Peng:2014oda,Aaij:2015xoa,Aaij:2019jot,Zupanc:2009sy,TheBABAR:2016gom,Asner:2012xb, LHCb:2021ykz} \footnote{Performed in the case of allowed CP violation.} reads\\
\begin{equation}
  x = \frac{\Delta M_D  }{  \Gamma_{D^0}} = (0.409^{+ 0.048}_{-0.049}) \% \, \,  , \quad 
  y = \frac{\Delta \Gamma_D }{2 \Gamma_{D^0}} = 0.615^{+0.056}_{-0.055} \% \, ,
  \label{eq:x-y}
\end{equation}\\
where $\Delta M_D$ is the mass difference of the neutral $D^0$ mesons mass eigenstates
and $\Delta \Gamma_D$ the corresponding decay rate difference. 
However, the theoretical predictions for $x$ and $y$ cover a vast range of values, which spread over 
several orders of magnitude, see e.g.\ Refs.~\cite{Nelson:1999fg,Petrov:2003un}.
Future measurements will not only increase the precision of $x$ and $y$, but also provide stronger bounds or even evidence for  CP violation in mixing \cite{Cerri:2018ypt}. 
It is clear that having a reliable range of potential SM predictions is necessary in order to benefit  from these experimental improvements.

In Section~\ref{sec:pheno1}  we discuss the study of the total decay width of charmed mesons and obtain theoretical predictions for the lifetimes of the $D^0$, $D^+$, and $D_s^+$ mesons and their ratios, as well as for the semileptonic branching fractions ${\rm Br} (D \to X e^+ \nu_e)$, and their ratios.
Furthermore, in Section~\ref{sec:pheno2} we present a possible solution for the discrepancy between previous HQE determinations of $D$-mixing with data. The content of this chapter closely follows the one of Ref.~\cite{King:2021xqp} and  Ref.~\cite{Lenz:2020efu}.


\section{Theoretical study of the total decay width of charmed mesons}
\label{sec:pheno1}

In the present section, we analyse the structure of the HQE in the charm sector, to try to shed further light
into the question, whether the expansion parameters
$\alpha_s (m_c)$, and $\Lambda_{\rm QCD} / m_c$, are small
enough in order to ensure meaningful theoretical predictions for the observables listed in Table~\ref{tab:exp-data}.
The Particle Data Group \cite{Zyla:2020zbs} quotes, respectively
for the pole and the $\overline{\rm MS}$ mass of the charm
quark, the values\\
\begin{equation}
m_c^{\rm Pole} = (1.67  \pm 0.07) \; {\rm GeV}
\; ,
\hspace{1cm}
\overline{m}_c(\overline{m}_c) = (1.27  \pm 0.02) \; {\rm GeV}\,,
\label{eq:Mass}
\end{equation}\\
while the dependence of $\alpha_s$ on both
the charm scale and the loop order, obtained using the
RunDec package \cite{Herren:2017osy}, is shown in
Table~\ref{tab:alphas}. We emphasise that in our numerical analysis we use 
the five-loop running result. 
\begin{table}[h]
\centering
\renewcommand{\arraystretch}{1.4}
\begin{tabular}{|c||c|c|c|}
\hline
$\alpha_s(m_c)$ &
$m_c = 1.67 \, {\rm GeV}$ & 
$m_c = 1.48 \, {\rm GeV}$ &
$m_c = 1.27 \, {\rm GeV}$
\\
\hline
\hline
\mbox{two-loop}     &  
$0.322$ &
$0.346$ &
$0.373$
\\
\hline
\mbox{five-loop}     &  
$0.329$ & 
$0.356$ &
$0.387$ 
\\
\hline
\end{tabular} 
\caption{ Numerical values of the strong coupling
$\alpha_s$ evaluated at different scales and loop
order, obtained using the RunDec
package~\cite{Herren:2017osy}.}
\label{tab:alphas}
\end{table}\\
The relation between the pole and $\overline{\rm MS}$ mass schemes, up to third order in the strong coupling, reads \cite{Chetyrkin:1999qi,Chetyrkin:1999ys,Melnikov:2000qh}\\
\begin{align}
  m_c^{\rm Pole}  & =
   \overline{m}_c(\overline{m}_c) 
   \left[ 1 
   + \frac43 \frac{\alpha_s(  \overline{m}_c)}{\pi} 
   + 10.43
  \left( \frac{\alpha_s(  \overline{m}_c)}{\pi} \right)^2 
   + 116.5
  \left( \frac{\alpha_s(  \overline{m}_c)}{\pi} \right)^3
  \right] 
  \nonumber
  \\[3mm]
  & = 
  \overline{m}_c(\overline{m}_c) 
   \left[ 1 
   + 0.1642
   + 0.1582 
   + 0.2176 
  \right]       \, ,
   \label{eq:MSpole}
\end{align}\\
where we have used the
five-loop value of $\alpha_s$,
at the scale 1.27 GeV. The strong dependence of $\Gamma_3$ on the charm pole mass, cf.\  Eq.~(\ref{eq:Gamma0}), leads to different
results according to how higher orders in
Eq.~(\ref{eq:MSpole}) are treated. 
Specifically, by truncating the expansion in
Eq.~\eqref{eq:MSpole} at first order in $\alpha_s$, and using 
$\overline{m}_c(\overline{m}_c)
= 1.27$ GeV, we obtain for the pole mass the value 
$m_c^{\rm Pole} = 1.479 $ GeV, which, respectively yields \\
\begin{equation}
\left( m_c^{\rm Pole} \right)^5 
= 
  \overline{m}_c(\overline{m}_c)^5 
  \left[ 1 +0.1642 \right]^5
= 2.14 \, \overline{m}_c(\overline{m}_c)^5,
\label{eq:prefactor1}
\end{equation}\\
by computing the fifth power of $m_c^{\rm Pole}$, and\\
\begin{equation}
 \left( m_c^{\rm Pole} \right)^5   \approx 
  \overline{m}_c (\overline{m}_c)^5 
  \left[ 1 + 5 \cdot 0.1642 \right]
  = 1.82 \, \overline{m}_c(\overline{m}_c)^5 
   \, ,
   \label{eq:MSpole5}
\end{equation}\\
by further expanding the fifth power up to the first 
order in $\alpha_s$. Note that the result in Eq.~(\ref{eq:MSpole5}) 
is about 15~$\%$ smaller than the one in 
Eq.~(\ref{eq:prefactor1}). Conversely, by including also all the higher order terms shown in 
Eq.~(\ref{eq:MSpole}), gives\\
\begin{equation}
\left( m_c^{\rm Pole} \right)^5 
 =
  \overline{m}_c(\overline{m}_c)^5 
  \left[  1 
  + 0.1642
   + 0.1582
   + 0.2176
   \right]^5
   =  8.66 \,
  \overline{m}_c(\overline{m}_c)^5 
   \, ,
   \label{eq:MSpole5b}
\end{equation}\\
which is roughly four times larger than the result in Eq.~(\ref{eq:prefactor1}). In order to deal with this numerical instability, in the following we investigate different scenarios:\\
\begin{enumerate}
\item Use Eq.~(\ref{eq:MSpole}) to first order in  
      $\alpha_s$, since this is the order at which most of
      the Wilson coefficients are known. In this case we fix
      $m_c^{\rm Pole} = 1.48$ GeV and $\alpha_s = 0.356$.
      A further possibility would be to use as input the pole mass value
      from the PDG, i.e.\  $m_c^{\rm Pole} = 1.67 $ GeV. However, in this case, our numerical analysis gives results for the decay rates which are roughly $30\%$ larger than the ones obtained in the $1S$ scheme, discussed below. Since we expect this enhancement to be compensated by missing NNLO
      corrections to the non-leptonic decay rates,
      we do not present explicit results for $m_c^{\rm Pole} = 1.67 $ GeV.
       
\item Express $m_c^{\rm Pole}$ in terms of the 
     $\overline{\rm MS}$ mass  \cite{Bardeen:1978yd}\\
     \begin{equation}
     m_c^{\rm Pole} = \overline m_c(\overline m_c) 
     \left[ 1 + \frac{4}{3} \frac{\alpha_s(\overline m_c) }{\pi} \right]\,,
     \label{eq:c-quark-mass-Pole-to-MS}
     \end{equation}\\
     using $\overline{m}_c(\overline{m}_c) = 1.27$ GeV~\cite{Zyla:2020zbs}, 
     and expand consistently up to order $\alpha_s$. Because
     of the dependence on the fifth power of the charm quark
     mass, in this case, $\Gamma_3$ receives large
     corrections  $\sim 5 \times (4/3)(\alpha_s /\pi)$.
     
\item Express $m_c^{\rm Pole}$ in terms of the kinetic mass \cite{Bigi:1994ga,Bigi:1996si}. The kinetic scheme has been introduced in order to obtain a short distance definition of the heavy quark mass which allows a faster convergence of the perturbative series and which is still valid at small scales $\mu \sim 1$ GeV. The relation between the kinetic scheme and the $\overline{\rm MS}$ and pole schemes can be found, up to N$^3$LO corrections, in Ref.~\cite{Fael:2020njb}. At order $\alpha_s$ we have\\
\begin{equation}
      m_c^{\rm Pole}  = m_c^{\rm Kin}
      \left[ 1 + \frac{4 \alpha_s}{3 \pi} 
      \left( 
      \frac43 \frac{\mu^{\rm cut}}{ m_c^{\rm Kin} }
      +
      \frac12 \left(\frac{\mu^{\rm cut}}{ m_c^{\rm Kin}}\right)^2
      \right)\right] \, ,
      \label{eq:Pole-Kin-scheme}
      \end{equation}\\
where $\mu^{\rm cut}$ is the Wilsonian cutoff 
separating the perturbative and non perturbative regimes.
In our numerical analysis we set $\mu^{\rm cut} = 0.5$~GeV, which gives, at NLO-QCD and using as an input $\overline{m}_c (3 \, {\rm GeV})$ \cite{Fael:2020njb} \\
 \begin{equation}
  m_c^{\rm kin} (0.5 \, {\rm GeV}) = 1.363 \, {\rm GeV} \,.
 \end{equation}
\item Express $m_c^{\rm Pole}$ in the $1S$-mass scheme, defined as
\cite{Hoang:1998ng,Hoang:1998hm}\\
\begin{equation}
m_c^{\rm Pole} = \frac{m_{J/\psi}}{2} \left(1 + \frac{(\alpha_s \, C_F)^2}{8} \right) \,,
\label{eq:mc-1S}
\end{equation}\\
where $C_F = 4/3$, and we use
$m_{J/\psi} = 3.0969$~GeV~\cite{Zyla:2020zbs},
so that $m_{J/\psi}/2 \approx 1.55$~GeV. Note that in Eq.~(\ref{eq:mc-1S}), the NLO correction actually starts at order $\alpha_s^2$, see Ref.~\cite{Hoang:1998ng}.
\end{enumerate}


\subsection{Description of the computation}

The non-leptonic decay of a charm quark $c \to q_1 \bar{q}_2 u$, with $q_1, q_2 =\{ u,d,s\}$, is described by the following effective  Hamiltonian, cf.\ Section~\ref{sec:Heff}, i.e.
\begin{align}
  {\cal H}_{\rm eff}^{\rm NL} & = 
  \frac{G_F}{\sqrt{2}} 
  \left[
  \sum \limits_{q_{1,2}=d,s} \lambda_{q_1 q_2}
  \Big(C_1  \, Q_1^{q_1q_2} + C_2 \, Q_2^{q_1q_2} \Big)
   - \lambda_b \sum \limits_{j=3}^6  C_j  Q_j 
   \right] + {\rm h.c.},
   \label{eq:Heff-NL}
\end{align}
where we have defined the CKM factors respectively as $\lambda_{q_1 q_2} = V_{cq_1}^* V_{uq_2} $
and $\lambda_{b} = V_{cb}^* V_{ub} $, and introduced the following notation for the tree-level $\Delta C = 1$ operators:\\
\begin{align}
Q_1^{q_1 q_2} & =   
\left(\bar{q}_1^i \gamma_\rho (1- \gamma_5) c^i \right)
\, \left(\bar{u}^j \gamma^\rho (1- \gamma_5) q_2^j \right)\,,
\label{eq:Q1}
\\[3mm]
Q_2^{q_1 q_2} & =  
\left(\bar{q}_1^i \gamma_\rho (1- \gamma_5) c^j \right)
\, \left(\bar{u}^j \gamma^\rho (1- \gamma_5) q_2^i \right)\,,
\label{eq:Q2}
\end{align}\\
while $Q_j,$  with $ j = 3, \ldots, 6 $, refer to the penguin operators, which can only arise in the singly Cabibbo 
suppressed decays $c \to s \bar{s} u$ and $c \to d \bar{d} u$. In Eq.~(\ref{eq:Heff-NL}), 
$C_i (\mu_1)$, with $i = 1, \ldots, 6$, denote the corresponding Wilson coefficients evaluated at the renormalisation scale $\mu_1 \sim m_c$. 
A comparison of their values respectively at NLO-(LO-)QCD and for different choices of $m_c$, is shown  
 in Table~\ref{tab:WCs}. \\
\begin{table}[th]
\renewcommand{\arraystretch}{1.3}
\centering
   \begin{tabular}{|c||C{1.7cm}|C{1.7cm}|C{1.7cm}|C{1.7cm}|C{1.7cm}|C{1.7cm}|}
   \hline
     $\mu_1  [{\rm GeV}]$  & 
     $1$     &  
     $1.27 $ &  
     $1.36 $ &  
     $1.48 $ &   
     $1.55 $ &
     $3 $ 
     \\
    \hline \hline
     $C_1 (\mu_1) $  & 
     $1.25$ \qquad $(1.34)$ & 
     $1.20$ \qquad $(1.27)$ & 
     $1.19$ \qquad $(1.26)$ & 
     $1.18$ \qquad $(1.24)$ & 
     $1.17$ \qquad $(1.23)$ & 
     $1.10$ \qquad $(1.15)$ 
     \\ 
     \hline
     $C_2 (\mu_1) $  & 
     $-0.48$ \qquad $(-0.62)$ & 
     $-0.39$ \qquad $(-0.50)$ & 
     $-0.40$ \qquad $(-0.53)$ & 
     $-0.37$ \qquad $(-0.48)$ & 
     $-0.36$ \qquad $(-0.47)$ & 
     $-0.24$ \qquad $(-0.32)$ 
     \\ 
     \hline
     $C_3 (\mu_1) $  & 
     $0.03$ \qquad $(0.02)$ &
     $0.02$ \qquad $(0.01)$ & 
     $0.02$ \qquad $(0.01)$ & 
     $0.01$ \qquad $(0.01)$ & 
     $0.01$ \qquad $(0.01)$ & 
     $0.00$ \qquad $(0.00)$ 
     \\
     \hline
     $C_4 (\mu_1) $  & 
     $-0.06$ \qquad $(-0.04)$ &
     $-0.05$ \qquad $(-0.03)$ &
     $-0.04$ \qquad $(-0.03)$ &
     $-0.04$ \qquad $(-0.02)$ &
     $-0.04$ \qquad $(-0.02)$ &
     $-0.01$ \qquad $(-0.01)$ 
     \\
     \hline
     $C_5 (\mu_1) $  & 
     $0.01$ \qquad $(0.01)$ &
     $0.01$ \qquad $(0.01)$ & 
     $0.01$ \qquad $(0.01)$ & 
     $0.01$ \qquad $(0.01)$ & 
     $0.01$ \qquad $(0.01)$ & 
     $0.00$ \qquad $(0.00)$ 
     \\
     \hline
     $C_6 (\mu_1) $  & 
     $-0.08$ \qquad $(-0.05)$ &
     $-0.05$ \qquad $(-0.03)$ &
     $-0.05$ \qquad $(-0.03)$ &
     $-0.04$ \qquad $(-0.03)$ &
     $-0.04$ \qquad $(-0.02)$ &
     $-0.01$ \qquad $(-0.01)$ 
     \\
     \hline 
    \end{tabular} 
    \caption{Comparison of
    the Wilson coefficients 
    at NLO-QCD (LO-QCD), for different values of $\mu_1 = m_c$.
    }
    \label{tab:WCs}
\end{table}\\
We see that the Wilson coefficients $C_j$, with $j = 3, \ldots, 6$, are very small and additionally strongly CKM suppressed because of the factor $\lambda_b \ll \lambda_{q_1 q_2}$. For these reasons, in the following, the contribution due to the penguin operators in Eq.~(\ref{eq:Heff-NL}) is neglected. However, the most general effective Hamiltonian describing all possible $c$-quark decays
is a sum of non-leptonic, semileptonic as well as radiative contributions, namely
\begin{align}
  {\cal H}_{\rm eff} & =   {\cal H}_{\rm eff}^{\rm NL} + {\cal H}_{\rm eff}^{\rm SL} + {\cal H}_{\rm eff}^{\rm rare} \, ,
  \label{eq:H-eff-tot-charm}
\end{align}\\
here, ${\cal H_{\rm eff}^{\rm NL}}$ is given in Eq.~\eqref{eq:Heff-NL},\\
\begin{equation}
{\cal H}_{\rm eff}^{\rm SL} 
= 
\frac{G_F}{\sqrt 2} \sum_{q = d, s} \sum_{\ell = e, \mu}
V_{cq}^* \, Q^{q \ell} + {\rm h.c.}\,,
\label{eq:Heff-SL}
\end{equation}\\
where we have introduced the semileptonic operator \\
\begin{equation}
Q^{q \ell} =\left(\bar{q} \gamma^\mu (1- \gamma_5) c \right)
\left(\bar{\nu}_\ell \gamma_\mu (1 - \gamma_5) \ell \right)\,, \qquad \ell = e, \mu\,,
\end{equation}\\
while ${\cal H}_{\rm eff}^{\rm rare}$ describes decays like
$D \to \pi \ell^+ \ell^-$, whose branching fraction is much smaller than those corresponding to the
tree-level transitions. Hence, in the following, we also neglect the presence of rare decays 
and omit to specify further the form of ${\cal H}_{\rm eff}^{\rm rare}$. 
Starting from the expression of the effective Hamiltonian in Eq.~(\ref{eq:H-eff-tot-charm}), the total decay width of the heavy charm mesons $ D^0 , D^+, D_s^+$, can be computed according to Eq.~(\ref{eq:Gamma-HQ}), where now we need to set $Q = c$. The structure of the HQE is schematically given in Eq.~(\ref{eq:HQE}), and its diagrammatic representation can be visualised as in Figure~\ref{fig:HQE-exp}. The lowest dimensional contributions, namely those due to two-quark operators up dimension-six, and to four-quark operators up dimension-seven, have been discussed in detail in Chapter~\ref{ch:HQE-ex} and Chapter~\ref{ch:Darwin}. Despite having considered  explicitly the case of the $B$ meson, almost all the expressions obtained, can be used also in the charm sector, taking into account the proper replacements i.e.\ $b \to c$, $c \to s$, etc..  However, we cannot directly use the coefficients of the Darwin operator listed in Eqs.~(\ref{eq:Ruud})-(\ref{eq:R22ccs}), since by setting $\rho = m_s^2/m_c^2$, it is straightforward to see that some of the functions diverge in the limit $m_s \to 0$. The presence of IR divergences  reflects the fact that now there are further contributions due to mixing of four-quark operators with external $s$-quarks and the Darwin operator, that must be included, whereas for the $b$-system, the corresponding operators with external $c$-quarks did not. This point will be discussed further later on, and we refer to Ref.~\cite{Breidenbach:2008ua } for more details. 

Following Ref.~\cite{King:2021xqp}, we try to analyse each of the contributions that enter the HQE of a $D$ meson, in order to identify the presence of possible cancellations which might affect the charm system. We start from the leading order term $\Gamma_3$, cf.\ Eq.~(\ref{eq:HQE}), which, including also NLO-QCD corrections to the short distance coefficients, can be schematically written as\\
 \begin{equation}
    \Gamma_3 = \Gamma_0 \,c_3 = \Gamma_0 \left[
    3 \, C_1^2    \, {\cal C}_{3,11} 
 +  2 \, C_1 C_2  \, {\cal C}_{3,12} 
 +  3 \, C_2^2    \, {\cal C}_{3,22} 
 +             {\cal C}_{3, \rm SL} 
    \right]  \, ,
\end{equation}\\
where a summation over all modes is implied and we stress that now $\Gamma_0$ is defined slightly differently compared  to Eq.~(\ref{eq:Gamma0}), i.e. \\
\begin{equation}
\Gamma_0 = \frac{G_F^2 m_c^5}{192 \pi^3} |V_{cs}|^2\,.
\end{equation}\\
At LO-QCD, the three non-leptonic coefficients ${\cal C}_{3,11}$, ${\cal C}_{3,12}$, and ${\cal C}_{3,22}$, for each of the $c \to q_1 \bar q_2 u$ modes, reduce to ${\cal C}^{(q_1 \bar q_2 u)}_0$, computed in Chapter~\ref{ch:Darwin}, cf.\ Eq.~(\ref{eq:Gamma-NL-res-scheme}) \footnote{Up to the CKM factor $\lambda_{q_1 q_2}^2 / |V_{cs}|^2$.}. 
The expressions for the QCD corrections to the non-leptonic coefficients ${\cal C}_{3,11}$ and ${\cal C}_{3,22}$,
as well as to ${\cal C}_{3,\rm SL}$, are obtained from Ref.~\cite{Hokim:1983yt}. In the latter, the computation has been performed  
for three arbitrary massive final states of the decaying quark, hence their results can be easily applied to all $c$-quark decay modes, by taking the appropriate mass limits.
For the coefficient ${\cal C}_{3,12}$, we use the results of Ref.~\cite{Bagan:1994zd}, respectively for
the $c \to s \bar d u$, $c \to d \bar s u$ and $c \to d \bar d u$ decay channels, 
and those of Ref.~\cite{Krinner:2013cja} in the case of two massive final states, e.g.\ $c \to s \bar s u$.
To compare the size of $c_3$ between the $b$- and $c$-system, it is interesting to consider the effect of the NLO corrections in the case of dominant CKM non-leptonic and semileptonic modes \footnote{For example $b \to c \bar u d $, $b \to c \ell \nu_\ell$ and $c \to s \bar d u $,  $c \to s \ell \nu_\ell $ transitions.}, neglecting for simplicity the mass of the final state particles. The result was determined in 1991 in Ref.~\cite{Altarelli:1991dx}, and reads \footnote{Note that the factor $|V_{ud}|^2 \approx 1$ has been omitted for simplicity.} 
\begin{align}
    c_3^{\rm NLO} -  c_3^{\rm LO}& =    8 \frac{\alpha_s}{4 \pi}
    \left[ 
    \underbrace{ \left( \frac{25}{4} - \pi^2\right)}_{< 0}
      \underbrace{  +  (C_1^2 + C_2^2)    \left( \frac{31}{4} - \pi^2\right)}_{< 0}
      \underbrace{   - \frac23 C_1 C_2    \left( \frac{7}{4} + \pi^2\right)}_{\geq 0}
    \right] 
    \, .
    \label{eq:G3_LO_nom}
\end{align}
The first term on the r.h.s.\ of Eq.~(\ref{eq:G3_LO_nom}), corresponds to the semileptonic mode while the remaining two terms to the non-leptonic one. For the $b$-quark decay, the NLO corrections are negative, while for the charm system, the third term
can dominate over the second one and lead to a positive correction to $c_3$. Moreover, there is a sizeable enhancement of the QCD corrections for non-leptonic $b$-quark decays due to finite charm quark mass effects \cite{Voloshin:1994sn,Bagan:1994zd,Bagan:1995yf,Krinner:2013cja}, whereas the corresponding increase for charm is much less pronounced since $m_c^2/m_b^2 \approx 0.1 \gg m_s^2/m_c^2 \approx 0.005$.
A comparison of $\Gamma_3$, both at LO- and NLO-QCD, for different $c$-quark mass schemes is shown in Table~\ref{tab:Gamma_3}.
\begin{table}
\centering
\renewcommand{\arraystretch}{1.5}
   \begin{tabular}{|c||c|c|}
   \hline
     &  
     $\Gamma_3^{\rm LO}$  [ps$^{-1}$]  &  
     $\Gamma_3^{\rm NLO}$ [ps$^{-1}$] 
     \\
     \hline  \hline
     $m_c^{\rm Pole} = 1.48$ GeV  &
     $1.45_{-0.14}^{+0.17}$  &
     $1.52_{-0.16}^{+0.20}$ 
     \\
     \hline
     $\overline{m}_c(\overline{m}_c) = 1.27$ GeV  &
     $0.69_{-0.09}^{+0.06}$  &
     $1.32_{-0.03}^{+0.06}$
     \\
     \hline
     $m_c^{\rm kin} (0.5 \, {\rm GeV}) = 1.363$ GeV  &
     $0.97_{-0.11}^{+0.10}$  &
     $1.47_{-0.30}^{+0.27}$ 
     \\
     \hline
     $m_c^{1S} = 1.548$ GeV  &
     $1.80_{-0.16}^{+0.24}$  &
     $2.12_{-0.30}^{+0.51}$ 
     \\
     \hline
    \end{tabular} 
\caption{ Numerical values of $\Gamma_3$ and LO- and NLO-QCD, using different schemes for the $c$-quark mass. The uncertainties are obtained by varying the renormalisation scale $\mu_1$ between 1 GeV and 3 GeV.}
\label{tab:Gamma_3}
\end{table}
The range of values between $1.3\, {\rm ps}^{-1}$ to $2.7\, {\rm ps}^{-1}$ 
for the  free charm-quark decay at NLO-QCD, is in good agreement with the experimental determinations in
Table~\ref{tab:exp-data}, and we find that the effect of a non-vanishing strange quark mass leads to small corrections ($<5 \% $). 
Interestingly, the NLO-QCD result is affected by strong cancellations.
We in fact observe a suppression of the non-leptonic contribution
because of the opposite sign between the NLO corrections to the $\Delta C = 1$ operators and to their Wilson coefficients.
Furthermore, a cancellation is present between the semileptonic
and the non-leptonic modes. In the ${\rm \overline{MS}}$ scheme an
additional NLO contribution arises from the conversion factor of
$m_c^5$, which is the origin of the large shift between the LO and
the NLO value. This is explicitly indicated in the following two equations\\
\begin{align}
{\rm (Pole)} \qquad \Gamma_3 & =   \Gamma_3^{\rm LO} \, \left[ 1 + \left(  \overbrace{\underbrace{1.84}_{\rm oper.} - \underbrace{0.74}_{\rm WC}}^{\rm NL} - \overbrace{0.67}^{\rm SL} \right) \, \frac{\alpha_s}{\pi} + {\cal O}\left(\frac{\alpha_s}{\pi}\right)^2 \right] \,, \label{eq:cancel1}
\\[3mm]
{\rm (\overline{MS})} \qquad \Gamma_3 & =   \Gamma_3^{\rm LO} \, \left[ 1 + \left(\overbrace{ \underbrace{2.10}_{\rm oper.} - \underbrace{0.70}_{\rm WC}}^{\rm NL} - \overbrace{0.71}^{\rm SL} + \overbrace{6.66}^{\rm conv. fac.} \right) \, \frac{\alpha_s}{\pi}  + {\cal O}\left(\frac{\alpha_s}{\pi}\right)^2 \right] \,.
\label{eq:cancel2}
\end{align}\\
To obtain a first indication of the behaviour of the QCD series for $\Gamma_3$ at higher orders, the authors of Ref.~\cite{King:2021xqp} have compared the results 
for the NNLO- \cite{Biswas:2009rb} and NNNLO- \cite{Fael:2020tow} 
QCD corrections to the semi leptonic $b$-quark decay and the preliminary NNLO-QCD corrections 
for the non-leptonic $b$-quark decay \cite{Czarnecki:2005vr}, concluding that higher order corrections seem to be crucial for a reliable determination of $\Gamma_3$ \footnote{Note that the results presented in Ref.~\cite{Czarnecki:2005vr} are not complete and hence cannot be used for phenomenological applications.}. 

\begin{figure}
    \centering
    \includegraphics[scale=1.0]{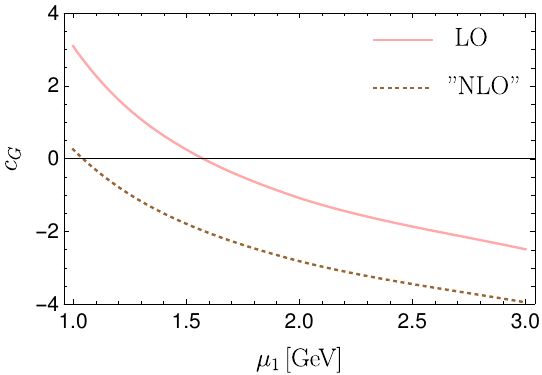}
    \caption{Scale dependence of the coefficient of the chromo-magnetic operator $c_G$.}
    \label{fig:cG}
\end{figure}

At order $1/m_c^2$ in the HQE, Eq.~(\ref{eq:HQE}), we find the contribution of the kinetic and the chromo-magnetic operators, defined in Eqs.~(\ref{eq:O-kin}), (\ref{eq:O-magn}), and respectively parametrised by the non perturbative input $\mu_\pi^2$ and $\mu_G^2$, see Eq.~(\ref{eq:dim-5-ME-parameters}). At this order and at LO-QCD, we can schematically write\\
\begin{equation}
    \Gamma_5 \frac{\langle {\cal O}_5 \rangle}{m_c^2} 
    = \Gamma_0 
    \left[
    c_{\mu_\pi} \frac{\mu_\pi^2}{m_c^2} 
    + c_G \, \frac{\mu_G^2}{m_c^2} 
    \right] \,,
    \label{eq:Gamma_5}
\end{equation}\\
where now, compared to the corresponding ones introduced in Eq.~(\ref{eq:Gamma-NL-res-scheme}), the short distance coefficients $c_{\mu_\pi}$, $c_{G}$, contain also the contribution due to the semileptonic modes as well as the dependence on the CKM factor $\lambda_{q_1 q_2}/|V_{cs}|^2$, due to the different definition of $\Gamma_0$. Their expressions can be then obtained from those of ${\cal C}_{G, mn}^{(q_1 \bar q_2 u)}$ and ${\cal C}_{0}^{(q_1 \bar q_2 u)}$ listed in Appendix~\ref{app:4}. Note that again a summation over all modes is implied. Specifically, we can decompose $c_G$ as\\
\begin{equation}
    c_G = 
    3 \, C_1^2    \, c_{G,11} 
 +  2 \, C_1 C_2  \, c_{G,12} 
 +  3 \, C_2^2    \, c_{G,22} + c_{G,SL} \, ,
\label{eq:CG}
\end{equation}\\
which leads to the following expression if we neglect the strange and muon masses and consider only the dominant CKM modes, i.e.\\
\begin{equation}
    c_G \approx -|V_{ud}|^2  \left[
    \frac92 \Bigl( C_1^2   + C_2^2 \Bigr)  + 19 \, C_1 C_2  \right]  - 3.
    \label{eq:cG}
\end{equation}\\
Because of the large coefficient in front of $C_1 C_2$ 
and of its negative value, Eq.~(\ref{eq:cG}) can be affected by cancellations. This can be visualised in Figure~\ref{fig:cG}, in which we show the dependence of the function $c_G$ in Eq.~(\ref{eq:CG}), on the
renormalisation scale $\mu_1$, for both LO- and NLO-QCD, $\Delta C =1 $ Wilson coefficients. Note that the latter case in Figure~\ref{fig:cG}, is indicated in quotation marks since it does not represent the complete NLO result, as corrections for non-leptonic modes are still missing and their effect could significantly reduce the strong scale dependence. In particular, from Figure~\ref{fig:cG}, we see that a change of sign occurs in the region between $1$ and $2$ GeV,  leading to a large uncertainty due to scale variation.
The numerical determinations of $\mu_\pi^2$ and $\mu_G^2$ are presented at the end of this section.

\begin{figure}[t]
    \centering
    \includegraphics[scale=1.0]{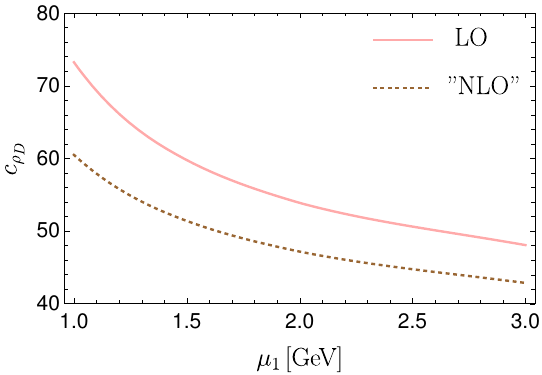}
    \caption{Scale dependence of the coefficient of the Darwin operator $c_{\rho_D}$.}
    \label{fig:c-rho-D}
\end{figure}

We now turn to analyse the contribution of the Darwin operator, which arises at order $1/m_c^3$ in Eq.~(\ref{eq:HQE}) and which has been discussed in detail in Chapter~\ref{ch:Darwin}. This can be compactly written as\\
\begin{equation}
\Gamma_6 \frac{\langle {\cal O}_6 \rangle}{m_c^3}
=
\Gamma_0 \, c_{\rho_D} \frac{\rho_D^3}{m_c^3}
 \, ,
\end{equation}\\
where again a summation over all modes is implied and the coefficient $c_{\rho_D}$ includes the effect of non-leptonic and semileptonic channels, namely\\
\begin{equation}
    c_{\rho_D} = 
    3 \, C_1^2 \, {\cal C}_{\rho_D,11} 
 +  2 \, C_1 C_2 \, {\cal C}_{\rho_D,12} 
 +  3 \, C_2^2 \, {\cal C}_{\rho_D,22} 
 +           {\cal C}_{\rho_D,SL} 
    \, .
\label{eq:crhoD}
\end{equation}\\
As already mentioned, the expressions for the non-leptonic coefficients obtained in Eqs.~(\ref{eq:Ruud})-(\ref{eq:R22ccs}), cannot be directly applied to the charm sector, since by naively replacing $m_b \to m_c $ and $m_c \to m_s$, some of the functions would develop infrared divergences in the limit $m_s \to 0$, whereas in the $b$-system the corresponding coefficients were finite functions of $\rho = m_c^2/m_b^2$. In fact, while we can assume $m_b \sim m_c \gg \Lambda_{\rm QCD}$, and neglect the effect of four-quark operators with external $c$-quarks in matrix elements between $B$-meson states \footnote{We recall that we do not consider the case of the $B_c$ meson.},  see e.g.\ Ref.~\cite{Breidenbach:2008ua},  in the charm sector, it is $m_c \gg m_s \sim \Lambda_{\rm QCD}$, and there are further contributions due to the mixing of four-quark operators with external $s$-quarks which must be additionally included. 
Specifically, this leads to a modification of the coefficients proportional to $C_1^2$ and $C_1 C_2$. 
Using the same procedure as discussed in Ref.~\cite{Lenz:2020oce}, 
the coefficients of the Darwin operator required for the study of $D$-meson decays have been computed in Ref.~\cite{King:2021xqp}, and the analytical expressions, including 
the full $s$-quark mass dependence, however finite in the limit $m_s \to 0$, are listed in Appendix~\ref{app:6},
for all non-leptonic modes. 
The results for ${\cal C}_{\rho_D,SL}$, can then be obtained by setting,
$N_c =1$, $C_1 = 1$, $C_2 = 0$ and $m_s \to m_\mu$ in the case of $c \to s \mu^+  \nu_\mu$ decay. By neglecting the strange and muon masses and 
by considering only the dominant CKM modes, we have\\
\begin{equation}
    c_{\rho_D} \approx 
     |V_{ud}|^2 \, \left(18 \, C_1^2 \, -\frac{68}{3} \, C_1 C_2 
     + 18 \, C_2^2\right) + 12        
    \, .
    \label{eq:Darwin-coeff}
\end{equation}\\
It is interesting to note that in this
combination all terms have the same sign and no
cancellations arise. In Figure~\ref{fig:c-rho-D} we show the dependence on the
renormalisation scale $\mu_1$ of the function $c_{\rho_D}$ in Eq.~(\ref{eq:crhoD}), where the quotation marks in the NLO-QCD result reflect again the fact that only corrections due to the $\Delta C =1 $ Wilson coefficients have been included, since also in this case a complete determination of the NLO corrections is still missing.
Estimates for the matrix element of the Darwin operator will be presented at the end of this section.

In order to discuss the contribution of four-quark operators at order $1/m_c^3$ in Eq.~(\ref{eq:HQE}), see Section~\ref{sec:4q-contr}, we introduce the following basis, in complete analogy to Eqs.~(\ref{eq:Ova-4q})-(\ref{eq:Tsp-4q}) \footnote{Note however that now, we do not use the tilde to denote the colour-octet operators. }, namely
\begin{align}
  O_1^q  
  & =   
  \big( \bar{c}\,\gamma_\mu (1-\gamma_5) q \big)\,\big(\bar{q}\,\gamma^\mu (1-\gamma_5) c\big) \,,
  \label{eq:O1} \\[3mm]
  O_2^q  
  & =  
  \big(\bar{c} (1 - \gamma_5) q\big)\,\big(\bar{q} (1 + \gamma_5) c\big) \,,
  \label{eq:O2} \\[3mm]
  O_3^q  
  & =  
 \big(\bar{c} \, \gamma_\mu (1-\gamma_5) t^a q\big) 
  \, \big(\bar{q} \, \gamma^\mu (1-\gamma_5) t^a  c\big) \,,
  \label{eq:T1} \\[3mm]
  O_4^q 
  & =  
  \big(\bar{c} (1-\gamma_5) t^a q\big)\,\big(\bar{q}(1 + \gamma_5) t^a c\big) \,,
  \label{eq:T2}
\end{align}\\
where a summation over colour indices is implied and we have replaced $c_v$ with $c$, cf.\ Eq.~(\ref{eq:Qx-HQE}). The parametrisation of the matrix element of the operators in Eqs.~(\ref{eq:O1})-(\ref{eq:T2}) is given in Appendix~\ref{app:7}. However, by evaluating them in the framework of the HQET, the dependence on the charm quark mass can be further extracted from the $c$-quark field and meson state cf.\ Section~\ref{sec:dim-7-four-q}, and in this case, the corresponding dimension-six operators read  \\
\begin{align}
  {\cal O}_1^q  
  & =   
  (\bar{h}_v\,\gamma_\mu (1-\gamma_5) q)\,(\bar{q}\,\gamma^\mu (1-\gamma_5) h_v) \,,
  \label{eq:O1-HQET} \\[3mm]
  {\cal O}_2^q  
  & =  
  (\bar{h}_v (1 - \gamma_5) q)\,(\bar{q} (1 + \gamma_5) h_v) \,,
  \label{eq:O2-HQET} \\[3mm]
  {\cal O}_3^q  
  & =  
  (\bar{h}_v \, \gamma_\mu (1-\gamma_5) t^a q) 
  \, (\bar{q} \, \gamma^\mu (1-\gamma_5) t^a  h_v) \,,
  \label{eq:T1-HQET} \\[3mm]
  {\cal O}_4^q 
  & =  
  (\bar{h}_v (1-\gamma_5) t^a q)\,(\bar{q}(1 + \gamma_5) t^a h_v) \,,
  \label{eq:T2-HQET}
\end{align}\\
where $h_v$ denotes the effective heavy quark field, see Section~\ref{sec:HQET}. The parametrisation in HQET of the matrix element of the operators in Eqs.~(\ref{eq:O1-HQET})-(\ref{eq:T2-HQET}), see also Appendix~\ref{app:7}, can be written as\\
\begin{align}
\langle {D}_q | {\cal O}_i^q \, | {D}_q \rangle 
& = 
F^2(m_c) \, m_{D_q} \left( {\tilde B}_i^q  + \tilde \delta^{q q}_i \right)
 = 
f_{D_q}^2 m_{D_q}^2 
\left(1 + \frac{4}{3} \frac{\alpha_s (m_c)}{ \pi} \right)
 \left( {\tilde B}_i^q  + \tilde \delta^{q q}_i \right) \,,
\label{eq:ME-dim-6-HQET-q-q}
\\[3mm]
\langle {D}_q | {\cal O}_i^{q^\prime} | {D}_q \rangle 
& = 
F^2 (m_c) \, m_{D_q} \tilde \delta^{q q^\prime}_i
= 
f_{D_q}^2 m_{D_q}^2 
\left(1 + \frac{4}{3} \frac{\alpha_s (m_c)}{\pi} \right)
\tilde \delta^{q q^\prime}_i, \quad q \neq q^\prime \,,
\label{eq:ME-dim-6-HQET-q-q-prime}
\end{align}\\
where $q, q^\prime = u, d, s$, 
${\tilde B}_i^q$ are the Bag parameters computed in HQET, while
$F (\mu)$ and $f_{D_q}$ correspond respectively to HQET and QCD decay constants,
defined, by \footnote{The subscript `QCD' or `HQET' on the states is usually omitted, however for clarity it is specified in the definition of the decay constant.}\\
\begin{equation}
\langle 0 | \bar q \gamma^\mu \gamma_5 c|{D_q (v)} \rangle_{\rm QCD} 
= i f_{D_q} \, p^\mu,
\label{eq:DecayConstQCD}
\end{equation}\\
with $p^\mu = m_D v^\mu$, and \\
\begin{equation}
\langle 0 | \bar q \gamma^\mu \gamma_5 h_v| {D}_q (v) \rangle_{\rm HQET} = i \, F (\mu) \, \sqrt{m_{D_q}} \, v^\mu.
\label{eq:DecayConstHQET}
\end{equation}\\
The relation between $f_D$ and $F(\mu)$ up to QCD and power corrections, can be found e.g.\ in Refs.~\cite{Neubert:1992fk, Kilian:1992cj}. At the scale $\mu = m_c$, it reads \\
\begin{align}
f_D = \frac{F (m_c)}{\sqrt{ m_D}} 
\left(1 - \frac{2}{3} \frac{\alpha_s (m_c)}{\pi} 
+ \frac{G_1 (m_c)}{m_c} + 6 \, \frac{G_2 (m_c)}{m_c}
- \frac 1 2 \frac{\bar \Lambda}{m_c} \right),
\label{eq:decay_constant-conversion}
\end{align}\\
where $\bar \Lambda = m_D - m_c$, and the parameters $G_1$ and $G_2$ characterise matrix elements of non-local operators.
Note that in Eqs.~\eqref{eq:ME-dim-6-HQET-q-q}, \eqref{eq:ME-dim-6-HQET-q-q-prime}, by expressing the HQET decay constant in terms of the one defined in QCD, we have included only corrections due to $\alpha_s$, which become part of the NLO-QCD contribution at dimension-six. In fact, as we will discuss, the power corrections can be absorbed in the contribution of some of the dimension-seven operators appearing in HQET. 

\begin{figure}[t]
\centering
\includegraphics[scale = 0.4]{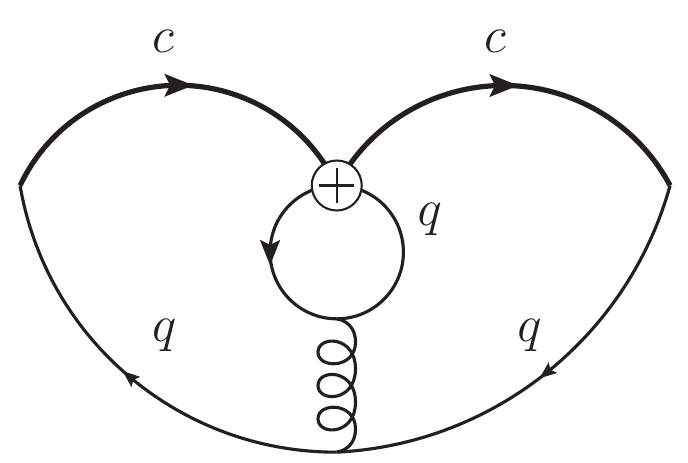}
\qquad 
\includegraphics[scale = 0.4]{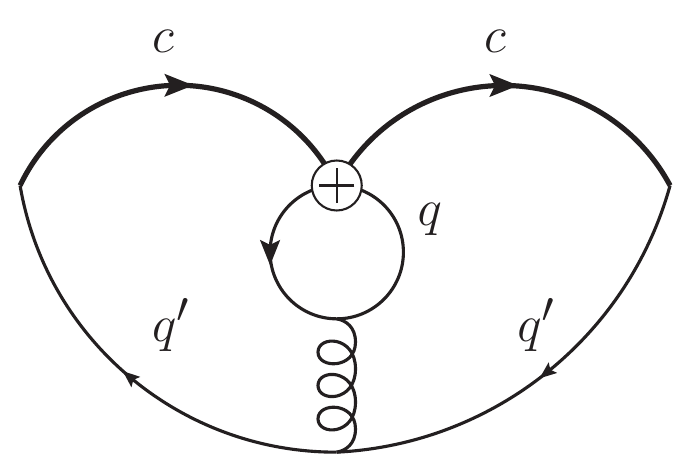}
\caption{Diagram describing the eye-contractions.}
\label{fig:eye-contractions}
\end{figure}

In vacuum insertion approximation (VIA), the Bag parameters of the colour-singlet operators are equal to one, i.e.\ ${\tilde B}_{1,2}^q = 1$, and those  of the colour-octet
operators vanish, i.e.\ $\tilde B^q_{3,4} = 0$. Note that we assume isospin symmetry, so that  ${\tilde B}_i^u = {\tilde B}_i^d $.
The parameters $\tilde \delta^{q q}_i$, $\tilde \delta^{q q^\prime}_i$, 
in \eqref{eq:ME-dim-6-HQET-q-q-prime}, describe subleading effects in the non perturbative matrix elements, compared to the dominant $\tilde B_i$, and correspond to the so called eye-contractions, shown in Figure~\ref{fig:eye-contractions}.
In VIA, the contribution of 
all eye-contractions vanish i.e.\ $\tilde \delta^{q q}_i = \tilde \delta^{q q'}_i = 0$. However, beyond vacuum insertion approximation, 
the matrix element of four-quark operators with external $q^\prime$ quarks, differ from zero even 
when the spectator quark $q$ in the $D_q$ meson does not coincide with 
$q^\prime$, as it is indicated in Eq.~\eqref{eq:ME-dim-6-HQET-q-q-prime} and in Figure~\ref{fig:eye-contractions}.
Again, due to isospin symmetry, we assume $\tilde \delta^{u q^\prime}_i = \tilde \delta^{d q^\prime}_i $ and $\tilde \delta^{q u }_i = \tilde \delta^{q d}_i $.
The Bag parameters ${\tilde B}_i$ and $ \tilde \delta_i^{q q^\prime}$ have been computed using HQET sum rules, specifically,
the formed were obtained for the $D^{+,0}$ 
mesons in Ref.~\cite{Kirk:2017juj}, while
corrections due to the strange quark mass as well as the
contribution of the eye-contractions,
have been determined for the first time in Ref.~\cite{King:2020}.
The numerical values of the HQET Bag parameters are listed in Table~\ref{tab:Bag-parameters}.

By considering only the dominant CKM modes and by neglecting the effect of the eye-contractions, at LO-QCD and at dimension-six, the contribution of four-quark operators 
to the decay rate of the $D^0, D^+$ and $D_s^+$ mesons, cf.\ Eq.~(\ref{eq:HQE}), reads\\ 
\begin{align}
16 \pi^2  \tilde{\Gamma}^{D^0}_6 \frac{\langle {\tilde O}_6\rangle^{D^0}}{m_c^3} 
 & = 
 \Gamma_0 |V_{ud}^*|^2 \, 16 \pi^2 \frac{M_{{D^0}} f_{D^0}^2}{m_c^3}(1- x_s)^2  \Bigg\{
 C_{\rm WE}^S \left[
  ({\tilde B}_2^{u} -  {\tilde B}_1^{u}) 
  \right]
\nonumber \\[3mm]
& + x_s \, \left(2 {\tilde B}_2^{u} - \frac{ {\tilde B}_1^{u}}{2} \right)   + C_{\rm WE}^O \Big[ ({\tilde B}_4^{u} -  \tilde B_3^{u}) 
  + x_s \, (2 \, {\tilde B}_4^{u} - \frac{ \tilde B_3^{u}}{2}) \Big] \Bigg\}  \, ,
  \label{eq:dim-6-4q-LO-D0}
  \end{align}\\
  which corresponds to the WE topology,\\
  \begin{align}
16 \pi^2  \tilde{\Gamma}^{D^+}_6 \frac{\langle {\tilde O}_6 \rangle^{D^+}}{m_c^3}
   =  
  \Gamma_0 |V_{ud}^*|^2 16 \pi^2 \,\frac{M_{{D^+}} f_{D^+}^2}{m_c^3} \, (1-x_s)^2
  \,\Big\{
  C_{\rm PI}^S\,  {\tilde B}_1^{d}
  + C_{\rm PI}^O\,  \tilde B_3^{d}  
  \Big\}
  \, ,
  \label{eq:dim-6-4q-LO-Dp}
  \end{align}\\
  describing the PI contribution, and \\
  \begin{align}
  \hspace*{-3mm}
 16 \pi^2  \tilde{\Gamma}^{D_s^+}_6 \frac{\langle {\tilde O}_6 \rangle^{D_s^+}}{m_c^3}
  & = 
  \Gamma_0 |V_{ud}^*|^2 16 \pi^2
  \frac{M_{{D_s^+}}f_{D_s^+}^2}{m_c^3}
  \Bigg\{
  \left( C_{\rm WA}^S + \frac{2}{|V_{ud}^*|^2 }  \right) 
  \left({\tilde B}_2^{s} -  {\tilde B}_1^{s} \right)
  \nonumber\\[2mm]
  &
  +C_{\rm WA}^O  \left(\tilde B_4^{s} -  \tilde B_3^{s} \right) 
  \Bigg\} \, ,
  \label{eq:dim-6-4q-LO-Ds}
\end{align}\\
due to WA. Here $x_s = m_s^2/m_c^2$, and we have introduced the following notation for the combinations of Wilson coefficients, cf.\  Eqs.~(\ref{eq:WC-WE}), (\ref{eq:WC-PI}), and (\ref{eq:WC-WA}), namely\\
\begin{align}
      & C_{\rm WE}^S = \frac 1 3 C_1^2 + 2 \, C_1 C_2 + 3 \, C_2^2, 
      & & C_{\rm WE}^O =  2 \, C_1^2,
      \label{eq:C-WE} \\[3mm]
      & C_{\rm PI}^S = C_1^2 + 6 \, C_1 C_2 +  C_2^2,
      & & C_{\rm PI}^O = 6 \, (C_1^2  + C_2^2)  \, ,
      \label{eq:C-PI} \\[3mm]
      & C_{\rm WA}^S = 3 \, C_1^2 + 2 \, C_1 C_2 + \frac 1 3 C_2^2, 
      & & C_{\rm WA}^O = 2 \, C_2^2  \, ,
      \label{eq:C-WA}
\end{align}\\
where the superscript ``S'' and ``O'' refers to coefficient in front of 
the colour-singlet and of colour-octet Bag parameters, respectively. 
Note that in Eq.~(\ref{eq:dim-6-4q-LO-Ds}), the contribution due to the muon mass in the semileptonic decay $c \to s \mu^+ \nu_\mu $, has been neglected. 
The expressions in Eqs.~(\ref{eq:dim-6-4q-LO-D0})-(\ref{eq:dim-6-4q-LO-Ds}) lead to some interesting numerical effects.
First, in the charm system, one expects that the contribution due to the spectator quark is of similar size compared to 
      the leading term
      $\Gamma_3$ in the HQE, unless some additional cancellations are present. Using the pole mass $m_c^{\rm Pole} = 1.48$ GeV and Lattice QCD values for the decay constants~\cite{Aoki:2019cca}, roughly yields\\
      \begin{align}
          16 \pi^2
        \frac{M_{{D^0}}f_{D^0}^2}{m_c^3}  =  {4.1} \approx 
        {\cal  O} (c_3) \, , \qquad
        16 \pi^2
         \frac{M_{{D_s^+}}f_{D_s^+}^2}{m_c^3}
           =  { 6.0}  \approx 
        {\cal O} (c_3) \, .
      \end{align}\\
This result has led the authors of Ref.~\cite{Mannel:2021uoz} to propose a different
way to rearrange the HQE series in the charm sector.
However, to investigate further the size of four-quark contributions at dimension-six, 
we consider the combinations of Wilson coefficients that appear in 
Eqs.~\eqref{eq:dim-6-4q-LO-D0} - \eqref{eq:dim-6-4q-LO-Ds}.
A comparison of these coefficients at LO- and NLO-QCD, for different values of the 
renormalisation scale $\mu_1$ is
shown in Table~\ref{tab:C-WE-PI-WA}.
\begin{table}[ht]
\centering
\renewcommand{\arraystretch}{1.25}
   \begin{tabular}{|c||C{1.5cm}|C{1.5cm}|C{1.5cm}|C{1.5cm}|C{1.5cm}|C{1.5cm}|}
   \hline
   $\mu_1$ [GeV] & 1 & 1.206 & 1.27 & 1.48 & 1.67 & 3 
   \\
   \hline 
   \hline
   $C_{\rm WE}^{\rm S} (\rm LO)$ &
   $\phantom{-}0.09 $ & 
   $\phantom{-}0.04 $ &
   $\phantom{-}0.03 $ &
   $\phantom{-}0.01 $ &
   $\phantom{-}0.00 $ &
   $\phantom{-}0.01 $ 
   \\
   $C_{\rm WE}^{\rm S} (\rm NLO)$ &
   $-0.03 $ &
   $-0.03 $ &
   $-0.03 $ &
   $-0.02 $ &
   $-0.02 $ &
   $\phantom{-}0.04$
   \\
   \hline
   $C_{\rm WE}^{\rm O} (\rm LO)$ &
   $\phantom{-}3.57 $ & 
   $\phantom{-}3.30 $ &
   $\phantom{-}3.24 $ &
   $\phantom{-}3.08 $ &
   $\phantom{-}2.98 $ &
   $\phantom{-}2.63 $
   \\
   $C_{\rm WE}^{\rm O} (\rm NLO)$ &
   $\phantom{-}3.11 $ & 
   $\phantom{-}2.93 $ &
   $\phantom{-}2.89 $ &
   $\phantom{-}2.77 $ &
   $\phantom{-}2.70 $ &
   $\phantom{-}2.44 $
   \\
   \hline
   $C_{\rm PI}^{\rm S} (\rm LO)$ &
   $-2.80 $ & 
   $-2.25 $ &
   $-2.12 $ &
   $-1.79 $ &
   $-1.57 $ &
   $-0.79 $
   \\
   $C_{\rm PI}^{\rm S} (\rm NLO)$ &
   $-1.74$ & 
   $-1.36 $ &
   $-1.28 $ &
   $-1.04 $ &
   $-0.88 $ &
   $-0.27 $
   \\
   \hline
   $C_{\rm PI}^{\rm O} (\rm LO)$ &
   $\phantom{-}13.0 $ & 
   $\phantom{-}11.7 $ &
   $\phantom{-}11.4 $ &
   $\phantom{-}10.6 $ &
   $\phantom{-}10.1 $ &
   $\phantom{-}8.50 $
   \\
   $C_{\rm PI}^{\rm O} (\rm NLO)$ &
   $\phantom{-}10.6 $ & 
   $\phantom{-}9.73 $ &
   $\phantom{-}9.55 $ &
   $\phantom{-}9.05 $ &
   $\phantom{-}8.72 $ &
   $\phantom{-}7.60 $
   \\
   \hline
   $C_{\rm WA}^{\rm S} (\rm LO)$ &
   $\phantom{-}3.82 $ & 
   $\phantom{-}3.65 $ &
   $\phantom{-}3.61 $ &
   $\phantom{-}3.51 $ &
   $\phantom{-}3.45 $ &
   $\phantom{-}3.24 $
   \\
   $C_{\rm WA}^{\rm S} (\rm NLO)$ &
   $\phantom{-}3.57 $ & 
   $\phantom{-}3.45 $ &
   $\phantom{-}3.42 $ &
   $\phantom{-}3.35 $ &
   $\phantom{-}3.31 $ &
   $\phantom{-}3.16 $
   \\
   \hline
   $C_{\rm WA}^{\rm O} (\rm LO)$ &
   $\phantom{-}0.77 $ & 
   $\phantom{-}0.59 $ &
   $\phantom{-}0.55 $ &
   $\phantom{-}0.46 $ &
   $\phantom{-}0.40 $ &
   $\phantom{-}0.21 $   
   \\
   $C_{\rm WA}^{\rm O} (\rm NLO)$ &
   $\phantom{-}0.41 $ & 
   $\phantom{-}0.32 $ &
   $\phantom{-}0.30 $ &
   $\phantom{-}0.24 $ &
   $\phantom{-}0.21 $ &
   $\phantom{-}0.10 $
   \\
   \hline
  \end{tabular} 
  \caption{Comparison of the combinations $C_{\rm WE, PI, WA}^{\rm S, O} $, respectively at LO- and NLO-QCD, for different values of the renormalisation scale $\mu_1$.}
  \label{tab:C-WE-PI-WA}
\end{table}
The first observation is that $C^{\rm S}_{\rm WE}$ is strongly suppressed. 
Moreover, in Eq.~(\ref{eq:C-WE}) the Bag parameters of the colour-singlet operators exactly cancel in VIA.
On the other side, the combination of Wilson coefficients in front of the colour-octet operators
is not suppressed for weak exchange, indicating that both colour structures might be equally important in this case.
For Pauli interference, the combinations of Wilson coefficients 
multiplying the colour-singlet operators are significantly
enhanced compared to those in WE, the same holds for the
colour-octet operators.
Note that $C_{\rm PI}^O$ and $C_{\rm PI}^S$ get
large modifications, and even a change of sign, compared to 
the case $C_1 = 1$ and $C_2 = 0$ revealing the importance of 
gluon radiative corrections.
Moreover $C_{\rm PI}^O$ is enhanced compared to $C_{\rm PI}^S$, again 
indicating that both colour structures might be equally important for PI.
In the case of weak annihilation, $C_{\rm WA}^S$ is large.
On the other hand, the Bag parameters of the colour-singlet operators exactly cancel in VIA.
 The above arguments show that by neglecting the effect of the colour-octet operators in VIA, one might be led to misleading conclusions, and therefore an accurate determination of the deviation 
of the Bag parameters from their VIA values, using non-perturbative methods like HQET sum rules or lattice simulations, is necessary.
Finally, by including all CKM modes as well as NLO-QCD corrections,
the contribution of four-quark operators to the total decay width at order $1/m_c^3$, schematically reads
\begin{table}[t]
    \centering
    \renewcommand{\arraystretch}{1.6}
    {
    \begin{tabular}{|c|c|c|c|}
    \hline
    scheme & $D^0$ & $D^+$ & $D^+_s$ 
    \\[2mm]
    \hline
    \multicolumn{4}{|c|}{VIA}
    \\
    \hline
    Pole 
    &
    $\underbrace{-0.06}_{\rm NLO}
    = \underbrace{0.00}_{\rm LO} \underbrace{-0.06}_{\rm \Delta NLO}$ 
    & 
    $\underbrace{-11.3}_{\rm NLO} 
    = \underbrace{-7.18}_{\rm LO} \underbrace{-4.13}_{\rm \Delta NLO}$ 
    & 
    $\underbrace{-0.85}_{\rm NLO} 
    = \underbrace{-0.51}_{\rm LO} \underbrace{-0.34}_{\rm \Delta NLO}$ 
    \\[5mm]
    \hline
    $\overline{\rm MS}$
    &
    $\underbrace{-0.09}_{\rm NLO} 
    = \underbrace{0.00}_{\rm LO} \underbrace{- 0.09}_{\rm \Delta NLO}$ 
    & 
    $\underbrace{-22.9}_{\rm NLO} 
    = \underbrace{-11.3}_{\rm LO} \underbrace{- 11.5}_{\rm \Delta NLO}$ 
    & 
    $\underbrace{-1.66}_{\rm NLO} 
    = \underbrace{-0.77}_{\rm LO} \underbrace{- 0.89}_{\rm \Delta NLO}$ 
    \\[5mm]
    \hline
    Kinetic
    &
    $\underbrace{-0.08}_{\rm NLO} 
    = \underbrace{0.00}_{\rm LO} \underbrace{- 0.08}_{\rm \Delta NLO}$ 
    & 
    $\underbrace{-16.3}_{\rm NLO} 
    = \underbrace{-9.18}_{\rm LO} \underbrace{- 7.14}_{\rm \Delta NLO}$ 
    & 
    $\underbrace{-1.21}_{\rm NLO} 
    = \underbrace{-0.64}_{\rm LO} \underbrace{- 0.57}_{\rm \Delta NLO}$ 
    \\[5mm]
    \hline
    $1S$
    &
    $\underbrace{-0.05}_{\rm NLO} 
    = \underbrace{0.00}_{\rm LO} \underbrace{-0.05}_{\rm \Delta NLO}$ 
    & 
    $\underbrace{-10.1}_{\rm NLO} 
    = \underbrace{-6.27}_{\rm LO} \underbrace{-3.82}_{\rm \Delta NLO}$ 
    & 
    $\underbrace{-0.76}_{\rm NLO} 
    = \underbrace{-0.45}_{\rm LO} \underbrace{-0.31}_{\rm \Delta NLO}$ 
    \\[5mm]
    \hline
    \multicolumn{4}{|c|}{HQET SR}
    \\
    \hline
    Pole 
    &
    $\underbrace{0.06}_{\rm NLO}
    = \underbrace{0.10}_{\rm LO} \underbrace{-0.04}_{\rm \Delta NLO}$ 
    & 
    $\underbrace{-12.3}_{\rm NLO} 
    = \underbrace{-7.97}_{\rm LO} \underbrace{-4.37}_{\rm \Delta NLO}$ 
    & 
    $\underbrace{-0.93}_{\rm NLO} 
    = \underbrace{-0.69}_{\rm LO} \underbrace{-0.24}_{\rm \Delta NLO}$ 
    \\[5mm]
    \hline
    $\overline{\rm MS}$
    &
    $\underbrace{0.23}_{\rm NLO} 
    = \underbrace{0.17}_{\rm LO} \underbrace{+0.06}_{\rm \Delta NLO}$ 
    & 
    $\underbrace{-24.9}_{\rm NLO} 
    = \underbrace{-12.6}_{\rm LO} \underbrace{- 12.3}_{\rm \Delta NLO}$ 
    & 
    $\underbrace{-1.78}_{\rm NLO} 
    = \underbrace{-1.06}_{\rm LO} \underbrace{- 0.72}_{\rm \Delta NLO}$ 
    \\[5mm]
    \hline
    Kinetic
    &
    $\underbrace{0.12}_{\rm NLO} 
    = \underbrace{0.13}_{\rm LO} \underbrace{- 0.01}_{\rm \Delta NLO}$ 
    & 
    $\underbrace{-17.8}_{\rm NLO} 
    = \underbrace{-10.2}_{\rm LO} \underbrace{- 7.61}_{\rm \Delta NLO}$ 
    & 
    $\underbrace{-1.31}_{\rm NLO} 
    = \underbrace{-0.87}_{\rm LO} \underbrace{- 0.43}_{\rm \Delta NLO}$ 
    \\[5mm]
    \hline
    $1S$
    &
    $\underbrace{0.06}_{\rm NLO} 
    = \underbrace{0.09}_{\rm LO} \underbrace{-0.03}_{\rm \Delta NLO}$ 
    & 
    $\underbrace{-11.0}_{\rm NLO} 
    = \underbrace{-6.97}_{\rm LO} \underbrace{-4.05}_{\rm \Delta NLO}$ 
    & 
    $\underbrace{-0.83}_{\rm NLO} 
    = \underbrace{-0.60}_{\rm LO} \underbrace{-0.23}_{\rm \Delta NLO}$ 
    \\[5mm]
    \hline
    \end{tabular}
    }
    \caption{Dimension-six contributions to the $D$-mesons decay width normalised by $\Gamma_0$ and split up into LO-QCD and NLO-QCD corrections within different mass schemes and using 
    both VIA and HQET SR values for the Bag parameters. }
    \label{tab:dim-6-NLO-vs-LO}
\end{table}
\begin{align}
\tilde{\Gamma}^{D_q}_6 \frac{\langle \tilde{\cal O}_6\rangle^{D_q}}{m_c^3}
& = 
\frac{\Gamma_0}{{|V_{cs}|^2}} \!\!\! \sum_{q_1, q_2 = d, s} \!\!\! 
\left| \lambda_{q_1 q_2} \right|^2
\sum_{i = 1}^4 
\Bigg[
A_{i, q_1 q_2}^{\rm WE} \frac{\langle D_q | {\cal  O}_i^u | D_q \rangle}{m_c^3}
+ A_{i, q_1 q_2}^{\rm PI} \frac{\langle D_q | {\cal  O}_i^{q_2} | D_q \rangle}{m_c^3}
\nonumber
\\[3mm]
&  + A_{i, q_1 q_2}^{\rm WA} \frac{\langle D_q | {\cal  O}_i^{q_1} | D_q \rangle}{m_c^3}
\Bigg] + 
\frac{\Gamma_0}{{|V_{cs}|^2}} \sum_{q_1 = d, s} |V_{c q_1}|^2
\sum_{\ell = e, \mu} \sum_{i = 1}^4 
\left[
A_{i, q_1 \ell}^{\rm WA} \frac{\langle D_q | {\cal  O}_i^{q_1} | D_q \rangle}{m_c^3}
\right]\,,
\label{eq:dim-6-4q-NLO-scheme}
\end{align}
where the matrix elements of the four-quark operators are given
in Eqs.~\eqref{eq:ME-dim-6-HQET-q-q}, \eqref{eq:ME-dim-6-HQET-q-q-prime},  and the short-distance
coefficients
for the WE, PI and WA topologies are denoted by $A_{i, q_1 q_2}^{\rm WE}$, 
$A_{i, q_1 q_2}^{\rm PI}$ and $A_{i, q_1 q_2}^{\rm WA}$, 
$A_{i, q_1 \ell}^{\rm WA}$, respectively. Their expressions at LO-QCD have been derived in Section~\ref{sec:T-4q-6}, while NLO corrections to
$A_{i, q_1 q_2}^{\rm WE}$ and $A_{i, q_1 q_2}^{\rm PI}$ have been
computed in Ref.~\cite{Franco:2002fc}.
The corresponding results for $A_{i, q_1 q_2}^{\rm WA}$ can be
obtained by using Eq.~(\ref{eq:WE-WA-Fierz}), since the Fierz symmetry is respected also at one-loop. For the semileptonic modes, the coefficients
$A_{i, q_1 \ell}^{\rm WA}$ have been determined in Ref.~\cite{Lenz:2013aua}.
Note that in our analysis, we treat the contribution of the $\tilde \delta_i^{q q^\prime}$ parameters effectively as a NLO effect, therefore their coefficients are included only at LO-QCD. 
Finally, in Table~\ref{tab:dim-6-NLO-vs-LO}, we compare the size of the LO- and NLO-QCD corrections in Eq.~(\ref{eq:dim-6-4q-NLO-scheme}), normalised by $\Gamma_0$, 
both in VIA and using HQET SR results for the Bag parameters. 
The NLO-QCD corrections turn out to have an essential
numerical effect for the contribution of four-quark at order $1/m_c^3$. In particular, in the case of
the $D^0$ and $D_s^+$ mesons, they lift the helicity
suppression present in the weak exchange and weak annihilation topologies 
at LO-QCD and in VIA. Note that
for the $D^+_s$ meson, in addition to the CKM dominant WA contribution,
there is a correction due to the CKM suppressed, but
nevertheless large PI topology. 
In the case of the $D^+$ meson the NLO corrections to Pauli interference are very large, $50 \% - 100 \%$ depending on the mass scheme. Already in the $B$ system they were found to be of the order of $30\%$ for the ratio $\tau (B^+)/\tau (B_d)$, in the pole scheme, see e.g. Ref.~\cite{Beneke:2002rj}. 
We conclude that, neglecting these contributions in the study of charm lifetime, as it has been previously done in Ref.~\cite{Cheng:2018rkz}, is clearly not justified and the determination of higher order corrections would be highly desirable.

We now consider the contribution of four-quark operators at order $1/m_c^4$ to the HQE in Eq.~(\ref{eq:HQE}), which have been   discussed in detail in Section~\ref{sec:dim-7-four-q}. By expanding $p^\mu = p_c^\mu \pm p_q^\mu$, only in the small momentum of the light spectator quark $p_q \sim \Lambda_{QCD}$, leads to the following basis for the dimension-seven operators 
 \footnote{
Note that the basis used e.g.\ in Ref.~\cite{Lenz:2013aua}, is redundant, since it contains also the additional operator denoted by $P_2^q$, related to $P_1^q$ by hermitean conjugation, namely
$P_2^q = m_q \, (\bar c (1 + \gamma_5) q) (\bar q (1 +  \gamma_5) c) 
= (P^q_1)^\dagger$. Leading to the same matrix element, we do not include this operator in our basis.
}\\
\begin{align}
P_1^q 
& =   
m_q\, (\bar c (1- \gamma_5) q) (\bar q (1- \gamma_5) c) \, ,
\label{eq:P1q}
\\[3mm]
P_2^q 
& =  
\frac{1}{m_c} (\bar c \overset{\leftarrow}{D_\nu} \gamma_\mu (1- \gamma_5)  D^\nu q) (\bar q \gamma^\mu (1 - \gamma_5) c) \, ,
\label{eq:P3q}
\\[3mm]
P_3^q
& = 
\frac{1}{m_c} (\bar c \overset{\leftarrow}{D_\nu} (1- \gamma_5)  D^\nu q) 
(\bar q (1 + \gamma_5) c)\, , 
\label{eq:P4q}
\end{align}\\
together with the corresponding $\tilde P_1^q,\tilde P_2^q, \tilde P_3^q$, containing the generators $t^a$. Due to the presence in Eqs.~(\ref{eq:P3q}), (\ref{eq:P4q}), of a covariant derivative acting on the charm quark field, 
which scales as $m_c$ at this order, there is no immediate power counting for these operators, contrary to those defined in HQET, cf.\  Eqs.~(\ref{eq:P3q-HQET}), (\ref{eq:P4q-HQET}). Moreover, note that this basis differs from the one 
used in Ref.~\cite{Gabbiani:2004tp} for the computation of the dimension-seven and dimension-eight contributions.
By evaluating the matrix elements of the dimension-seven four-quark operators using the framework of the HQET, we have to further expand the charm quark momentum, according to $p_c^\mu = m_c v^\mu + k^\mu$, see Section~\ref{sec:HQET}, as well as to include $1/m_c$ corrections to the effective heavy quark field and to the HQET Lagrangian, cf.\ Section~\ref{sec:dim-7-four-q}.
In this case, we obtain the following basis, in complete analogy to Eqs.~(\ref{eq:P1-d7})-(\ref{eq:M4-d7}), namely\\
\begin{align}
  {\cal P}_1^q 
 & =  
 m_q \, (\bar h_v (1- \gamma_5) q) (\bar q (1- \gamma_5) h_v)\, , 
 \label{eq:P1q-HQET}
\\[3mm]
 {\cal P}_2^q 
& =
(\bar h_v \gamma_\mu (1- \gamma_5)  (i v \cdot D) q) (\bar q \gamma^\mu (1- \gamma_5) h_v)\, ,
\label{eq:P3q-HQET}
\\[3mm]
 {\cal P}_3^q
& =  
 (\bar h_v(1- \gamma_5)  (i v \cdot D) q) (\bar q  (1+ \gamma_5) h_v)\, ,
\label{eq:P4q-HQET}
\end{align}\\
due to the contribution of the light spectator quark momentum, \\
\begin{align}
 {\cal R}_1^q 
& =  
 (\bar h_v \gamma_\mu (1- \gamma_5)  q) (\bar q \gamma^\mu (1- \gamma_5) (i \slashed D) h_v) \, ,
\label{eq:P5q-HQET}
\\[3mm]
 {\cal R}_2^q
& = 
 (\bar h_v (1- \gamma_5) q) (\bar q  (1+ \gamma_5)  (i \slashed D)  h_v)\, ,
\label{eq:P6q-HQET}
\end{align}\\
due to $1/m_c$ corrections to the effective heavy quark field $h_v$, and \\
\begin{align}
{\cal M}_{1}^{q}
& =
  i \int d^4 y \,  {\rm T}
\left[ 
 {\cal O}_1^q (0), 
(\bar h_v (i D)^2 h_v) (y)
\right],
\label{eq:M1-pi} 
\\[3mm]
{\cal M}_{2}^{q}
& =
 i \int d^4 y \, {\rm T}
\left[ 
 { \cal O}_1^q (0), 
\frac{1}{2} g_s \left(\bar h_v \sigma_{\alpha \beta} G^{\alpha \beta} h_v \right) (y)
\right],
\label{eq:M1-G}
\\[3mm]
{\cal M}_{3}^{q}
& =
 i \int d^4 y \, {\rm T}
\left[ 
 {\cal  O}_2^q (0), 
(\bar h_v (i D)^2 h_v) (y)
\right],
\label{eq:M2-pi}
\\[3mm]
{\cal M}_{4}^{q}
& =
 i \int d^4 y \, {\rm  T}
\left[ 
 { \cal O}_2^q (0), 
\frac{1}{2} g_s \left(\bar h_v \sigma_{\alpha \beta} G^{\alpha \beta} h_v \right) (y)
\right],
\label{eq:M2-G}
\end{align}\\
due to $1/m_c$ corrections to the HQET Lagrangian, which we have explicitly indicated, cf.\ Eq.~(\ref{eq:O-kin-HQET}), (\ref{eq:O-magn-HQET}). Moreover, the set of operators in Eqs.~(\ref{eq:P1q-HQET})-(\ref{eq:M2-G}), are supplemented by the corresponding colour octet ones. 
To parametrise the matrix element of the dimension-seven operators in HQET, we use VIA and account for deviations from it by including the corresponding Bag parameters, as it is explicitly shown in Appendix \ref{app:7}. However, since for these matrix elements, does not exist a non perturbative evaluation available yet, in our analysis we have to rely only on VIA.
It follows that, at LO-QCD the matrix element of the dimension-seven operators listed above, can be expressed in terms of the HQET
non perturbative parameters $F (\mu)$, $G_1 (\mu)$, 
$G_2 (\mu)$, and $\bar \Lambda$, so far determined only with large uncertainties.
For this reason, we prefer to use as an input the QCD decay
constant $f_D$, which is computed very precisely using Lattice QCD
\cite{Aoki:2019cca}. In doing so, we obtain that in VIA and at 
the matching scale $\mu = m_c$,  
the contribution of the local operators 
${\cal R}_{1,2}^q$, as well as that of the non-local 
${\cal M}_{1}^q$, 
${\cal M}_{2}^q$, 
${\cal M}_{3}^q$ and
${\cal M}_{4}^q$, 
can be entirely absorbed in the QCD decay constant $f_D$, cf.\ Eq.~\eqref{eq:decay_constant-conversion}, 
more precisely, in the QCD matrix element of the dimension-six operators in Eqs.~(\ref{eq:O1}), (\ref{eq:T2}), which are proportional to $f_D$, and the only remaining $1/m_c$
contribution is due to the operators $ {\cal P}_{1,2,3}^q$, analogously to the QCD case \footnote{In the matrix element of  $\tilde P_{1,2,3}^q$ one can replace the HQET decay constant with the QCD one, up to higher order corrections.}.
To make this point more clear, we consider as an example the contribution due to Pauli interference at LO-QCD and up to order $1/m_c^4$, in the case of $c \to s \bar d u $ transition, which constitutes the dominant correction to $\Gamma(D^+)$, namely\\
\begin{align}
\hspace*{-2.5mm}
{\rm Im} \, {\cal T}^{\rm PI} 
& =  
\Gamma_0 \, |V_{ud}^*|^2 \, 
\frac{32 \pi^2}{m_c^3} (1 - x_s)^2  
\Biggl[ 
C_{\rm PI}^S \left( {\cal O}_1^d +  \frac{{\cal R}_1^d}{m_c} + 
\frac{{\cal M}_{1}^d}{m_c} + \frac{{\cal M}_{2}^d }{m_c}
+ 2 \frac{1 + x_s}{1 - x_s}\frac{ {\cal P}_3^q }{m_c} \right) 
\nonumber \\[3mm]
& + ( {\rm singlet} \to {\rm octet} ) 
\Biggr]\,.
\label{eq:Gamma-PI-c-to-s-d-u}
\end{align}\\
By evaluating the matrix element of ${\rm Im} {\cal T} ^{\rm PI}$ in VIA, the contribution due to the colour-octet operators vanishes. Moreover, using the parametrisation for the matrix elements of the four-quark operators given in Eq.~\eqref{eq:ME-dim-6-HQET-q-q} and in Appendix~\ref{app:7}, we obtain in VIA and setting $\mu  = m_c$, that\\
\begin{align}
\langle  {\cal O}_1^d +  \frac{{\cal R}_1^d}{m_c} + \frac{{\cal M}_{1}^d}{m_c} + \frac{{\cal M}_{2}^d}{m_c}  \rangle_{\rm HQET}
& =  
F^2 (m_c) \, m_{D^+} 
\left[1 - \frac{\bar \Lambda}{m_c} + \frac{2 \,G_1 (m_c)}{m_c} + \frac{12 \, G_2 (m_c)}{m_c} \right]
\nonumber \\[3mm]
& =  f_D^2 \, m_{D^+}^2 = \langle O_1^d \rangle_{\rm QCD},
\label{eq:PI-HQET-QCD}
\end{align}
where in the second line we have used the conversion between the QCD and HQET decay constants given in Eq.~\eqref{eq:decay_constant-conversion}. From Eq.~(\ref{eq:PI-HQET-QCD}) we see that the contribution of the local operators ${\cal R}_1^q$ and
non-local operators ${\cal M}^q_{1}$ and
${\cal M}^q_{2}$, is entirely absorbed by using the QCD decay
constant. Note that, by neglecting the effect due to the strange quark mass and using
VIA we reproduce the result in Eq.~(19) of Ref.~\cite{Mannel:2021uoz}.
The same argument applies also to the remaining topologies i.e.\  WE and WA. 
However, it is worth remarking that in VIA and neglecting the strange quark mass, the contribution of WE and WA exactly vanishes at LO-QCD, due to the helicity suppression.
This suppression 
is lifted once the $s$-quark mass or perturbative
gluon corrections are included, and in this case it becomes again manifest that the contribution
of ${\cal R}_i^q$, and $ {\cal M}^q_{i}$ in HQET, can be completely absorbed in $f_D$ by evaluating the matrix elements in VIA.
\footnote{Note, that for the operator ${ O}_2^q$ the contribution
of $ { \cal R}_2^q$ is absorbed by the combination $(m_D \, f_D/m_c)^2
\approx (1 + 2 \, \bar \Lambda/m_c) \, f_D^2$.}.
We stress that in our numerical analysis, we employ this argument also when using the results of the Bag parameters determined from HQET SR, by neglecting the small deviation from their corresponding VIA values. 
Note that a detailed analysis of the dimension-seven
contributions within the HQET has been performed in
Ref.~\cite{Kilian:1992cj} for the case of $B - \bar B$-mixing.
Specifically, it was found that in VIA, subleading power
corrections due to non-local operators can be entirely absorbed
in the definition of the QCD decay constant,  and that the
residual $1/m_b$ corrections, due to the running of the local
dimension-seven operators from the scale $m_b$ to
$\mu \sim$~1~GeV, is numerically small ($\sim 5 \%$ for
Ref.~\cite{Kilian:1992cj}).\footnote{By neglecting the effect of
running down to a lower scale, from Ref.~\cite{Kilian:1992cj} one
can see that in VIA the QCD decay constant entirely absorbs all
the $1/m_b$ contributions.}
Finally, by summing over all the CKM
modes, at LO-QCD, the dimension-seven contribution can 
 be presented as\\
\begin{align}
16 \pi^2 \tilde{\Gamma}^{D_q}_7 \frac{\langle \tilde{\cal O}_7\rangle^{D_q}}{m_c^4}
& = 
\frac{\Gamma_0}{{|V_{cs}|^2}} \!\!\! \sum_{q_1, q_2 = d, s} \Bigg\{
\left| \lambda_{q_1 q_2} \right|^2 \!
\sum_{i = 1}^3 
\Big[G_{i, q_1 q_2}^{\rm WE} \frac{\langle D_q | { \cal  P}_i^u | D_q \rangle}{m_c^4} 
+ G_{i, q_1 q_2}^{\rm PI} \frac{\langle D_q | {\cal  P}_i^{q_2} | D_q \rangle}{m_c^4}
\nonumber
\\[3mm]
& + G_{i, q_1 q_2}^{\rm WA} \frac{\langle D_q | { \cal P}_i^{q_1} | D_q \rangle}{m_c^4}
\Big] +  |V_{c q_1}|^2
\sum_{\ell = e, \mu} \sum_{i = 1}^3 
\left[
G_{i, q_1 \ell}^{\rm WA} \frac{\langle D_q | { \cal P}_i^{q_1} | D_q \rangle}{m_c^4}
\right] 
\nonumber \\[3mm]
& + (\mbox{colour-octet part}) \Bigg\} \,,
\label{eq:dim-7-4q-NLO-scheme}
\end{align}\\
and we confirm the results for the short-distance coefficients $G_{i, q_1 q_2}^{\rm WE}$, $G_{i, q_1 q_2}^{\rm PI}$
and $G_{i, q_1 q_2}^{\rm WA}$, $G_{i, q_1 \ell}^{\rm WA}$ presented in Ref.~\cite{Lenz:2013aua}. 
Note that, due to the current accuracy of the analysis, at dimension-seven we include only the contribution of the valence-quark therefore e.g.\ $\langle D^0 | {\cal  P}_i^s | D^0 \rangle = 0$.

Having presented each of the contributions that enter the HQE of $D$-mesons, we now turn to discuss the numerical evaluation of the corresponding matrix elements. For most of the non perturbative parameters in the charm-sector there is no determination available yet, contrary to the $b$-system, where e.g.\ the value of $\mu_G^2$, $\mu_\pi^2$, and $\rho_D^3$, has been extracted performing fits to experimental data for $B^0$ and $B^+$ semileptonic decays \cite{Alberti:2014yda}. Specifically, for $\mu_G^2$ they find  \cite{Alberti:2014yda}\\
\begin{equation}
 \mu_G^2 (B) =    (0.332 \pm 0.062) \,  {\rm GeV}^2
  \, .
  \label{eq:muG_B_fit}
 \end{equation}\\
By using the heavy quark symmetry, we could expect the corresponding
parameter in the $D$ system to have a similar size. The value of $\mu_G^2$ can also be obtained taking into account the spectroscopy relation~\cite{Uraltsev:2001ih}\\
\begin{equation}
\mu_G^2 (D_{(s)}) = 
\frac{3}{2} m_c \, \left(M_{D_{(s)}^*} - M_{D_{(s)}}\right) \, ,
\label{Eq:mupi_spec}
\end{equation}\\
which holds up to power corrections.
Using the value for the meson masses given in the PDG \cite{Zyla:2020zbs} and setting $m_c = 1.27 \, {\rm GeV}$, we obtain the following estimates\\
\begin{equation}
\mu_G^2 (D) = (0.268 \pm  0.107) \,  {\rm GeV}^2, 
\qquad
\mu_G^2 (D_{s}) = (0.274 \pm  0.110) \,  {\rm GeV}^2, 
\label{eq:mu_G}
\end{equation}\\
where we have conservatively added an uncertainty of $40 \%$ 
due to unknown power corrections of order~$1/m_c$. 
The values in Eq.~(\ref{eq:mu_G}) are roughly 19$\%$ smaller than those obtained from experimental fits for semileptonic $B$-meson decays, 
see Eq.~(\ref{eq:muG_B_fit}).
Moreover, Eq.~(\ref{Eq:mupi_spec}) leads to a tiny amount of
$SU(3)_f$-symmetry breaking of 
$ \approx 2 \%$, which might, however,
be enhanced by the neglected power
corrections.
In the literature instead of Eq.~(\ref{Eq:mupi_spec}) it is often adopted the relation \cite{Falk:1992wt, Neubert:1993mb}\\
 \begin{equation}
 \mu_G^2 (D_{(s)})  
  =  \frac 3 4 \left(M_{D_{(s)}^*}^2 - M_{D_{(s)}}^2 \right)\,,
 \label{Eq:mupi_spec2}
 \end{equation}\\
 which coincides with Eq.~(\ref{Eq:mupi_spec}) up to corrections of order $1/m_c$.
 Numerically we find that
 Eq.~(\ref{Eq:mupi_spec2}) yields \\
 \begin{equation}
 \mu_G^2 (D)  =  0.41 \, {\rm GeV}^2
 \, , 
  \qquad 
 \mu_G^2 (D_s^+)  =  0.44 \, {\rm GeV}^2 \, ,
 \label{eq:mu_G2}
 \end{equation}\\
 which are roughly 23$\%$ higher than the values in 
 Eq.~(\ref{eq:muG_B_fit}).
 In our numerical analysis we take the average of the two determinations in  Eq.~(\ref{eq:mu_G})
and Eq.~(\ref{eq:mu_G2}). This gives
\begin{equation}
\mu_G^2 (D)  =  (0.34 \pm 0.10) \, {\rm GeV}^2,
\qquad 
\mu_G^2 (D_s^+)  =  (0.36 \pm 0.10) \, {\rm GeV}^2 \, ,
\label{eq:mu_G3}
\end{equation}
\begin{table}[ht]
\centering
\renewcommand{\arraystretch}{1.6}
\begin{tabular}{|c||C{2cm}|C{2cm}|C{2cm}|C{2cm}|C{2cm}|}
  \hline
  {\rm Source}  
  & LQCD \cite{FermilabLattice:2018est}
  & LQCD \cite{Gambino:2017vkx}
  & Exp.~fit \cite{Alberti:2014yda} 
  & QCD~SR \cite{Neubert:1996wm} 
  & QCD~SR \cite{Ball:1993xv} 
  \\
  \hline
  $\mu_\pi^2 [{\rm GeV}^2]$   
  & 0.05(22)  
  & 0.314(15) 
  & 0.465(68)
  & 0.10(5) 
  & 0.6(1) \\
  \hline
\end{tabular}
\caption{Different determinations of $\mu_\pi^2 (B)$ available in the literature.}
\label{tab:mu-pi-sq}
\end{table}
which agrees  
well with the one in Eq.~(\ref{eq:muG_B_fit}).
From Eq.~(\ref{eq:Gamma_5}), we expect corrections to the total
decay rate due to the chromo-magnetic operator,
$c_G \, \mu_G^2/(c_3 \, m_c^2)$ ranging between $-6 \%$ and $+8 \%$
with respect to the leading free-quark decay contribution.
A large part of the sizeable uncertainty derives from the
cancellations in  the coefficient $c_G$, as shown in Figure~\ref{fig:cG}, which could be
reduced with a complete determination of the NLO-QCD corrections.
For semileptonic rates the contribution of the chromomagnetic operator
can be even of the order of $20\%$.
An experimental determination of $\mu_G^2(D)$ from inclusive
semileptonic $D$-meson decays could further reduce the uncertainties
and could in particular give some insight into the numerical size of
$SU(3)_f$ breaking.

For the matrix element of the kinetic operator, there are several predictions of $\mu_\pi^2$, available in the literature 
for the  $B$-meson, which cover a large range of values, see Table~\ref{tab:mu-pi-sq}.
Assuming heavy quark symmetry we can again use the determination  in Ref.~\cite{Alberti:2014yda}
\begin{equation}
  \mu_\pi^2 (B)   =  
  (0.465 \pm 0.068) \, {\rm GeV}^2 \, ,
 \end{equation}
to obtain the following estimate in the case of $D$-meson\\
\begin{equation}
  \mu_\pi^2 (D)   =   (0.465  \pm 0.198) \,  {\rm GeV}^2,
 \end{equation}\\
where we have added a conservative uncertainty of $40 \% $ in order to account for the breaking of the heavy 
quark symmetry. This value fulfils the theoretical bound $\mu_\pi^2 \geq \mu_G^2$, see e.g. the review \cite{Bigi:1997fj}.
We then expect, from Eq.~(\ref{eq:Gamma_5}), corrections 
due to the kinetic  operator of the order of $-10\%$.
The $SU(3)_f$ breaking effects for the matrix element of the kinetic operator 
have been estimated in Refs.~\cite{Bigi:2011gf, Lenz:2013aua}, i.e.\\
\begin{equation}
\mu_\pi^2 (D_s^+) - \mu_\pi^2 (D^0)  \approx   0.09 \,  {\rm GeV}^2 \, ,
\end{equation}\\
leading to the following estimate in the case of the $D_s$ meson\\\
\begin{equation}
\mu_\pi^2 (D_s^+) = (0.555 \pm 0.232) \, {\rm GeV}^2.
\end{equation}\\
Again a more precise experimental determination of $\mu_\pi^2$ from
fits to semileptonic $D^+$, $D^0$ and $D_s^+$ meson decays, as it has been
done for the $B^{+}$ and $B^{0}$ decays, would be very desirable.

For the matrix element of the Darwin operator, we can again 
assume the validity of the heavy quark symmetry and use the corresponding value
obtained from fits of the semileptonic $B$ decays \cite{Alberti:2014yda}, namely\\
\begin{equation}
  \rho_D^3 (B)  =   (0.170 \pm 0.038) \, {\rm GeV}^3 \, ,
  \label{eq:rho_d-fit_B}
 \end{equation}\\
which, by adding quadratically an uncertainty of {$40\%$} to account for the breaking of the heavy-quark symmetry,  leads to a first estimate of
\begin{align}
  \rho_D^3 (D)^{\rm I}  =   (0.17 \pm 0.07) \, {\rm GeV}^3 \, .
 \label{eq:Darwin_MEI}
 \end{align}
Alternatively the Darwin parameter can be
related to the matrix elements of the
dimension-six four-quark operators
through the equation of motion 
for the gluon field.
At leading order in $1/m_Q$, we obtain\\
\begin{align}
\rho_D^3 (H_Q) &= 
\frac{g_s^2}{18} f_{H_Q}^2 \, m_{H_Q} 
\Bigg[
2 \, \tilde B_2^{q} -  \tilde B_1^{q} 
+ \frac{3}{4}  \tilde B_3^{q}
- \frac{3}{2}  \tilde B_4^{q} 
\nonumber \\[3mm]
&
+ \sum_{q^\prime = u, d, s} 
\left(
2 \tilde \delta_2^{q^\prime q} -  \tilde \delta_1^{q^\prime q} 
+ \frac{3}{4}  \tilde \delta_3^{q^\prime q} - \frac{3}{2}  \tilde \delta_4^{q^\prime q}
\right)
\Bigg]\,,
\label{eq:EoM-Darwin}
\end{align}\\
where $H_Q$ is a heavy hadron with the mass $m_{H_Q}$ and the decay constant $f_{H_Q}$,
$q = u, d, s,$ is the light valence quark in $H_Q$, 
and the Bag parameters
have bee introduced in Eqs.~(\ref{eq:ME-dim-6-HQET-q-q}).
The strong coupling $g^2_s = 4 \pi \alpha_s$, should be evaluated at a non perturbative scale and e.g.\ Ref.~\cite{Bigi:1993ex} suggests to set $\alpha_s = 1$.
\begin{table}[ht]
\renewcommand{\arraystretch}{1.5}
\centering
\begin{tabular}{|c||c|c||c|c||c|c|}
\hline
&
\multicolumn{2}{|c||}{$\mu = 1.5$ GeV} &
\multicolumn{2}{|c||}{$\mu = 1.0$ GeV} &
\multicolumn{2}{|c|}{$\alpha_s = 1$} 
\\
\hline
$\rho_D^3 [{\rm GeV^3}]$
& VIA & \mbox{ HQET} 
& VIA & \mbox{ HQET}
& VIA & \mbox{ HQET}
\\
\hline \hline
$B^+, B_d$ &  
0.048  & 0.047 &  
0.066  & 0.064 & 
0.133  & 0.129
\\
\hline
$B_s$ &  
0.072  & 0.070 &  
0.098  & 0.095 & 
0.199  & 0.193
\\
\hline
$D^+, D^0 $
& 0.021 & 0.020  
& 0.027 & 0.026 
& 0.059 & 0.056 
\\
\hline
$D_s^+ $
& 0.030 & 0.029 
& 0.040 & 0.038 
& 0.086 & 0.082 
\\
\hline
\end{tabular}
\caption{Values of $\rho_D^3 (H)$ for $B$- and $D$-mesons
in VIA and using HQET SR for the Bag parameters for three different choices of $\alpha_s$ in Eq.~\eqref{eq:EoM-Darwin}.}
\label{tab:rhoD}
\end{table}
From the input listed in Table~\ref{tab:input} and the expression in Eq.~(\ref{eq:EoM-Darwin}), we can  estimate the size of $\rho_D^3$ for both the $B$- and $D$-mesons and using the VIA 
as well as the HQET SR values for the Bag parameters. 
The results are summarised in Table~\ref{tab:rhoD}
for the three different choices, namely $\alpha_s(\mu = 1.5 \, {\rm GeV})$, $\alpha_s(\mu = 1 \, {\rm GeV})$ and $\alpha_s = 1$.
By setting $\alpha_s = 1$ in Eq.~(\ref{eq:EoM-Darwin}), we obtain values for $\rho_D^3(B)$ that are close to the one in Eq.~(\ref{eq:rho_d-fit_B}), indicating $1/m_b$-corrections in Eq.~(\ref{eq:EoM-Darwin}) of the order of $+30 \%$.
Moreover, the difference between using VIA and HQET sum rule is small.
We emphasise that because of the sizeable $SU(3)_F$ breaking in the decay constants, Eq.~(\ref{eq:EoM-Darwin}) leads also to a sizeable $SU(3)_F$ breaking for the non-perturbative parameters $\rho_D^3(D)$, $\rho_D^3(D_s^+)$. 
By setting $\alpha_s = 1$ and using HQET SR results for the Bag parameters we arrive at the second estimate, cf.\ last column in Table~\ref{tab:rhoD}\\
\begin{align}
  \rho_D^3 (D)^{II}  =  (0.056 \pm 0.022) \, {\rm GeV}^3 \, ,
  \quad
  \rho_D^3 (D_s^+)^{II}  =  (0.82 \pm 0.033) \, {\rm GeV}^3 \, ,
 \label{eq:Darwin_MEII}
 \end{align}\\
where we have again added $40\%$ uncertainty.
Finally, another possibility to extract $\rho_D^3(D)$ is to substitute in Eq.~(\ref{eq:EoM-Darwin}) the values of the Bag parameters in VIA, which gives\\
\begin{equation}
\rho_D^3 (H_Q) \approx \frac{g_s^2}{18} f_{H_Q}^2 \, m_{H_Q}.
\label{eq:EoM-Darwin-VIA}
\end{equation}\\
Assuming the strong coupling to have a  similar size for both the $B$- and $D$-meson matrix elements, from Eq.~(\ref{eq:EoM-Darwin-VIA}) we obtain\\
\begin{align}
\rho_D^3 (D) 
\approx 
\frac{f_D^2 \, m_D}{f_B^2 \, m_B} \, \rho_D^3 (B) \, ,
 \quad 
\rho_D^3 (D_s) 
 \approx 
\frac{f_{D_s}^2  m_{D_s}}{f_B^2 \, m_B} \, \rho_D^3 (B) \, .
\end{align}\\
Using the most precise determination of the decay constants from Lattice QCD \cite{Aoki:2019cca}, and of the meson masses from PDG \cite{Zyla:2020zbs} and taking into account the value of $\rho_D^3 (B)$ in Eq.~(\ref{eq:rho_d-fit_B}), leads to the third estimate\\
\begin{align}
\rho_D^3 (D)^{III} 
 = 
(0.075 \pm 0.034) \, {\rm GeV^3} \, ,
 \quad 
\rho_D^3 (D_s)^{III} 
 = 
(0.110 \pm 0.050) \, {\rm GeV^3} \, ,
\label{eq:Darwin_ME}
\end{align}\\
where we again assign in addition a conservative $40 \%$
uncertainty due to missing power corrections.
These values are consistent with the numbers shown in Table~\ref{tab:rhoD} for $\alpha_s = 1$.
Contrary to the case of the dimension-five matrix elements, in Eq.~(\ref{eq:Darwin_ME}) we observe a large $SU(3)_f$ symmetry breaking of $\approx 46 \% $, and similarly of $\approx 49 \%$ for the $B_{(s)}$-mesons, as already stated above, mostly stemming from the ratios $f_{B_s}/f_{B_d}$ and  $f_{D_s^+}/f_{D^0}$.
In our numerical analysis we use the values shown in Eq.~(\ref{eq:Darwin_ME}), which lie between the estimates 
obtained in Eq.~(\ref{eq:Darwin_MEI}) and
Eq.~(\ref{eq:Darwin_MEII}).
Again, a more precise experimental determination of $\rho_D^3$ from 
fits to semileptonic $D^+$, $D^0$ and $D_s^+$ meson decays, as it has been
done for the $B^{+}$ and $B^{0}$ decays , would be very desirable
and could have a significant effect on the phenomenology of inclusive charm decays.

Finally, the dimension-six Bag parameters of the $D^+$ and $D^0$ mesons
have been determined using HQET sum rules in Ref.~\cite{Kirk:2017juj}. Corrections due to the inclusion of the
strange quark mass, needed in the case of
the  $D_s^+$  meson, as well as the effect of the eye-contractions, have been
computed for the first time in
Ref.~\cite{King:2020}, again using HQET sum rules.
The results, collected in Table~\ref{tab:Bag-parameters}, show only a small deviation from the corresponding VIA values.
For the dimension-seven Bag parameters we use only 
 VIA. In HQET the matrix
 elements of dimension-seven operators depend also on the
 parameters $\bar \Lambda_{(s)} = m_{D_{(s)}} - m_c $, for which
 we use the following range of values \cite{King:2020}\\
\begin{equation}
\bar \Lambda = (0.5 \pm 0.1 ) \, {\rm GeV}, \qquad 
\bar \Lambda_s = (0.6 \pm 0.1 ) \, {\rm GeV}\,.
\end{equation}


\subsection{Numerical results}
\label{sec:numer}

\begin{table}
\centering
\renewcommand{\arraystretch}{1.6}
\begin{tabular}{|C{2.6cm}|C{4.3cm}|C{7.6cm}|}
\hline
Parameter & Value & Source \\
\hline
\hline 
$\alpha_s (M_Z)$ & $0.1179 \pm 0.0010$ & PDG~\cite{Zyla:2020zbs} \\
\hline
$|V_{us}|$ & $0.224834^{+0.000252}_{-0.000059}$  
& \multirow{4}{*}{CKMfitter \cite{Charles:2004jd}} \\
$|V_{ub}|/|V_{cb}|$ & $0.088496^{+0.001885}_{-0.002244}$ 
&  \\
$|V_{cb}|$ & $0.04162^{+0.00026}_{-0.00080}$ &   
 \\
$\delta$ & $\left(65.80^{+0.94}_{-1.29}\right)^\circ $ & 
 \\
\hline
$\overline{m}_c (\overline{m}_c)$ & $(1.27 \pm 0.02) \, {\rm GeV}$ & 
PDG \cite{Zyla:2020zbs}
\\  
$m_c^{\rm kin} (0.5 \, {\rm GeV})$ 
& 
$1.306 \, {\rm GeV}$
& 
\cite{Fael:2020njb} 
\\
$m_{J/\psi}$ & $3.0969$ GeV & PDG \cite{Zyla:2020zbs} \\
\hline
$m_s$ & $\left(93^{+11}_{-5}\right) \, {\rm MeV}$ 
& PDG \cite{Zyla:2020zbs} \\
\hline
$M_{D^0}$ & $1.86493 \, {\rm GeV}$ &  \\
$M_{D^+}$ & $1.86965 \, {\rm GeV}$ & PDG \cite{Zyla:2020zbs} \\
$M_{D_s^+}$ & $1.96834 \, {\rm GeV}$ &   \\
\hline
 $f_D$ & $(0.2120 \pm 0.0007)$ GeV
       & \multirow{2}{*}{Lattice QCD \cite{Aoki:2019cca}} \\
 $f_{D_s}$ & $(0.2499 \pm 0.0005)$ GeV 
       & \\
\hline
$\mu_\pi^2 (D)$ 
    & $(0.465 \pm  0.198)$ GeV$^2$
    & Exp. fit \cite{Alberti:2014yda} 
      and HQ symmetry \\
$\mu_\pi^2 (D_s)$     
    & $(0.555 \pm  0.232)$ GeV$^2$
    & $SU(3)_f$-breaking \cite{Bigi:2011gf} 
      and HQ symmetry \\
\hline
$\mu_G^2 (D)$
     & $(0.339 \pm  0.098)$ GeV$^2$ 
     & \multirow{2}{*}{Spectroscopy relations~\cite{Uraltsev:2001ih, Falk:1992wt}} \\
$\mu_G^2 (D_s)$ 
     & $(0.357 \pm  0.104)$ GeV$^2$
     & \\
     \hline
$\rho_D^3 (D)$ 
     & $(0.075 \pm  0.034)$ GeV$^3$
     & \multirow{2}{*}{Exp. fit \cite{Alberti:2014yda} and 
       E.O.M relations} \\
$\rho_D^3 (D_s)$ 
     & $(0.110 \pm  0.050)$  GeV$^3$
     & \\             
\hline       
\end{tabular}
\caption{Numerical input used in the numerical analysis.}
\label{tab:input}
\end{table}
\begin{table}\centering
\renewcommand{\arraystretch}{1.6}
\begin{tabular}{|c||c|c|c|c|}
\hline
${\rm HQET}$    
&  $ \tilde B_1$ 
&  $ \tilde B_2$ 
& $ \tilde B_3$ 
& $ \tilde B_4$ 
\\
\hline
\hline
    $D^{+,0}$ 
     & $1.0000^{+0.0020}_{-0.0006}$ 
     & $\phantom{-}1.0000^{+0.0007}_{-0.0000}$ 
     & $-0.0161^{+0.0115}_{-0.0206}$ 
     & $-0.0007^{+0.0104}_{-0.0170}$
\\
\hline
     $D_s^+$  
     & $1.0000^{+0.0014}_{-0.0003}$ 
     & $\phantom{-}1.0000^{+0.0007}_{-0.0000}$ 
     & $-0.0094^{+0.0103}_{-0.0171}$ 
     & $-0.0001^{+0.0104}_{-0.0169}$
\\
\hline
\end{tabular}
\begin{tabular}{|c||c|c|c|c|}
\hline
${\rm HQET}$    
& $ \tilde \delta_1$
& $ \tilde \delta_2$ 
& $ \tilde \delta_3$ 
& $ \tilde \delta_4$ 
\\
\hline
\hline
$\langle D_q | \tilde O^q | D_q \rangle $
& $0.0026^{+0.0004}_{-0.0009}$ 
& $-0.0018^{+0.0005}_{-0.0002}$ 
& $-0.0004^{+0.0001}_{-0.0001}$ 
& $\phantom{-}0.0003^{+0.0000}_{-0.0001}$
\\
\hline
$\langle D_s |  \tilde O^q | D_s \rangle$ 
& $0.0025^{+0.0004}_{-0.0008}$ 
& $-0.0018^{+0.0005}_{-0.0002}$ 
& $-0.0004^{+0.0001}_{-0.0001}$ 
& $\phantom{-}0.0003^{+0.0000}_{-0.0001}$
\\
\hline
$\langle D_q | \tilde O^s | D_q \rangle$ 
& $0.0017^{+0.0005}_{-0.0009}$ 
& $-0.0012^{+0.0005}_{-0.0003}$ 
& $-0.0003^{+0.0001}_{-0.0001}$ 
& $\phantom{-}0.0002^{+0.0001}_{-0.0001}$
\\
\hline
$\langle D_s | \tilde O^s | D_s \rangle$ 
& $0.0023^{+0.0005}_{-0.0009}$ 
& $-0.0017^{+0.0005}_{-0.0002}$ 
& $-0.0004^{+0.0002}_{-0.0001}$ 
& $\phantom{-}0.0003^{+0.0000}_{-0.0001}$
 \\
\hline
\end{tabular}
\caption{Numerical values of the HQET Bag parameters \cite{Kirk:2017juj,King:2020}, at the renormalisation scale $\mu_0 = 1.5 \, {\rm GeV}$.}
\label{tab:Bag-parameters}
\end{table}
In this section, using all the ingredients described above, we present the theoretical predictions
for the total and semileptonic decay rates of the $D^0$, $D^+$ and $D_s^+$ mesons, and for their ratios. 
All the input included in our numerical analysis are collected in Table~\ref{tab:input}.
For each observable we investigate several quark mass schemes, using as default the kinetic and the $1S$ scheme, and compare the corresponding results with both VIA and HQET SR
values for the dimension-six Bag parameters. 
The uncertainties quoted are obtained by varying all the input parameters
within their intervals.
For the renormalisation scales, we fix the central values to
$\mu_1 = \mu_0 = 1.5 \, {\rm GeV}$ \footnote{The renormalisation scale $\mu_0$ enters in the NLO-QCD corrections to the dimension-six coefficients as well as in the running of the Bag parameters.}, and vary both of them independently between 1 and 3 GeV.
Moreover, we add an estimated uncertainty due to missing higher power and QCD corrections. 
\begin{table}[t]\centering
\renewcommand{\arraystretch}{1.7}
\begin{tabular}{|c|C{1.5cm}|C{1.5cm}|C{1.5cm}|C{1.5cm}||c|}
\hline
\multicolumn{6}{|c|}{VIA} 
\\
\hline
Observable 
& 
Pole
&
$\overline{\rm MS}$
&
Kinetic
&
$1S$
& Exp. value  \\
\hline
\hline
$\Gamma (D^0) [{\rm ps}^{-1}]$ 
& $1.68 $ & $1.47 $ & $1.56 $ & $2.31 $
& $2.44 $ \\
\hline
$\Gamma (D^+) [{\rm ps}^{-1}]$ 
& $0.19 $ & $-0.03 $ & $0.09 $ & $0.56 $
& $0.96 $ \\
\hline
$\bar \Gamma (D_s^+) [{\rm ps}^{-1}]$ 
& $1.72 $  & $1.48 $ & $1.58 $ & $2.34 $
& $1.88 $ \\
\hline
\hline
$\tau (D^+)/\tau(D^0) $ 
& $2.55 $ & $2.56 $ & $2.53 $ & $2.82 $
& $2.54 $ \\
\hline
$\bar \tau (D_s^+)/\tau(D^0) $ 
& $0.99 $ & $1.00 $ & $0.99 $ & $0.99 $
& $1.30 $ \\
\hline
\hline
$B_{sl}^{D^0} [\%]$ 
& $5.31 $ & $6.46 $ & $6.03 $ & $8.48 $
& $6.49 $ \\
\hline
$B_{sl}^{D^+} [\%]$ 
& $13.5$ & $16.4$ & $15.3$ & $21.5$
& $16.07$ \\
\hline
$B_{sl}^{D^+_s} [\%]$ 
& $6.88 $ & $8.24 $ & $7.74 $ & $10.8 $
& $6.30 $ \\
\hline
\hline
$\Gamma_{sl}^{D^+}/\Gamma_{sl}^{D^0}$
& $1.000 $ & $1.000 $ & $1.000 $ & $1.000 $
& $0.985 $ \\
\hline
$\Gamma_{sl}^{D^+_s}/\Gamma_{sl}^{D^0}$
& $1.04$ & $1.04$ & $1.04$ & $1.05$
& $0.790$ \\
\hline
\end{tabular}
\caption{Central values of the charm observables in different
quark mass schemes using VIA for the matrix elements of the
four-quark operators compared to the corresponding experimental
values (last column).}
\label{tab:summary-diff-schemes-VIA}
\end{table}

\begin{figure}[t]
    \centering
    \includegraphics[scale=1.0]{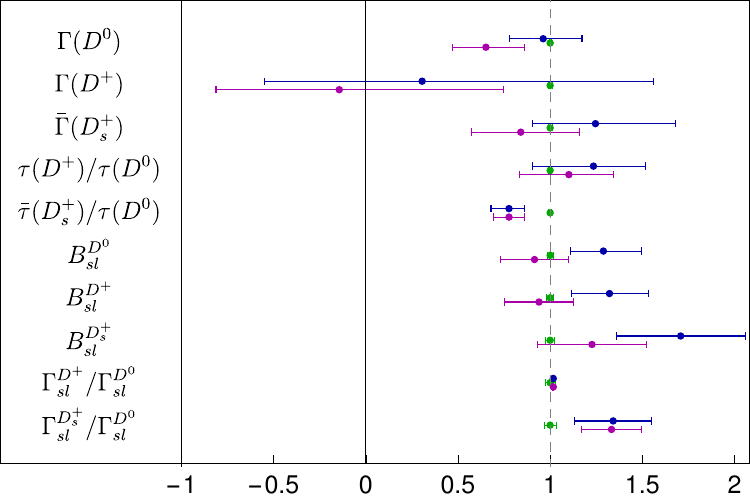}
    \caption{A comparison of the HQE prediction for the charm observables
    in the kinetic scheme (magenta) and in the $1S$ scheme (blue),  
    with the corresponding experimental data (green). Note that all the quantities are normalised to the corresponding experimental central values. }
    \label{fig:summary-comparison}
\end{figure}

We start by considering the total decay rates, 
which are expected to have sizeable uncertainties due to the dependence of the
free quark decay on the fifth power of the charm quark mass and due to large 
perturbative and power corrections. 
A comparison of the central values for the HQE prediction of the decay widths in several mass
schemes is shown in the three first rows of 
Table~\ref{tab:summary-diff-schemes-VIA},
using VIA for the Bag parameters and of
Table~\ref{tab:summary-diff-schemes-HQET-SR} 
using the HQET sum rules results.
In Table~\ref{tab:summary-with-uncertainties} we present the complete theoretical
prediction including the corresponding uncertainties, using the $1S$ and
kinetic scheme for the quark masses and the HQET SR values for the
dimension-six Bag parameters, the same results can be visualised also  in
Figure~\ref{fig:summary-comparison}. In each table, 
the corresponding experimental determinations are listed in the last column. 
For the $D_s^+$ meson there is an additional subtlety due to the
large branching fraction of the leptonic decay $D_s^+ \to \tau^+ \nu_\tau$, which however is not included in the HQE, since the tau lepton is more massive than the charm quark. 
\begin{table}[t]\centering
\renewcommand{\arraystretch}{1.7}
\begin{tabular}{|c|C{1.5cm}|C{1.5cm}|C{1.5cm}|C{1.5cm}||c|}
\hline
\multicolumn{6}{|c|}{HQET SR} 
\\
\hline
Observable 
& 
Pole
&
$\overline{\rm MS}$
&
Kinetic
&
$1S$
& Exp. value  \\
\hline
\hline
$\Gamma (D^0) [{\rm ps}^{-1}]$ 
& $1.71 $ & $1.50 $ & $1.59 $ & $2.34 $
& $2.44 $ \\
\hline
$\Gamma (D^+) [{\rm ps}^{-1}]$ 
& $-0.05 $ & $-0.25$ & $-0.14 $ & $0.29 $
& $0.96 $ \\
\hline
$\bar \Gamma (D_s^+) [{\rm ps}^{-1}]$ 
& $1.70 $  & $1.46 $ & $1.56 $ & $2.32 $
& $1.88 $ \\
\hline
\hline
$\tau (D^+)/\tau(D^0) $ 
& $2.83 $ & $2.83 $ & $2.80$ & $3.14 $
& $2.54 $ \\
\hline
$\bar \tau (D_s^+)/\tau(D^0) $ 
& $1.01 $ & $1.02 $ & $1.01 $ & $1.01 $
& $1.30 $ \\
\hline
\hline
$B_{sl}^{D^0} [\%]$ 
& $5.18 $ & $6.37 $ & $5.93 $ & $8.34 $
& $6.49 $ \\
\hline
$B_{sl}^{D^+} [\%]$ 
& $13.2$ & $16.2$ & $15.1$ & $21.2$
& $16.07$ \\
\hline
$B_{sl}^{D^+_s} [\%]$ 
& $6.79 $ & $8.19 $ & $7.67 $ & $10.7 $
& $6.30 $ \\
\hline
\hline
$\Gamma_{sl}^{D^+}/\Gamma_{sl}^{D^0}$
& $1.002 $ & $1.001 $ & $1.001 $ & $1.002 $
& $0.985 $ \\
\hline
$\Gamma_{sl}^{D^+_s}/\Gamma_{sl}^{D^0}$
& $1.05$ & $1.05$ & $1.05$ & $1.05$
& $0.790$ \\
\hline
\end{tabular}
\caption{Central values of the charm observables in different
quark mass schemes using  HQET sum rule results
\cite{Kirk:2017juj,King:2020} for the matrix elements of the
four-quark operators compared to the corresponding experimental
values (last column).}
\label{tab:summary-diff-schemes-HQET-SR}
\end{table}
Using the experimental value of the leptonic branching ratio 
\cite{Zyla:2020zbs} (online update), we obtain\\
\begin{equation}
{\rm Br} (D_s^+ \to \tau^+ \nu_\tau)  =  (5.48 \pm 0.23)\% \, ,
\end{equation}\\
accordingly, we define the reduced decay rate $\bar{\Gamma} (D_s^+)$, as\\
\begin{equation}
    \bar{\Gamma} (D_s^+)  \equiv  
    {\Gamma} (D_s^+) - \Gamma (D_s^+ \to \tau^+ \nu_\tau) 
     =  
    (1.88 \pm 0.02) \, \mbox{ps}^{-1} \, ,
\end{equation}\\
which leads to the reduced lifetime ratio\\
\begin{equation}
    \frac{\bar{\tau} (D_s^+)}{\tau (D^0)}   =   1.30 \pm 0.01 \, .
\end{equation}\\
The first and main result we can derive from
Table~\ref{tab:summary-with-uncertainties} 
and from Figure~\ref{fig:summary-comparison}, is that the HQE can reproduce
the experimental values of $\Gamma (D^0)$, $\Gamma (D^+)$, and 
$\Gamma (D_s^+)$, within very large uncertainties. 
\begin{table}[t]\centering
\renewcommand{\arraystretch}{1.7}
{
\begin{tabular}{|c|c|c|c|}
\hline
Observable & Kinetic scheme & $1S$-scheme & Exp. value \\
\hline
\hline
$\Gamma (D^0) [{\rm ps}^{-1}]$ 
& $1.590 \pm 0.242^{+0.451 \, +0.002}_{-0.365 \, -0.002}$
& $2.348 \pm 0.247^{+0.651 \, +0.002}_{-0.489 \, -0.001}$
& $2.44 \pm 0.01 $
\\
\hline
$\Gamma (D^+) [{\rm ps}^{-1}]$ 
& $-0.138 \pm 0.572^{+0.581 \, +0.252}_{-0.273 \, -0.102}$
& $0.293 \pm 0.664^{+0.939 \, +0.360}_{-0.453 \, -0.161}$
& $0.96 \pm 0.01$
\\
\hline
$\bar \Gamma (D_s^+) [{\rm ps}^{-1}]$ & 
$1.572 \pm 0.309^{+0.508 \, +0.018}_{-0.399 \, -0.004}$ & 
$2.330 \pm 0.349^{+0.734 \, +0.027}_{-0.540 \, -0.009}$ & 
$1.88 \pm 0.02$ 
\\
\hline
\hline
$\tau (D^+)/\tau(D^0) $ 
& $2.798 \pm 0.606^{+0.001 \, +0.109}_{-0.135 \, -0.263} $
& $3.137 \pm 0.691^{+0.023 \, +0.169}_{-0.299 \, -0.376} $
& $2.54 \pm 0.02$
\\
\hline
$\bar \tau (D_s^+)/\tau(D^0) $ 
& $1.010 \pm 0.105^{+0.018 \, +0.003}_{-0.030 \, -0.010} $
& $1.010 \pm 0.118^{+0.027 \, +0.006}_{-0.044 \, -0.015} $
& $1.30 \pm 0.01 $
\\
\hline
\hline
$B_{sl}^{D^0} [\%]$ 
& $5.94 \pm 1.15^{+0.33}_{-0.28}$
& $8.36 \pm 1.31^{+0.23}_{-0.08}$
& $6.49 \pm 0.11$
\\
\hline
$B_{sl}^{D^+} [\%]$ 
& $15.1 \pm 2.91^{+0.83}_{-0.72}$
& $21.2 \pm 3.32^{+0.58}_{-0.19}$
& $16.07 \pm 0.30$
\\
\hline
$B_{sl}^{D^+_s} [\%]$ 
& $7.73 \pm 1.80^{+0.45}_{-0.40}$
& $10.76 \pm 2.18^{+0.33}_{-0.04}$
& $6.30 \pm 0.16$
\\
\hline
\hline
$\Gamma_{sl}^{D^+}/\Gamma_{sl}^{D^0}$
& $1.002 \pm 0.002 \pm 0.001$
& $1.002 \pm 0.003 \pm 0.001$
& $0.985 \pm 0.028$
\\
\hline
$\Gamma_{sl}^{D^+_s}/\Gamma_{sl}^{D^0}$
& $1.053 \pm 0.130^{+0.006}_{-0.007} $
& $1.060 \pm 0.164^{+0.006}_{-0.007} $
& $0.790 \pm 0.026 $
\\
\hline
\end{tabular}
}
\caption{ HQE predictions for all the ten observables  
in the kinetic (second column) and in the $1S$-schemes (third column),
using HQET SR results for the Bag parameters.
The first uncertainty is parametric one, 
second and third uncertainties are due to $\mu_1$- and $\mu_0$-scales variation, respectively.
The results are compared with the corresponding experimental measurements
(fourth column).}
\label{tab:summary-with-uncertainties}
\end{table}
Moreover, we find that in the 1$S$ scheme we obtain larger values for the decay rates, while the kinetic and the $\overline{\rm MS}$ scheme typically result in smaller values, close to the pole scheme. 
Within the uncertainties the predictions in the different mass schemes are compatible with each other, however, to a large extent.
Given the current precision then, to consider only one
quark mass scheme might lead to considerably underestimate the uncertainties.
Due to the fact that the values of the HQET Bag parameters
\cite{Kirk:2017juj,King:2020} are close to the corresponding ones in VIA,
the predictions shown in Table~\ref{tab:summary-diff-schemes-VIA} 
and in Table~\ref{tab:summary-diff-schemes-HQET-SR} do not differ much. 
A peculiar role is played by the $D^+$ meson, for which we obtain huge
theoretical uncertainties because of the large negative value of the Pauli
interference contribution at dimension-six. This term actually dominates the 
total decay rate. Furthermore, the large negative value is enhanced by the
NLO-QCD corrections, but partly compensated by the dimension-seven contribution.
In this respect, having an independent determination of the HQET sum rule results, e.g.\ with a lattice QCD computation,
as well as higher order QCD corrections to dimension-six and seven might could significantly bring more insights.
For the determination of the lifetime ratios, in order to eliminate the contribution of the free-quark decay, we use\\
{\begin{equation}
\frac{\tau (D^+_{(s)})}{\tau (D^0)} = 1 + 
\left( \Gamma^{\rm HQE} (D^0) - \Gamma^{\rm HQE} (D^+_{(s)}) \right) \tau^{\rm exp} (D^+_{(s)})\,.
\label{eq:lifetime-ratio}    
\end{equation}\\
In Eq.~(\ref{eq:lifetime-ratio}), $\Gamma_3$ cancels exactly and $\Gamma_5$ and 
$\Gamma_6 $ cancel up to isospin or $SU(3)_f$ breaking corrections 
in the corresponding non-perturbative matrix elements. The lifetime ratios should then be dominated by the contribution of four-quark operators.
The results for the HQE prediction of the lifetime ratios, in several mass schemes, 
are shown in the fourth and fifth rows of
Table~\ref{tab:summary-diff-schemes-VIA},
Table~\ref{tab:summary-diff-schemes-HQET-SR},
Table~\ref{tab:summary-with-uncertainties}
as well as in
Figure~\ref{fig:summary-comparison}. We observe that the large lifetime ratio 
$\tau (D^+) / \tau (D^0)$ is well reproduced in all schemes considered,
while in the case of 
$\tau (D_s^+) / \tau (D^0)$ the HQE result lies closer to one,
compared to the experimental value. In the latter case, the theoretical prediction is dominated
by $SU(3)_f$-symmetry breaking effects in the non-perturbative parameters 
$\mu_\pi^2$, $\mu_G^2$ and $\rho_D^3$, which are only very roughly known. Future and more precise
determinations of their values can considerably improve our conclusion 
for these lifetime ratios.
\begin{figure}[t]
    \centering
    \includegraphics[scale=1.0]{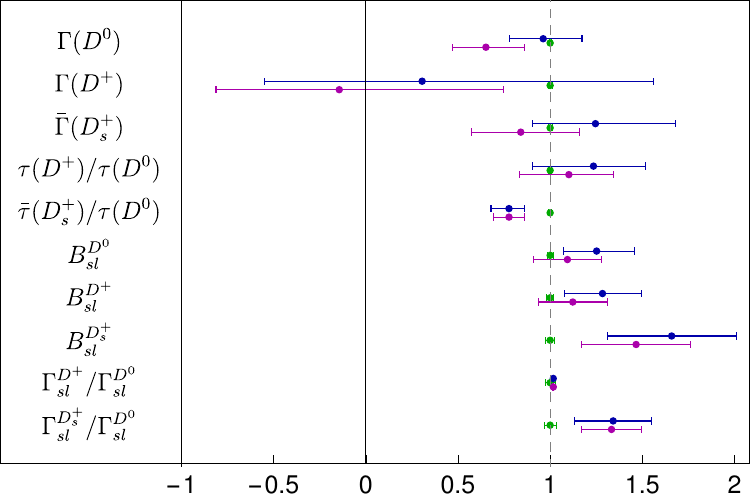}
    \caption{A comparison of the HQE prediction for the charm observables
    in the kinetic scheme (magenta) and in the $1S$ scheme (blue)  
    with the corresponding experimental data (green).
    All the quantities are normalised to the corresponding experimental central values.
    In comparison to Fig.~\ref{fig:summary-comparison}  now the NNLO corrections to the semileptonic branching fractions are included, taken from the talk of Matteo Fael at the CHARM-2020 conference.
}
    \label{fig:summary-comparison2}
\end{figure}
In the case of the inclusive semileptonic decays, we introduce the
shorthand notations $\Gamma_{sl}^D \equiv \Gamma (D \to X e^+ \nu_e)$
and $B_{sl}^D \equiv {\rm Br} (D \to X e^+ \nu_e)$.
The theoretical predictions are then obtained as
\begin{equation}
 B_{sl}^{D, \rm HQE}    =  
 \Gamma_{sl}^{D, \rm HQE}  
 \cdot \tau (D)^{\rm Exp.}  \, .
\end{equation}\\
The HQE results 
in several mass schemes are shown in the sixth, seventh and eighth row of
Table~\ref{tab:summary-diff-schemes-VIA},
Table~\ref{tab:summary-diff-schemes-HQET-SR}
and
Table~\ref{tab:summary-with-uncertainties}, as well as in 
Figure~\ref{fig:summary-comparison}.
In the kinetic scheme all HQE predictions for the semi-leptonic branching
fractions cover the experimental values, while the results in the 1$S$ scheme tend to be too large. It is interesting to note, that by adding NNLO-QCD corrections
to the semileptonic decays, the difference between the 
two quark mass schemes is considerably reduced, see Figure~\ref{fig:summary-comparison2}.
Using the experimental values respectively for the $D^0$ lifetime and the semileptonic branching fraction, we determine the semileptonic ratios in the following way\\
\begin{align}
\frac{\Gamma_{sl}^{D^+} }{\Gamma_{sl}^{D^0} }
 = 
1 + \left[\Gamma_{sl}^{D^+}  - \Gamma_{sl}^{D^0}\right]^{\rm HQE}
\left[\frac{\tau(D^0)}{B_{sl}^{D^0}}\right]^{\rm exp} 
 \, ,
\\[3mm]
\frac{\Gamma_{sl}^{D_s^+}}{\Gamma_{sl}^{D^0}}
 = 
1 + \left[\Gamma_{sl}^{D^+_s}  - \Gamma_{sl}^{D^0}\right]^{\rm HQE}
\left[\frac{\tau(D^0)}{B_{sl}^{D^0}}\right]^{\rm exp} 
\, .
\end{align}\\
The HQE results for these ratios are shown in the ninth and tenth row of
Table~\ref{tab:summary-diff-schemes-VIA},
Table~\ref{tab:summary-diff-schemes-HQET-SR}
and
Table~\ref{tab:summary-with-uncertainties}
and in
Figure~\ref{fig:summary-comparison}. In agreement with the experimental data, the 
HQE leads to values for $\Gamma_{sl}^{D^+}/\Gamma_{sl}^{D^0}$
very close to one. Also for $\Gamma_{sl}^{D_s^+}/\Gamma_{sl}^{D^0}$ the corresponding theoretical prediction is close to one,
however, the corresponding experimental number is as low as 0.79, confirming the necessity of having better control over the $SU(3)_f$-symmetry  breaking effects in the non-perturbative parameters $\mu_G^2$, $\mu_\pi^2$ and $\rho_D^3$ for the $D$ mesons.


\section{Charm mixing}
\label{sec:pheno2}

In Section~\ref{sec:pheno1} we have shown that the HQE is able to reproduce, within large theoretical uncertainties, the experimental pattern for the lifetime of charmed mesons. However, a naive application of the HQE yields results for the
decay rate difference of neutral $D$ mesons that are four orders of magnitude smaller than the experimental ones. It is well known that this huge suppression results from severe Glashow-Iliopoulos-Maiani (GIM) cancellations \cite{Glashow:1970gm}. Following Ref.~\cite{Lenz:2020efu}, we discuss a possible explanation for the large discrepancy between the theoretical prediction for $D$-mixing and experimental data. We stress though, that we do not present a detailed derivation of the fundamentals of the theory of mixing, for which, instead, we refer to the comprehensive reviews \cite{Proceedings:2001rdi, Nierste:2009wg, Silvestrini:2019sey}.


\subsection{GIM cancellations in $D$-mixing}
Because of the weak interaction, neutral mesons, here we consider the case of the $D^0$ meson, can mix with their corresponding antiparticles through the box diagrams shown in Figure~ \ref{fig:bix-diag}. The process is described by a $2 \times 2$ Hamiltonian matrix with non vanishing off-diagonal entries $M_{12}$ and $\Gamma_{12} $. By diagonalising the mixing matrix of the $D^0$ and the $\bar{D}^0$ mesons, we can obtain the two eigenstates with definite mass and decay width. The corresponding observables $\Delta M_D$ and $\Delta \Gamma_D$, denoting respectively the mass and decay width difference between the two eigenstates, are functions of $\Gamma_{12}$ and $M_{12}$,. Moreover we define \\
\begin{equation}
x = \frac{\Delta M_D}{ \Gamma_{D^0}}\,, \qquad y = \frac{\Delta \Gamma_D}{2 \Gamma_{D^0}}\,,
\label{eq:mix-obs}
\end{equation}\\
and\\
\begin{equation}
    x_{12} = \frac{2 \, |M_{12}|}{\Gamma_{D^0}} \,, \qquad 
    y_{12} = \frac{2 \, |\Gamma_{12}|}{\Gamma_{D^0}}\,, \qquad 
   \phi_{12} = \arg \left( \frac{M_{12}}{\Gamma_{12}} \right) \,.
   \label{def_phase}
\end{equation}\\
In the following we only discuss the computation of $\Gamma_{12}$ and, by taking into account the bound $\Delta \Gamma_D \leq 2 \, | \Gamma_{12}| $, see e.g.\ Refs.~\cite{Nierste:2009wg,Jubb:2016mvq}, we derive the theoretical prediction for $\Delta \Gamma_D$. In fact, we do not consider the calculation of $M_{12}$, and hence we can only determine one contribution to the mixing phase $\phi_{12}$ in Eq.~(\ref{def_phase}). 
\begin{figure}
\centering
\includegraphics[scale = 0.5]{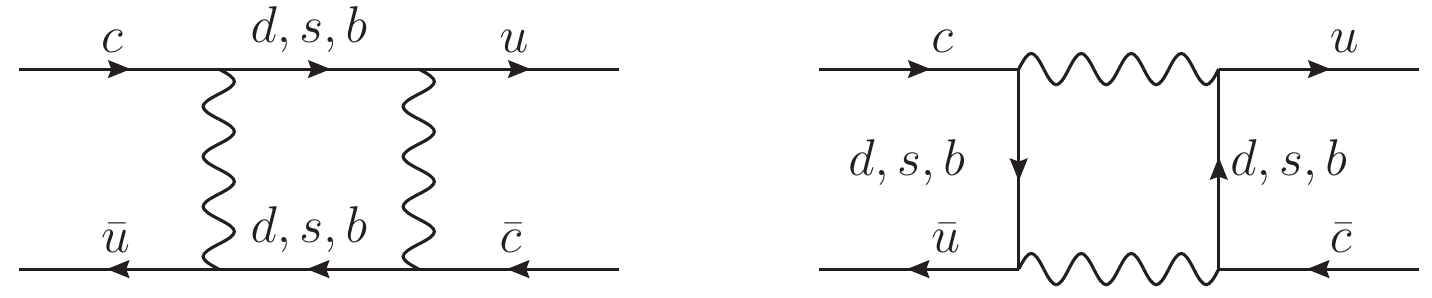}
\caption{Box diagrams contributing to mixing of neutral $D$-mesons.}
\label{fig:bix-diag}
\end{figure}
$\Gamma_{12}$ corresponds to the absorptive part of the mixing amplitude of the $D^0$ - $\bar D^0$ system, and it is then obtained by computing the imaginary part of the matrix element of the effective Hamiltonian describing the $c$-quark decay, between the $D^0$ and $\bar D^0$ states. Using the formalism described in Section~\ref{sec:HQE}, $\Gamma_{12}$ can be expanded in inverse powers of the heavy $c$-quark mass, leading to\\
\begin{equation}
  \Gamma_{12} = \left[ \Gamma_6^{(0)} + \frac{\alpha_s}{4 \pi}\,  \Gamma_6^{(1)}
    + \ldots \right] 
  \frac{\langle Q_6 \rangle}{m_c^3}
   + \ldots \, ,
\label{eq:HQE-mixing}
\end{equation}\\
where the ellipsis stand for terms of higher order and we have explicitly shown the perturbative expansion of the short distance coefficient $\Gamma_6$, cf.\ Eq.~(\ref{eq:pert-Gamma}). Eq.~(\ref{eq:HQE}),  is diagrammatically represented in Figure~\ref{fig:mixing}. The product of $\Delta C = 1$ operators in the effective Hamiltonian, i.e.\ in the ``full" theory, cf.\ Eq.~(\ref{eq:Heff-NL}), 
is now matched into a series of local $\Delta C = 2$ operators $Q_d$ of increasing dimension $d \geq 6$, with the short distance coefficients denoted by $\Gamma_{d}$. 
The expressions for $\Gamma_6^{(0)}$ can be easily derived in complete analogy to what it has been done in Section~\ref{sec:4q-contr}, while those for  $\Gamma_6^{(1)}$ can be obtained from the corresponding ones for $B$-mixing determined in Refs.~\cite{Beneke:1996gn,Beneke:1998sy,Dighe:2001gc,Beneke:2003az,Ciuchini:2003ww,Lenz:2006hd} by replacing $m_b \to m_c$, $m_c \to m_s$, etc. Furthermore,
the matrix elements of the dimension-six operators have been computed e.g.\ in Refs.~\cite{Kirk:2017juj,Bazavov:2017weg}.
The experimental value of the decay rate difference reads \footnote{Note that for consistency we present the numbers used in the analysis of Ref.~\cite{Lenz:2020efu} based on the previous determination $y = 0.68^{+0.06}_{-0.07} \,\% $, however, using the new value quoted in Eq.~(\ref{eq:x-y}), would not lead to any significant difference, and we would have instead $\Delta \Gamma_D^{\rm Exp} \geq 0.027$ ps$^{-1}$. }\\
\begin{equation}
    \Delta \Gamma_D^{\rm Exp} = 2 y / \tau (D^0) 
= (0.032 \pm 0.003) {\, \rm ps}^{-1},
\end{equation}\\
which, at one standard deviation, leads to the following bound \\
\begin{equation}
    \Delta \Gamma_D^{\rm Exp} \geq 0.028 \, {\rm ps}^{-1}\,.
    \label{eq:Omega}
\end{equation}\\
In order to compare the theoretical predictions with the experimental determinations, we investigate the quantities\\
\begin{equation}
 \alpha = - \arg (\Gamma_{12}) \,, \qquad \Omega = \frac{2 \, | \Gamma_{12}|^{\rm SM}}{  0.028 \, \mbox{ps}^{-1}} \,,
 \label{eq:alpha}
\end{equation}\\
where $\alpha$ contributes to CP violation in mixing and values of $\Omega$ smaller than one signal a discrepancy between the  theoretical and experimental description of $D$-mixing, within the one sigma range. A naive application of the HQE 
leads to $\Omega = 3.4 \cdot 10^{-5}$ at LO-QCD 
and to $6.2 \cdot 10^{-5}$ at NLO-QCD, showing that the theoretical prediction for the decay rate difference is more than four orders of 
magnitude smaller than the corresponding experimental number. 
Moreover, the phase $\alpha$ is very large, i.e.\ $\alpha =  93^\circ $ at LO-QCD
and $\alpha =  99^\circ $ at NLO-QCD.
By default in our numerical analysis we have used PDG \cite{Tanabashi:2018oca}
values for the quark, in the $\overline{\mbox{MS}}$ scheme, and meson masses, as well as for 
the strong coupling, CKM input from Ref.~\cite{Charles:2004jd},
the results of Ref.~\cite{Kirk:2017juj} for the non-perturbative matrix elements and Ref.~\cite{Aoki:2019cca} for the $D^0$ decay constant.
\begin{figure}
\centering
  \includegraphics[width=0.3\textwidth]{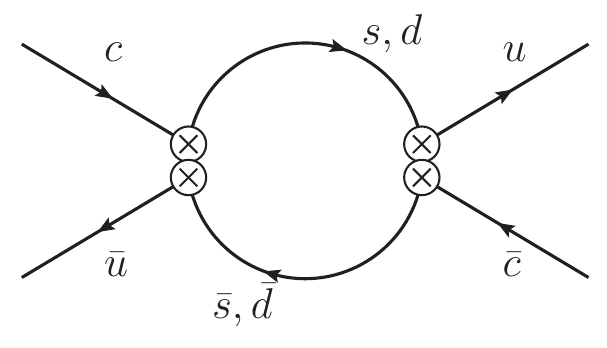}\qquad
  \includegraphics[width=0.3\textwidth]{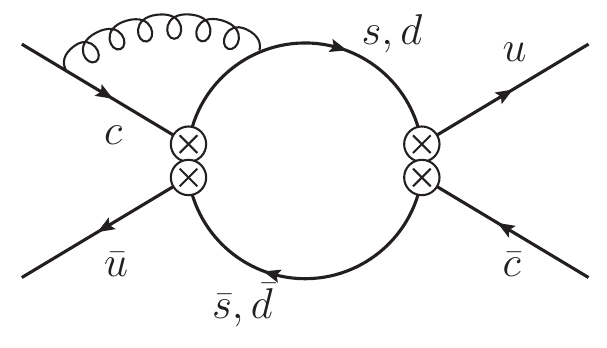}\qquad
  \includegraphics[scale = 0.4]{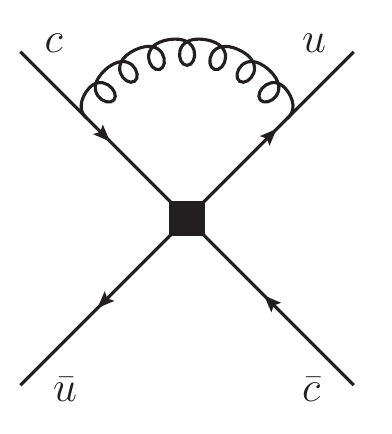}
  \caption{ Diagrams describing mixing of neutral $D$ mesons via intermediate $s \bar{s}$,
  $s \bar{d}$, $d \bar{s}$, and $d \bar{d}$, states
  in the``full" theory at LO-QCD (left) and NLO-QCD (center) and at NLO-QCD in the HQE (right).
  The crossed circles denote the insertion of $\Delta C=1$ operators of the effective Hamiltonian describing the charm-quark decay, while  
  the full dot indicates the insertion of $\Delta  C = 2$ operators in the HQE. 
 }
  \label{fig:mixing}
\end{figure}
In order to analyse the peculiarities of $D$-mixing, we decompose $\Gamma_{12}$
according to the flavour of the internal quark pair, cf.\ Figure~\ref{fig:mixing}. We denote the corresponding three contributions by $\Gamma_{12}^{ss}$, $\Gamma_{12}^{dd}$, and $\Gamma_{12}^{sd}$, i.e.\\
\begin{align}
  \Gamma_{12}  &=  - \Big( \lambda_s^2 \, \Gamma^{ss}_{12}   + 2 \,  \lambda_s \lambda_d \, \Gamma^{sd}_{12} 
  +  \lambda_d^2 \, \Gamma^{dd}_{12}  \Big)
  \nonumber \\[3mm]
 &=  - \,  \lambda_s^2  \Big( \Gamma^{ss}_{12} - 2   \Gamma^{sd}_{12} + \Gamma^{dd}_{12}  \Big)
  + \, 2 \lambda_s \lambda_b \Big( \Gamma^{sd}_{12} - \Gamma^{dd}_{12}                        \Big)
  -  \lambda_b^2 \Gamma^{dd}_{12}\,,
  \label{eq:GIM}
\end{align}\\
where the CKM factors are defined as $\lambda_q = V_{cq} V_{uq}^*$, and we have used the unitarity relation 
$\lambda_d+\lambda_s+\lambda_b=0$, to eliminate $\lambda_d$ in the second line of Eq.~(\ref{eq:GIM}). Taking into account the numerical value of the CKM elements, we see that Eq.~(\ref{eq:GIM}) shows the presence of a very pronounced 
hierarchy, namely
  \begin{align}
  -\lambda_s^2  =  -4.791 \cdot 10^{-2} + 3.094 \cdot 10^{-6} i,
  \\[3mm]
  + 2 \lambda_s \lambda_b  = 
  +2.751  \cdot 10^{-5}  +
  6.121  \cdot 10^{-5}  i,
  \label{eq:lambdasb}
  \\[3mm]
  -\lambda_b^2  =  +1.560 \cdot 10^{-8} - 1.757 \cdot 10^{-8} i.
\end{align}
The CKM factor in the first term of Eq.~(\ref{eq:GIM}) has considerably the largest real part, whereas the second term
has the largest imaginary part and it should then be important for the determination of the potential size
of CP violation in $D$-mixing.
The relative size between the imaginary and real part, is much larger in $\lambda_b$ than 
 in $\lambda_s$ and we therefore suggest to include all terms in Eq.~(\ref{eq:GIM}). 
Moreover, extreme GIM cancellations \cite{Glashow:1970gm} affect
the short distance coefficients of the CKM elements in Eq.~(\ref{eq:GIM}). 
By expanding in the small mass parameter 
$z =  m_s^2/  m_c^2$, we obtain at LO- (top line)
and at NLO-QCD (lower line), respectively \\
\begin{align}
  { \Gamma^{ss}_{12}}& =   
  \left\{ 
  \begin{array}{ll}
  { 1.62 - 2.34 \, z - 5.07 \, z^2 + \ldots } 
  \, ,
  \\[2mm]
  { 1.42 - 4.30 \, z - 12.45 \, z^2  + \ldots } 
  \, ,
  \end{array}
  \right.
  \label{noGIM}
  \\[3mm]
  { \Gamma^{sd}_{12}  - \Gamma^{dd}_{12}}  & =  \left\{ 
  \begin{array}{ll}
  { - 1.17 \, z - 2.53 \, z^2 + \ldots  } 
  \, , \\[2mm]
  { -2.15 \, z - 6.26 \, z^2 + \ldots }
  \, ,
  \end{array}
  \right.
  \label{littleGIM}
  \\[3mm]
  {\Gamma^{ss}_{12}  - 2  \Gamma^{sd}_{12} + 
  \Gamma^{dd}_{12}}  & =  
  \left\{ 
  \begin{array}{ll}
  { - 13.38 \, z^3 + \ldots } 
  \, , 
  \\[3mm]
  { 0.07 \, z^2 - 29.72 \, z^3 + \ldots }
  \, .
  \end{array}
  \right.
\label{crazyGIM}
\end{align}\\
Note that in the NLO result in Eq.~(\ref{crazyGIM}), the GIM suppression is lowered by one power of $z$, as it has been observed before \cite{Golowich:2005pt,Bobrowski:2010xg}.
We conclude that the peculiarity of Eq.~(\ref{eq:GIM}) lies in the fact that
the CKM dominant factor $\lambda_s^2$ multiplies the extremely GIM suppressed term given in
Eq.~(\ref{crazyGIM}), the CKM suppressed factor $\lambda_s \lambda_b$ multiplies the GIM suppressed term 
given in Eq.~(\ref{littleGIM}), while the very CKM suppressed factor $\lambda_b^2$ multiplies $\Gamma_{12}^{dd}$, obtained taking the limit $z \to 0$ in Eq.~(\ref{noGIM}), in which no GIM suppression is present. Therefore, the three
contributions in Eq.~(\ref{eq:GIM}) have actually a similar size, in fact\\
\begin{align}
\Gamma_{12} & =  \left( 2.08 \cdot10^{-7} - 1.34 \cdot 10^{-11} i \right)
\mbox{(1st term)}
\nonumber
\\[3mm]
&  - \left( 3.74 \cdot 10^{-7} + 8.31 \cdot 10^{-7} i \right)
\mbox{(2nd term)}
\nonumber
\\[3mm]
&+ \left( 2.22 \cdot 10^{-8} - 2.5 \cdot 10^{-8} i \right)
\mbox{(3rd term)}.
\end{align}\\
Because of Eq.~(\ref{eq:lambdasb}), it also follows that a sizeable contribution to the mixing phase can only arise if the slightly GIM suppressed term in Eq.~(\ref{littleGIM}) is enhanced. 
In order to explain the mismatch between the HQE prediction and experimental determination, in the literature different solutions have been proposed.
\begin{itemize}
\item[i)] At higher order in the HQE, the GIM suppression could be less pronounced
\cite{Georgi:1992as,Ohl:1992sr,Bigi:2000wn}. First estimates of the dimension-nine contribution to $D$-mixing, performed in Ref.~\cite{Bobrowski:2012jf}, show indeed such an enhancement, but not on a scale sufficient to reproduce 
the experimental result. For a final statement about this possibility, the complete determination of the 
dimension-nine and twelve contributions, would be necessary.
\item[ii)] The discrepancy is a signal of the violation of quark hadron duality. However, while it was originally suggested that  deviations of quark hadron duality should be as large as $10 ^{5}$, because of $\Omega \approx 10^{-5} $, this seems unlikely given
the many successful tests
of the HQE.
In fact in Ref.~\cite{Jubb:2016mvq} it was shown that violations as small as 20 per cent could be sufficient
to explain the experimental data.
\item[iii)] The HQE is not applicable and we have to consider different methods, like to sum over all the exclusive decays channels 
contributing to the decay rate difference, see e.g.\
Refs.~\cite{Falk:2001hx, Cheng:2010rv,Jiang:2017zwr}. 
\end{itemize}


\subsection{Alternative scale setting}

$\Gamma_{12}$ depends on the two scales $\mu_1$ and $\mu_2$. The former denotes the renormalisation scale of the $\Delta C = 1$ operators, and it is explicitly present in the expressions of the corresponding Wilson coefficients in the effective Hamiltonian, the latter is the renormalisation scale of the $\Delta C = 2$ operators that arise in the HQE and it appears also in their short distance coefficients. Up to higher order terms, the dependence on $\mu_1$ and $\mu_2$ must cancel between the local matrix elements and the corresponding short distance functions.
Without discussing the dependence on $\mu_2$, for which the cancellation is 
very effective, in the following we consider only that on $\mu_1$.
In the $B_s$ system the cancellation is
numerically only weakly realised when 
moving from LO- to NLO-QCD, see Refs.~\cite{Asatrian:2017qaz,Asatrian:2020zxa}, indicating the importance of higher order 
corrections. First steps in this direction show 
indeed large NNLO-QCD effects
\cite{Asatrian:2017qaz,Asatrian:2020zxa}. 
In the $D$ system a reduction of the $\mu_1$-dependence, when
moving from LO- to
NLO-QCD, is present in the individual
contributions $\Gamma^{ss,sd,dd}_{12}$, but not in 
$\Gamma_{12}$, see Figure~\ref{fig:Omega-lo-nlo}, which seems to be again consequence of the 
severe GIM cancellations.
\begin{figure}
\centering
    \includegraphics[scale = 1.]{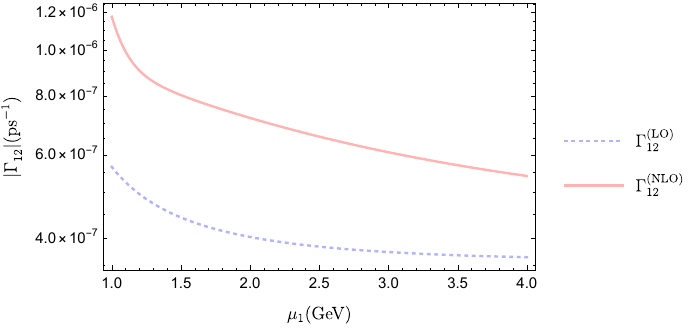}
    \caption{Comparison of $\mu_1$-dependence of $|\Gamma_{12}|$ at LO-QCD (dotted blue) and NLO-QCD (solid pink).}
    \label{fig:Omega-lo-nlo}
\end{figure}
By explicitly showing the scale dependence in $\Gamma_{12}$, we can write\\
\begin{equation}
\Gamma_{12} = 
\sum \limits_{q_1q_2 =ss,sd,dd} 
\Gamma_6^{q_1q_2} (\mu_1^{q_1q_2},\mu_2^{q_1q_2}) \langle Q_6 \rangle (\mu_2^{q_1q_2})
\frac{1}{m_c^3} + \ldots \,.
\end{equation}\\
In general different internal quark pairs contribute to
different decay channels of the $D^0\, (\bar D^0)$ meson
e.g.\  $s \bar{s}$ 
to a $K^+ K^-$  
final state and $s \bar{d}$ to a $\pi^+ K^-$ 
final state.
For each of these observables the choice of the renormalisation scales is a priori arbitrary, nevertheless typically one fixes
$\mu_x^{ss} = \mu_x^{sd} = \mu_x^{dd} = \mu$, 
which is then chosen to be equal to the mass of the decaying heavy 
quark, i.e.\ $\mu = m_Q$ for $Q$ quark decays,
in order to minimise the effect of the logarithmic terms $\alpha_s(\mu) \log (\mu^2/m_Q^2)$. 
Uncertainties due to unknown 
higher order corrections are estimated varying $\mu$
between $m_Q/2$ and $2\, m_Q$ and in the case of the 
charm quark 
we fix the lower bound to $1$ GeV 
to still ensure 
reliable perturbative results.

We consider two alternative possibilities to
treat the renormalisation 
scale $\mu_1^{q_1 q_2}$, which both allow to reduce
the discrepancy between the theoretical and the experimental determination of $D$-mixing,
while leaving the other HQE predictions unchanged, namely
\begin{itemize}
    \item[A)] We fix the central value of the three scales $\mu_1^{ss}$,  
$\mu_1^{sd}$, and $ \mu_1^{dd}$, to $m_c$, but we vary them independently 
between $1$ GeV and $2 m_c$. 
\item[B)] We choose different central values for the three scales $\mu_1^{ss}$,  
$\mu_1^{sd}$, and $ \mu_1^{dd}$, 
according to the
size of the available phase
space. 
In particular we evaluate $\Gamma^{ss}_{6}$  at the scale
$\mu_1^{ss} = \mu - 2 \epsilon$,
$\Gamma^{sd}_{6}$ at the scale
$\mu_1^{sd} = \mu -  \epsilon$, and $\Gamma^{dd}_{6}$ at 
the scale $\mu_1^{dd} = \mu$, where $\epsilon$ is an unknown parameter,
related to the kinematics of the decays.
\end{itemize}
If $\epsilon$ is not too large, both methods yield results for the individual
$\Gamma^{ss}_{6}$, $\Gamma^{sd}_{6}$ and $\Gamma^{dd}_{6}$, which lie within the usually 
quoted theory uncertainties obtained
following the prescription stated above,
but they affect in a sizeable way the severe GIM cancellations in Eqs.~(\ref{littleGIM}) and (\ref{crazyGIM}).
The first method gives
a considerably enhanced range 
of values for $\Omega$ in Eq.~(\ref{eq:Omega}), i.e.\\
\begin{equation}
\Omega \in [4.6 \cdot 10^{-5} , 1.3] \, ,
\label{sol1}
\end{equation}\\
which nicely covers also the experimental determination 
of the decay rate difference. 
Scanning independently over $\mu_1^{ss}$, 
$\mu_1^{sd}$, and $\mu_1^{dd}$, in 11 equidistant
steps we find that out of the 1331 points only 14 give a value of 
$\Omega < 0.001$, while 984 give a value of 
$\Omega > 0.1$.
The large discrepancy between the theoretical and the experimental
determinations, seems then to be
an artefact of fixing the scales $\mu_1^{ss}$,  $\mu_1^{sd}$, and 
$\mu_1^{dd}$, to be the same.
The range of values shown in Eq.~(\ref{sol1}) does not change significantly if we use the pole scheme for the quark masses, lattice results instead of the HQET sum rule results, or a different $\Delta C = 2$ operator basis. In all these cases we can obtain $\Omega \geq 1$.
For $\alpha$ in Eq.~(\ref{eq:alpha}), 
we observe that the results lie in the range $[-\pi, \pi]$.
A closer look however, shows
that for $\Omega  > 0.5$
only values of $\alpha  <  0.1^\circ $ are allowed, and conversely 
large values of $\alpha$ correspond to results for $\Omega$ inconsistent with the experimental data.
\\[2mm]
The second method for the scale setting
requires the introduction of
a mass scale $\epsilon$. 
A possible estimate for the size of this parameter
could be the
strange quark mass $\epsilon = m_s \approx 0.1$~GeV or 
the phase space difference of the
corresponding exclusive
decay channels, specifically, by  
comparing the energy release of $D^0 \to K^+ K^-$, $M_{D^0} - 2 M_{K^+} = 0.88$ GeV, with that of $D^0 \to \pi^+ \pi^-$, $M_{D^0} - 2 M_{\pi^+} = 1.59$ GeV we might
expect that $\epsilon \approx 0.35$~GeV.
\begin{figure}
    \centering
    \includegraphics[scale=0.7]{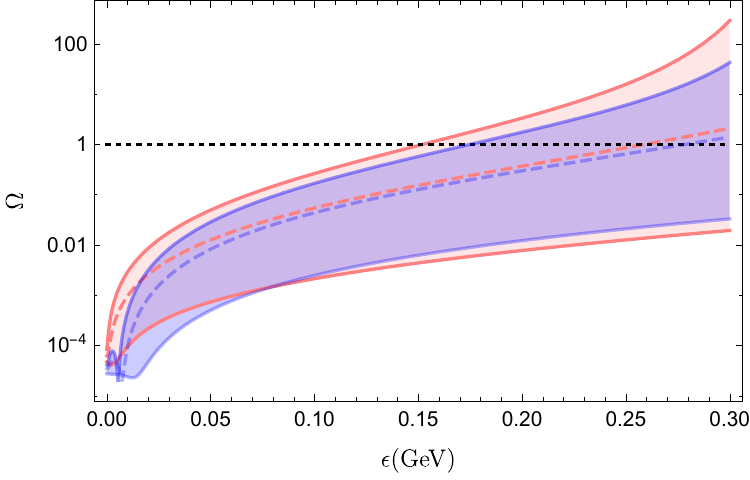}
    \caption{Comparison of the $\epsilon$ dependence of $\Omega$ at LO-QCD (blue) and  NLO-QCD (pink) for different values of $\mu$: the dashed line corresponds to $\mu =~m_c$ while the two solid lines to $\mu = 1$~GeV and $\mu = 2 m_c$.}
    \label{fig:Omega-epsilon-variation}
\end{figure}
In Figure~\ref{fig:Omega-epsilon-variation} it is shown how the 
HQE prediction of $\Omega$ would be affected in this
scenario. 
Also in this case, it appears possible that the theoretical prediction could reach the experimental
value for $\epsilon \approx 0.2$~GeV.   

Finally, we have to consider how other HQE predictions would be affected by choosing a different scale setting procedure.
In the case of observables in which GIM-like cancellations are not present, like the lifetime, both in the charm and bottom system, and the decay rate difference 
$\Delta \Gamma_s$, there is no significant change, but only a shift within the usually quoted 
theory uncertainties.
However, the semileptonic CP
asymmetries are governed 
by the weakly GIM suppressed contribution in $B_s$-mixing. The SM predictions read\\
\begin{align}
{\rm Re} \left( 
\frac{\Gamma_{12}^q}{M_{12}^q}\right)^{\rm SM} 
& =  
- \frac{\Delta \Gamma_q}{\Delta M_q} = 
\left\{
\begin{array}{ll}
- (49.9 \pm 6.7) \cdot 10^{-4} & q = s 
\\[3mm]
- (49.7 \pm 6.8) \cdot 10^{-4} & q = d 
\end{array}
\right.
\,  ,
\nonumber
\\[5mm]
{\rm Im} \left( \frac{\Gamma_{12}^q}{M_{12}^q} 
\right)^{\rm SM} 
& = 
a_{sl}^q = 
\left\{
\begin{array}{ll}
(+2.2 \pm 0.2) \cdot 10^{-5} & q = s
\\[3mm]
(-5.0 \pm 0.4) \cdot 10^{-4} & q = d
\end{array}
\right.
\,  ,
\end{align}
\\
while in the scenario B we obtain\\
\begin{displaymath}
\begin{array}{|c||c|c|}
\hline
\epsilon \,( {\rm GeV} ) & \Gamma_{12}^s / M_{12}^s & \Gamma_{12}^d / M_{12}^d
\\
\hline \hline
0.    & {\bf -0.00499 + 0.000022 i}  & {\bf -0.00497 - 0.00050 i}
\\ \hline
0.2.  & {\bf -0.00494 + 0.000023 i}& {\bf -0.00492 - 0.00053 i}
\\ \hline
0.5.  &  {\bf -0.00484 } + 0.000026 i&  {\bf -0.00482} - 0.00059 i
\\ \hline
1.0   &  {\bf -0.00447 } + 0.000037 i&  {\bf -0.00448} - 0.00084 i
\\ \hline
1.5.  & -0.00287 + 0.000091 i    &                -0.00309 - 0.0021 i
\\ \hline
\end{array}
\end{displaymath}\\
We see that for values of $\epsilon \leq 1$ GeV, the
predictions for the real part lie within the usually quoted
theory uncertainties, indicated in bold type. However, they would be increased by almost $100 \%$ in correspondence of larger values of $\epsilon$. 

Our conclusion is that, by modifying the usually
adopted scale setting,  the theoretical uncertainty
of $y$, within the HQE,  becomes larger than previously thought and it
can cover the experimental value. However, this does not represent a complete solution and
more precise estimates of higher
power corrections to the HQE, as well as full NNLO-QCD corrections to the leading dimension-six term, could bring further insights.
The alternative scale setting procedure shows 
that  a small contribution to CP violation in mixing stemming from 
the decay rate can be up to one per mille within in the 
SM, which agrees with estimates made in Refs.~\cite{Kagan:2020vri,Li:2020xrz}. For a prediction of 
CP violation in mixing, the contribution
coming from $M_{12}$ needs to be determined in addition. 
This might be done in future via the help of dispersion 
relations, see e.g.\  Refs.~\cite{Falk:2004wg, Cheng:2010rv,Li:2020xrz}.
We would like to note that our suggested procedure
still respects the GIM mechanism, because for
vanishing internal strange quark mass, also the parameter
$\epsilon$ is zero.
Finally this alternative scale setting does not affect quantities like $\tau (D^+)/ \tau(D^0)$, $b$-hadron lifetimes and  $\Delta \Gamma_s$ outside the range of their quoted theoretical errors, but it affects the semileptonic CP asymmetries and we get enhanced SM ranges \\
\begin{equation}
a_{sl}^d  \in  [-9.2;-4.6] \cdot 10^{-4}\,, \qquad 
a_{sl}^s  \in  [2.0; 4.0] \cdot 10^{-5} \, .
\end{equation}\\
Note that in Ref.\cite{Alonso-Alvarez:2021qfd}, the CP violating effects responsible for creating the baryon asymmetry stem actually from $a_{sl}^d$ and  $a_{sl}^s$.


\chapter*{Conclusion}
\addcontentsline{toc}{chapter}{Conclusion}

As discussed at the beginning this work, indirect BSM searches with quark flavour 
observables, represent a promising route to improve the current understanding of the fundamental laws of physics. However, this strategy strongly relies on the ability to systematically increase both the experimental and the theoretical precision. In the present thesis we have 
analysed the theoretical status for the study of the inclusive decay widths of heavy hadrons, which define observables of primary phenomenological importance in heavy flavour physics. In particular, we have discussed the computation of higher power corrections to the HQE, and tested its applicability in the charm sector, for the case of inclusive quantities like lifetimes, semileptonic branching fractions and mixing observables. Specifically,
the first part of this thesis has been dedicated to presenting the main ingredients required for the computation.
We have started by introducing the weak effective Hamiltonian and the heavy quark effective theory, which constitute the two effective theories that allow to disentangle a multi-scale problem like the weak decay of heavy hadrons, by progressively integrating out the heavier degrees of freedom, respectively, the $W$-boson and the massive component of the heavy quark field. 
We stress that our exposure, far from being exhaustive, has only covered the aspects relevant for the subsequent discussions and has mostly followed the excellent reviews available in the literature. We have then performed a detailed derivation of the expansion of the quark propagator in the external gluon field using the Fock-Schwinger gauge, which provides a gauge covariant parametrisation of the soft interaction with the non perturbative QCD field, for a quark propagating with large momentum inside the hadronic state. The corresponding expressions, up to terms proportional to one covariant derivative of the gluon field strength tensor, have been computed both in momentum and in coordinate space.  Finally, we have presented a pedagogical introduction to the heavy quark expansion, the theoretical framework in which all of the remaining computations and discussions are embedded. 
In the second part of this work, we have shown the explicit calculation of the lowest dimensional contributions to the HQE of a $B$ meson, namely those of two-quark operators up to order $1/m_b^2$ and of four-quark operators up to order $1/m_b^4$. For the former, the computation has been performed using the coordinate representation of the quark propagator and considering the single decay mode $b \to c \bar u d$. Moreover, we have reproduced the results given in the literature and in some cases derived more general expressions.  Also in this case we have tried to provide a very comprehensive and detailed presentation. The most important results, from a technical point of view, have been discussed in the third part of this work. Here, we have outlined the detailed computation of the contribution of two-quark operators up to order $1/m_b^3$, for arbitrary non-leptonic decays of the heavy $b$-quark and using the momentum representation of the quark propagator in the external gluon field. Particular emphasis has been put in describing the mixing between four-quark operators and the Darwin operator at dimension-six, which ensures the cancellations of the infrared divergences otherwise present in the coefficients of the Darwin operator, due to the emission of a soft gluon from a light quark propagator. The contribution of the Darwin operator for non-leptonic $b$-quark decays, has been only recently determined and found to be sizeable, hence its effect, previously neglected, might have important consequences for $b$-physics phenomenology. In the last part of this work, we have considered two phenomenological applications of the HQE in the charm sector, specifically the study of the inclusive decay width of $D$-mesons and of the Glashow-Iliopoulos-Maiani (GIM) cancellations in $D$-meson mixing. Due to the value of its mass, the charm quark sits at the boundary between the heavy- and light-quark region, and the applicability of the HQE in the charm sector is a priori questionable. In fact, both the perturbative and the power corrections might not describe a well converging series. For this reason, the charm system can be considered an important testing ground for the theoretical framework here discussed. 
We have then performed a comprehensive study of the structure of the HQE for the inclusive decay width of charmed mesons, including for the first time the contribution of the Darwin operator and in addition clarified some inconsistencies related to the contribution of dimension-seven four-quark operators. Our predictions appear to be consistent with the corresponding experimental pattern, albeit with large theoretical uncertainties. Given the current poor knowledge of many of the non perturbative input in the charm sector, as well as the absence of determinations of higher order perturbative and power corrections, we conclude that our numerical analysis does not show signals for a breakdown of the HQE in the charm system. Finally, we have also attempted to clarify the long standing puzzle due to the big discrepancy between the experimental determination
for neutral $D$-meson mixing and the corresponding HQE prediction, which might naively point at a complete failure of the HQE in the charm sector. In this respect, we
have proposed a novel procedure to treat the renormalisation scale for observables affected by GIM cancellations. Our results show that the experimental value can be accommodated within the HQE, albeit again with very large theoretical uncertainties.  

In conclusion, we have improved the current theoretical status of the HQE, by computing the contribution of the Darwin operator to non-leptonic decays of heavy quarks and by testing its applicability in the charm system. 
The same framework can in future be applied to the $B$ system, to improve the theoretical prediction for lifetimes and mixing observables. A precise determination of e.g.\  $\tau(B_s) / \tau (B_d)$, could in fact increase the bounds on the size of potential new physics contributions in the decay $b \to s \tau \tau$, which are predicted by some of the current BSM models that explain the $B$-anomalies, or could further constrain the baryogengesis model discussed in~Ref.~\cite{Alonso-Alvarez:2021qfd}.

\newpage
\chapter*{Acknowledgements}
\addcontentsline{toc}{chapter}{Acknowledgements}
First and outmost I thank Prof.\ Dr.\ Alexander Lenz for his precious guidance and the inestimable support provided during the last three years. I also thank him for being an inexhaustible source of energy and motivation and I consider myself very lucky to have had the possibility to learn from him. 
\\[3mm]
A special thanks goes to my collaborator and dear friend Dr.\ Aleksey Rusov, he has deeply contributed to my technical growth and I am grateful for the chance had to work together. Moreover, I sincerely thank him for the patience and the constant support offered in the last few months while writing this thesis.
\\[3mm]
I warmly thank Prof.\ Dr.\ Michael Spannowsky and Prof.\ Dr.\ Frank Krauss for giving me the opportunity to start my PhD program at the IPPP in Durham and Dr.\ Daniel Maitre for the supervision during the first year there.
\\[3mm]
A big thanks to everyone met in the past four years, first in Durham and now in Siegen, for each of them has contributed in some way to this journey.
\\[3mm]
Arrived at this point, I feel obliged to also thank all the teachers and professors I have had the fortune to find on my lifelong learning path, many of them have certainly inspired me and motivated me to continue studying.
\\[3mm]
Last but not least, I thank my family, my friends, and Peter, for understanding the many times I have been absent and for taking care of me despite the distance.

\begin{appendix}
\pagestyle{plain}

\chapter{Expansion of operators containing the heavy quark momentum}
\label{app:1}

Here we discuss a general procedure to generate operators of higher dimension, bilinear in the heavy quark field, starting from expressions containing lower dimensional operators with coefficients proportional to the heavy quark momentum. For definiteness we consider the case of the $b$-quark in order to make the connection to Chapter~\ref{ch:HQE-ex} and Chapter~{\ref{ch:Darwin}} easier.
We start by recalling that inside a heavy hadron, the momentum of the heavy quark can be conveniently parametrised as $p_b^\mu = m_b v^\mu + k^\mu$. Correspondingly the rescaled heavy quark field $b_v(x)$ is defined by\\
\begin{equation}
b(x) = e^{-i m_b v \cdot x}\,  b_v(x)\,,
\label{eq:bv-def}
\end{equation}\\
where the phase factor removes the large `kinetic' part of the heavy quark momentum so that a derivative acting on $b_v(x)$ returns only the residual component $k$. Consider now the following expression $p_b^\mu\, \bar b b$, with the $b$-fields evaluated at the origin $x= 0$. By taking into account Eq.~(\ref{eq:bv-def}), we can write, see Refs.~\cite{Blok:1992hw, Blok:1992he}\\
\begin{align}
p_b^\mu \bar b b= \lim_{x \to 0} \bar b(x) i \partial^\mu b(x) = \lim_{x \to 0} \bar b_v(x) (m_b v^\mu + i \partial^\mu) b_v(x)\,,
\label{eq:App-2}
\end{align}\\
here the partial derivative is acting on the right. Using that in the FS gauge, see Section~\ref{sec:FS}, the gauge field satisfies the useful property $A^\mu(0) = 0$, cf.\ Eq.~(\ref{eq:A_mu}), it follows that on the r.h.s.\ of Eq.~(\ref{eq:App-2}), the partial derivative acting on $b_v(x)$ can be replaced by the corresponding covariant derivative in the limit $x \to 0$, namely\\
\begin{equation}
p_b^\mu \bar b b = \lim_{x \to 0} \bar b_v (x) \big(m_b v^\mu + i \partial^\mu + A^\mu(x) \big) b_v(x) = \bar b_v (m_b v^\mu + i D^\mu) b_v\,,
\label{eq:App-3}
\end{equation}\\
which generates the dimension-four operator $\bar b_v i D^\mu b_v$. 
If the original expression contains more than one power of the heavy quark momentum, to keep track of the order of the covariant derivatives we can symmetrise the action of the partial derivatives, e.g.\ the case of $p_b^\mu p_b^\nu \, \bar b b$ gives\\
\begin{align}
&p_b^\mu p_b^\nu \, \bar b b = \frac12 \lim_{x\to 0} \bar b(x) \big ( i \partial^\mu i \partial^\nu + i \partial^\nu i \partial^\mu \big) b(x) 
\nonumber \\[3mm]
 = \frac12 \lim_{x \to 0} \bar b_v(x)\Big( (m_b v^\mu &+ i \partial^\mu) (m_b v^\nu + i \partial^\nu) +  (m_b v^\nu + i \partial^\nu) (m_b v^\mu + i \partial^\mu) \Big) b_v(x)
\,.
\label{eq:App-4}
\end{align}\\
Here again we would like to replace the action of the partial derivative with that of the covariant derivative. Note that in this case, apart from terms containing the gauge field on the most left, which vanish when taking the limit $x \to 0$, we introduce also terms with derivatives of the gauge field evaluated at the origin, that are in general non zero. From Eq.~(\ref{eq:A_mu}), it follows that $\partial^\mu A^\nu (0) =(1/2) G^{\mu \nu} (0)$, however, due to the antisymmetry of the gluon field strength tensor, these contributions  cancel in the symmetric combination in Eq.~(\ref{eq:App-4}), namely\\
\begin{align}
& \hspace{4.5cm}  \frac12 \, \bar  b_v \Big( i D^\mu  i D^\nu +  i D^\nu  i D^\mu  \Big) b_v
\nonumber \\[3mm]
& = \frac12 \lim_{x \to 0} \bar b_v(x)\Big( \big(  i\partial^\mu i\partial^\nu + A^\mu (x) i \partial^\nu + i \partial^\mu A^\nu(x)+ A^\nu (x) i \partial^\mu  + A^\mu(x) A^\nu (x) \big)
\nonumber \\[3mm]
& \quad \quad +
 \big( i\partial^\nu i\partial^\mu + A^\nu (x) i \partial^\mu + i \partial^\nu A^\mu(x) + A^\mu (x) i \partial^\nu + A^\nu(x) A^\mu (x) \big) \Big) b_v(x)
\nonumber \\[3mm]
&\hspace{2.2cm} = \frac12 \, \bar b_v  \big ( i \partial^\mu i \partial^\nu + i \partial^\nu i \partial^\mu  + \frac{i}{2} G^{\mu \nu} (0) + \frac{i}{2} G^{\nu \mu} (0) \big) b_v  
\nonumber \\[3mm]
& \hspace{4.3cm} =  \frac12  \bar b_v \big ( i \partial^\mu i \partial^\nu + i \partial^\nu i \partial^\mu \big) b_v
 \,.
\end{align}\\
We then obtain that\\
\begin{equation}
p_b^\mu p_b^\nu \, \bar b b = \frac12 b_v \Big\{ (m_b v^\mu + i D^\mu), (m_b v^\nu +  i D^\nu) \Big\}  b_v\,,
\end{equation}\\
where the curly brackets denote the anticommutator. Notice that we generate operators of dimension-four and dimension-five with respectively one and two covariant derivatives. In the case of three powers of the heavy quark momentum, $p_b^\mu p_b^\nu p_b^\rho \, \bar b b$, in rewriting this expression in terms of partial derivatives we consider all the permutations in the three Lorentz indices $\mu \nu \rho$. The replacement $i \partial^\mu \to i D^\mu$ again follows from the symmetric combination of antisymmetric tensors, since Eq.~(\ref{eq:A_mu}) now gives $\partial^\mu \partial^\nu A^\rho(0) = (1/2) \partial^\mu G^{\nu \rho}(0) + (1/3 ) D^\mu G^{\nu \rho}(0)$. For $n$-powers of the heavy quark momentum, we use the general expression\\
\begin{align}
 p_b^{\mu_1} \ldots p_b^{\mu_n} \,  \bar b b
=  \frac{1}{n!} \sum_{\sigma \in S_n}  
  \bar b_v  \,  \big( m_b v +  i D \big)^{\sigma(\mu_1)} \, \ldots\,  \big( m_b v + i D \big)^{\sigma(\mu_n)}\, b_v \,,
 \label{eq:expansion-momentum}
\end{align}\\
where $S_n$ is the group of all permutations of $n$ elements. Furthermore consider the following function \\
\begin{equation}
f(r) = a + b (r)\, r + c (r)\, r^2 + d(r) \, r^3 + e(r) \, r^4 \,,
\label{eq:fp-not-expanded}
\end{equation}\\
with the dependence on the heavy quark momentum contained in the argument $r = m^2/p_b^2$. In this case, we can use that $p_b^2 = m_b^2 \, (1 + X)$, where \\
\begin{equation}
 X = 2  \, \frac{v \cdot k}{m_b} + \frac{k^2}{m_b^2}  \ll 1\,,
\end{equation}\\
and expand Eq.~(\ref{eq:fp-not-expanded}) in series, i.e.\\
\begin{equation}
f(r) =   \sum \limits_{n= 0}^{\infty}  \, g_n(\rho)\, (- X) ^n \,,
\label{eq:fr-series}
\end{equation}\\
here $g_0(\rho) = f(\rho)$ and the dimensionless mass parameter $\rho =  m^2/m_b^2$.
The series in Eq.~(\ref{eq:fr-series}) can be truncated at a certain value of $n$, leading to corrections up to order $1/m_b^{2n}$. Expressing $X$ in Eq.~(\ref{eq:fr-series}) back in terms of the four-momentum, we then obtain\\
\begin{equation}
\nonumber \\
f(r) = \sum _{n= 0}^{\infty} \, g_n (\rho) \, \Bigg(1- \frac{p_b^2}{m_b^2} \Bigg)^n \,,
\end{equation}\\
from which it follows that $ f(r) \, \bar b b $ can be expanded according to Eq.~(\ref{eq:expansion-momentum}). Finally, for the computation of the coefficient of the Darwin operator discussed in Chapter~\ref{ch:Darwin}, we need to expand also the structure $p_b^\rho\,  \bar b \, G^{\mu \nu} b$. In this case we write\\
\begin{align}
p_b^\rho \, \bar b \, G^{\mu \nu} b &= \frac12 \, \lim_{x \to 0} \Big( \bar b(x) (- i \overset{\leftarrow}{\partial^\rho}) G^{\mu \nu}(x) b(x) + \bar b(x) G^{\mu \nu}(x) (i \overset{\rightarrow}{\partial^\rho}) b(x) \Big) 
\nonumber\\[3mm]
&=  \frac12 \, \lim_{x \to 0} \Big( \bar b_v(x) (m_b v^\rho - i \overset{\leftarrow}{\partial^\rho}) G^{\mu \nu}(x) b_v(x) + \bar b_v(x) G^{\mu \nu}(x) ( m_b v^\rho + i \overset{\rightarrow}{\partial^\rho}) b_v(x) \Big) 
\nonumber \\[3mm]
&=  \frac12 \, \lim_{x \to 0} \Big( \bar b_v(x) (m_b v^\rho +  i \overset{\rightarrow}{\partial^\rho}) G^{\mu \nu}(x) b_v(x) + \bar b_v(x) G^{\mu \nu}(x) (m_b v^\rho + i \overset{\rightarrow}{\partial^\rho}) b_v(x) \Big) 
\nonumber \\[3mm]
& =  \frac12 \,  \Big( \bar b_v (m_b v^\rho +  i D^\rho ) G^{\mu \nu} b_v + \bar b_v G^{\mu \nu} (m_b v^\rho + i D^\rho) b_v \Big)   \,, 
\end{align}\\
where in the third step we have applied the chain rule for the derivative operator acting on the left and dropped the corresponding   term with a total derivative since it does not contribute in forward matrix elements with zero momentum transfer.

In general, by expressing powers of the heavy quark momentum in terms of operators acting on the heavy quark field, we generate a set of higher dimensional operators with multiple covariant derivatives. Specifically, in order to compute the contributions discussed in Chapter~\ref{ch:HQE-ex} and Chapter~\ref{ch:Darwin}, it is sufficient to consider the expansions up to three covariant derivatives. The operators obtained in this way, must be then evaluated between external $B$ meson states for the calculation of the total decay width. This can be easily done by taking into account the results presented in Ref.~\cite{Dassinger:2006md}, in which the complete parametrisation of these matrix elements, up to order $1/m_b^4$ was derived.

\chapter{Results for one-loop integrals in dimensional regularisation}
\label{app:2}
Here we list some useful results for the computation of one-loop integrals in dimensional regularisation \cite{tHooft:1972tcz, Bollini:1972ui, Cicuta:1972jf, Ashmore:1972uj } with $D = 4 - 2 \epsilon$. All the expressions presented can be found in standard QFT textbooks like Refs.~\cite{Peskin:1995ev, Itzykson:1980rh, Schwartz:2014sze}. Note that in the following we use the notation \\
\begin{equation}
 \mu^{2 \varepsilon }\int \frac{d^D l}{(2 \pi)^D} \equiv \int_l\,.
 \label{eq:in-l}
\end{equation}\\
The scalar tadpole and bubble integrals are respectively given by\\
\begin{equation}
  \int_l  \frac{1 }{ (l^2 - m^2 + i \varepsilon ) }   = {\cal A}_0 (m^2)\,,
 \label{eq:int-tadpole}
 \end{equation}\\
 and
 \begin{equation}
\int_l  \frac{1 }{ (l^2 - m_1^2 + i \varepsilon) ((l-p)^2 - m_2^2 + i \varepsilon)}   = {\cal B}_0 (p^2, m_1^2, m_2^2)\,,
\label{eq:int-bubble}
\end{equation}\\
with
\begin{align}
{\cal A}_0(m^2) =  \frac{i \, m^2}{16 \pi^2} \, \Bigg( \frac{1}{\epsilon} -\gamma_E + \log(4\pi) + 1 - \log \left( \frac{m^2}{\mu^2} \right) \Bigg) + {\cal O} (\epsilon) \,, 
\label{eq:A0}
\\[3mm]
{\cal B}_0 (p^2, m_1^2, m_2^2) = \frac{i}{16 \pi^2} \Bigg(  \frac{1}{\epsilon} -\gamma_E + \log(4\pi)  + f(p^2, m_1^2, m_2^2) \Bigg)  + {\cal O} (\epsilon)  \,, 
\label{eq:B0}
\end{align} \\
where $\gamma_E$ is the Euler-Mascheroni constant and \\
\begin{equation}
f(p^2, m_1^2, m_2^2) = - \int \limits_0^1 dx \, \log \left( \frac{(1-x) \, m_1^2 + x\,  m_2^2 - x\, (1-x )\, p^2 - i \varepsilon }{\mu^2} \right)\,,
\label{eq:fp}
\end{equation}\\
Note that as $\varepsilon \to 0$, the function $f(p^2, m_1^2, m_2^2)$ is real and analytic in $s \in {\mathbb C}$, with ${\rm Re}(s) = p^2 $, only for $p^2 <  \big( m_1 + m_2 \big)^2 \vee 0 \leq x \leq 1$, while it exhibits a branch point when the argument of the logarithm in the integrand on the r.h.s\ of Eq.~(\ref{eq:fp}) vanishes. This leads to a cut in the complex $s$-plane, in correspondence of the points $p^2 > \big( m_1 + m_2 \big)^2 $ in which the argument of logarithm becomes negative. By introducing $\varepsilon$, the computation of the integral, performed above and below the branch cut, gives see e.g.\ the lecture notes~\cite{Kniehl:1996rh, Soldati:QFT2} \\
\begin{equation}
{\rm Disc} \,  f(p^2, m_1^2, m_2^2) =  2i \pi \,   \frac{\sqrt{\lambda(p^2, m_1^2, m_2^2)}}{p^2} \,\, \theta \left( p^2 - \big( m_1 + m_2 \big)^2  \right) \,,
\label{eq:disc-fp}
\end{equation}\\
and $\theta (x)$ is the Heaviside function. From Eqs.~(\ref{eq:A0})-(\ref{eq:disc-fp}) it follows that\\
\begin{equation}
{\rm Im} \,  {i \cal A}_0(m^2)  = 0 \,,
\label{eq:iA0}
\end{equation}
while\\
\begin{align}
{\rm Im} \,  {i \cal B}_0(p^2, m_1^2, m_2^2)  = - \frac{1}{16 \pi^2} \, {\rm Im} f(p^2, m_1^2, m_2^2) 
 = - \frac{1}{16 \pi} \,   \frac{\sqrt{\lambda(p^2, m_1^2, m_2^2)}}{p^2} \,,
 \label{eq:iB0}
\end{align}\\
for  $p^2 \geq \big( m_1 + m_2 \big)^2$. The corresponding one- and two-point one-loop tensor integrals of rank $r$ can be computed, using 
the Passarino-Veltman reduction algorithm \cite{Passarino:1978jh}, as a linear combination of tensors of the same rank, built from the metric $g^{\mu \nu}$ and the external momentum $p^\mu$, with coefficients proportional to the scalar integrals in Eqs.~(\ref{eq:int-tadpole}), (\ref{eq:int-bubble}). Let us consider explicitly the cases $r =1, 2$.
Rank-1 integrals can only depend on the four-vector $p^\mu$, namely\\
\begin{equation}
\int_l \, \frac{l^\rho }{ (l^2 - m_1^2 + i \varepsilon) ((l-p)^2 - m_2^2 + i \varepsilon)}  = B_{11} \, p^\rho\,,
\label{eq:r1-bubble}
\end{equation}\\
where the coefficient $B_{11}$ is obtained contracting both sides of Eq.~(\ref{eq:r1-bubble}) with $p_\rho$ i.e.\\
\begin{equation}
B_{11} = \frac{1}{p^2} \, \int_l \,  \frac{l \cdot p }{ (l^2 - m_1^2 + i \varepsilon) ((l-p)^2 - m_2^2 + i \varepsilon)}\,.
\label{eq:B11-bubbe}
\end{equation}\\
By substituting the identity\\
\begin{equation}
l \cdot p = \frac12 \Big( l^2 + p^2 - (l-p)^2 + m_1^2 - m_1^2 + m_2^2 - m_2^2 \Big) \,,
\label{eq:ldotp}
\end{equation}\\
into Eq.~(\ref{eq:B11-bubbe}), evidently yields\\ 
\begin{align}
B_{11} & = \frac{1}{2 p^2} \int_l \frac{ (l^2 - m_1^2) - ((l-p)^2 - m_2^2) + (p^2 + m_1^2 - m_2^2)}{(l^2 - m_1^2 + i \varepsilon) ((l-p)^2 - m_2^2 + i \varepsilon)}
\nonumber \\[3mm]
& = \frac{1}{2 p^2} \, \Bigg( {\cal A}_0(m_2^2) - {\cal A}_0(m_1^2) + \Big( p^2 +m_1^2 - m_2^2 \Big)\, {\cal B}_0 (p^2, m_1^2, m_2^2) \Bigg)\,,
\label{eq:B11}
\end{align}\\
where we have taken into account Eqs.~(\ref{eq:int-tadpole}), (\ref{eq:int-bubble}). In the case of rank-2 integrals, we can build two independent tensors, $g^{\rho \sigma}$ and $p^\rho p^\sigma$, namely \\
\begin{equation}
\int_l \, \frac{l^\rho l^\sigma }{ (l^2 - m_1^2 + i \varepsilon) ((l-p)^2 - m_2^2 + i \varepsilon)}  = B_{21} \, p^\rho p^\sigma + B_{22} \, p^2 g^{\rho \sigma}\,,
\label{eq:r2-bubble}
\end{equation}\\
here, the coefficients $B_{21}$, $B_{22}$, are the solutions of the system of two equations obtained contracting both sides of Eq.~(\ref{eq:r2-bubble}) respectively with $p_\rho $ and $g_{\rho \sigma}$, i.e.\\
\begin{align}
\int_l \, \frac{l^2 }{ (l^2 - m_1^2 + i \varepsilon) ((l-p)^2 - m_2^2 + i \varepsilon)} &  = B_{21} \, p^2  + \, p^2 \, B_{22}  \, D\,,
\label{eq:linear-equations-1}
\\[4mm]
\int_l \, \frac{ (l \cdot p ) \, l^\sigma  }{ (l^2 - m_1^2 + i \varepsilon) ((l-p)^2 - m_2^2 + i \varepsilon)} & = B_{21} \, p^2 \, p^\sigma + \, p^2 \, B_{22} \, p^\sigma \, .
\label{eq:linear-equations-2}
\end{align}\\
Adding and subtracting $m_1^2$ in the numerator of Eq.~(\ref{eq:linear-equations-1}) and substituting Eq.~(\ref{eq:ldotp}) into Eq.~(\ref{eq:linear-equations-2}), we obtain \footnote{Note that integrals of odd functions of $l^\mu$  vanish due to parity.}\\
\begin{align}
&\Big( B_{21}  + D \, B_{22} \Big)\,  p^2 =   {\cal A}_0(m_2^2) + m_1^2 \,  {\cal B}_0(p^2, m_1^2, m_2^2) \,,
\\[4mm]
&\Big( B_{21}  +  B_{22} \Big) \, p^2  =  \displaystyle{\frac12} \Big( {\cal A}_0(m_2^2) + (p^2 + m_1^2 - m_2^2) \, B_{11} \Big) \,,
\end{align}\\
solved by\\
\begin{align}
B_{21} = & \displaystyle{\frac{1}{D -1}\, \frac{1}{2 p^2}}  \Bigg( \, {\cal A}_0(m_2^2) (D-2) - 2 m_1^2\, {\cal B}_0(p^2, m_1^2, m_2^2)
\nonumber \\[2mm]
& + \displaystyle{ \frac{D\, (m_1^2 - m_2^2 +p^2) \Big({\cal A}_0(m_2^2) - {\cal A}_0(m_1^2) + ( p^2 +m_1^2 - m_2^2 ) {\cal B}_0(p^2, m_1^2, m_2^2) \Big)}{2p^2} } \Bigg)\,,
\label{eq:B21}
\end{align}
and\\
\begin{align}
B_{22} = & \displaystyle{\frac{1}{D -1}\, \frac{1}{2 p^2}} \Bigg( \, {\cal A}_0(m_2^2) + 2 m_1^2\, {\cal B}_0(p^2, m_1^2, m_2^2)
\nonumber \\[2mm]
&- \displaystyle{ \frac{ (m_1^2 - m_2^2 +p^2) \Big( {\cal A}_0(m_2^2) - {\cal A}_0(m_1^2) + ( p^2 +m_1^2 - m_2^2) {\cal B}_0(p^2, m_1^2, m_2^2) \Big)}{2p^2} } \Bigg)\,,
\label{eq:B22}
\end{align}\\
where Eq.~(\ref{eq:B11}) has been used. The expressions in Eq.~(\ref{eq:int-tadpole})-(\ref{eq:B22}) are needed for the computation of the four-quark operators contribution discussed in Chapter~\ref{ch:HQE-ex}. 

Finally we derive some useful results used in the calculation of the one-loop diagram in Figure~\ref{fig:Mixing}, describing the mixing of four-quark operators with two-quark operators at dimension-six, discussed in Chapter~\ref{ch:Darwin}. Consider the following scalar integrals\\
\begin{equation}
\int_l  \frac{1}{(l^2 - m^2 + i \varepsilon)^2}  = {\cal I}_0(m^2)
 \,,
 \label{eq:tadpole-order2}
\end{equation}\\
 and\\
 \begin{equation}
\int_l \frac{l^2}{(l^2 - m^2 + i \varepsilon)^3} = \frac{D}{4} \, {\cal I}_0(m^2) \,,
\label{eq:lsq-tadpole}
\end{equation}\\
with\\
\begin{equation}
 {\cal I}_0(m^2) = \frac{i}{16 \pi^2} \Bigg( \frac{1}{\epsilon} - \gamma_E + \log(4 \pi) + \log \left( \frac{\mu^2}{m^2} \right) \Bigg)  + {\cal O}(\epsilon)\,.
 \label{eq:I0-tadpole}
 \end{equation}\\
Rank-2 integrals, since now there is no external momentum they can depend on, can be only parametrised by the metric tensor i.e. \\
 \begin{equation}
\int_l  \frac{l^\mu l^\nu}{(l^2 - m^2 + i \varepsilon)^{3}} = g(m^2)\, g^{\mu \nu} \,,
\label{eq:r2-tensor-integral-tadpole}
\end{equation}\\
where the coefficient $g(m^2)$ is obtained by contracting both sides of Eq.~(\ref{eq:r2-tensor-integral-tadpole}) with $g_{\mu \nu}$. This gives\\
\begin{equation}
\int_l  \frac{l^2}{(l^2 - m^2 + i \varepsilon)^{3}} = D \, g(m^2) \,,
\end{equation}\\
and from Eq.~(\ref{eq:lsq-tadpole}) it then follows that\\
\begin{equation}
g(m^2) = \frac{1}{4} \, {\cal I}_0(m^2)\,.
\label{eq:g-msq}
\end{equation}


\chapter{Decomposition of tensor integrals}
\label{app:3}
In the following section we describe how to reduce rank-$r$ integrals to a linear combination of tensors of rank-$r$ with coefficients given by scalar integrals. We consider explicitly integrals of the sunset type encountered in Chapter~\ref{ch:Darwin} although the same procedure can be generalised to any tensor integral without loss of generality. We work again in dimensional regularisation with $D = 4 - 2 \varepsilon$ and to simplify the notation we define\\
\begin{equation}
 \int_{l_1 l_2} \equiv  \int_{l_1}\, \int_{l_2} \,  \frac{1}{(l_1^2 - m_1^2 + i \varepsilon )\,  (l_2^2 - m_2^2+ i \varepsilon)((l_1 +l_2-p)^2 - m_3^2+ i \varepsilon)} \,,
\end{equation}\\
with $\int_l$ defined as in Eq.~(\ref{eq:in-l}).
Rank-$1$ integrals can only be parametrised in terms of the external momentum $p^\mu$, hence\\
\begin{equation}
\int_{l_1  l_2 } l_j^\mu= a_j(p^2, m_i^2)\,  p^\mu \,,
\label{eq:r1-int}
\end{equation} \\
where $j = 1,2$ and the coefficient $a_j(p^2, m_i^2)$ is obtained by contracting both sides of Eq.~(\ref{eq:r1-int}) with $p_\mu$ i.e.\\
\begin{equation}
a_j(p^2, m_i^2) = \frac{1}{p^2} \int_{l_1 l_2} (l_j \cdot p) \, .
\label{eq:r1-int-coef}
\end{equation}\\
In the case of rank-$2$ integrals, we can build two independent tensors, namely $p^2 g^{\mu \nu}$ and $p^\mu p^\nu$, hence\\
\begin{equation}
\int_{l_1 l_2} l_j^\mu l_k^\nu= b_{ jk}(p^2, m_i^2) \, p^2 g^{\mu \nu}  + c_{jk}(p^2,m_i^2) \, p^\mu p^\nu\, , 
\label{eq:r2-int}
\end{equation}\\
where $j, k = 1,2$ and the coefficients $b_{ jk}(p^2, m_i^2)$, $c_{jk}(p^2,m_i^2)$ are the solutions of the system of two equations obtained by contracting both sides of Eq.~(\ref{eq:r2-int}) respectively with $g_{\mu \nu}$ and $p_\mu p_\nu$ i.e.\\
\begin{align}
b_{ jk} (p^2,m_i^2) &=  \frac{1}{D -1 }\, \frac{1}{p^4}\, \int_{l_1 l_2}   \Big( (l_j \cdot l_k) p^2 - (l_j \cdot p) ( l_k \cdot p) \Big) \,,
\label{eq:r2-int-coef-1}
\\[4mm]
c_{ jk} (p^2,m_i^2) &=   \frac{1}{D -1}\, \frac{1}{p^4}\, \int_{l_1 l_2}  \Big(  D\,  (l_j \cdot p) (l_k \cdot p)  - ( l_j \cdot l_k) p^2  \Big) \, .
\label{eq:r2-int-coef-2}
\end{align}\\
In the case of rank-$3$ integrals, there are four independent tensors $p^2 g^{\{ \mu \nu} p^{\rho \}}$ with the curly brackets denoting all possible permutation of the Lorentz indices, hence\\
\begin{align}
\int_{l_1 l_2}  l_j^\mu l_k^\nu l_m^\rho & =   d_{1, jkm}(p^2, m_i^2) \, p^2  \, g^{\mu \nu}  p^\rho +   d_{2, jkm}(p^2, m_i^2) \, p^2  \, g^{\rho \mu}  p^\nu 
\nonumber \\[3mm]
& +    d_{3,jkm}(p^2, m_i^2)\,  p^2  \, g^{\nu \rho}  p^\mu  + e_{jkm}(p^2, m_i^2)\, p^\mu p^\nu p^\rho \, ,
\nonumber \\
\label{eq:r3-int}
\end{align}
where $j,k,m = 1,2$ and the coefficients $d_{r,jkm}(p^2, m_i^2)$, $e_{jkm}(p^2, m_i^2)$ are the solutions of the system of four equations obtained  by contracting both sides of Eq.~(\ref{eq:r3-int}) respectively with $g_{\mu \nu} p_{\rho }$, $g_{\rho \mu} p_{\nu }$, $g_{\nu \rho} p_{\mu }$ and $p_\mu p_\nu p_\rho$. Defining for simplicity\\
\begin{align}
 d_{jkm}(p^2,m_i^2) & =   \frac{1}{D-1}\, \frac{1}{p^6} \,\int_{l_1 l_2}  \Big(  (l_j \cdot l_k) (l _m\cdot p) p^2 - (l_j \cdot p) (l_k \cdot p) ( l_m\cdot p)  \Big)\,,
 \label{eq:cjkm}
 \end{align}\\
leads to
 \begin{align}
 d_{1, jkm} (p^2,m_i^2)=  d_{jkm} (p^2,m_i^2)\,, 
 \label{eq:r3-int-coef-1}
  \\[2mm]
 d_{2, jkm} (p^2,m_i^2)=  d_{kmj} (p^2,m_i^2)\,,
 \label{eq:r3-int-coef-2}
 \\[2mm]
  d_{3, jkm} (p^2,m_i^2) =  d_{mjk} (p^2,m_i^2)\,,
 \label{eq:r3-int-coef-3}
 \end{align}\\
 while\\
\begin{align}
 e_{jkm}(p^2,m_i^2) & =    \frac{1}{D-1 }\, \frac{1}{p^6}\, \int_{l_1 l_2}  \Big( (D + 2)\, (l_j \cdot p) (l_k \cdot p) (l_m \cdot p) 
\nonumber \\[2mm]
 &  - ( l_j \cdot l_k)( l_m \cdot p)\,  p^2   -(l_k \cdot l_m)(l_j \cdot p)\, p^2   -   (l_m \cdot l_j)(l_k \cdot p)\, p^2  \Big) \, .
\label{eq:r3-int-coef-4}
\end{align}\\
Finally in the case of rank-$4$ integrals there are $10$ independent tensor structures we can build, namely $p^4 \, g^{\{ \mu \nu} g^{\rho \sigma\}} $,  $p^2 g^{\{\mu \nu} p^\rho p^{\sigma\}}$, and $p^\mu p^\nu p^\rho p^\sigma$, hence \\
 \begin{align}
\int_{l_1 l_2} l_j^\mu l_k^\nu l_m^\rho l_n^\sigma &= f_{1, jkmn}(p^2, m_i) \, p^4 \, g^{ \mu \nu} g^{\rho \sigma}  +  f_{2, jkmn}(p^2, m_i) \, p^4 \, g^{ \mu \rho} g^{\nu \sigma}
\nonumber \\
&+ f_{3, jkmn}(p^2, m_i) \, p^4 \, g^{ \mu \sigma} g^{\nu \rho}  +  g_{1, jkmn}(p^2, m_i) \, p^2 \, g^{ \mu \nu} p^\rho p^\sigma
\nonumber \\[2mm]
&+  g_{2, jkmn}(p^2, m_i) \, p^2 \, g^{ \mu \rho} p^\nu p^\sigma +  g_{3, jkmn}(p^2, m_i) \, p^2 \, g^{ \mu \sigma} p^\nu p^\rho
\nonumber \\[2mm]
&+  g_{4, jkmn}(p^2, m_i) \, p^2 \, g^{ \nu \rho} p^\mu p^\sigma +  g_{5, jkmn}(p^2, m_i) \, p^2 \, g^{ \nu \sigma} p^\mu p^\rho
\nonumber \\[2mm]
& +  g_{6, jkmn}(p^2, m_i) \, p^2 \, g^{ \rho \sigma} p^\mu p^\nu  + h_{jkmn}(p^2, m_i) \, p^{\mu} p^\nu p^\rho p^\sigma   \, ,
\label{eq:r4-int}
\end{align}\\
where the coefficients $f_{r,jmk}(p^2, m_i)$, $g_{s,jmk}(p^2, m_i)$ and $h_{jmk}(p^2, m_i)$ are the solutions of the system of $10$ equations obtained by contracting both sides of Eq.~(\ref{eq:r4-int}) with each of the tensor structures. Defining 
\begin{align}
f_{jkmn}(p^2, m_i) &= \frac{1}{(D^2 - 1)(D-2)} \, \frac{1}{p^8} \, \int_{l_1l_2} \, \Big[ D\, \Big( (l_j \cdot p)(l_k \cdot p)(l_m \cdot p)(l_n \cdot p)
\nonumber \\[2mm]
& +  (l_j \cdot l_k) \Big( (l_m \cdot l_n) -  (l_m \cdot p)(l_n \cdot p) \Big) - (l_j \cdot p)(l_k \cdot p) (l_m \cdot l_n)
\Big)
\nonumber \\[2mm]
& + (l_j \cdot l_m)(l_k \cdot p)(l_n \cdot p) + (l_j \cdot l_n)(l_k \cdot p)(l_m \cdot p) 
\nonumber \\[3mm]
& + (l_j \cdot p )(l_n \cdot p)(l_k \cdot l_m) + (l_j \cdot p )(l_m \cdot p)(l_k \cdot l_n)
\nonumber \\[3mm]
& - \Big( (l_j \cdot l_m)(l_k \cdot l_n) + (l_j \cdot l_n)(l_k \cdot l_m) + 2\,  (l_j \cdot p)(l_k \cdot p)(l_m \cdot p)(l_n \cdot p)\Big) \Big]
\,,
\end{align}
and
\begin{align}
g_{jkmn}(p^2, m_i) &= -  \frac{1}{(D^2 - 1)(D-2)} \, \frac{1}{p^8} \, \int_{l_1l_2} \, \Big[ - D^2 (l_j \cdot l_k) (l_m \cdot p )(l_n \cdot p)
\nonumber \\[2mm]
& + D \Big( D\,  (l_j \cdot p)(l_k \cdot p)(l_m \cdot p)(l_n \cdot p) + (l_j \cdot l_k)(l_m \cdot l_n)   - (l_j \cdot p )(l_k \cdot p)(l_m \cdot l_n) \Big)
\nonumber \\[2mm]
& + 2\,  (l_j \cdot l_k) (l_m \cdot p )(l_n \cdot p) 
 + (l_j \cdot l_m)(l_k \cdot p)(l_n \cdot p)  + (l_j \cdot l_n)(l_k \cdot p)(l_m \cdot p)   
\nonumber \\[2mm]
&  + (l_j \cdot p )(l_n \cdot p)(l_k \cdot l_m)  + (l_j \cdot p )(l_m \cdot p)(l_k \cdot l_n) 
\nonumber \\[2mm]
& - \big( (l_j \cdot l_m)(l_k \cdot l_n) + (l_j \cdot l_n)(l_k \cdot l_m)  
 + 4 \,  (l_j \cdot p)(l_k \cdot p)(l_m \cdot p)(l_n \cdot p) \big) \Big] \,,
\end{align}
it follows that\\
\begin{align}
f_{1, jkmn}(p^2, m_i) & = f_{jkmn}(p^2, m_i)\,,
\\[2mm]
f_{2, jkmn}(p^2, m_i) & = f_{jmkn}(p^2, m_i)\,,
\\[2mm]
f_{3, jkmn}(p^2, m_i) & = f_{jnkm}(p^2, m_i)\,,
\end{align}\\
\begin{align}
g_{1, jkmn}(p^2, m_i) & = g_{jkmn}(p^2, m_i)\,,
\\[2mm]
g_{2, jkmn}(p^2, m_i) & = g_{jmkn}(p^2, m_i)\,,
\\[2mm]
g_{3, jkmn}(p^2, m_i) & = g_{jnkm}(p^2, m_i)\,,
\\[2mm]
g_{4, jkmn}(p^2, m_i) & = g_{kmjn}(p^2, m_i)\,,
\\[2mm]
g_{5, jkmn}(p^2, m_i) & = g_{knjm}(p^2, m_i)\,,
\\[2mm]
g_{6, jkmn}(p^2, m_i) & = g_{mnjk}(p^2, m_i)\,,
\end{align}\\
while\\
\begin{align}
h_{jkmn}(p^2, m_i) &= -  \frac{1}{D^2 - 1} \, \frac{1}{p^8} \, \int_{l_1l_2} \, \Bigg[ - (D + 4)(D+2)  \, (l_j \cdot p)(l_k \cdot p)(l_m \cdot p)(l_n \cdot p) 
\nonumber \\[2mm]
& + (D + 2) \Big(  (l_j \cdot l_k)(l_m \cdot p)(l_n \cdot p) + (l_j \cdot l_m)(l_k \cdot p)(l_n \cdot p) 
\nonumber \\[3mm]
& + (l_j \cdot l_n)(l_k \cdot p)(l_m \cdot p) + (l_k \cdot l_m)(l_j \cdot p)(l_n \cdot p) 
\nonumber \\[3mm]
& +  (l_k \cdot l_n)(l_j \cdot p)(l_m \cdot p) + (l_m \cdot l_n)(l_j \cdot p)(l_k \cdot p)   \Big)
\nonumber \\[2mm]
& - \Big( (l_j \cdot l_k)(l_m \cdot l_n)  +  (l_j \cdot l_m)(l_k \cdot l_n)  
 +  (l_j \cdot l_n)(l_k \cdot l_m)  \Big) \Bigg] \,.
\end{align}\\


\chapter{Expressions for the coefficients ${\cal C}_0^{(q_1 \bar q_2 q_3)}$ and ${\cal C}_{G, mn}^{(q_1 \bar q_2 q_3)}$}
\label{app:4}
Here we present the analytic expressions for the coefficients ${\cal C}_0^{(q_1 \bar q_2 q_3)}$ and ${\cal C}_{G, mn}^{(q_1 \bar q_2 q_3)}$ with $mn = 11, 12, 22$, introduced in Eq.~(\ref{eq:Gamma-NL-res-scheme}). They read respectively\\
\begin{align}
 {\cal C}_{0}^{(u \bar u d)} = 1\,,
\qquad {\cal C}_{G,11}^{(u \bar u d)} 
= - \frac{3}{2}   = {\cal C}_{G,22}^{(u \bar u d)}   \,,
 \qquad
{\cal C}_{G,12}^{(u \bar u d)}  =  -\frac{19}{2},
\end{align}
\begin{align}
{\cal C}_{0}^{(u \bar c s)} 
& =  
1 - 8 \rho - 12 \rho^2 \log (\rho) + 8 \rho^3 - \rho^4\,,
\\[3mm]
{\cal C}_{G,11}^{(u \bar c s)}  
& = 
-\frac{1}{2} \left(3 - 8 \rho + 12 \rho^2 \log( \rho) 
+ 24 \rho ^2 - 24 \rho^3 + 5 \rho^4 \right)  = {\cal C}_{G,22}^{(u \bar c s)} \,,
\\[3mm]
{\cal C}_{G,12}^{(u \bar c s)} & =  
-\frac{1}{2} 
\left(19 + 16 \rho + 12 \rho (\rho + 4) \, \log(\rho) - 24 \rho ^2 - 16 \rho ^3 
+ 5 \rho^4  \right)\,,
\end{align}
\begin{align}
{\cal C}_{0}^{(c \bar u d)} & =  
1 - 8 \rho - 12 \rho^2 \log (\rho) + 8 \rho^3 - \rho^4\,,
 \\[3mm]
{\cal C}_{G,11}^{(c \bar u d)}  & =  
-\frac{1}{2} \left(3 - 8 \rho + 12 \rho^2 \log( \rho) + 24 \rho ^2   
- 24 \rho^3 + 5 \rho^4 \right) = {\cal C}_{G,22}^{(c \bar u d)} \,,
\label{eq:C0ud} \\
{\cal C}_{G,12}^{(c \bar u d)} & = 
-\frac{1}{2} 
\left(19 - 56 \rho + 12 \rho^2 \log(\rho) + 72 \rho ^2 - 40 \rho ^3 +5 \rho^4  \right)\,,
\end{align}
\begin{align}
{\cal C}_{0}^{(c \bar c s)} & =  
\sqrt{1 - 4 \rho} \left(1  - 14 \rho - 2 \rho^2 - 12 \rho^3 \right) 
+ 24 \rho^2 (1 - \rho^2)  
\log \left(\frac{1 + \sqrt{1 - 4 \rho^{\phantom{\! 1}}}}
{1 - \sqrt{1 - 4 \rho^{\phantom{\! 1}}}} \right) \! \,,
 \\[3mm]
\! \! \! \! {\cal C}_{G,11}^{(c \bar c s)} & =  
- \frac{1}{2} \Biggl[
\sqrt{1 - 4 \rho} \left(3  - 10 \rho + 10 \rho^2 + 60 \rho^3 \right) 
\nonumber \\[3mm]
& \qquad \qquad \qquad- 24 \rho^2 (1 - 5 \rho^2)  
\log \left(\frac{1 + \sqrt{1 - 4 \rho^{\phantom{\! 1}}}}
{1 - \sqrt{1 - 4 \rho^{\phantom{\! 1}}}} \right) \Biggr] = {\cal C}_{G,22}^{(c \bar c s)}  \,,
 \\[3mm]
{\cal C}_{G, 12}^{(c \bar c s)} & = 
- \frac{1}{2} \Biggl[
\sqrt{1 - 4 \rho} \left(19  - 2 \rho + 58 \rho^2 + 60 \rho^3 \right) 
\nonumber\\[3mm]
& \qquad \qquad \qquad -  24 \rho \, (2 + \rho - 4 \rho^2 - 5 \rho^3)  
\log \left(\frac{1 + \sqrt{1 - 4 \rho^{\phantom{\! 1}}}}
{1 - \sqrt{1 - 4 \rho^{\phantom{\! 1}}}} \right) \Biggr]\,,
\end{align}


\chapter{Expressions for the divergent functions ${\cal D}_{mn}^{(q_1 \bar q_2 q_3)}$}
\label{app:5}
Here we list the expressions for the divergent functions ${\cal D}_{nm}^{( q_1 \bar q_2 q_3)}$ given in Eq.~(\ref{eq:lim-Dnm}). \\
\begin{align}
{\cal D}_{11}^{(u \bar u d)}  & =  8\,  \log \left( \frac{m_u^2}{m_b^2} \right)\,,
\\[1mm]
{\cal D}_{12}^{( u \bar u d)} & = 8 \,\left[ \log \left( \frac{m_d^2}{m_b^2} \right) - \log \left( \frac{m_u^2}{m_b^2} \right) \right]\,,
 \\[1mm]
 {\cal D}_{22}^{(u \bar u d)} & =  8 \,  \log \left( \frac{m_d^2}{m_b^2} \right)\,,
\\[5mm]
{\cal D}_{11}^{(u  \bar c s)} & =  8 \, (1- \rho)^2 \, (1 + \rho ) \log \left( \frac{m_u^2}{m_b^2} \right)\,,
\\[1mm] 
{\cal D}_{12}^{(u  \bar c s)} & =  8 \, (1- \rho)^2 \, (1 + \rho )  \left[ \log \left( \frac{m_u^2}{m_b^2} \right) + \log \left( \frac{m_s^2}{m_b^2} \right)\right]\,,
\\[1mm]
{\cal D}_{22}^{(u \bar c s)} & = 
8 \, (1- \rho)^2 \, (1 + \rho ) \log \left( \frac{m_s^2}{m_b^2} \right)\,,
\\[5mm]
{\cal D}_{12}^{(c \bar u d)} & =  - 16 \, (1 - \rho)^2             \log \left( \frac{m_u^2}{m_b^2} \right) + 
		        	  8 \, (1 - \rho)^2 (1 + \rho)  \log \left( \frac{m_d^2}{m_b^2} \right)\,,
\\[1mm]
{\cal D}_{22}^{(c \bar u d)} & =  8 \, (1 - \rho)^2 (1 + \rho) \log \left( \frac{m_d^2}{m_b^2} \right)\,,
\\[5mm]
{\cal D}_{12}^{(c \bar c s)} & =   
8 \, \sqrt{1 - 4 \rho} \, \log \left( \frac{m_s^2}{m_b^2} \right)\,,
\\[1mm]
{\cal D}_{22}^{(c \bar c s)} & =  
8 \, \sqrt{1 - 4 \rho} \, \log \left( \frac{m_s^2}{m_b^2} \right)\,.
\end{align}


\chapter{Coefficients of the Darwin operator for the charm system}
\label{app:6}

The coefficients ${\cal C}_{\rho_D, mn}^{(q_1q_2)}$ of the Darwin operator corresponding to the $c \to q_1 \bar q_2 u$ decya, needed for the analysis of $D$-meson decays, including the full $s$-quark mass dependence with $\rho = m_s^2/m_c^2$, are given by the following expressions\\ 
%
%
\begin{equation}
{\cal C}_{\rho_D, 11}^{(d \bar d)} 
= 6 \,,
\qquad
{\cal C}_{\rho_D, 12}^{(d \bar d)} 
 =  
-\frac{34}{3}\,,
\qquad
{\cal C}_{\rho_D, 22}^{(d \bar d)} 
 = 
6\,,
\label{eq:CrhoDdd22}
\end{equation}
%
%
\begin{align}
{\cal C}_{\rho_D, 11}^{(d \bar s)} 
& =  
\frac{2}{3} (1 - \rho) \biggl[ 9  + 11 \rho - 12 \rho ^2 \log (\rho ) 
 - \, 24 \left(1 - \rho^2 \right) \log (1-\rho )- 25 \rho ^2  + 5 \rho^3 \biggl]\,,
\label{eq:CrhoDds11} 
\\[3mm]
{\cal C}_{\rho_D, 12}^{(d \bar s)} 
& = 
- \frac{2}{3} \biggl[17 + 12 \rho \left(5 + 2 \rho - 2 \rho ^2 \right) \log(\rho) 
+ \, 48 (1 - \rho) (1 - \rho^2) \log (1-\rho)
\nonumber \\[3mm]
& 
\quad \quad  - 26  \rho + 18 \rho^2 - 38 \rho^3 + 5 \rho ^4 
\biggl]\,,
\label{eq:CrhoDds12} 
\\[3mm]
{\cal C}_{\rho_D, 22}^{(d \bar s)} 
& = 
\frac{2}{3} (1 - \rho) \biggl[ 9  + 11 \rho - 12 \rho ^2 \log (\rho )  - \, 24 \left(1 - \rho^2 \right) \log (1-\rho )- 25 \rho ^2  + 5 \rho^3 \biggr] \,,
\label{eq:CrhoDds22}
\end{align}
%
%
\begin{align}
{\cal C}_{\rho_D, 11}^{(s \bar d)} 
& =  
\frac{2}{3} \biggl[9 - 16 \rho - 12 \rho ^2 + 16 \rho ^3 - 5 \rho ^4 
\biggr]\,,
\label{eq:CrhoDsd11} 
\\[3mm]
{\cal C}_{\rho_D, 12}^{(s \bar d)} 
& = 
- \frac{2}{3} \biggl[17 + 12 \, \rho^2 \left(3 - \rho \right) \log(\rho) 
\nonumber \\[3mm]
& - \, 24 (1 - \rho)^3 \log (1-\rho)  -  50 \rho + 90 \rho ^2 - 54 \rho ^3 + 5 \rho ^4
\biggr],  
\label{eq:CrhoDsd12} 
\\[3mm]
{\cal C}_{\rho_D, 22}^{(s \bar d)} 
& = 
\frac{2}{3} (1 - \rho) \biggl[ 9  + 11 \rho - 12 \rho ^2 \log (\rho ) 
\nonumber\\[3mm]
& - \, 24 \left(1 - \rho^2 \right) \log (1-\rho)
 - 25 \rho^2  + 5 \rho^3 
\biggr]\,,
\label{eq:CrhoDsd22}
\end{align}
%
%
\begin{align}
{\cal C}_{\rho_D, 11}^{(s \bar s)} & =  
\frac{2}{3} \Biggl[
\sqrt{1 - 4 \rho} \left(17 + 8 \rho - 22 \rho^2 - 60 \rho^3 \right)
-4 \left(2 - 3 \rho + \rho^3 \right) +
\nonumber \\[3mm]
& 
\quad \quad - \, 12 \left(1 - \rho - 2 \rho^2 + 2 \rho^3 + 10 \rho^4 \right) 
\log \left(\frac{1 + \sqrt{1 - 4 \rho^{\phantom{\! 1}}}}
{1 - \sqrt{1 - 4 \rho^{\phantom{\! 1}}}} \right) 
\nonumber \\[3mm]
&  
\quad \quad - \, 12 \, (1 - \rho)(1 - \rho^2) 
\log (\rho)
\Biggr],
\label{eq:CrhoDss11} 
\\[3mm]
{\cal C}_{\rho_D, 12}^{(s \bar s)} & = 
\frac{2}{3} \Biggl[ \sqrt{1 - 4 \rho} \left(- 33 
+ 24 \, \log(\rho) - 24 \, \log (1 - 4 \rho)  
+ 46 \rho - 106 \rho^2 - 60 \rho^3 \right)
\nonumber \\[3mm]
&  
\quad \quad + \, 12 \left(3 - 2 \rho + 4 \rho^2 - 16 \rho^3 - 10 \rho^4 \right) 
\log \left(\frac{1 + \sqrt{1 - 4 \rho^{\phantom{\! 1}}}}
{1 - \sqrt{1 - 4 \rho^{\phantom{\! 1}}}} \right)  
\nonumber \\[3mm]
&  
\quad \quad + \, 4 \left(1 -\rho \right)^2 
\left(4 + 3 (1-\rho) \log(\rho) - \rho \right) 
\Biggr],  
\label{eq:CrhoDss12} 
\\[3mm]
{\cal C}_{\rho_D, 22}^{(s \bar s)} 
& =
\frac{2}{3} \Biggl[ \sqrt{1 - 4 \rho} \left(9 + 24 \, \log(\rho) - 24 \, \log (1 - 4 \rho)
+ 22 \rho - 34 \rho^2 - 60 \rho^3 \right)
\nonumber \\[3mm]
& 
\quad \quad  + \, 24 \left(1 - 2 \rho - \rho^2 - 2 \rho^3 - 5 \rho^4 \right) 
\log \left(\frac{1 + \sqrt{1 - 4 \rho^{\phantom{\! 1}}}}
{1 - \sqrt{1 - 4 \rho^{\phantom{\! 1}}}} \right) 
\Biggr].
\label{eq:CrhoDss22}
\end{align}


\chapter{Parametrisation of matrix elements of four-quark operators}
\label{app:7}
The matrix elements of the dimension-six operators in QCD are parametrised as \\
\begin{align}
\langle {D}_q | O_i^q \, | {D}_q \rangle 
& = 
A_i \, f_{D_q}^2 m_{D_q}^2 B_i^q \,, 
\label{eq:ME-dim-6-QCD-q-q}
\\[4mm]
\langle {D}_q | O_i^{q^\prime} | {D}_q \rangle 
& = 
A_i \, f_{D_q}^2 m_{D_q}^2 
\, \delta^{q q^\prime}_i, \qquad q \neq q^\prime 
\label{eq:ME-dim-6-QCD-q-q-prime}
\end{align}
where 
\begin{equation}
A_1^q = A_3^q = 1\,, \qquad A_2^q = A_4^q = \frac{m_D^2}{(m_c + m_q)^2} \,.
\end{equation}\\
In  VIA the Bag parameters reduce to $B_1^q = B_2^q = 1$, $B_3^q = 0$, $B_4^q = 0$ and $\delta^{q q'}_i = 0$.
The matrix elements of the dimension-seven four-quark operators in Eqs.~\eqref{eq:P1q-HQET} - \eqref{eq:M2-G} in HQET 
are parametrised in the following way\\
\begin{align}
\langle D_q |  {\cal P}_1^q | D_q \rangle 
& = 
- m_q F^2 (\mu_0) \, m_D \, \tilde B_{P, 1}^q \, , 
\label{eq:ME-dim-7-P1}
\\[3mm]
\langle D_q |  {\cal P}_2^q |D_q \rangle 
& =  
- F^2 (\mu_0) \, m_D \, \bar \Lambda \, \tilde B_{P, 2}^q\,, 
\label{eq:ME-dim-7-P3}
\\[3mm]
\langle D_q |  {\cal P}_3^q |D_q \rangle 
& = 
- F^2 (\mu_0) \, m_D \, \bar \Lambda \, \tilde B_{P, 3}^q\,, 
\label{eq:ME-dim-7-P4}
\\[3mm]
\langle D_q |  {\cal R}_1^q |D_q \rangle 
& =  
- F^2 (\mu_0) \, m_D \, (\bar \Lambda - m_q) \, \tilde B_{R, 1}^q\,, 
\label{eq:ME-dim-7-P5}
\\[3mm]
\langle D_q |  {\cal R}_2^q |D_q \rangle 
& =  
F^2 (\mu_0) \, m_D  \, (\bar \Lambda - m_q) \, \tilde B_{R, 1}^q\,,
\label{eq:ME-dim-7-P6}
\end{align}\\
with $ \bar \Lambda = m_D - m_c$, and \\
\begin{align}
\langle D_q |  {\cal M}_{1}^q | D_q \rangle 
& =
2 \, F^2 (\mu_0) \, m_D \, G_1 (\mu_0) \, \tilde B_{M,1}^q \, ,
\label{eq:ME-M1-pi}
\\[3mm]
\langle D_q |  {\cal M}_{2}^q | D_q \rangle 
& = 12 \, F^2 (\mu_0) \, m_D \, G_2 (\mu_0) \, \tilde B_{M,2}^q \, ,
\label{eq:ME-M2-pi}
\\[3mm]
\langle D_q |  {\cal M}_{3}^q | D_q \rangle 
& = 
2 \, F^2 (\mu_0) \, m_D \, G_1 (\mu_0) \, \tilde B_{M,3}^q \, ,
\label{eq:ME-M1-G}
\\[3mm]
\langle D_q |  {\cal M}_{4}^q | D_q \rangle 
& = 
12 \, F^2 (\mu_0) \, m_D \, G_2 (\mu_0) \, \tilde B_{M,4}^q \, ,
\label{eq:ME-M1-pi}
\end{align}\\
and similar expressions for the colour-octet operators.
Again, in VIA, the dimension-seven Bag parameters are 
$\tilde B_{P, i}^q = 1$, $\tilde B_{R, i}^q = 1$, and
$\tilde L_{M, i}^q = 1$, while the corresponding colour-octet Bag parameters vanish.
The expressions in Eqs.~\eqref{eq:ME-dim-7-P1} - \eqref{eq:ME-dim-7-P6} have been obtained using the general parametrisation in HQET of the matrix element of heavy-quark currents between a heavy pseudo-scalar meson and the vacuum,  
see e.g. Ref.~\cite{Neubert:1993mb}, namely
\begin{align}
\langle 0 | \bar q \,  \Gamma \, h_v| M(v)\rangle 
& =  
\frac{i}{2} F(\mu) \, {\rm Tr} [\Gamma {\cal M}(v)]\, ,
\\[3mm]
\langle 0 | \bar q \,  \Gamma \,i D_\alpha h_v| M(v) \rangle 
& = 
-\frac{i}{6} (\bar \Lambda -m_q) F(\mu) \, {\rm Tr} [(v_\alpha + \gamma_\alpha) \Gamma {\cal M}(v)]\, ,
\\[3mm]
\langle 0 | \bar q (-i \overset{\leftarrow}{D}_\alpha) \,  \Gamma \, h_v| M(v)\rangle 
& = 
- \frac{i}{6} F(\mu) \, {\rm Tr}[((4 \bar \Lambda - m_q) v_\alpha +  (\bar \Lambda - m_q) \gamma_\alpha) \Gamma {\cal M}(v)],
\end{align}\\
while for the non-local operators
\begin{align}
\langle 0 |
i \int d^4 y \, {\rm T}
\left[ 
(\bar q \, \Gamma \, h_v) (0), 
(\bar h_v (i D)^2 h_v) (y)
\right]
| {\cal M}(v) \rangle
& =  
F(\mu) \, G_1 (\mu) \, {\rm Tr} [\Gamma {\cal M}(v)],
\\[3mm]
\langle 0 |
i \int d^4 y \, {\rm T}
\left[ 
(\bar q \, \Gamma \, h_v) (0), 
\frac{1}{2} g_s \left(\bar h_v \sigma_{\alpha \beta} G^{\alpha \beta} h_v \right) (y)
\right]
| {\cal M}(v) \rangle
& =
6 \, F(\mu) \, G_2 (\mu) \,  {\rm Tr} [\Gamma {\cal M}(v)],
\end{align}\\
where $\Gamma$ is a generic Dirac structure, and\\
\begin{equation}
{\cal M}(v)  =  - \sqrt{m_D} \, \frac{(1+\slashed v)}{2} \gamma_5\,.
\end{equation}
%

\end{appendix}

\pagestyle{plain}

\bibliographystyle{hieeetr}
\bibliography{References}

\end{document}